\documentclass{aa}

\usepackage[colorlinks=true,citecolor=blue, urlcolor=blue, linkcolor=blue]{hyperref}
\usepackage[varg]{txfonts}

\makeatletter
\renewcommand*\aa@pageof{, page \thepage{} of \pageref*{LastPage}}
\makeatother

\usepackage{amsmath,amssymb,listings,upgreek,nicefrac}
\usepackage{grffile}
\usepackage{xcolor,xspace}
\usepackage[normalem]{ulem}
\usepackage[T1]{fontenc}
\usepackage{textcomp}
\usepackage{pdfcolfoot}

\def\entspr{\,\widehat{=}\,}                           %
\def\mH{m_\mathrm{H}}                              %
\def\mp{m_\mathrm{p}}                              %
\def\me{m_\mathrm{e}}                              %
\def\kB{k_\mathrm{B}}                              %
\def\Rydinfty{\mathrm{R}_\infty}                  %
\def\MJ{M_{\mathrm{J}}}                            %
\def\RJ{R_{\mathrm{J}}}                            %

\newcommand{\MSonne}{{M_{\odot}}}

\newcommand{\LSonne}{{L_{\odot}}}

\def\Ha{H\,$\alpha$\xspace}                               %
\def\HaM{\textrm{H}\,\alpha}                              %
\def\Hb{H\,$\beta$\xspace}                                %
\def\HepsM{\mathrm{H}\,\varepsilon}                       %
\def\Paa{Pa\,$\alpha$\xspace}                             %
\def\Pab{Pa\,$\beta$\xspace}                              %
\def\PabM{\textrm{Pa}\,\beta}                             %
\def\Pae{Pa\,$\varepsilon$\xspace}                        %
\def\Bra{Br\,$\alpha$\xspace}                             %
\def\BraM{\textrm{Br}\,\alpha}                            %
\def\Brb{Br\,$\beta$\xspace}                              %
\def\Brg{Br\,$\gamma$\xspace}                             %
\def\BrgM{\textrm{Br}\,\alpha}                            %
\def\Pfa{Pf\,$\alpha$\xspace}                             %
\def\Pfb{Pf\,$\beta$\xspace}                              %
\def\Pfg{Pf\,$\gamma$\xspace}                             %
\def\Pfd{Pf\,$\delta$\xspace}                             %
\def\Pfe{Pf\,$\varepsilon$\xspace}                        %
\def\HeI{He\,\textsc{i}\xspace}                             %

\def\nl{\ensuremath{n_\ell}\xspace}                       %
\def\nupp{\ensuremath{n_\mathrm{u}}\xspace}               %

\def\aHa{\ensuremath{a_{\HaM}}\xspace}                      %
\def\bHa{\ensuremath{b_{\HaM}}\xspace}                      %
\def\aBra{\ensuremath{a_{\BraM}}\xspace}                    %
\def\bBra{\ensuremath{b_{\BraM}}\xspace}                    %

\def\PDS{PDS\,70\xspace}                                  %
\def\PDSA{PDS\,70\,A\xspace}                              %
\def\PDSb{PDS\,70\,b\xspace}                              %
\def\PDSc{PDS\,70\,c\xspace}                              %
\def\PDSbc{PDS\,70\,b and~c\xspace}                       %
\def\PDSd{PDS\,70\,d\xspace}                              %
\def\Dlrmb{Delorme\,1\,(AB)b\xspace}                      %
\def\WISPA{WISPIT\,2\,A\xspace}                           %
\def\WISPb{WISPIT\,2\,b\xspace}                           %
\def\WISPc{WISPIT\,2\,c\xspace}                           %
\def\WISPbc{WISPIT\,2\,bc\xspace}                         %

\def\kms{\ensuremath{\mathrm{km}\,\mathrm{s^{-1}}}\xspace}             %
\def\MPkt{\ensuremath{\dot{M}}\xspace}             %
\def\MPktkumul{\ensuremath{\dot{M}^{\textrm{cml}}}\xspace}             %
\def\MPktPdir{\ensuremath{\dot{M}_{\textrm{direct}}}\xspace}      %
\def\MPktvekst{\ensuremath{\dot{M}_{\textrm{growth}}}\xspace}     %
\def\MPktzpSchin{\dot{M}_{\mathrm{inw}}}           %
\def\MPktempir{\ensuremath{\dot{M}_{\textrm{empirical}}}\xspace}    %
\def\MPktEmiss{\ensuremath{\dot{M}_{\textrm{emit}}}\xspace}    %
\def\vKep{\ensuremath{v_{\mathrm{Kepler}}}\xspace}       %
\def\vkrit{\ensuremath{v_{\mathrm{crit}}}\xspace}        %
\def\rvkrit{\ensuremath{r_{v_{\mathrm{crit}}}}\xspace}   %
\def\rin{\ensuremath{r_{\mathrm{in}}}\xspace}      %
\def\rout{\ensuremath{r_{\mathrm{out}}}\xspace}    %
\def\MP{\ensuremath{M_{\mathrm{p}}}\xspace}        %
\def\RP{\ensuremath{R_{\mathrm{p}}}\xspace}        %
\def\MzpSch{\ensuremath{M_{\textrm{CPD}}}\xspace}  %
\newcommand{\FOb}{\ensuremath{\mathcal{F}}\xspace} %
\newcommand{\LAkk}{\ensuremath{{L_{\textrm{acc}}}}\xspace}  %
\newcommand{\LAkkPdir}{\ensuremath{{L_{\textrm{acc,\,direct}}}}\xspace}  %
\def\LHa{\ensuremath{{L_{\HaM}}}\xspace}    %
\def\FHa{{F_{\textnormal{H}\,\alpha}}}             %
\def\FBra{{F_{\BraM}}}           %
\def\LBra{\ensuremath{L_{\BraM}}\xspace}
\def\LBrakumul{\ensuremath{L_{\BraM}^{\textrm{cml}}}\xspace}  %
\def\LBratotPl{\ensuremath{L_{\BraM}^{\textrm{tot,\,pl}}}\xspace}
\def\ABra{{A_{\BraM}}}           %
\def\FPab{{F_{\PabM}}}           %
\def\LPab{\ensuremath{L_{\PabM}}\xspace}
\def\FBrg{{F_{\BrgM}}}           %
\def\LBrg{\ensuremath{L_{\BrgM}}\xspace}
\def\LLinie{\ensuremath{L_{\mathrm{line}}}\xspace} %
\newcommand{\RHill}{{R_{\textnormal{Hill}}}}       %
\newcommand{\Rtap}{{R_{\textnormal{tap}}}}         %
\newcommand{\Teff}{\ensuremath{T_{\textnormal{eff}}}\xspace}   %
\newcommand{\fPhot}{\ensuremath{f_{\textnormal{phot}}}\xspace} %
\def\cs{c_{\mathrm{s}}}                            %
\def\AV{A_V}                              %
\def\RV{R_V}                            %
\def\vtherm{v_{\textnormal{th}}}                   %
\def\varthmaxP{\ensuremath{\vartheta_\mathrm{max,\,p}}\xspace}   %
\def\varmu{\ensuremath{\tilde{\mu}}}               %
\def\varmumin{\ensuremath{\varmu_\mathrm{min}}\xspace}    %
\def\thetaPolebene{\ensuremath{\theta_{\varphi=\phi}}\xspace}   %

\newcommand{\Rzent}{\ensuremath{{R_{\textrm{cent}}}}\xspace}
\newcommand{\fzent}{\ensuremath{{f_{\textrm{cent}}}}\xspace}    %
\newcommand{\vFfinfty}{\ensuremath{{\varv_{\textrm{ff},\,\infty}}}\xspace}

\newcommand{\thzpSch}{\ensuremath{\theta_{\textrm{CPD}}}\xspace}
\newcommand{\hzpSch}{\ensuremath{h_{\textrm{CPD}}}\xspace}

\def\tBeob{\ensuremath{t_\textrm{obs}}\xspace}
\def\FlmbdEmpf{\ensuremath{F_\lambda^{\mathrm{sens}}}\xspace}

\def\relDvBreite{\ensuremath{\mathfrak{w}}\xspace}  %
\def\fkrit{\ensuremath{f_{\mathrm{crit}}}\xspace}
\def\vZerf{\ensuremath{v_{\mathrm{brkp}}}\xspace}
\def\FWHM{\ensuremath{\mathrm{FWHM}}\xspace}

\def\WVerhaeltnis{\ensuremath{\mathfrak{r}}\xspace}
\def\SNR{\ensuremath{\mathrm{S/N}}\xspace}  %
\def\nuabschn{\ensuremath{\tilde{\nu}_{\mathrm{cut}}}\xspace}
\def\vSyst{\ensuremath{v_{\mathrm{syst}}}\xspace}
\def\RVwegenErde{\ensuremath{\mathrm{RV}_{\mathrm{E,\,orb}}}\xspace}

\def\fdown{\ensuremath{f_{\mathrm{down}}}\xspace}

\usepackage{etoolbox}

\makeatletter

\patchcmd{\NAT@citex}
  {\@citea\NAT@hyper@{%
     \NAT@nmfmt{\NAT@nm}%
     \hyper@natlinkbreak{\NAT@aysep\NAT@spacechar}{\@citeb\@extra@b@citeb}%
     \NAT@date}}
  {\@citea\NAT@nmfmt{\NAT@nm}%
   \NAT@aysep\NAT@spacechar\NAT@hyper@{\NAT@date}}{}{}

\patchcmd{\NAT@citex}
  {\@citea\NAT@hyper@{%
     \NAT@nmfmt{\NAT@nm}%
     \hyper@natlinkbreak{\NAT@spacechar\NAT@@open\if*#1*\else#1\NAT@spacechar\fi}%
       {\@citeb\@extra@b@citeb}%
     \NAT@date}}
  {\@citea\NAT@nmfmt{\NAT@nm}%
   \NAT@spacechar\NAT@@open\if*#1*\else#1\NAT@spacechar\fi\NAT@hyper@{\NAT@date}}
  {}{}

\makeatother

\def\degr{{\mbox{\textdegree}}}

\newcommand{\mum}{\ensuremath{\upmu\mathrm{m}}\xspace}
\newcommand{\mumM}{\upmu\mathrm{m}}

\makeatletter
\let\jnl@style=\rm
\def\ref@jnl#1{{\jnl@style#1}}

\def\aj{\ref@jnl{AJ}}                   %
\def\actaa{\ref@jnl{Acta Astron.}}      %
\def\araa{\ref@jnl{ARA\&A}}             %
\def\apj{\ref@jnl{ApJ}}                 %
\def\apjl{\ref@jnl{ApJ}}                %
\def\apjs{\ref@jnl{ApJS}}               %
\def\ao{\ref@jnl{Appl.~Opt.}}           %
\def\apss{\ref@jnl{Ap\&SS}}             %
\def\aap{\ref@jnl{A\&A}}                %
\def\aapr{\ref@jnl{A\&A~Rev.}}          %
\def\aaps{\ref@jnl{A\&AS}}              %
\def\azh{\ref@jnl{AZh}}                 %
\def\baas{\ref@jnl{BAAS}}               %
\def\bac{\ref@jnl{Bull. astr. Inst. Czechosl.}}
\def\caa{\ref@jnl{Chinese Astron. Astrophys.}}
\def\cjaa{\ref@jnl{Chinese J. Astron. Astrophys.}}
\def\icarus{\ref@jnl{Icarus}}           %
\def\jcap{\ref@jnl{J. Cosmology Astropart. Phys.}}
\def\jrasc{\ref@jnl{JRASC}}             %
\def\memras{\ref@jnl{MmRAS}}            %
\def\mnras{\ref@jnl{MNRAS}}             %
\def\na{\ref@jnl{New A}}                %
\def\nar{\ref@jnl{New A Rev.}}          %
\def\pra{\ref@jnl{Phys.~Rev.~A}}        %
\def\prb{\ref@jnl{Phys.~Rev.~B}}        %
\def\prc{\ref@jnl{Phys.~Rev.~C}}        %
\def\prd{\ref@jnl{Phys.~Rev.~D}}        %
\def\pre{\ref@jnl{Phys.~Rev.~E}}        %
\def\prl{\ref@jnl{Phys.~Rev.~Lett.}}    %
\def\pasa{\ref@jnl{PASA}}               %
\def\pasp{\ref@jnl{PASP}}               %
\def\pasj{\ref@jnl{PASJ}}               %
\def\qjras{\ref@jnl{QJRAS}}             %
\def\rmxaa{\ref@jnl{Rev. Mexicana Astron. Astrofis.}}%
\def\rnaas{\ref@jnl{RNAAS}}             %
\def\skytel{\ref@jnl{S\&T}}             %
\def\solphys{\ref@jnl{Sol.~Phys.}}      %
\def\sovast{\ref@jnl{Soviet~Ast.}}      %
\def\ssr{\ref@jnl{Space~Sci.~Rev.}}     %
\def\zap{\ref@jnl{ZAp}}                 %
\def\nat{\ref@jnl{Nature}}              %
\def\natas{\ref@jnl{Nat.~Ast.}}         %
\def\iaucirc{\ref@jnl{IAU~Circ.}}       %
\def\aplett{\ref@jnl{Astrophys.~Lett.}} %
\def\apspr{\ref@jnl{Astrophys.~Space~Phys.~Res.}}
\def\bain{\ref@jnl{Bull.~Astron.~Inst.~Netherlands}} 
\def\fcp{\ref@jnl{Fund.~Cosmic~Phys.}}  %
\def\gca{\ref@jnl{Geochim.~Cosmochim.~Acta}}   %
\def\grl{\ref@jnl{Geophys.~Res.~Lett.}} %
\def\jcp{\ref@jnl{J.~Chem.~Phys.}}      %
\def\jgr{\ref@jnl{J.~Geophys.~Res.}}    %
\def\jqsrt{\ref@jnl{J.~Quant.~Spec.~Radiat.~Transf.}}
\def\memsai{\ref@jnl{Mem.~Soc.~Astron.~Italiana}}
\def\nphysa{\ref@jnl{Nucl.~Phys.~A}}   %
\def\physrep{\ref@jnl{Phys.~Rep.}}   %
\def\physscr{\ref@jnl{Phys.~Scr}}   %
\def\planss{\ref@jnl{Planet.~Space~Sci.}}   %
\def\procspie{\ref@jnl{Proc.~SPIE}}   %

\def\rprphys{\ref@jnl{Rep.~Prog.~Phys\@}}   %
\def\ptp{\ref@jnl{Prog.~Th.~Phys.}}   %
\def\natas{\ref@jnl{NatAs}}           %
\def\amjm{\ref@jnl{AmJM}}             %
\def\jatis{\ref@jnl{J.\ Astron.\ Tel.\ Instr.\ Syst.\@}} %

\makeatother

\defcitealias{goli04}{G04}
\defcitealias{marl07}{M07}
\defcitealias{scvh}{SCvH}
\defcitealias{bl94}{BL94}
\defcitealias{mc14}{MC14}
\defcitealias{ensman94}{E94}
\defcitealias{lever81}{LP81}
\defcitealias{Aoyama+2020}{A20}

\newcommand{\Vekt}[1]{\mathbf{#1}} 

\newcommand{\eqsep}{\;\;\;}

\newcommand{\MPktEinhJ}{\MJ\,\mbox{yr}^{-1}}
\newcommand{\MMPktEinhJ}{{M_{\mathrm{J}}}^2\,\mbox{yr}^{-1}}
\newcommand{\MPktEinhS}{\MSonne\,\mbox{yr}^{-1}}

\def\fpg{\ensuremath{f_{\textrm{d/g}}}\xspace}
\def\Lphot{\ensuremath{L_{\mathrm{phot}}}\xspace}  %
\def\Hi{\textnormal{H\,\textsc{i}}\xspace}

\newcommand{\MStern}{\ensuremath{M_{\star}}\xspace}

\newcommand{\RStern}{\ensuremath{R_{\star}}\xspace}
\newcommand{\vFf}{{v_{\textnormal{ff}}}}
\newcommand{\rmin}{\ensuremath{r_{\mathrm{min}}}\xspace}
\newcommand{\rmax}{\ensuremath{r_{\mathrm{max}}}\xspace}

\def\FlmbdPhot{\ensuremath{F_{\mathrm{\lambda,\,phot}}}\xspace}
\def\FlmbdSchock{\ensuremath{F_{\mathrm{\lambda,\,shock}}}\xspace}

\usepackage{ulem}

\begin{document}

\title{Detectability of resolved hydrogen lines from the accretion shock at gas giants and their circumplanetary discs}
\titlerunning{Hydrogen-line profiles from accreting gas giants and CPDs}

\author{%
Gabriel-Dominique~Marleau\inst{\ref{MPIA},\ref{UDE},\ref{Bern}}\thanks{Current main affiliation: Universit\"at Duisburg--Essen\inst{\ref{UDE}}.}\and  %
Thomas~Henning\inst{\ref{MPIA}}\and
Roy van~Boekel\inst{\ref{MPIA}}\and
Myriam~Benisty\inst{\ref{MPIA}}\and
Yuhiko~Aoyama\inst{\ref{SYSU}}\and
Inga~Kamp\inst{\ref{Kapteyn}}
}

\authorrunning{G.-D.\ Marleau et al.}

\institute{%
Max-Planck-Institut f\"ur Astronomie,
K\"onigstuhl 17,
69117 Heidelberg, Germany\\
\email{gabriel.marleau@uni-due.de}
\label{MPIA}%
\and
Fakult\"at f\"ur Physik,
Universit\"at Duisburg--Essen,
Lotharstra\ss{}e~1,
47057 Duisburg, Germany
\label{UDE}
\and
Division of Space Research \&\ Planetary Sciences,
Physics Institute, University of Bern,
Sidlerstr.~5,
3012 Bern, Switzerland
\label{Bern}
\and
School of Physics and Astronomy,
Sun Yat-sen University,
Guangdong 519082,
People's Republic of China
\label{SYSU}%
\and
Kapteyn Astronomical Institute,
University of Groningen,
PO Box 800,
9700 AV Groningen,
The Netherlands
\label{Kapteyn}%
}

\date{Received 16 November 2025 / Accepted 10 August 2026}

\abstract%
{%
Many mature gas giants are known but only a few have been observed in their accretion phase. However, Extremely Large Telescope (ELT) instruments will have a higher sensitivity and a smaller inner working angle than the tools used up to now. This should increase search yields and allow for detailed characterisation of such objects.
}%
{%
We examine what
METIS, the first-generation ELT spectrograph, can reveal about accreting gas giants. We focus on the accretion-tracing hydrogen 
recombination lines accessible at a resolution of $R\sim10^5$, mainly \Bra and Pfund series lines. Our approach is general, but we took PDS\,70\,b as a fiducial case, which is very similar to WISPIT\,2\,b.
}%
{%
To calculate high-resolution line profiles, we combined a semi-analytical multi-dimensional description of the flow onto an accreting planet and its circumplanetary disc (CPD) with local non-equilibrium shock-emission %
models. %
We assumed the limiting scenario of no extinction, which is appropriate for gas giants in gaps, and of negligible contribution from magnetospheric accretion columns.
We used simulated detector sensitivities to compute the required observing times.
}%
{%
Both the planet surface and the CPD surface shocks contribute to the total line profile, which has a Gaussian core but wider and asymmetrical wings. The line is much narrower than the free-fall velocity, and in fact it has a constant width of $\FWHM\approx30$--40~km\,s$^{-1}$ for a large part of parameter space.
For the adopted baseline accretion rate onto PDS\,70\,b, 
the \Bra line peak excess is as strong as  %
the photospheric continuum, which is modulated mostly by water features. %
However, the rotation of the planet broadens the features, helping the shock excess stand out.  %
At \Bra, already the continuum of PDS\,70\,b should yield a per-bin $\SNR=12$ in 4~h.
Observing the peak excess at $\SNR\approx3$ should require only about 10~min.  %
With ProDiMo, we estimate that the CPD does not hinder the detection of the line emission.
}%
{%
\Bra is a potent planet formation tracer accessible to METIS that requires little integration time.
For pure shock emission, the line shape is barely sensitive to the planetary or system parameters. A complex profile would indicate that magnetospheric accretion contributes significantly. In any case, the high spectral resolving power of METIS will help reveal the line shapes of even faint accretors with great fidelity.%
}


\maketitle

\section{Introduction}
In the past decade, our understanding of the formation of gas giants has considerably improved.
One important factor has been the discovery and study of low-mass accretors through their accretion-shock-tracing hydrogen line emission (e.g., \citealp{schmidt08,bowler11,Haffert+2019,eriksson20,demars23,luhman23c,close25a}). Exquisitely precise indirect studies of the formation phase are possible for Jupiter based on analyses of present-day properties \citep{batygin25,knierim26a}. However, in general, exoplanets need to be studied while they are in a formation or at least an accretion phase.

Only a handful of planetary-mass accretors are known despite several searches \citep{Cugno+2019,xie20,Zurlo+2020,floresrivera23,follette23,plunkett25},
and most are isolated or nearly so. Thus, they might not probe the same physics as the bulk of the planet population. Also, planets forming in protoplanetary discs are especially interesting because of complex physics, including interaction with the disc leading to radial migration, or their effect on the spatial and size distribution of dust or on the disc chemistry. Further, young disc-bearing systems might show signposts of the forming gas giants they harbour \citep[e.g.,][]{bae23ppvii}, making their detection easier. Also, the age of the parent star can be used, with appropriate caution \citep[taking the formation delay into account; e.g.,][]{fort05,zhang24}, to derive information on the physical properties of the detected forming companion.

Most planetary-mass accretors have been detected at large separations from their primary or host star, but many more planets should be found closer to their star, with the giant-planet mass function peaking at a few astronomical units (\citealp{fernandes19,wittenmyer20,fulton21}; but see also \citealp{lagrange23}).
The Extremely Large Telescope (ELT), thanks to its 39-m %
primary-mirror diameter \citep{ramsay18}, will offer a high angular resolution capable of probing down to such scales at the typical distance of nearby star-forming regions: $a\approx(2.2\lambda/D)d=0.7~(\lambda/4~\mum)(d/150~\mathrm{pc})$~au, where $a$ is the semimajor axis, $D$ is the primary-mirror diameter, and $d$ is the distance to the system. In its integral field unit (IFU) mode, the first-generation Mid-infrared ELT Imager and Spectrograph (METIS) on the ELT will offer high-contrast, diffraction-limited observations at a spectral resolving power of $R\approx10^5$ at $\lambda\approx2.9$--5.3~$\mum$, covering the $L$ and $M$ bands \citep{brandl21,brandl22,feldt24}, as summarised in Figure~\ref{Abb:Linien}.
This will enable in-depth studies of the physical and chemical properties of forming planets and their environments, amongst other topics (e.g., \citealp{takami25}).
In a pioneering study, \citet{oberg23metis} carefully simulated observations of circumplanetary discs (CPDs) with METIS at several $^{12}$CO transitions at $\lambda\approx5$~\mum, and they find that CPDs should be easily detectable.

In this work, we look at neutral-hydrogen (\Hi) line emission from the shocks on the surface of the planet and on its CPD.
We study the brightness of the shock and the shape of the emission line for a range of parameters and also estimate the importance of the confounding photospheric signal from the planet or its CPD, or from the PPD.
Line shapes have the potential of revealing the physical mechanisms governing accretion, even though the interpretation of the profiles can be challenging for any accretor mass (e.g., \citealp{edwards94,currie25b}).

We take as an obvious example \PDSb, the best-known accreting planetary-mass gas giant found in a gap in a transition disc \citep{mueller18,keppler18,Haffert+2019,zhou21,zhou25,close25a}.
It is a fair representative of its class.
\PDSc has essentially the same physical properties within the large errorbars \citep{shibaike24}, also as \citet{faruqi26} derive with very different methods.
Atacama Large Millimeter Array (ALMA) observations provide evidence of accretion onto \PDSc \citep{dom25}, with variability even at the timescale of an hour \citep{casassus26}.
However, \PDSc is partially obscured by protoplanetary disc (PPD) material \citep{Haffert+2019}.
Recently, \WISPbc \citep{vancapelleveen25a,close25b,lawlor26} and J16120\,b \citep{ginski25,li25,sierra26} were added to the short list of planetary-mass (possible\footnote{At \WISPc, only weakly constraining upper limits at \Ha \citep{close25a} and \Brg (Fig.~4 of \citealp{lawlor26}) exist, but it might only be a matter of time before detections are obtained, for example, with JWST/NIRSpec \citep{balmer26jwst}. However, \WISPc might presently not be fed any material \citep{benisty26}.}) accretors in a PPD gap.
\WISPA is a solar-mass spectroscopic binary ($0.97~\MSonne$ and $0.33~\MSonne$ on a 4.8-day orbit, corresponding to roughly 0.07~au; \citealp{buergy26}) that is 133.4~pc away \citep{bj21}; thus it is only slightly more distant than \PDS{} (see Table~\ref{Tab:Par}). The (non-dereddened) \Ha luminosity of \WISPb is somewhat higher than that of \PDSbc{} \citep{close25b}. We return to this in the discussion.

The structure of the paper is as follows.
In Section~\ref{Th:Mod}, we introduce the physical picture we consider and summarise our computation of the line emission.
We present the parameter range we cover, with fiducial values guided by \PDSb.
In Section~\ref{Th:EmLiProf}, we present our theoretical line profiles and their dependence on the parameters.
In Section~\ref{Th:Detektierbarkeit}, we estimate how much the planetary photosphere and other `noise' sources can mask the accretion, and we compute the needed observation time given the instrumental sensitivity.
In Section~\ref{Th:Disk}, we discuss different aspects within or beyond our study,
and we summarise and conclude in Section~\ref{Th:Zus}.
Appendices~\ref{Th:FlbeimBeob}--\ref{Th:Linienlisteneffekt} present additional or supporting material.
In particular, in Appendix~\ref{Th:zpSchProDiMo} we present estimates of the CPD emission with ProDiMo \citep{woitke16}.

\section{Model set-up}
 \label{Th:Mod}

\subsection{Physical picture}

We consider planets around stars that still host a PPD. We assume a `super-thermal' planet mass \MP, such that the planet has opened a deep gap (e.g., \citealp{fung19}), which is a rapid process (e.g., \citealp{malik15}). This implies that the gas flow towards the planet is supersonic
and that consequently a shock forms on---in fact, defines---the surface of the planet and of at least a part of the CPD. Another consequence is that the in-system extinction can be assumed to be moderate at $L$ and $M$ thanks to the orders-of-magnitude reduction in the surface density of the PPD \citep{kanagawa17,okuzumi26}, especially if the opacity decreases with wavelength (e.g., \citealp{woitke16}).
Nevertheless, it would be easy to fold in an arbitrary amount of extinction since we focus on a narrow spectral range. This could be the extinction by the accreting gas and dust at least approximately (e.g., \citealp{maea21}).

We model the accretion-flow streamline within the planetary Hill sphere using the ballistic infall model of \citet{m24expeditus}. It extends the model
of \citet{ab22,ab25} and \citet{taylor24,taylor25}, which is based on \citet{ulrich76} in the stellar context, and use the result that the streamlines responsible for the line emission originate close to the pole \citep{m22Schock}.
As discussed in Section~\ref{Th:MagAkk},
we do not include any accretion from the CPD to the planet by magnetospheric accretion \citep{fendt03,lovelace11,hartmann16}, which would be a source of line emission.
The flow model yields the preshock velocity and density at the surface of the planet and of the CPD. Being supersonic, the gas flow, to first order, does not depend on the temperature distribution on the surface of the planet or of the CPD \citep{m22Schock}, which reduces the number of free parameters.

One parameter is the centrifugal radius \Rzent, which we parametrise by
$\fzent=\Rzent/\RHill=\ell^2/3$, where $\RHill$ is the Hill radius and $\ell$ is the so-called angular momentum bias (e.g., \citealp{ward10}; \citealp{ab25}).
Other parameters of our framework are introduced in the following sections.
In \citet{m24expeditus}, we find that $\fzent\approx0.03$ reproduces well the simulations in \citet{m22Schock} in terms of the fraction of the influx at the Hill sphere that lands directly on the planet. While this is smaller than the canonical estimate $\fzent=1/3$ (e.g., \citealp{quillen98}),  %
it does match hydrodynamical simulations \citep{ward10}, as reported in \citet[][see their Eq.~1]{shibaike24},
especially in the limit of very super-thermal masses (Hill radius much larger than the Bondi radius).
Also, \citet{ab22} argue that \fzent is uncertain but likely smaller than the classical estimate and \citet{ab25} mention that this reduction is expected to be in the range of $\ell^2\approx1/9$--$1/4$, with smaller values towards higher planet masses. Our adoption of $\fzent=0.03\approx1/9\times1/3$ is thus in line with this. Nevertheless, we show in Section~\ref{Th:Parvar} that it does not play an important role for our results. Also, \fzent should not be confused with the outer edge of the CPD; in a classical viscous disc, for instance, there is a wide decretion region between $\Rzent=\fzent\RHill$ and $\RHill$ (e.g., Fig.~1 in \citealt{ab25}). Thus a small \fzent does not imply a small CPD, as discussed in more detail in Sect.~2.4.3 of \citet{m24expeditus}.

\subsection{Accretion geometry and emission geometry}

We assume that the accreting planet is surrounded by a CPD.
Neither theory nor observations have yet constrained tightly the thickness of CPDs around forming super-Jupiters. As studied by several authors, the thickness  depends on the thermodynamics of the gas, especially its ability to cool (e.g., \citealp{ab09b,szul16,fung19,schulik20,krapp24}) but we surmise and assume that at sufficiently high masses, CPDs will be flatter, rotationally supported, rather than distended envelopes. The latter have been found in large-scale simulations that had to smooth the gravitational potential, out to 1--10\,\%\ of the Hill sphere or tens of Jupiter radii at best (e.g., \citealp{schulik19,schulik20,szul20,lega24,sagynbayeva25}). On the other hand, true CPDs, even if thick, around super-thermal mass objects have been seen in smoothing-free, high-resolution simulations (e.g., \citealt{b19b,takasao21,m22Schock} and overview table in the latter work).
Thus we assumed a rather thin CPD, with the thickness a free parameter.
We define \thzpSch as the polar angle of the surface of the CPD. With this, the aspect ratio is
\begin{equation}
\label{Gl:hzpSch}
\hzpSch=\tan\left(90\degr-\thzpSch\right).
\end{equation}
Since the emission is dominated by the regions close to the planet,
it is not necessary to consider the flaring of the CPD.

By conservation of angular momentum, most gas coming from large scales falls onto the CPD, not the planetary surface (e.g., \citealp{tanigawa12,schulik20,ab22,ab25}). We assume azimuthal symmetry around the planet, as in the simulations of \citet{m22Schock}.
\citet{taylor24} also make this choice, justifying it by the typically large timescale ratio between the infall and orbits in the CPD, which they quote respectively as $\sim1$~Myr versus $\sim1$~yr.
Within this assumption, the flux density at the surface of the planet or the CPD depends only on the polar angle or the radial distance from the planet, respectively.

In Appendix~\ref{Th:FlbeimBeob}, we detail the geometry needed to integrate the emission towards the observer, considering separately the visible portions of the planet and CPD surfaces. The derivation relies on geometry and trigonometry. The quantitative results are obtained by numerical integration of the flux per projected area over the visible surface at every wavelength of interest, which we perform with \texttt{IDL}/\texttt{GDL}\footnote{The GNU Data Language (\texttt{GDL}) is a mature, open-source, actively developed drop-in alternative to \texttt{IDL} (\url{https://github.com/gnudatalanguage/gdl}).}. This yields line profiles, from which we obtain line-integrated fluxes, and for control and visualisation purposes we can also calculate the spatially resolved emission from the planet and CPD surfaces. 

\subsection{Local emission}
 \label{Th:lokEm}

To compute the local emission---a shock is always a local, 1D process---, we use the non-equilibrium radiation-hydro\-dynamic\-al models of \citet{aoyama18}, developed specifically for planetary accretion and used in several works (see references thereto).
The model computes line profiles and therefore intensities, which depend essentially only on the preshock quantity $n_0=X\rho_0/\mH$ and on the preshock velocity $v_0$, where $\mH$ is the atomic mass and $n_0$ has the units of a number density but, for fixed hydrogen mass fraction $X$, is directly proportional to the mass density $\rho_0$.

Each spectral line can be thought of as a sum of Gaussians from the receding postshock gas in the cooling layer \citep{aoyama18,Aoyama+2020}.
The emitting gas has a temperature of $T > 10^4$~K, so no spectral feature narrower than a resolution
$R = c/(2\surd[{2\ln2}]\vtherm) \approx \textrm{11,000}$ is expected,
where $\vtherm=\surd[{8\kB T/(\pi\mu\mH)}]\approx10~\kms$ is the thermal speed \citep{aoyama18},  %
with $\mu=2.29$ denoting the mean molecular weight,
$\kB$ Boltzmann's constant, and $c$ the speed of light.
Depending on the preshock velocity, which is $v_0\sim100$--180~\kms for planets, but also on the preshock density, the full width at half maximum (\FWHM) of the line ranges between between 10\,\%\ and 130\,\%\ of preshock velocity, perhaps surprisingly, and is in fact nearly constant at $\FWHM\approx30$--40~\kms for low preshock densities (Appendix~\ref{Th:DvundmuAo18}).
An important fact is that below $v_0\approx25$--30~\kms, if the hydrogen arrives in molecular form at the shock, the emission drops precipitously to zero because there is not enough energy to excite electrons to higher levels \citep{aoyama18,AMIM21L}.
Therefore, planet masses of a least roughly one %
Jupiter mass
are needed to generate any emission \citep{Aoyama+2020}.

\begin{figure}[t] %
 \centering
 \includegraphics[width=0.45\textwidth]{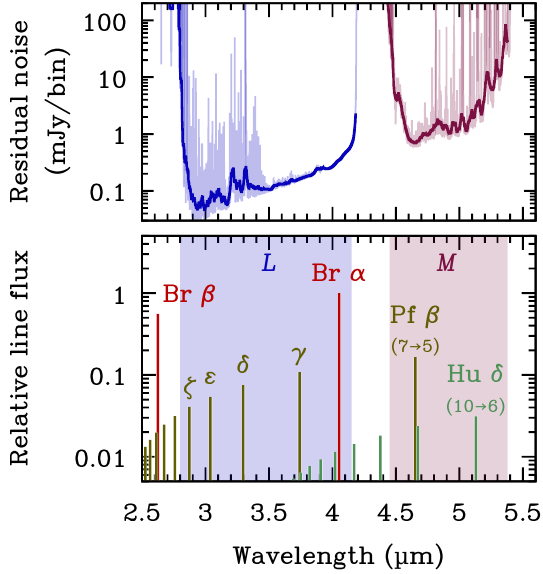}
\caption{%
\Hi{} transitions accessible to METIS.
\textit{Bottom panel}: Line fluxes from the \citet{aoyama18} model for typical shock conditions relative to the \Bra flux
in the $L$ and $M$ bands of METIS.
We show the Brackett (Br; lower level $\nl=4$), Pfund (Pf; $\nl=5$), and
Humphries (Hu; $\nl=6$) series, partially labelled.
\textit{Top panel}:
Residual 5$\sigma$ noise per spectral bin for classical ADI post-processing of a 1-h integration (`sensitivity curve'; details in Section~\ref{Th:Detektorempf}),  %
box-car averaged over 100 (pale), and 2000 (dark) detector bins.
}
\label{Abb:Linien}
\end{figure}

The calculations presented below apply in the same way to all lines, but for convenience, Figure~\ref{Abb:Linien} gives an overview of the lines accessible to METIS in IFU mode, that is, at $L$ and $M$. The strongest are
\Bra ($n=5{\rightarrow4}$; in vacuum\footnote{
The values quoted here are from the NIST database \citep{NIST_ASD20230624} and represent an average over finer levels. As detailed in Appendix~\ref{Th:DvundmuAo18}), this can deviate from the Rydberg formula (Eq.~(\ref{Gl:Rydberg})) but only up to 2.5~\kms. The respective air values are at $\Delta v=-84.3$~\kms.}:
$\lambda_0=4.052279~\mum$),
\Pfb ($n=7{\rightarrow}5$; 4.65378~\mum),
\Pfg ($n=8{\rightarrow}5$; 3.74058~\mum), and
\Pfd ($n=9{\rightarrow}5$; 3.2970~\mum).
\Brb (2.62587~\mum) is not accessible, and the brightest Pfund line, \Pfa (7.45990~\mum), is just outside of the $N$ band (covering approximately $\lambda=7.5$--13.5~\mum;
not shown, which will only have imaging capabilities. Higher-order transitions have been used as accretion tracers \citep[e.g.,][]{rigliaco15,tofflemire25}
Thus \Bra is the main transition of interest, and \Pfb\ is a
secondary candidate, first studied (in the stellar domain) by \citet{salyk13}.

\subsection{The accretion rate}
 \label{Th:MPkt}

`The accretion rate' towards a planet has different meanings in different contexts and works. In multi-dimensional global hydrodynamics PPD simulations, it is often defined as the (net) mass inflow into the Hill sphere. In low-dimensional calculations of planet formation, the accretion rate is taken as the rate of growth of the planet, which is fed by direct infall onto the planet and the mass transfer rate from the CPD, whether through magnetospheric or boundary-layer accretion. Observationally, the accretion rate is the line-emitting, that is, observable, portion of the mass influx towards the planet and CPD. These different definitions are partially `orthogonal', partially overlapping.

One possibility is that most of the gas first landing on the CPD migrates radially inward quickly, which requires a sufficiently high viscosity for the CPD (see e.g., \citealp{papnel05}). Then, the gas will generate lines at the footpoints of magnetospheric-accretion columns on the surface of the planet. In this case, all definitions more or less agree: Even the gas hitting the CPD surface too slowly to emit lines will ultimately reach the planetary surface at velocities on the same order of magnitude as in free-fall from infinity, even if reduced because of the finite truncation radius (e.g., \citealp{demars23}). At the other extreme, the CPD could be a decretion disc directly connected to the planet \citep{dong21} or not \citep{batygin20}, so only direct infall onto the planet surface would contribute to its growth, while the line-emitting rate would be larger. The total rate of infall onto the CPD would be even greater but relevant neither for line-emission nor for planet growth.

\begin{figure}[t] %
 \centering
 \includegraphics[width=0.45\textwidth]{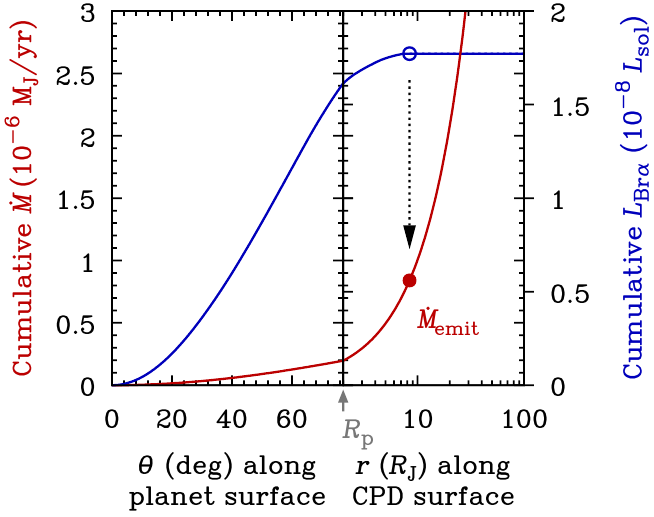}
\caption{%
Definition of the line-emitting accretion rate \MPktEmiss, shown here for the fiducial planet and CPD parameters as an example (details in Section~\ref{Th:Par}).  %
This rate is the value of the cumulative accretion rate \MPktkumul (Equation~(\ref{Gl:MPktkumul}); \textit{red curve, left axis}) at the location (\textit{filled/empty circle}) where the cumulative \Bra luminosity \LBrakumul (Equation~(\ref{Gl:Lkum}); \textit{blue line, right axis}) reaches the total \LBra. %
At that location, the preshock velocity drops below $v_0\approx30~\kms$ (Eq.~(\ref{Gl:rvkrit})).
In the left panel, \MPktkumul and \LBrakumul,
which start at zero at the pole,
are plotted against
the angle along the planet surface, and the curves continue in the right panel but running against the distance along the CPD surface. %
}%
\label{Abb:MPktEmiss}
\end{figure}

Given that the motion of the gas in the CPD depends on very unknown parameters such as the viscosity, we take a simpler, pragmatic approach in this work.
First, we defined the cumulative mass inflow rate, letting it begin at zero at the pole and integrating first along the surface of the planet down to the CPD and then along the CPD surface:
\begin{align}
\label{Gl:MPktkumul}
\MPktkumul = \left\{\!\!\!
\begin{array}{ll}
\int_{0}^\theta 4\pi\RP^2 \rho v \sin\theta'\,{\rm d}\theta'  & \mbox{planet surface} \\
\int_{\RP}^r                4\pi r'   \rho v \sin\thzpSch\,{\rm d}r' + 
   \MPktPdir & \textrm{CPD surface,} \\
\end{array}
\right.
\end{align}
where \MPktPdir is the infall rate reaching the planet surface directly,
that is, the first line of Equation~(\ref{Gl:MPktkumul}) evaluated at $\theta=\thzpSch$ (see the discussion also in Section~4.4 of \citealt{m24expeditus}).
Then, we defined \MPktEmiss as  %
the cumulative accretion rate \MPktkumul evaluated at the position
where %
the total line-integrated \Bra luminosity is emitted.
For numerical reasons, we used a fraction $f=0.9999$ but this yielded almost exactly the same result as with $f=0.999$, with the corresponding \MPkt greater only by some tens of percent than the result with $f=0.99$ (not shown).
As illustrated in Figure~\ref{Abb:MPktEmiss},
while the integrated accretion rate grows `indefinitely' with distance from the pole, the luminosity asymptotes because the emission drops at some point essentially to zero, once the preshock velocity falls below $\vkrit\approx30$~\kms \citep{aoyama18} as mentioned in Section~\ref{Th:lokEm}.
That location \rvkrit along the CPD surface is easily determined from Equation~(4b) of \citet{m24expeditus}:
\begin{equation}
 \label{Gl:rvkrit}
 \rvkrit =7.6~\left(\frac{\MP}{5~\MJ}\right)\left(\frac{\vkrit}{30~\kms}\right)^{-2}\left(\frac{\sin\left[{\thzpSch/(2\times77\degr)}\right]}{0.623}\right)^2~\RJ.
\end{equation}

Therefore, contrary to the total accretion rate, \MPktEmiss is defined by the flow close to the planet, where the flow is well understood within the no-magnetospheric-accretion assumption; \MPktEmiss does not depend on the outer radius of the CPD or its large-scale structure.
Also, \MPktEmiss is simply an infall rate and thus conceptually detached from the fate of the gas once it has hit the planet or CPD surface, whether the gas flows in- or outwards in the latter case.
For small masses, the location corresponding to \MPktEmiss is on the planet, while normally, it is on the CPD but within a modest distance of the planet.

Only for the purpose of defining the accretion rate \MPktEmiss, we computed the line-integrated cumulative luminosity \LBrakumul by summing the fluxes along the surfaces, as in Equation~(\ref{Gl:MPktkumul}):
\begin{align}
\label{Gl:Lkum}
\LBrakumul \approx \left\{\!\!\!
\begin{array}{ll}
\int_{0}^\theta 4\pi\RP^2 \FBra\sin\theta'\,{\rm d}\theta'  & \mbox{planet surface} \\
\int_{\RP}^r                4\pi r'   \FBra\sin\thzpSch\,{\rm d}r' + 
   \LBratotPl & \textrm{CPD surface,} \\
\end{array}
\right.
\end{align}
where $\FBra=\FBra(n_0,\,v_0)$ is the local line-integrated flux, $\theta'$ and $r'$ are the polar angle along the surface of the planet and the radial coordinate along the surface of the CPD (see similarly Equation~11 in \citealt{m22Schock}),
and \LBratotPl is the total \Bra luminosity from the free planetary surface, that is, from the pole down to \thzpSch.
Equation~\ref{Gl:Lkum} thus sums the contributions from both hemispheres and we proceed the same way for the accretion rate.
Equation~(\ref{Gl:Lkum}) is approximate because it effectively assumes an `isotropic observer', without any reference to a viewing geometry. As we show below (Section~\ref{Th:Parvar}), the inclination does modify the total flux reaching the observer.

The caveat to be kept in mind with Equation~(\ref{Gl:Lkum}) is that it traces the line-emitting mass inflow, while the mass growth rate can be larger or smaller. The relation between the two depends on the fate of the gas once it has reached the CPD and also on the structure of the CPD close to the planet, as mentioned at the beginning of this section. We return to the different accretion rates in Section~\ref{Th:MPktausLLinie}.

Internally (as in \citealt{ab22} or \citealt{m22Schock}), we take as a free parameter the gas surface density $\Sigma$ at the Hill sphere, while assuming a vertical stratification \citep{m24expeditus}.
However, this means that at a given $\Sigma$ the mass infall rate %
depends on the planetary mass since the Hill radius depends on \MP.   %
Therefore, it is easier, also for relating to observations, to work with the line-emitting accretion rate as defined above.
The surface density determines linearly the incoming mass flux.

As mentioned and we shall see below, the planet surface dominates the line emission, and by angular momentum conservation the gas landing there comes predominantly from the polar regions above the planet \citep{tanigawa12,m22Schock}.
Thus the density across the gas-providing region is relatively constant, independently of the angular distribution of the inflow across the hemisphere. \citet{taylor24,taylor25}, and \citet{ab25} considered a range of distributions and carefully studied their impact on different quantities such as the surface density of the CPD or its evolution. However, this angular dependence does not matter very much here because we are examining only the line emission, which comes from the innermost regions.

\subsection{Parameter values}
 \label{Th:Par}

We considered a range of parameter values for the mass, accretion rate (see definition in Section~\ref{Th:MPkt}), 
and other properties.
The fiducial values are inspired by the \PDS planets and listed in Table~\ref{Tab:Par}.
Within the large observational uncertainties, the two planets are relatively similar, but for definiteness we select \PDSb as a reference because it is less likely to be affected by extinction from the inner edge of the PPD \citep{benisty21}. %
The surface density $\Sigma$ is an important and uncertain parameter.
To obtain a fiducial value, we took an empirical approach by calculating the total line flux for different values of $\Sigma$ and using the observational constraints to select a value approximately consistent with the data, which yields $\Sigma\approx0.2$~g\,cm$^{-2}$ (Section~\ref{Th:Sigmakalibrierung}; Figure~\ref{Abb:LLinie}). We used this as the fiducial value.
The corresponding accretion rate %
$\MPktEmiss\approx7.8\times10^{-7}~\MPktEinhJ$
is within the range of values found in the literature \citep{shibaike24} but
higher than some estimates \citep{close25a}.
Part of the reason is that planetary emission might be less efficient than in the stellar case \citep{AMIM21L,m24expeditus}. %

\begin{table}
\caption{Fiducial parameter values considered in this work.}
\label{Tab:Par}
\centering
\begin{tabular}{cc}
\hline
\hline
\multicolumn{2}{c}{\textit{Planet intrinsic parameters}} \\  %
\hline
  Mass    &   $\MP  = 5~\MJ$ \\  %
  Radius  &   $\RP  = 2~\RJ$ \\ %
  Effective temperature\tablefootmark{a}  & $\Teff=1400$~K \\
  Surface gravity\tablefootmark{a}        & $\log g=3.5$ \\
  Projected rotation speed\tablefootmark{a}  & $v\sin i=10$~\kms \\
\hline
\multicolumn{2}{c}{\textit{CPD parameters}} \\
\hline
  Relative centrifugal radius  & $\fzent=\Rzent/\RHill = 0.03$ \\ %
  CPD opening angle\tablefootmark{b}  & $\thzpSch =  77$\degr \\ %
\hline
\multicolumn{2}{c}{\textit{System and observer parameters}} \\
\hline
  Stellar mass             & $\MStern   =     {0.9}~\MSonne$ \\ %
  PPD surface density in gap\tablefootmark{c}   & $\Sigma =   0.2$~g\,cm$^{-2}$ \\  %
  Distance to observer\tablefootmark{d}      & $d=113.4$~pc   \\
  Viewing inclination      & $i  = 50$\degr \\  %
\hline
\end{tabular}
\tablefoot{
Fiducial values, inspired by observations or modelling of \PDSb, based in part on the detailed compilation of \citet{shibaike24}. The mass also agrees with \citet{wahhaj24}, \citet{doi24}, and the results of \citet{trevascus25} or \citet{taylor26b}.
\tablefoottext{a}{Discussed in Section~\ref{Th:PlanEm} and used only there and onwards. We multiplied the flux in the atmospheric model with a factor of $\fPhot=1.036$ (see Section~\ref{Th:PlanEm}) in addition to the $(\RP/d)^2$ conversion.}
\tablefoottext{b}{Measured from the pole. The CPD aspect ratio is given by Equation~(\ref{Gl:hzpSch}).}
\tablefoottext{c}{Instead, one could choose the unperturbed surface density in the PPD and a reduction factor based on, for example, \citet{kanagawa17} or \citet{okuzumi26}. The accretion rate (not indicated; see Section~\ref{Th:MPkt}) scales linearly with $\Sigma$.}
\tablefoottext{d}{The updated value from \textit{Gaia} DR3 is $d=112.32$~pc
\citep{gDR3}, but for simplicity we keep the value from \citet{gDR2}, which is barely 1\,\%\ larger.}
}
\end{table}

\section{Emission line profiles}
 \label{Th:EmLiProf}

\subsection{Fiducial case}
 \label{Th:EmLiProfNormalfall}

\begin{figure*}[t] %
 \centering
  \includegraphics[height=0.37\textwidth]{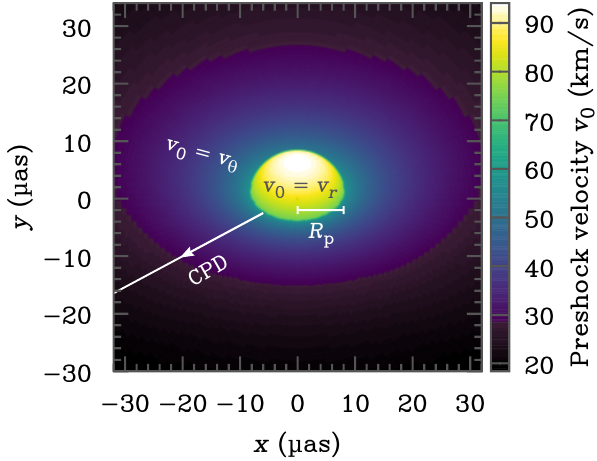}~~~
  \includegraphics[height=0.37\textwidth]{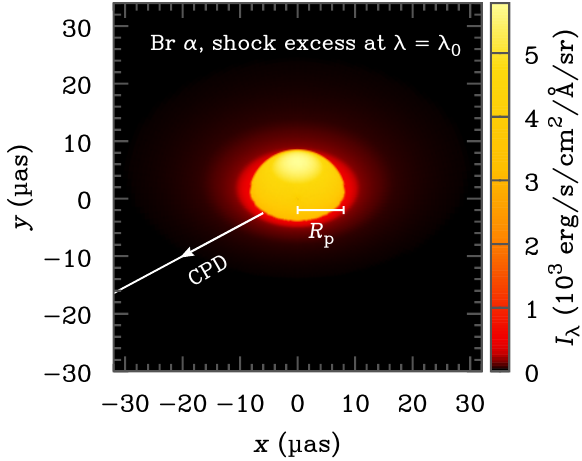}~~~~~~  %
\caption{%
Results for the shock excess line emission (i.e., without photospheric emission) in the fiducial case (for \PDSb-like parameters, including $d=113.4$~pc and $i=50\degr$; see Table~\ref{Tab:Par}). The planet is in the centre, and the inner part of the CPD is visible.
The diffraction-limited beam of the ELT is $26$~mas, about 1000~times larger than the panels.
(\textit{Left})~Preshock velocity $v_0$ at the surface of the planet ($v_r$ velocity component; the bright central parts) or of the CPD (polar component $v_\theta$).
The colour scale switches to grey at the minimum preshock velocity for line emission \vkrit{} (see Eq.~\ref{Gl:rvkrit}).
(\textit{Right})~Intensity of the emission at the central wavelength of \Bra.
The planet and CPD rotation are neglected for the line emission because it originates from postshock layers that are assumed to have not yet joined the Keplerian rotation.  %
The shock is much brighter per area at the planet surface than at the CPD, and only the closest regions of the CPD, out to $r\approx(3$--$4)\RP$, contribute somewhat to the total flux.  %
The sharp red transition at $r\approx1.2\RP$ is a minor effect coming from the linear interpolation of the $F_\lambda(v_0)$ tables.  %
}
\label{Abb:Bild}
\end{figure*}

We show in Figure~\ref{Abb:Bild} the preshock velocity, which largely controls the local emission, and the spatially resolved intensity of the line excess at the planet and CPD surfaces.
The first panels serves also to understand the origin of the emission and to verify the plausibility of our calculation of the geometry.
As expected, the highest preshock velocity is found at the pole, with $v_0=94~\kms\approx\vFfinfty$, where
\begin{equation}
 \label{Gl:vFfinfty}
\vFfinfty\equiv\sqrt{\frac{2G\MP}{\RP}}=59.5~\sqrt{\frac{\MP}{\MJ}\frac{\RJ}{\RP}}~\kms
\end{equation}
is the free-fall velocity from infinity,
and $v_0$ drops towards the equator almost by a factor of $\surd{2}$ (Equation~(3) of \citealt{m24expeditus}) because angular-momentum conservation fans out the flow.

The planet surface is much brighter per area than the CPD thanks to the higher preshock velocity: at the planet surface, the radial\footnote{In this work, unless otherwise noted (as in Appendix~\ref{Th:wosindbundc}), by `radial' or `radial velocity' we mean `towards the planet' in the reference frame of the planet, not `towards the observer'.} component is the %
preshock velocity, whereas for the CPD emission, the polar ($\theta$) velocity component is the preshock velocity within the assumption of a purely radial, or constant-aspect-ratio (flaring-free), CPD surface, as discussed in \citet{m22Schock}.
Equation~(3) of \citet{m24expeditus} gives the velocity components explicitly.
We do not include the possible scattering of photons from the planetary-surface shock off the CPD\footnote{Scattering can increase the total flux from the planet and CPD near 1--2~\mum while decreasing it past 5~\mum in a modest but noticeable way (Fig.~7 of \citealp{sun26}), with the details likely depending on the dust shape and size distribution.}.
Figure~\ref{Abb:Bild} shows only a small region of the CPD, approximately out to $r=3\RP$, because near this the preshock velocity drops below $v_0\approx25$--30~\kms, and therefore the emission effectively goes to zero \citep{aoyama18}, as Figure~\ref{Abb:MPktEmiss} reflects.
The quantities of Figure~\ref{Abb:Bild} are plotted spatially resolved, over a region 60~$\upmu$as across, but in reality this is 2000~times smaller than the diffraction limit of the ELT in the $L$ band, $\theta\approx26$~mas, set by its 39-m primary mirror.

\begin{figure}[tph] %
 \centering
\includegraphics[width=0.4\textwidth]{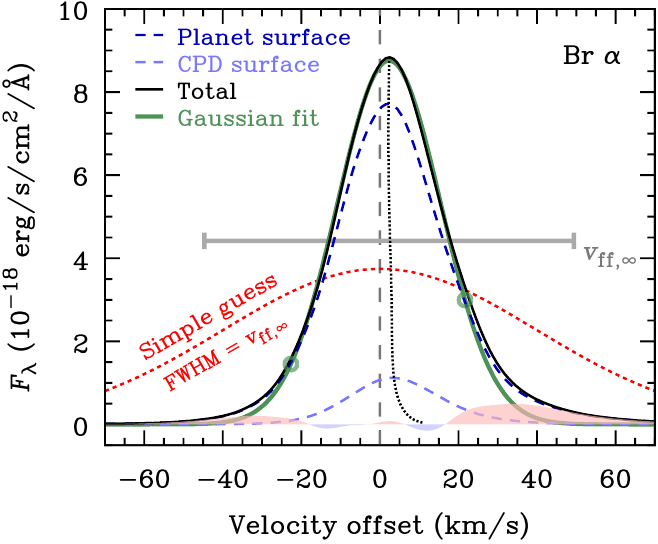}
\caption{%
Total emission line (\textit{thick black line}), which sums the planetary- and CPD-surface contributions (\textit{dashed dark and pale blue}, respectively) corresponding to Figure~\ref{Abb:Bild} (parameters in Table~\ref{Tab:Par}).
At this mass, the planet surface dominates the sum.
The bisector (\textit{black dotted}) emphasises the asymmetry of the line.
A best-fitting Gaussian (\textit{thick mint curve}) is also shown,
with circles delimiting the region within which the fit is better than 10\,\%, and the residual (total line minus fit) is shown as a shaded region.
The line is  %
clearly narrower ($\FWHM=31~\kms$)
than $\vFfinfty=94~\kms$ (\textit{thick grey bar}),
and thus has a much higher peak than a Gaussian with $\FWHM=\vFfinfty$ and the same total flux (`simple guess'; \textit{red dashed line}).
}
\label{Abb:Prof}
\end{figure}

Figure~\ref{Abb:Prof} displays the resulting line profile for \Bra, with a peak flux of $F_\lambda=8.8\times10^{-18}$~erg\,s$^{-1}$\,cm$^{-2}$\,\AA$^{-1}$. For our fiducial parameter values, even the line core is not saturated, that is, is not optically thick.
This implies that all other lines observable by METIS, which are optically even thinner because they are higher-order transitions (Fig.~\ref{Abb:Linien}), will have a similar shape.  %
The shock at the planet surface contributes almost all of the line flux.
The whole line is redshifted by about $\Delta v=+2$~\kms, which is easily measureable with METIS given the typical line-centroid positioning accuracy of one tenth of a resolution element.   %
The line is %
mostly symmetrical, but the bisector shows that the wings, especially for flux levels below approximately ten percent of the flux peak,
are asymmetrical. This is because the line is formed in the postshock region, where the gas is systematically receding away from the observer at a significant fraction of the thermal velocity \citep{aoyama18}. The far wings come from the highest-temperature postshock layers, which have the highest velocities.
As \citet{AMIM21L} note, this is typical for shock emission, whereas funnel emission (magnetospheric accretion) can have a broader blue wing \citep[e.g.,][]{demars23,demars26,viswanath26,hashimoto26}.

For the fiducial case, the line is much narrower than the free-fall velocity from infinity, with a relative width of
\begin{equation}
 \label{Gl:relDvBreite}
\relDvBreite \equiv\frac{\FWHM}{\vFfinfty}\approx0.33. %
\end{equation}
For comparison, \citet{thanathibodee19} adopted $\MP=6~\MJ$ and $\RP=1.3~\RJ$, which implies $\vFfinfty=128$~\kms, and used their magnetospheric accretion model to fit the \Ha line observed with the MUSE instrument of the Very Large Telescope (VLT) at a spectral resolution\footnote{\citet{thanathibodee19} used a higher approximate value $R=2800$, while $R=2516$ was derived by \citet{eriksson20} from the user manual.} $R=2516$ \citep{Haffert+2019}. \citet{thanathibodee19} found $\FWHM = 57~\kms$ (see their Fig.~3),
that is, $\relDvBreite_{\mathrm{H}\,\alpha}=0.46$.
For GSC\,06214\,B, whose \Pab line shape is consistent with both a pure shock or magnetospheric-accretion origin, \citet{demars23} found a maximum deconvolved $\FWHM\approx100$--$130~\kms\approx(0.5$--$0.7)\vFfinfty$, which will be somewhat an overestimate if the intrinsic line shape is not Gaussian (D.~Demars 2025, priv.~comm.). At TWA\,27\,B, which seems to be accreting through magnetospheric accretion\footnote{The results of \citet{patapis25}, who find that a transitional CPD model of \citet{sun24} matches well the mid-infrared spectrum, seem to be consistent with this, with the inner cavity conceivably cleared by the magnetic field of the planet (e.g., \citealp{zhu15}). Of course, alternative explanations nevertheless exist.} \citep{aoyama24twa}, the average of several NIR Balmer lines yields $\relDvBreite\approx0.6$, as discussed in \citet{m23alois}.

The upshot of this is that irrespective of the exact accretion mechanism,
these relatively weak accretors have
a narrow line width $\relDvBreite\lesssim0.5$ and not unity as one might assume. %
There are two aspects to consider:
\begin{enumerate}
 \item In the magnetospheric accretion model \citep[e.g.,][]{hartmann94}, the base of the wing is of the order of the free-fall velocity \vFfinfty because the velocity of the accretin gas causes the broadening; the local line width is narrow compared to \vFfinfty.

 \item In the postshock emission model, the \FWHM is coincidentally of the same order as, but smaller by a significant factor than, the pre-shock velocity $v_0$, which is at most \vFfinfty.
\end{enumerate}
\citet{zhu15} assumed Point~1 (the line being a box of width $\vFfinfty$, broadened by the infall).
Here, since we are considering postshock emission rather than the accretion-flow emission, one should compare the \FWHM to \vFfinfty (Equation~(\ref{Gl:relDvBreite})). Interestingly, under their set of assumptions, in the context of supernova remnants, \citet[][and works cited therein]{heng07} found the width of Balmer lines to be proportional to the preshock velocity. Their Fig.~7 shows $\FWHM\approx0.8v_0$, flattening towards higher preshock velocities $v_0\gtrsim2000$~\kms.

In Appendix~\ref{Th:DvundmuAo18}, we show that at moderate to low preshock densities $n_0\lesssim10^{12}$~cm$^{-3}$, the \citet{aoyama18} models predict a remarkably constant line width $\FWHM\approx30$--40~\kms for a wide range of $v_0\approx30$--200~\kms. Thus all regions of the planet surface and the CPD emit with approximately the same line shape in terms of width.
In Figure~\ref{Abb:Prof}, the contribution from the planet surface has $\FWHM=31.3$~\kms and the CPD component has $\FWHM=27.8$~\kms, with a width of $\FWHM=30.9$~\kms for the total line.
We show the $(n_0,v_0)$ values on the planet and CPD surfaces in Figure~\ref{Abb:DvAo18}. The CPD has a larger area than the planet surface but also lower velocities, with correspondingly quickly decreasing flux. Therefore, the $(n_0,v_0)$ values at the planet surface dominate, where $\FWHM/v_0\approx0.3$--0.4. Figure~\ref{Abb:DvAo18} suggests that preshock densities two orders of magnitude higher would be needed to obtain lines with significantly different relative width.
For a given total flux, a narrow line has a line peak higher above the photospheric level, and is thus easier to detect, than would be expected from a simple estimate. We illustrate the converse in Figure~\ref{Abb:Prof} by a Gaussian of width \vFfinfty.

At first sight, the line profile seems Gaussian but this holds only for the line core.
In Figure~\ref{Abb:Prof}, we compare the curve with the fit to a Gaussian of mean $\mu$ and standard deviation $\sigma$. We used the built-in \texttt{fit} function of \texttt{gnuplot} and a $1/F_\lambda$ weighting to match the peak better.
This yields $\sigma=13.2$~\kms, implying a fitted $\FWHM = 2\surd({2\ln 2})\,\sigma=31.1~\kms$, which agrees with the actual \FWHM.
The horizontal offset is %
$\mu=2.1$~\kms,   %
also matching the bisector above the ten-percent flux level.
We comment on this shift in Appendix~\ref{Th:DvundmuAo18}.
From $\Delta v\approx-25$ to $+25~\kms$, the Gaussian reproduces the profile to better than 10\,\%\ but away from this, the line becomes non-Gaussian, %
with a flux that remains much higher.
Indeed, Figure~\ref{Abb:W10relGauAo18} shows that the ten-percent width $W_{10}$ of the shock emission, at the $(n_0,v_0)$ appropriate for the fiducial case, is larger than for a Gaussian by roughly ten percent. Towards higher preshock densities, $W_{10}$ becomes narrower than for a Gaussian. %

\subsection{PDS 70 b: Comparison with the other line fluxes}
 \label{Th:Sigmakalibrierung}

\begin{figure} %
 \centering
\addvspace{1em}
 \includegraphics[width=0.45\textwidth]{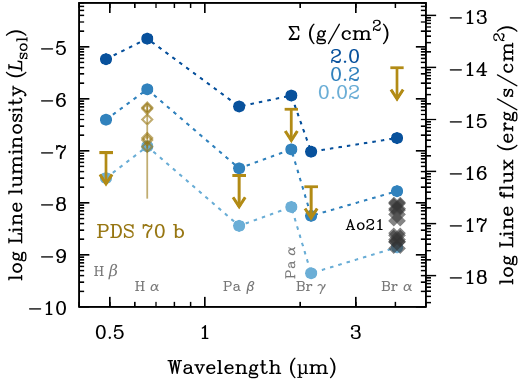}
\caption{%
Line-integrated luminosity in our models (\textit{blue circles}) compared to the observational constraints at \PDSb (\textit{gold symbols}), not correcting for extinction.
We used the fiducial parameters (Table~\ref{Tab:Par}) and varied only the gas surface density at the Hill sphere (\textit{hues of blue}), a proxy for the mass accretion rate.
The golden diamonds and arrows show detections and 3$\sigma$ upper limits, and the thick crosses at \Bra are the corresponding predictions using the \citet{AMIM21L} scalings (see text).
}
\label{Abb:LLinie}
\end{figure}

We repeated the computation for several hydrogen lines, and 
in Figure~\ref{Abb:LLinie}, we compare the spatially and spectrally integrated line fluxes
to the available data. These are the \Ha measurements (using for simplicity only \citealt{sanghi22,close25a,zhou25}) and upper limits at \Hb, \Pab, \Paa, \Bra, and \Brg (\citealp{christiaens19b, %
hashimoto20,Stolker+20b,wang21vlti,uyama21b,christiaens24}).
We did not correct for possible extinction, which might be $\AV\approx0.9$~mag \citep{uyama21b} and thus only $\ABra\approx0.06$~mag \citep{chiar06}.
Varying the mass to $\MP=7~\MJ$ (or roughly equivalently, considering instead a radius $\RP=\sqrt{5/7}\times2~\RJ=1.4~\RJ$; Equation~\ref{Gl:vFfinfty}) does not change our line flux estimates.

At face value, a surface density $\Sigma\approx0.2$~g\,cm$^{-2}$ is roughly consistent with the line detections and the 3$\sigma$ upper limits.
This is why we adopted this $\Sigma$ as the fiducial value, as reported in Table~\ref{Tab:Par}.
We recall that our $\Sigma$ is not meant as a realistic estimate of the true average surface density around the Hill sphere, which for gap-opening planets will not be an azimuthally symmetric quantity. Rather, $\Sigma$ should be seen as a convenient internal parameter controlling linearly the accretion rate.

A slight outlier is the \Hb flux, with an upper limit 0.5~dex fainter than the model flux for $\Sigma\approx0.2$~g\,cm$^{-2}$. However, besides \Ha, \Hb is the most likely to be affected by extinction, and consistent combinations of fluxes and extinction values are not easy to determine \citep{Aoyama+Ikoma2019}.
Recent studies \citep{close25a,zhou25} have clearly shown that \PDSbc are variable at \Ha by a factor of a few over a span of roughly five years.
Thus the comparison is only indicative, especially since the data in Figure~\ref{Abb:LLinie} were obtained at different epochs.
Similarly, we can tolerate that the \Ha flux with the nominal $\Sigma$ is slightly above the measurements so far; this might be slightly optimistic but a non-zero extinction, for which we do not correct, would make this approach conservative again.

Our detailed model, with this $\Sigma$ calibration, predicts an integrated \Bra flux $\FBra\approx4\times10^{-17}$~erg\,s$^{-1}$\,cm$^{-2}$.
As a very mild consistency check, we used the \citet{AMIM21L} fit to derive the accretion luminosity \LAkk from the \Ha measurements at different epochs and used the inverse of the fit  %
to obtain \Bra fluxes from the \LAkk values:
\begin{subequations}
\label{Gl:LBravonLHa}  %
\begin{align}
 \log \LBra/\LSonne &= \frac{\aHa}{\aBra}\log\LHa/\LSonne+\frac{\bHa-\bBra}{\aBra}\\
            &= 1.01 \log\LHa/\LSonne - 1.82,
\end{align}
\end{subequations}
where $\aHa=0.95$, $\aBra=0.94$, $\bHa=1.61$, and $\bBra=3.32$.
This approach,\footnote{For classical T~Tauri stars (CTTSs), only \citet{Komarova+Fischer2020} provide an $\LBra(\LAkk)$ relationship. For Class~I protostars, \citet{testi25} present a relationship but it is not calibrated with UV excess measurements. Several works study $\LLinie(\LAkk)$ for other lines; recent ones include \citet{rogers24}, \citet{fiorellino25}, and \citet{shridharan26}.} $\LLinie\rightarrow\LAkk\rightarrow\LBra$,
can be used to estimate signal strengths.  %
This yields a range of values, shown as thick crosses in Figure~\ref{Abb:LLinie}, 
extending one order of magnitude below the line flux of our detailed model, down to $\FBra\approx4\times10^{-18}$~erg\,s$^{-1}$\,cm$^{-2}$.
This agreement is almost by construction, since the underlying physical model is the same in \citet{AMIM21L}, who use the  %
 calculations of \citet{aoyama18} for single parameter combinations, as here, where we consider a spatial distribution of emitting patches.
The agreement is good because the emitting regions are not too optically thick. If they were, line saturation would lead to a departure from the fits of \citet{AMIM21L},
also for some lines including \Ha and \Pab, for example.
Our fiducial parameters place us in the regime of low preshock densities.

\subsection{Varying the parameters}
 \label{Th:Parvar}

\begin{figure*} %
 \centering
 \includegraphics[height=0.25\textwidth]{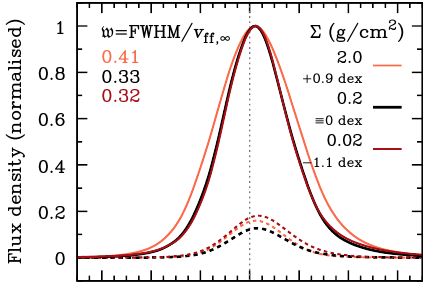}%
~%
 \includegraphics[height=0.25\textwidth]{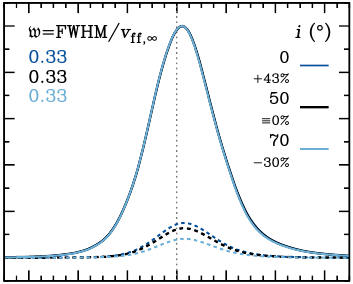}%
~%
 \includegraphics[height=0.25\textwidth]{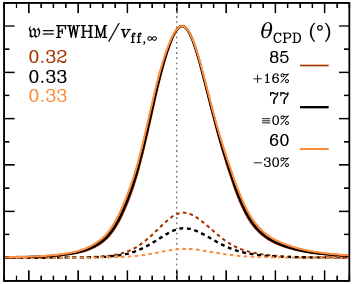}%
\\[0.1em]  %
 \includegraphics[height=0.3033\textwidth]{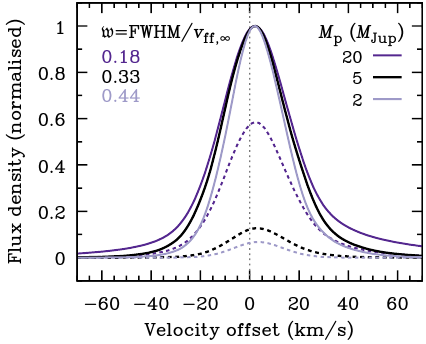}%
~%
 \includegraphics[height=0.3033\textwidth]{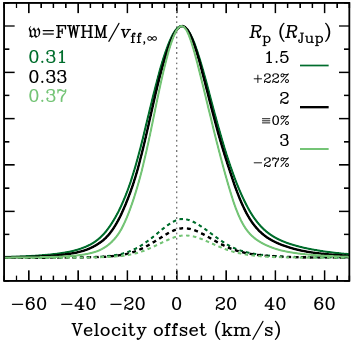}%
~%
 \includegraphics[height=0.3033\textwidth]{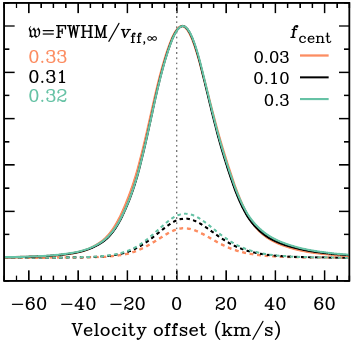}
\caption{%
Shock excess line shapes at \Bra (\textit{solid}: total, i.e., planet and CPD; \textit{dashed}: CPD contribution) when varying respectively (across and down)
 the gas surface density $\Sigma$,
 inclination $i$, 
 opening angle of the CPD \thzpSch, 
 planet mass \MP,
 planet radius \RP,
 and
 $\fzent\equiv\Rzent/\RHill$.
Parameter values are given in the respective legends.
Each CPD-contribution curve is normalised to the peak total flux.
The reference case (\textit{black line}, same in each panel) is as in Table~\ref{Tab:Par}. %
The legend reports the change in total flux relative to the fiducial case for all panels except where \MP or \fzent is varied.
Numbers on the left give the corresponding \relDvBreite values (Eq.~(\ref{Gl:relDvBreite})) of the total line.
}
\label{Abb:Varpar}
\end{figure*}

\begin{figure}
 \centering
\addvspace{1em}
 \includegraphics[height=0.3033\textwidth]{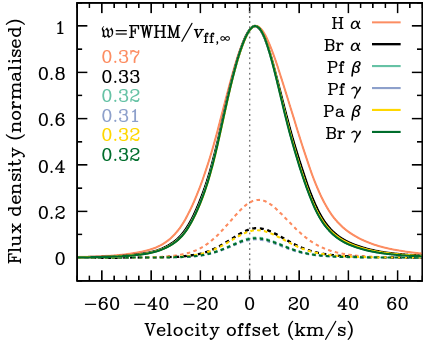}  %
\caption{%
As in Figure~\ref{Abb:Varpar} but comparing different hydrogen-line transitions.
}
\label{Abb:VarparLinie}
\end{figure}

In Figure~\ref{Abb:Varpar}, we compare the \Bra line shape resulting from varying the parameters \MP, \RP, \thzpSch, $\Sigma$, $i$, and \fzent in turn. In Figure~\ref{Abb:VarparLinie}, we compare the line shapes of \Ha, \Bra, \Pfb, and \Pfg. For the purposes of Section~\ref{Th:MICADO}, where we discuss prospects for MICADO, we add \Pab and \Brg. The left part of each panel reports \relDvBreite, the relative \FWHM (Eq.~\ref{Gl:relDvBreite}).

The behaviour is mostly intuitive. At higher preshock densities (through a higher $\Sigma$ or \MP, or a lower \RP), the \Bra line core is closer to being saturated, making the line broader. The contribution of the CPD shock to the total line flux increases with planet mass, as in \citet{m22Schock}, as well as for a thinner CPD (larger preshock velocity $v_\theta$). The same holds at \Ha (Fig.~\ref{Abb:VarparLinie}) because it is an optically thicker transition, which affects the CPD-surface contribution less than the planet-surface one. Conversely, for optically thinner lines such as as \Pfb and \Pfg, the CPD contributes less, even though the overall line shape is very similar.

In Figure~\ref{Abb:Varpar}, we also indicate the absolute change in flux when varying the parameters. We do not do this for the planet mass because it also changes the size of the Hill sphere and the mass influx towards the planet and CPD,
impeding a parameter-sensitivity analysis. Similarly, varing \fzent changes the large-scale mass influx. For all other parameters, however, the analysis is straightforward because they do not affect the flow geometry. We discuss the effect of each parameter in turn.

\begin{itemize}
\item The surface density $\Sigma$ only changes the mass loading but even with a factor-ten increase, we remain in the roughly linear regime for the flux response. The line starts to broaden only for the highest $\Sigma$ value. Figure~\ref{Abb:DvAo18} shows the more general view of the absolute and relative \FWHM, namely the dependence of the local emission line width on the preshock density $n_0$, itself proportional to $\Sigma$, and on $v_0$. For the fiducial parameters, $n_0\approx(3$--$4)\times10^{11}$~cm$^{-3}$ over the planet surface and emitting CPD surface.
\item Viewing the same system pole-on ($i=0\degr$) makes the shock 40\,\%\ brighter at the observer because the high-preshock-velocity, polar regions are better visible, whereas going towards a grazing view above the CPD surface ($i=70\degr\approx\thzpSch$) conversely reduces the flux, in reality probably more than the 30\,\%\ we find here because extinction might become important.
\item Increasing the thickness of the CPD (decreasing \thzpSch) does not change the supersonic inflow but it has a double effect: the streamlines that would have landed in the lowest-latitude regions on the planet instead shock on the CPD surface close to the planet, with a preshock velocity given by their polar component $v_\theta$ instead of the higher radial-velocity $v_r$ component. The second effect is that the streamlines hitting the CPD at some distance now do so higher up above the midplane, less deeply in the potential and thus at a lower velocity.
\item Changing the radius \RP, finally, has a subtlety: while a smaller radius leads to a higher preshock velocity on the planet, it also reduces the amount of mass hitting the planet because the flow field again remains the same but now some streamlines do not shock anymore on the planet but rather on the CPD, where again their $v_\theta$ is smaller than their $v_r$. However, the line flux depends strongly on the preshock velocity (at these relatively low densities,
the local emission $F\sim v_0^3$),
so that the contribution from the CPD remains minor, at the 10--20\,\%\ level. Therefore,
the change in total flux, with $+22\,\%$ and $-27\,\%$ for $\RP=1.5$~and 3~$\RJ$, is smaller than what the $\vFfinfty\propto1/\surd{\RP}$ scaling would naively imply, because of the reduction in area ($\propto\RP^2$) or, equivalently, in the fraction of the accretion rate hitting the planet surface.
\item Increasing the centrifugal radius by setting $\fzent=0.1$ or $\fzent\approx1/3$ (\citealt{quillen98} but see discussion in \citealt{m24expeditus})
changes the overall line profile barely noticeably, similarly to the effect of \thzpSch, and increases the contribution of the CPD minimally. This is still intuitive because a larger \fzent means that the gas is falling further away from the planet and thus more on the CPD.
\end{itemize}
Thus each parameter affects the total flux and also, to different degrees, the line shape. As Figure~\ref{Abb:Varpar} shows, only the highest $\Sigma$ values and otherwise the mass and the radius change noticeably the line shape.
The fiducial case is in the regime where the single-$(n_0,v_0)$ line width is nearly constant at $\FWHM\approx30~\kms$, which implies $\relDvBreite=0.33$. For a tenfold increase in \MPkt, \relDvBreite~rises to $\approx0.4$, as for a reduction of the mass to $\MP\approx2~\MJ$ or a larger radius $\RP\approx3~\RJ$.

Nevertheless, in general (not shown) there is no single profile from the grid of \citet{aoyama18} which matches the total profile coming from the spatial integral of the different regions. This means that explaining an observed line profile requires the kind of modelling presented here. %
At the same time, the line shape cannot be expected to yield tight constraints on the parameter values. We come back to this in Section~\ref{Th:MagAkk}.

\section{Detectability of the resolved accretion lines}
  \label{Th:Detektierbarkeit}

So far, we have calculated only the accretion-line emission.
Here, we estimate its detectability.
As mentioned above, we take the limit of negligible extinction in the $L$ and $M$ bands since opacity decreases with wavelength and we are considering planets which have opened gaps in their natal PPDs.
Therefore, we only need to look at the different competing signals and the noise.
Specifically, we estimate for the former
the photospheric signal of the planet (Section~\ref{Th:PlanEm}),
the brightness of photosphere of the CPD (Section~\ref{Th:zpSchEm}),
and
the thermal and scattered-light emission from the PPD directly at the location of the planet (Section~\ref{Th:zstSchEm}).
For the true noise, we could use the \texttt{ScopeSim} package\footnote{See \url{https://scopesim.readthedocs.io}.} \citep{leschinski20} with the settings for METIS, a combination formerly known as \texttt{SimMETIS} (e.g., \citealp{chen22,oberg23metis}), and fold in a detailed observational strategy.
Instead, we look separately at
the non-speckle noise (residual from the thermal background, dark current, and read-out noise; Section~\ref{Th:Detektorempf})
and the speckle noise (Section~\ref{Th:RauschenvonPrimEm}).
Putting this together, in Section~\ref{Th:LiniemitRauschen} we use the sensitivity curve to simulate the post-processed, that is, background-subtracted, accretion-line spectrum.
Finally, in Section~\ref{Th:alleszusammen} we generalise the calculation to other masses and accretion rates.
\subsection{Emission from the planetary photosphere}
 \label{Th:PlanEm}

To assess the possible signal from the planetary photosphere, we use resolved spectra because the resolution of METIS ($R\sim10^5$) is sufficiently high to sample at better than the Nyquist rate typical atmospheric features of objects down to a \Teff of several hundred kelvin.
Two approaches can be used to estimate the absolute flux from the photosphere, here specifically of \PDSb.
One is to consider the literature best fits to the global spectral energy distribution (SED), yielding a radius and \Teff, amongst others, and to use their predictions as-is.
The other is to consider a somewhat wider range of models (in terms of \Teff but also surface gravity, chemistry, etc.) but to scale them so that the synthetic photometry over the NB4.05 filter%
\footnote{The 
correct width is $\Delta\lambda=0.061640$~\mum and not 0.02~\mum as reported in several works; see Footnote~28 of \citet{maea21}.%
} match the observed photometry \citep{Stolker+20b}.
We prefer this more robust approach because even the global ($\lambda\approx1$--5~\mum) fits do not reproduce the data completely satisfactorily, which raises doubts about their reliability over very small spectral ranges.

We searched in publicly-available data and found only\footnote{Fortunately, in the last stages of reviewing of this paper, the situation has nearly changed thanks to the release of the cloudy and clear ExoREM `k26' grid \citep{radcliffe26} at $R=500$ and $R=10^4$ resolution, with the paper on $R=\textrm{200,000}$ in preparation.} %
the CIFIST models \citep{allard12philtrans,allard13} from the Spanish Virtual Observatory (SVO)\footnote{See \url{http://svo2.cab.inta-csic.es/svo/theory/newov2}.} to be
fully resolved at $L$ and $M$, where they have $R\approx(3$--$5)\times10^5$.
These models are cloud-free but in general, clouds mute rather the large-scale ($R\sim10$) and not the small-scale spectral features  %
(e.g., Figure~3 of \citealt{radcliffe26}). Thus the non-inclusion of clouds in the models might not affect too much our predictions. %
Only the solar-metallicity grid is available but this is also appropriate given the $K$-band analysis of \citet{hsu24c}.  %
Nevertheless, one should keep in mind that over a very small spectral window, the true spectrum could happen to be more similar to a model with a different metallicity or C/O ratio, for example.

\begin{figure}[!t]
 \centering
\includegraphics[width=0.47\textwidth]{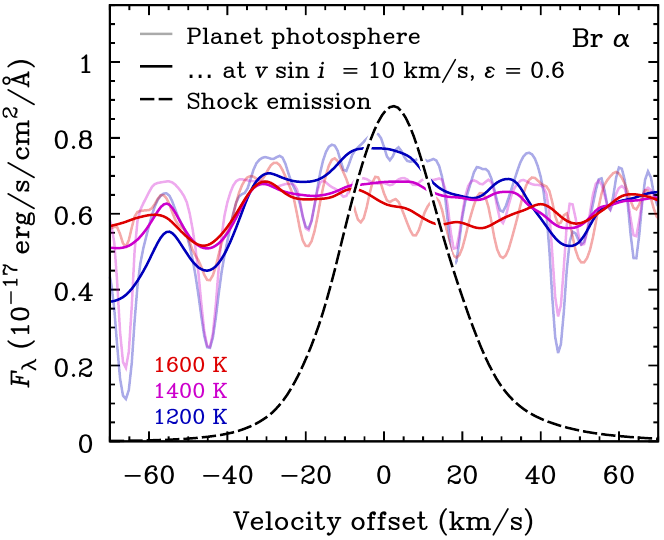}  %
\caption{%
Resolved continuum at \Bra for $\Teff=1200$, 1400, 1600~K (\textit{blue to red}) from the CIFIST models for $\log g=3.5$ and solar metallicity,
scaled to match, integrated over the filter, the NB4.05 photometry (\citealt{Stolker+20b}; see text),  %
and without rotational broadening (\textit{pale lines}) or broadened to $v\sin i\approx10~\kms$ assuming a standard limb darkening coefficient $\varepsilon$ (\textit{darker}). %
The \textit{black dashed line} is the total shock emission for the fiducial case as in Figure~\ref{Abb:Prof}. In Figure~\ref{Abb:AtmbeitragnormmitgCMCRT}, we make a brief comparison to a spectrum from gCMCRT \citep{lee22}.
}
\label{Abb:Atmbeitragnorm}
\end{figure}

Figure~\ref{Abb:Atmbeitragnorm} shows the possible contribution of the planetary atmosphere. We consider $\Teff=1200$--1600~K, noting that lower \Teff values are not available in the CIFIST models. 
This range is wider than given by the formal uncertainties on the global-SED fits (for instance in \citealt{wang21vlti} or \citealt{blakely25}; however, they use other families of models). As mentioned above, we choose this wider span for greater robustness.
We multiplied the model surface flux density by $(\RP/d)^2$ with \RP and $d$ from Table~\ref{Tab:Par}, %
and matched the NB4.05 photometry by additionally multiplying the model spectrum by a factor of
$\fPhot=1.188$, 1.036, 0.848 for $\Teff=1200$, 1400, 1600~K, respectively.
The fits with BT-Settl models in \citet{wang21vlti} yield a surface gravity $\log g\approx3.5$--4 and our fiducial mass and radius values correspond to $\log g=3.49$.
Thus we adopt $\log g=3.5$ for Figure~\ref{Abb:Atmbeitragnorm} and afterwards.
In higher-gravity models, the spectral features are less pronounced, which could have given an overly optimistic estimate of the observability.

Figure~\ref{Abb:Atmbeitragnorm} compares the photospheric signal to the accretion line. The maximum continuum\footnote{This is a pseudocontinuum but we shall use `continuum' for brevity. Also, while young planetary-mass objects and spectral-type LTY brown dwarfs can be variable \citep[e.g.,][]{radigan14,biller15,biller17,vos22,zhangmoll25,adamszhou25,cushing26,zhou26}, this variability is of at most a few percent in amplitude, and will therefore not substantially affect our ability to subtract it to reveal accretion signatures.} level turns out to be similar to the peak accretion-line flux density.
This is an encouraging result but since the line is not much brighter than the continuum, we need to look next also at the root mean square (RMS) deviation of the continuum. %
Indeed, with little information about the photosphere, one can always approximately remove the continuum through a low-order polynomial and treat the residuals as noise that could mask the line.

Ideally, forward models of the photosphere or atmospheric retrievals could provide an even better description of the spectral features but in practice there might still be large residuals, especially with high-resolution data. Incompleteness in line lists, choices for the line-wing truncation, and other uncertainties in the microphysics have an observable effect on the spectrum and can even limit the accuracy of the inferred basic parameters \citep{baudino17,deregt25}.
Approaches to deal with this have been proposed \citep{rotman25}
but at a simpler level, the RMS provides a useful and robust upper limit on the amount of photospheric noise that could hide the line.

Figure~\ref{Abb:Atmbeitragnorm} shows that the photosphere exhibits a few narrow absorption features with widths $\approx5$--7~\kms, but the deep features (50--80\,\%\ of the continuum) are fortunately $\Delta v=-65$~and $\pm45$~\kms away from the centre of the \Bra line.
Within the 10-percent width of the shock line, the strongest photospheric features are at most 20\,\%\ deep. Features are more pronounced towards low \Teff,
but are qualitatively similar for the range of \Teff considered.
For reference, we show in Appendix~\ref{Th:Phothelligkeitmehr} both the average photospheric flux and the RMS in the CIFIST models.

Given this limited dependence on \Teff, we adopt as the fiducial atmospheric effective temperature $\Teff=1400$~K with the flux correction factor of $\fPhot=1.036$ and report this in Table~\ref{Tab:Par}.
In other words, we are effectively using a photospheric radius of $R=\sqrt{1.036}\RP=2.04~\RJ$ but keep using separately $\RP=2~\RJ$ for the other parts of the calculations. Conveniently, \WISPb likely has a similar \Teff, as we discuss in Section~\ref{Th:Disk}.

One helpful potential physical effect is the rotation of the planet, which will smooth only the atmospheric features and not the accretion line.
Young planets might be spinning at $\fkrit\approx5$--30\,\%\ of their break-up velocity (\citealp{bryan20,snellen25}; \citealp{hsu26}),  %
\begin{equation}
\vZerf=\sqrt{\frac{G\MP}{\RP}},
\end{equation}
which translates into projected rotation velocities of
\begin{equation}
v\sin i\lesssim \fkrit\vZerf\approx 10 \left(\frac{\fkrit}{0.2}\right)\sin\left(\frac{i}{50\degr}\right) \sqrt{ \frac{M_5}{R_2} }~\kms,
\end{equation}
where $M_5\equiv\MP/(5~\MJ)$ and $R_2\equiv\RP/(2~\RJ)$.  %
With the Keck Planet Imager and Characterizer (KPIC), \citet{hsu24c} found a non-detection of spin broadening for \PDSb, deriving  %
a \mbox{95\%} (i.e., roughly 2$\sigma$) upper limit of $v\sin i<29$~\kms assuming a constant linear limb-darkening coefficient $\varepsilon=0.6$ \citep{gray92,claret00}   %
and no differential rotation\footnote{%
In general, there are degeneracies between the coefficient of differential rotation $\delta$ and the $v\sin i$; for example, a curve broadened with $v\sin i=22~\kms$ and $\delta=0.675$ (a $\delta$ value proposed by \citealp{smith94}, as mentioned in the comments of the \texttt{rotBroadInt.py} routine; \citealp{carvalho23}) is essentially identical to the result of $v\sin i=10~\kms$ and $\delta=0$.} \citep{hsu21,hsu21zndo}.

In Figure~\ref{Abb:Atmbeitragnorm}, we broaden the spectrum to $v\sin i=10~\kms$ with the \texttt{rotBroadInt} routine\footnote{See \url{https://github.com/Adolfo1519/RotBroadInt}.} of \citet{carvalho23}, with the same (default) value for the limb-darkening parameter %
and also without differential rotation as in \citet{hsu24c}.
The broadened spectra have much weaker fluctuations with now only roughly 10\,\%\ of the accretion-line peak, making a detection of the accretion line easier.
With rotational broadening, the shape of the continuum is even less sensitive to \Teff.

In summary, for the fiducial parameter values, the \Bra line should show up as a clear excess above the continuum,
but some care will be needed to remove the latter. %
We are in the intermediate regime where the line is neither an order of magnitude brighter than the photospheric continuum nor an order of magnitude fainter.
For comparison, at \Pab the resolved line peak outshines the continuum by an appreciable factor (this is discussed in Fig.~\ref{Abb:LinienfuerMICADO}).
If \PDSb is caught in an episode of low accretion and thus has an accretion-line flux weakened by a factor of a few, detection will be challenging, and an accurate determination of the continuum will be needed.
\subsection{Emission from the CPD}
 \label{Th:zpSchEm}
Recent works have presented fits to the SED of \PDSb from approximately 1~to 5~\mum using atmospheric models with or without additional extinction, and adding a second component or not to model approximately the CPD (e.g., \citealp{wang21vlti,blakely25}).
In their fits of the global spectrum of \PDSb, both \citet{Stolker+20b} and \citet{wang21vlti} fit blackbody components to approximate the CPD contribution,
with \citet{Stolker+20b} also considering some fits with the \citet{isella19} ALMA Band 7 non-detection as an upper limit.
In the absence of higher-resolution data, these studies yield rough but useful constraints on the emission from the CPD.
However, this is not suitable to describe the contribution over a narrow spectral range such as the width of an emission line.
In such a spectral window, realistically, several regions of the CPD might contribute to the emission.

Even more important here is the fact that while these various fits suggest different system parameters, they do robustly find that the measured VLT/NACO photometry 
\citep{Stolker+20b}
in the NB4.05 filter
is 1--2~$\sigma$ below the model predictions, even when fitting only for a planetary atmosphere.
The expected excess indicative of a CPD is seen only at $\lambda\approx4.8$~\mum consistently in VLT/NACO \citep{Stolker+20b}, JWST/NIRCAM \citep{christiaens24,christiaens25}, and JWST/NIRISS \citep{blakely25} data.
Therefore, currently, the limiting case that the CPD does not contribute at all to the NB4.05 photometry can be justified.
Because the thermal structure of the CPD, and therefore its emission, is very poorly constrained, we further assume that the CPD does not contribute, not only to the whole NB4.05 band but also over only the width of the \Bra line.

A priori, this is not a trivial assumption since there could happen to be strong spectral emission features over such a narrow interval, which would leave the narrow-filter-integrated flux consistent with the observed photometry.
Verifying this assumption by modelling the CPD while varying the several parameters \citep[e.g.,][]{k24,k24b,sun26} to generate high-resolution spectra would be outside the scope of this work, also because of the current lack of constraints.
Nevertheless, as a reference, we computed a few models with ProDiMo \citep{woitke16,kamp17} and show this in Appendix~\ref{Th:zpSchProDiMo}. It seems to indicate that the contribution from the CPD is modest and certainly that the CPD does not add fine spectral features at the position of \Bra.

\subsection{Emission from the protoplanetary disc at the location of the planet}
 \label{Th:zstSchEm}

To estimate the astrophysical noise from the PPD at \Bra and the other lines, we consider the JWST/NIRCAM F480M ($L$-band) image from \citet{christiaens24}.
We recall that the angular resolution of JWST %
is $185$~mas at NIRCAM/F480M,
whereas for the ELT it is $26$~mas.
\citet{christiaens24} obtain that the PPD has an average surface brightness $\mathfrak{B}=0.1$--0.5~GJy/sr  %
in the F480M filter ($\lambda_{\mathrm{pivot}}/\Delta\lambda=16$) depending on the location in the disc.
Over the beam of the ELT, this corresponds to
$\langle F_\lambda\rangle\approx(0.2$--$1)\times10^{-19}$~erg\,s$^{-1}$\,cm$^{-2}$\,\AA$^{-1}$.
This is much smaller than the accretion-line signal within $|\Delta v|\approx30$--$50~\kms$ of the line centre, and thus not a source of concern. %
\subsection{Sensitivity in the background-limited regime}
 \label{Th:Detektorempf}

\begin{figure}[!tp]
 \centering
\includegraphics[width=0.47\textwidth]{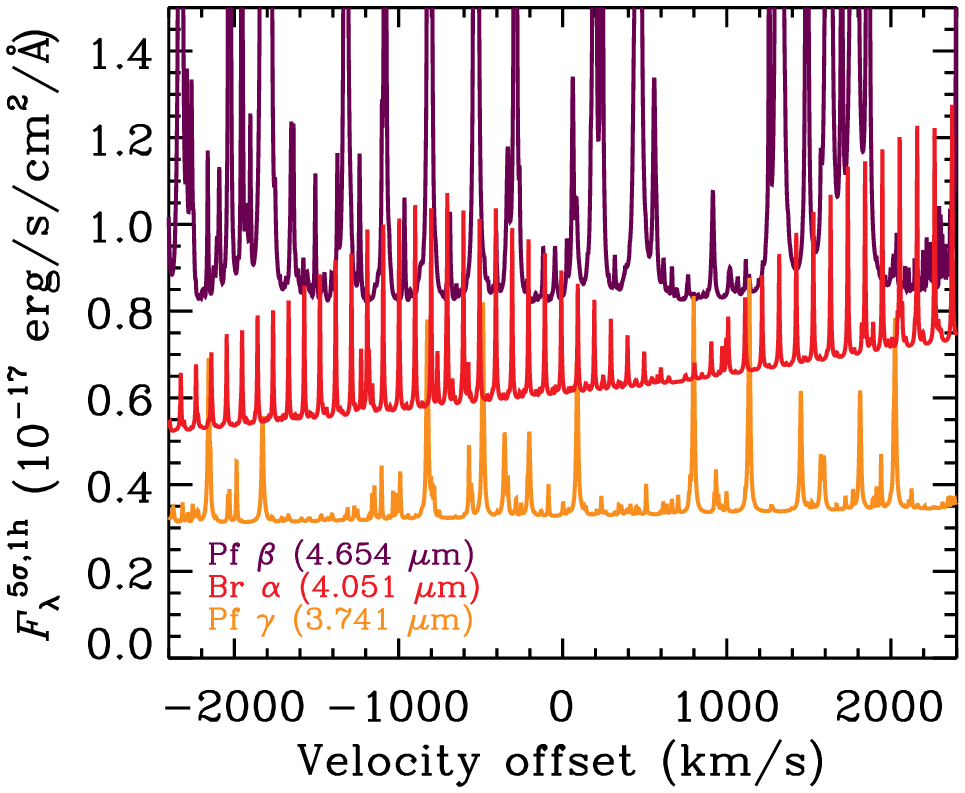}
\caption{%
Sensitivity of METIS (that is, residual noise per bin, at 5~$\sigma$ in 1~h; scalable with Eq.~(\ref{Gl:Empf})) around three of the \Hi emission lines.
Figure~\ref{Abb:tauAtmundDetektorempf} plots the optical depth of the Earth atmosphere over a narrower range around all \Hi lines, highlighting the similarities.
}
\label{Abb:DetektorempfpaarL}
\end{figure}

METIS is designed to have a high sensitivity.
Here, it suffices to use spatially averaged sensitivity curves without taking the exact spatial dependence of the detector characteristics into account.
Figure~\ref{Abb:DetektorempfpaarL} shows the sensitivity per spectral resolution element for spectrally resolved point sources (i.e., for the continuum or lines wider than the native resolution of $c/R\approx3~\kms$) 
in integral-field-unit (IFU) mode,
for median conditions in the background-limited\footnote{Of course, this `background' actually comes from the telescope and sky, which are in the foreground relative to the source.} regime
\citep{lovis22}, as calculated by one of us (RvB).
We calculate the sensitivity curve by
assuming that 
the target is in the one half of the IFU for 50\%\ of the time and in the other half for 50\%\ of the time. At each location the respective other half of the observing time is used to measure the background and subtract it.
The noise budget contains all fundamental noise sources. We use a radiometric model of the ELT and the METIS instrument, accounting for the thermal emission of the six warm telescope mirrors and the cryostat entrance window, the transmission of the telescope and all instrument-internal optical elements, and for the detector quantum efficiency and readout noise.
The dark current is not significant for the Hawaii-2RG detector.
For the telluric transmission and emission profile we adopt a \texttt{SkyCalc}\footnote{See \url{https://www.eso.org/observing/etc/bin/gen/form?INS.MODE=swspectr+INS.NAME=SKYCALC}.} model \citep{noll12,jones13} for median conditions of precipitable water vapour $\mathrm{PWV}=2.5$~mm and an airmass $x=1.3$.

On small scales, the sensitivity curves have periodic features caused mainly by resonances in N$_2$O (at \Bra, with a beating pattern), CH$_4$ and H$_2$O (at \Pfg), and N$_2$O and H$_2$O (at \Pfb) in Earth's atmosphere, as Figure~\ref{Abb:tauAtmundDetektorempf} shows.
At \Bra, outside of these narrow regions where it worsens by less than a factor of two (\textit{caveat lector}: a higher `sensitivity' value is worse), the sensitivity per bin is approximately
\begin{equation}
\label{Gl:Empf}
 \FlmbdEmpf(\BraM) \approx 6\times10^{-18}\,\frac{N_\sigma}{5}\sqrt{\frac{1~\mathrm{h}}{\tBeob}}\sqrt{\frac{R}{10^5}}~\textrm{erg\,s}^{-1}\,\textrm{cm}^{-2}\,\textrm{\AA}^{-1},
\end{equation}
where 
\tBeob is the time on source, assuming Poisson statistics,
$R=\lambda/\Delta \lambda$ is the bin spectral bin size expressed as a resolving power,
and $N_\sigma=\SNR$ 
is the detection significance per spectral bin.
Also in the $L$-band, the sensitivity prefactor for \Pfg is %
$\approx3\times10^{-18}$, and in the $M$-band, at \Pfb we have %
$\approx8\times10^{-18}$, briefly dropping the units and other factors, kept the same as in Equation~(\ref{Gl:Empf}).
The peaks of worse sensitivity at \Pfg are also a factor of two higher than the baseline, while at \Pfb they can be orders of magnitude larger.

\subsection{Speckle noise from the primary star and central regions}
 \label{Th:RauschenvonPrimEm}

Both the central star and the inner rim of the PPD emit at 4~\mum \citep{gaidos24} and will be unresolved even with the ELT. Therefore, we need to estimate whether this central emission is a problem for studying accretion at \PDSb at its separation of 115~mas (minimum around 2034) to 120~mas (around first light of the ELT; Appendix~\ref{Th:wosindbundc}). This corresponds to $(5$--$6)\lambda/D$, where the raw contrast\footnote{To about $\pm0.5$~dex, the raw contrast in Figure~10a of \citet{carlomagno20} matches the upper envelope of the Airy pattern, $F/F_0=[2J_1(x\pi)/(x\pi)]^2\leqslant8/(\pi^4x^3)$
(Digital Library of Mathematical Functions, Sect.~10.7(ii); \citealp{dlmf126}) for $x\equiv\lambda/D\gtrsim0.7$, where $J_1(z)$ is the Bessel function of the first kind.}, that is the flux relative to the central peak value, is already $10^{-4}$--$10^{-3}$ \citep{carlomagno20,carlomagno20err}. However, what is relevant is the speckle noise, which is the significantly smaller leftover signal from the central source (star and inner rim) after post-processing.

\begin{figure}
 \centering
 \includegraphics[width=0.47\textwidth]{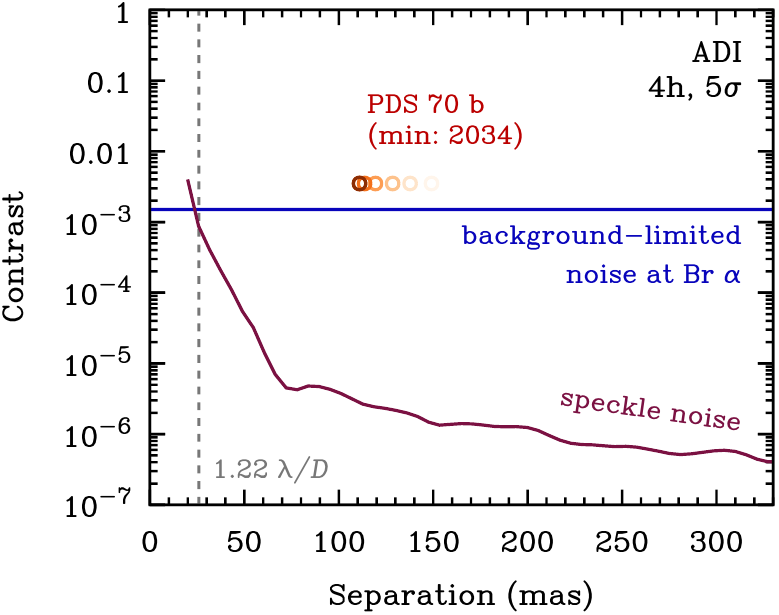}
\caption{%
Contrast curve (speckle noise remaining after post-processing) at $5\sigma$ for the speckle noise only in the CVC mode with the ELT for a 4-h observational sequence post-processed with standard ADI (\textit{purple curve}).
It stops at the `inner working angle' (IWA) of coronagraphic observations, about 20~mas for the $L$-band CVC.
It was computed with \texttt{HEEPS} and is provided courtesy of O.~Absil (priv.\ comm.\ 2026).
To the first order, the curve is independent of the stellar brightness for $L\lesssim8$~mag and applies both to imaging (IMG) and spectroscopy (LMS). Observing without a coronagraph would increase the speckle noise by a factor of five or less (see text), leaving \PDSb (\textit{circles}: separation over epochs, from Fig.~\ref{Abb:wosindbundc}) in the regime where the background noise (\textit{blue curve}) dominates.
}
\label{Abb:ELT-Kontrascht}
\end{figure}

To estimate the speckle noise, we show in Figure~\ref{Abb:ELT-Kontrascht} a 5$\sigma$ contrast curve kindly provided by O.\ Absil (2026, priv.\ comm.).
A contrast curve shows the remaining speckle noise from the central source after some form of post-processing has removed most of the central light. It was computed with \texttt{HEEPS} \citep{carlomagno20,carlomagno20err,delacroix22} and takes into account all possible types of wavefront perturbations, including non-common path aberrations (NCPA) and water vapour turbulence,
with the exception of misaligned M1 segments,
whereas the corresponding curve in Fig.~10a of \citet{carlomagno20} includes only the single-conjugate adaptive optics (SCAO) residuals.
The speckle-noise curve of Figure~\ref{Abb:ELT-Kontrascht} conservatively assumes classical angular differential-imaging (ADI) post-processing (median subtraction; \citealp{marois06a}). The background (shot noise) was estimated in Section~\ref{Th:Detektorempf} and is therefore not included here.
To first order, the speckle-noise curve is independent of stellar magnitude for $L\lesssim8$~mag and applies to both the imaging (IMG) and $L$- and $M$-band spectroscopy (LMS) sub-system; it was computed at 3.8~\mum but it applies to an excellent approximation also to \Bra.%

The curve in Figure~\ref{Abb:ELT-Kontrascht} is shown for the classical vortex coronagraph (CVC) mode, while it would likely be more advantageous to observe \PDS without a coronagraph because it is sufficiently faint.  %
Comparing with \citet{carlomagno20}, the speckle-limited contrast performance (again, this is not the raw stellar light but rather what remains once post-processed by a classical approach) should not degrade by more than a factor of five if the vortex coronagraph is removed.  
We return to this once the contrast has been calculated, that is, the brightness of the residual background.%

To compare the speckle noise and the thermal background, we need to express the latter, currently a physical flux value, also as a contrast. The simplest approach to obtain the brightness of the central source is to use the VLT/NACO $\mathrm{NB}4.05=7.77$~mag measurement \citep{Stolker+20b}. This is appropriate because the PPD within the VLT beam but outside the ELT beam does not contribute at those wavelengths (Section~\ref{Th:zstSchEm}). The NB4.05 brightness corresponds to an average flux of $F_\lambda=2.9\times10^{-15}$~erg\,s$^{-1}$\,cm$^{-2}$\,\AA$^{-1}$ over the NB4.05 filter \citep{lenzen03,rousset03}, of effective resolution $R\approx65$.
As a check, this agrees well with $F_\lambda=(3.5\pm0.1)\times10^{-15}$~erg\,s$^{-1}$\,cm$^{-2}$\,\AA$^{-1}$ (from $L'=7.86$ or $7.91$~mag; \citealp{keppler18,Stolker+20b}, respectively)
and the fit of a stellar photosphere and a second blackbody performed for the exoGRAVITY $K$-band library \citep{kammerer25},
using 2MASS/$JHK_{\mathrm{s}}$ and WISE/$W1W2$ photometry only.
It also agrees with the star plus inner-rim modelling of \citet{gaidos24} and \citet{jang24}.
\citet{keppler18} find a different fit, fainter past $\lambda\approx2$~\mum (fainter by about a factor of two at \Bra), %
but they were fitting only the NIR data. 
As we verified, high-resolution model spectra of the stellar photosphere near the relevant $\Teff\approx4000$--4500~K  %
do not exhibit emission lines, while the inner rim seems to contribute at the 50\%\ level \citep{gaidos24,jang24}. Therefore, it is likely that the NB4.05 brightness is representative for a narrow region around the \Bra line. There could be line emission from the inner rim but this is a small caveat.

One more serious consideration is that the estimate up to now considers only the emission from the photosphere and the inner rim but no potential line emission. However, \PDSA has been shown to be a weak accretor \citep{Thanathibodee+2020,skinner22,campbell23}, with $\MPkt\sim10^{-10}~\MSonne\,\mbox{yr}^{-1}$.
This implies \citep{Thanathibodee+2020}
\begin{equation}
\LAkk=\frac{G\MStern\MPkt}{\RStern}\times\left(1-\frac{1}{5}\right)\sim0.0015~\LSonne,
\end{equation}
for which the extrapolated \citet{Komarova+Fischer2020} relationship suggests $\LBra\approx10^{-5}~\LSonne$
(apparent flux 
$\FBra\approx2\times10^{-14}$~erg\,s$^{-1}$\,cm$^{-2}$),
while that of \citet{testi25} gives $\LBra\approx10^{-6}~\LSonne$.
The free-fall velocity is 
$\vFf\approx415$~\kms.
Assuming that the FWHM is half of this ($\relDvBreite\approx0.5$)  %
  implies a peak flux
$F_\lambda\approx4\times10^{-20}$~erg\,s$^{-1}$\,cm$^{-2}$\,\AA$^{-1}$ with the \citet{testi25} estimate.
Therefore, this would be completely negligible.

Actually, \citet{donati24} report the \Brg and \Pab lines to be narrow, within the stellar $v\sin i = 19.5\pm0.5$~\kms. They also find that the luminosities of these lines, combined with the scaling from \citet{alcal17}, would imply $\MPkt=10^{-11}~\MPktEinhS$, which is even less than others find (e.g., \citealt{campbell23}).  %
Thus the line emission from the central regions can be disregarded.

We are now in a position to compare the speckle noise to the background-limited contrast (the shot noise). Figure~\ref{Abb:ELT-Kontrascht} shows that outside of roughly $2\lambda/D$, the thermal background dominates the speckle noise by a factor of 100--1000.
Therefore, updates to the observation strategy---for example not using a coronagraph\footnote{%
As O.~Absil (priv.\ comm.\ 2026) points out,
non-coronagraphic imaging performs better in the background-limited regime.
In fact, it will probably not be %
possible
to operate the CVC on such a faint target with the LMS, because controlling the pointing onto the vortex phase mask relies on the 10\%\ of the flux that will be diverted to the IMG during LMS/IMG parallel observations. This means that there is effectively a star of $L > 10$~mag on the IMG side when observing \PDS with the LMS, and simulations suggest that pointing control will not work anymore on such a faint target.%
}, which should raise the speckle noise level by at most a factor of five---, will not modify the conclusion that only the residual background-limited noise needs to be taken into account. This applies also securely to the \WISPbc planets \citep{vancapelleveen25a,close25b,lawlor26}, at 320~and 100~mas, respectively, or J16120\,b some 140~mas away from its primary.

\subsection{Resulting accretion-line signal with noise} %
 \label{Th:LiniemitRauschen}

\begin{figure*}[!tp]
 \centering
\def\HFh{0.187}  %
 \includegraphics[width={0.96}\textwidth]{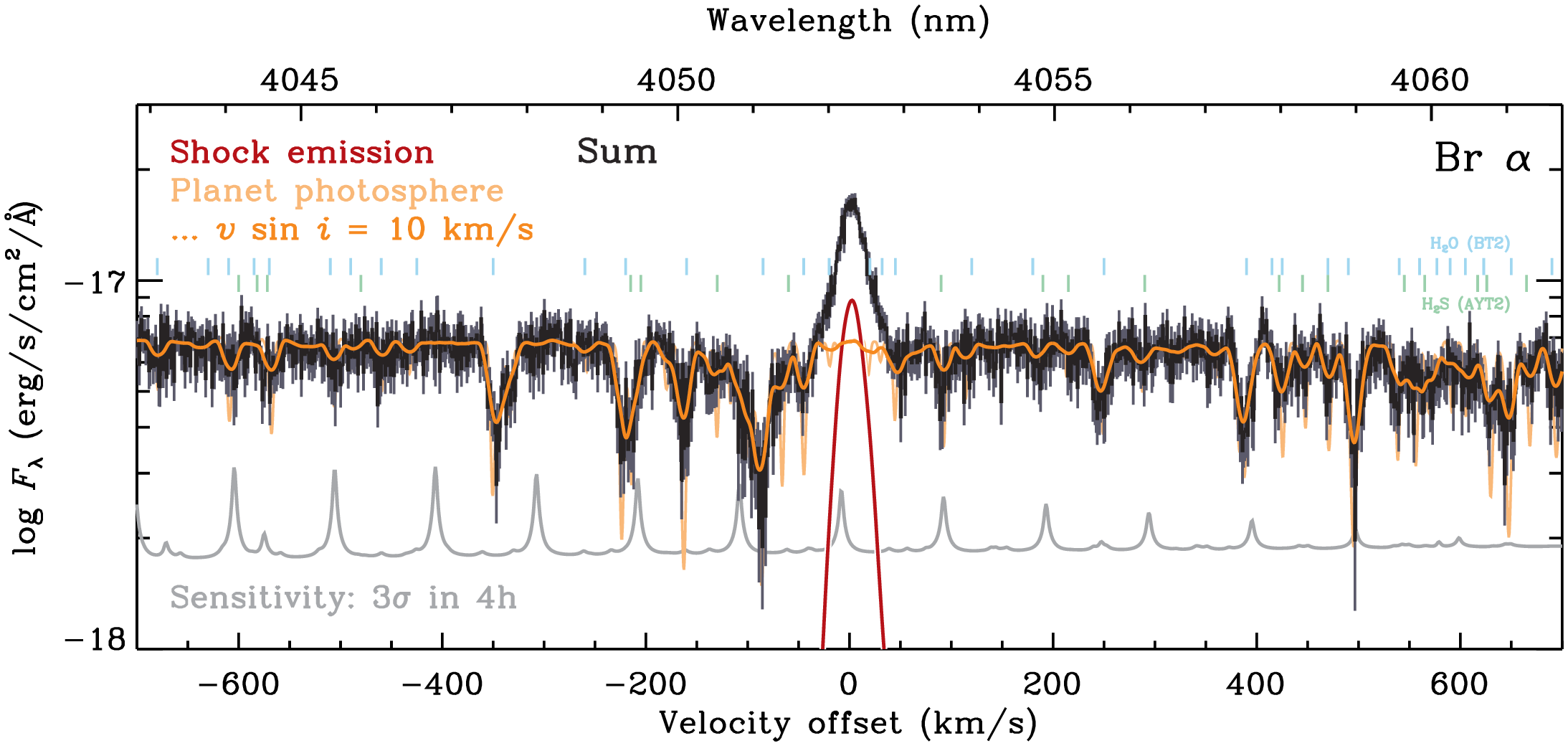}\\
~~~~\includegraphics[{height=\HFh\textheight}]{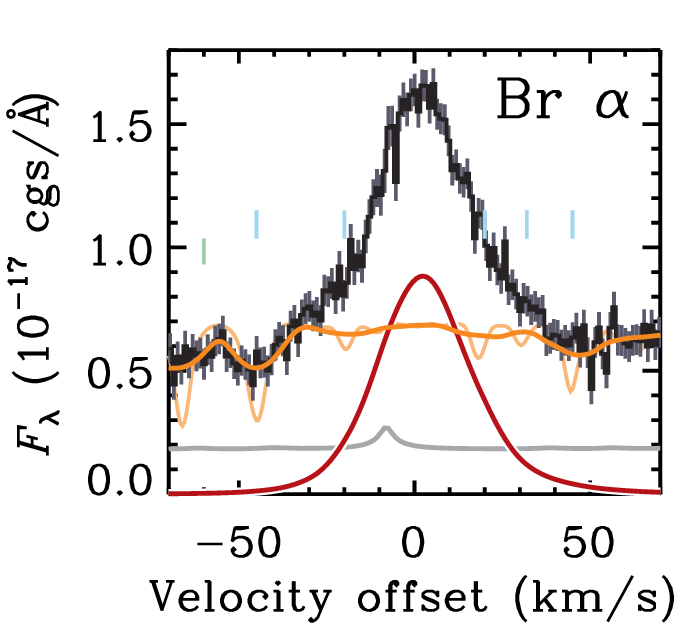}
 \includegraphics[{height=\HFh\textheight}]{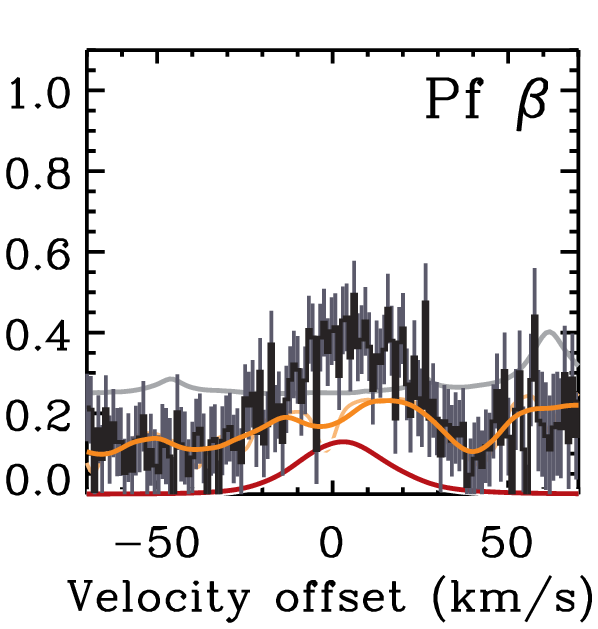}
 \includegraphics[{height=\HFh\textheight}]{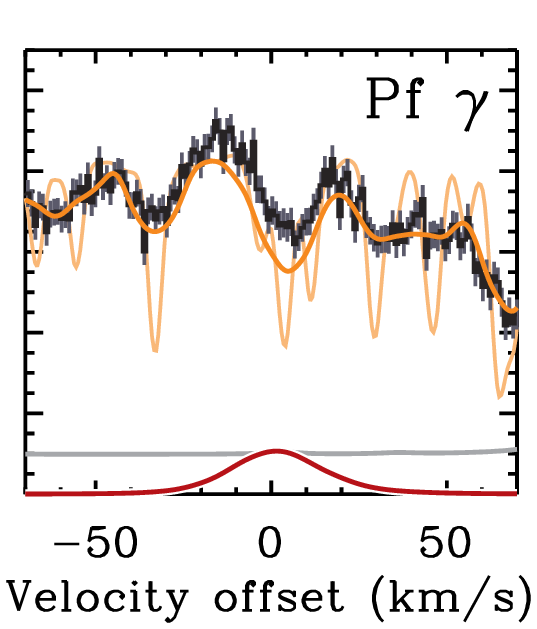}
 \includegraphics[{height=\HFh\textheight}]{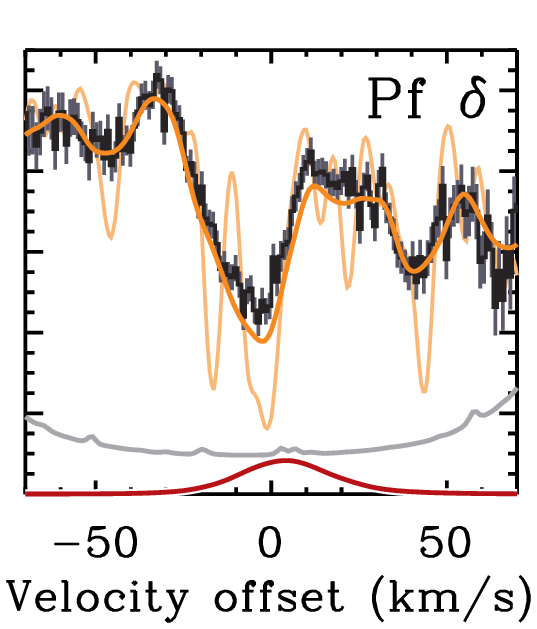}
\caption{%
Predicted spectrum of \PDSb at \Bra  %
and noise for a 4-h integration (\textit{black line} with 1$\sigma$ \textit{grey} errorbars) convolved to the resolution of METIS and sampled on its wavelength grid.
We combine the photosphere (CIFIST with $\Teff=1400$~K, $\log g=3.5$ broadened to $v\sin i=10$~\kms; \textit{pale and dark orange}, respectively)
and the shock emission (as in Figure~\ref{Abb:Prof}; \textit{dark red}).
The \textit{grey line} shows the 4-h, 3$\sigma$ METIS sensitivity curve.
\textit{Top panel}: One-third, centred on \Bra, of the range available at once.
Ticks indicate H$_2$O (\textit{pale blue}; \citealp{barber06}: `BT2') and H$_2$S (\textit{pale green}; \citealp{azzam16}: `AYT2') resonances.
\textit{Bottom panels}: Vignettes around accretion lines (see labels).
The $y$ range of the \Pfb ($\lambda\,4654$) panel applies 
also to \Pfg and \Pfd ($\lambda\lambda\,3741,\,3297$).
}
\label{Abb:Komboalles}
\end{figure*}

We are now ready to combine the predictions for the photosphere and the accretion shock to these sensitivity curves, yielding an estimate of the post-processed spectrum.
In Figure~\ref{Abb:Komboalles}, we re-interpolate the sum of the rotationally broadened photosphere and the line profile onto the detector wavelength grid (with $\Delta\lambda=0.144$~and 0.166~\AA\ at \Bra and \Pfb respectively, for example).
We chose an exposure time of 4~h (half a night),
and for every wavelength bin, we drew randomly from a Gaussian with a unity standard deviation and multiplied by the 1$\sigma$, 4-h $\FlmbdEmpf$ value at that wavelength.
We add this positive or negative noise realisation to the modelled signal
and plot as the symmetric errorbars the 1$\sigma$, 4-h sensitivity value.

The top panel of Figure~\ref{Abb:Komboalles} gives the spectral context of the accretion line. The displayed range of 700~\kms on either side of the \Bra line represents one third of the range which can be obtained in one exposure, $\Delta\lambda\approx567$~\AA.
Already without the additional flux from the accretion line, \PDSb should be detectable by its continuum around \Bra at a per-bin
$\SNR\approx12$ in 4~h.  %
The shock excess emission is clearly visible and contrasts in shape and amplitude with the photospheric features. Its \SNR is discussed three paragraphs down.

The bottom panels show zoom-ins around \Bra, \Pfb, \Pfg, and \Pfd with a linear flux scale.
At the line centres, the sensitivity of METIS turns out to be at a local best value or nearly so (compare to Figure~\ref{Abb:Linien}).
In Appendix~\ref{Th:tauAtm}, we compare the sensitivity curves to the absorption in the Earth's atmosphere.  %
As Figure~\ref{Abb:tauAtmundDetektorempf} shows, if around \Pfd and \Pfe the Earth's atmospheric features had been $\Delta v\approx100$~\kms blue- or redwards of their actual position, these lines would have been effectively unobservable from the ground.
These higher-order transitions are too faint to be relevant for \PDSb but they could be detectable at objects that are more massive or accreting more vigorously.

We have not taken the systemic radial velocity (RV) velocity of \PDS nor the proper RV of \PDSb into account but they are respectively only $\vSyst\approx3.1\pm1.4$~\kms \citep{gDR2}
and $|\mbox{RV}_{\mathrm{proper}}|\lesssim2$~\kms (Appendix~\ref{Th:wosindbundc}).  %
Thus both radial velocities are negligible here.
There remains only the apparent radial velocity \RVwegenErde due to the orbital motion of the Earth;
for \PDS, at an ecliptic latitude of $-26.6\degr$,
this is at most
\begin{equation}
 |\RVwegenErde|=\cos(-26.6\degr)\sqrt{\frac{G\MSonne}{1~\mathrm{au}}}=26.6~\kms.
\end{equation}
This will never move the hydrogen lines into wavelengths of significantly different sensitivity.

\begin{figure} %
 \centering
 \includegraphics[width=0.45\textwidth]{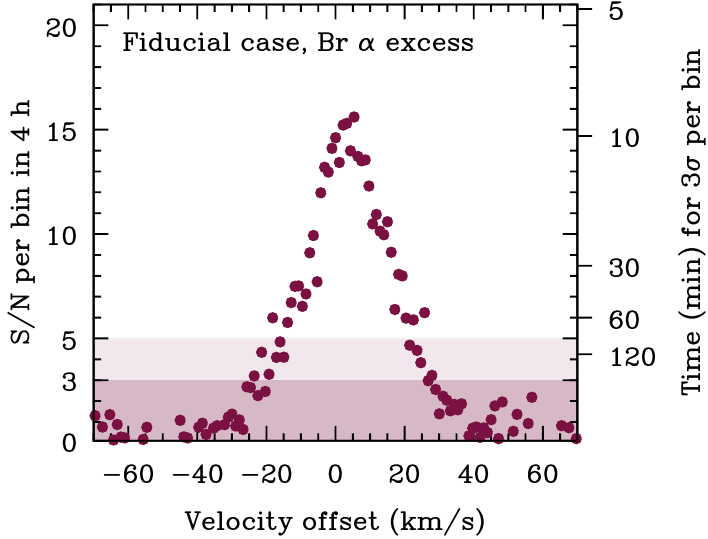}%
\caption{%
Signal-to-noise ratio per spectral bin for the excess shock emission in the fiducial case, from Figure~\ref{Abb:Komboalles}.
The right axis shows the time needed to reach $3\sigma$ per bin (scaled with Eq.~(\ref{Gl:Empf})).
Only the Gaussian core of the line (within $\Delta v \approx \pm25~\kms$; Figure~\ref{Abb:Prof}) is detectable at $\SNR>5$.%
}
\label{Abb:SRV}
\end{figure}

In Figure~\ref{Abb:SRV}, we show the signal-to-noise ratio \SNR per bin of the shock excess, and equivalently the time needed for a per-bin $3\sigma$ detection. With the fiducial parameters,
the peak flux excess (Fig.~\ref{Abb:Prof}) compared to the sensitivity (Equation~(\ref{Gl:Empf})) yields $\SNR=15$ for a 4-hour integration,
and conversely,
only around 10~min are required %
to reach a peak per-bin $\SNR=3$. %
These numbers are somewhat idealised because they assure perfect photosphere subtraction, but they suggest that \PDSb will be easy to observe in its shock excess.
The non-Gaussian wings, discussed in Figure~\ref{Abb:Prof}, have at best $\SNR\approx5$ and closer to~2. Therefore, the asymmetry of the line probably cannot be seen. Nevertheless, the Gaussian core, and indeed its shape, can be detected reliably. In particular, METIS can reveal narrow shock-dominated emission lines at low preshock density (Fig.~\ref{Abb:DvAo18}), but also more complex profiles as magnetospheric accretion is capable of generating.

\Pfb$\lambda46538$, not to be confused with \Pab$\lambda12822$, is intrinsically the brightest of the other hydrogen lines accessible to METIS. However, its peak flux---and also its integrated flux since the shape is essentially identical (Fig.~\ref{Abb:Varpar})---is 
$\approx7$~times fainter than that of \Bra, while the photosphere at \Pfb, despite the CO absorption beginning at 4.3~\mum, is only $\approx4$ times fainter. Thus the contrast is less favourable, making it more difficult to subtract the photosphere to isolate the line.
This conclusion holds more generally as clouds (e.g., Fig.~11 of \citealt{morley24}) or CPD emission will increase the photospheric emission at \Pfb relative to \Bra. The line ratio \Pfb/\Bra is roughly constant for low to moderate $n_0$ \citep{aoyama18}.  %
The \SNR even of the photosphere (bottom row of Figure~\ref{Abb:Komboalles}) is even lower because of the slightly reduced sensitivity at \Pfb (see Figure~\ref{Abb:DetektorempfpaarL} or after Equation~\ref{Gl:Empf}).

At \Pfg and \Pfd, which like \Bra are also in the $L$-band, the continuum is even more easily detectable than at \Bra, with a per-bin $\SNR\approx20$ in 1~h, thanks to a somewhat stronger signal and a sensitivity higher by a factor of two (see Figure~\ref{Abb:DetektorempfpaarL}). There is a wider range of \SNR at \Pfd than at \Pfg. Around both lines, even without rotational broadening of the photospheric features, the \SNR does not drop below $\approx5$.
The per-bin shock excess, however, only has $\SNR\approx3$ at its peak, making its detection very questionable.  %
These predictions for the photosphere need to be taken with a grain of sodium chloride. As detailed in Appendix~\ref{Th:Linienlisteneffekt}, most troughs at \Bra are due to H$_2$O or otherwise hydrogen sulphide, H$_2$S, which is indicated in Figure~\ref{Abb:Komboalles}.
Around \Pfg, a brief, non-exhaustive look suggests that CH$_4$ might cause some of the blue-side troughs.
For the water, the CIFIST models used the BT2 line list \citep{barber06}. We compare this in Figure~\ref{Abb:kappaundFlLinienlisten} to other, more recent work, namely POKAZATEL \citep{polyansky18} and the HITRAN~2020 release \citep{gordon22}.
The troughs are very different in position, depth, and even width.
\citet{polyansky18} note that the older line list of \citet{partridge97} is more accurate than BT2 past $\lambda\approx1$~\mum, and the latest HITRAN release \citep{tennyson24} recommends POKAZATEL over the others.
Thus, continued modelling and experimental efforts could change the quantitative and to some extent qualitative appearance of the continuum.

\subsection{Detectability over a range of masses and accretion rates}
 \label{Th:alleszusammen}

\begin{figure}[t!]  %
 \centering
 \includegraphics[width=0.45\textwidth]{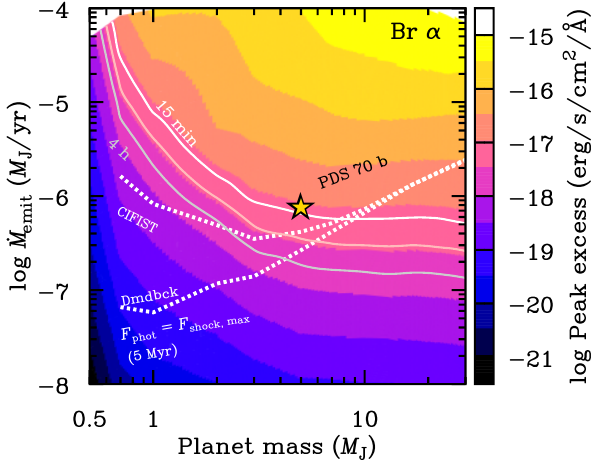}
\caption{%
Peak observed excess flux at \Bra for a range of planet masses and accretion rates \MPktEmiss (defined in Section~\ref{Th:MPkt}), for $d=113.4$~pc. The radius is fixed at $\RP=2~\RJ$, and the other parameters (\thzpSch, $i$, \fzent) are as in Table~\ref{Tab:Par}. We highlight the fiducial case for \PDSb{} (\textit{star symbol}).
The contours indicate integration times for a 3$\sigma$ detection of the shock excess (i.e., continuum-subtracted flux) with METIS in 15~min, 1~h, and 4~h  %
from Equation~(\ref{Gl:Empf}) (\textit{solid lines, top to bottom}).
The accretion rates for which the photospheric flux \FlmbdPhot equals the peak shock excess \FlmbdSchock are shown for the CIFIST and the cloud-free Diamondback models (\textit{dotted lines}; see text for details).
The results for \WISPb should be   %
similar (see begin of Section~\ref{Th:Disk}).
}
\label{Abb:Haupt}
\end{figure}

Figure~\ref{Abb:Haupt} presents a wider view of the parameter space. 
We report the peak flux excess at \Bra for masses from $\MP=0.5$~to 30~$\MJ$ and accretion rates (as defined in Section~\ref{Th:MPkt}) $\MPktEmiss\approx10^{-8}$~to $10^{-4}~\MPktEinhJ$,  %
 keeping the radius fixed at $\RP=2~\RJ$ for simplicity and the distance to $d=113.4$~pc for convenience.
The peak shock excess reaches $F_\lambda\approx3\times10^{-16}$~erg\,s$^{-1}$\,cm$^{-2}$\,\AA$^{-1}$ for $\MP\gtrsim5~\MJ$ and the highest accretion rates.
For masses above $\MP\approx3~\MJ$ and up to moderate accretion rates ($\MPktEmiss\lesssim3\times10^{-6}~\MPktEinhJ$), the peak flux is relatively independent of mass. At higher accretion rates, or at lower masses $\MP\lesssim3~\MJ$ and for all \MPktEmiss, the peak flux quickly decreases with decreasing mass.
We checked (not shown) that setting the radius to a constant $\RP=1.5~\RJ$ does not change the results significantly over the whole grid, as Figure~\ref{Abb:Varpar} suggested.

Contours in Figure~\ref{Abb:Haupt} indicate the integration time needed to reach a 3$\sigma$ detection of the peak excess, using the reference value and scaling of Equation~(\ref{Gl:Empf}).
Towards lower masses $\MP\approx1$--$2~\MJ$, the required time increases quickly into the several-hours regime, and below %
$\MP\approx1~\MJ$, the accretion rate and other system parameters %
would need to be extremely favourable to allow for a clear detection of the peak.
However, higher masses $\MP\gtrsim3~\MJ$ are well detectable even down to $\MPktEmiss\approx3\times10^{-7}~\MPktEinhJ$.

For context, Figure~\ref{Abb:Haupt} also shows where the photospheric contribution at \Bra is similar to the peak shock flux (dotted curves). Specifically, we combined the COND03 \citep{baraffe03} $t=5$~Myr `corrected' isochrone $\Teff'(M)$ (see next sentence) 
with the CIFIST flux at \Bra $\FBra(\Teff')$, averaged over $\Delta v=2200$~\kms around the line centre for robustness, and, for each mass, identified the \MPktEmiss value for which this photospheric flux \FlmbdPhot is equal to %
the peak of the shock excess flux density \FlmbdSchock.
The isochrone correction we applied was to scale \Teff to have the same bolometric luminosity %
with our choice $\RP=2~\RJ$, which turns out always to be larger than the radius in the isochrones ($R\approx1.5~\RJ$). Thus we took $\Teff'=\Teff\times(R/[2~\RJ])^{1/2}$, where $R$ is the radius in the isochrone tables,
and did not apply the flux correction factor of 1.086 introduced in Section~\ref{Th:PlanEm}.
We repeated this with the Sonora Diamondback isochrones\footnote{See \url{https://zenodo.org/records/12735103}.} \citep{morley24} for solar-metallicity evolution tracks, also using the `corrected' $\Teff'(M)$ values.
The use of `cloud-free' or `hybrid' evolutionary tracks made here essentially no difference.

Both the COND03 and the Diamondback cooling tracks assume `hot starts', which are a limiting case and are somewhat arbitrary (that is, the post-formation luminosity as a function of mass is not derived from for instance from a formation model; e.g., \citealt{fort05,marl07,sb12,mc14,mordasini17,m18hip}) but are acceptable for this comparison. The isochrones also predict different effective temperature for $\MP=5~\MJ$ ($\Teff\approx1380$~and 1260~K, respectively, or $\Teff'\approx1180$~and 1230~K) than our fiducial choice $\Teff=1400$~K.
This is also likely subdominant compared to the systematic uncertainties.
The CIFIST and Diamondback models essentially converge for $\MP\gtrsim5~\MJ$ but differ towards $\MP\approx1~\MJ$ by about 1~dex in flux or 1.3~dex in \MPktEmiss. This should be kept in mind as a kind of theoretical errorbar, especially if applying this to other systems.

For a given mass, the line $\FlmbdPhot=\FlmbdSchock$ indicates what minimum accretion rate is needed for the shock at its maximum to be as bright as the photosphere. For that and somewhat lower accretion rates, it should be possible to extract the shock signal, at least as long as the photosphere does not exhibit spectral features of order-unity amplitude, as argued in the discussion of the RMS in Section~\ref{Th:PlanEm} and Appendix~\ref{Th:Phothelligkeitmehr}. Unfortunately, at \Pfg and \Pfd, for example, the photospheric features are strong in that relative sense, but this is likely not an issue at \Bra (see Figure~\ref{Abb:Komboalles}).
The other significance of the $\FlmbdPhot=\FlmbdSchock$ line is relative to the flux (colour) scale, especially as delimited by the exposure-time contours. Figure~\ref{Abb:Haupt} shows that for $\MP\gtrsim2$--$4~\MJ$ at 5~Myr, depending on the cooling and atmospheric models, even a non-accreting planet should be detectable in 4~h at a per-bin $\SNR=3$ (at the native resolution of METIS) if located $d=113.4$~pc away.
\section{Discussion}
 \label{Th:Disk}

We took \PDSb as a fiducial case but our estimates apply similarly to \WISPb. Its mass is better constrained with $\MP=5\pm1~\MJ$, and the system age of 3.8--7.5~Myr \citep{vancapelleveen25a} implies a radius $\RP\approx1.5$--$1.7~\RJ$ and $\Teff\approx1300$--1500~K as estimated quickly from the clear to cloudy evolutionary tracks of \citet{morley24}.
Also, at least at the epoch of its discovery, the \Ha luminosity $\LHa\approx7\times10^{-7}~\LSonne$ of \WISPb, not correcting for any possible extinction, was remarkably close to that of \PDSb at different epochs (Figure~\ref{Abb:LLinie}). Therefore, our results apply almost directly also to \WISPb, with the small distance correction which decreases the fluxes by 40\,\%\ and thus increases the exposure-time estimates by 18\,\%\ at fixed \SNR, which is negligible. The separation of 320~mas also puts \WISPb in the background-dominated regime.

Before concluding, we discuss our neglect of magnetic fields (Section~\ref{Th:MagAkk}) and of extinction (Section~\ref{Th:ohneExtinkt}).
We relate our work to that of \citet{takasao21}, who were dealing with the \Ha line (Section~\ref{Th:vglTakasao21}) and to that of \citet{dom25}, concerned with the emission at radio frequencies (Section~\ref{Th:vglDom25}).
In Section~\ref{Th:MPktausLLinie}, we calibrate the scaling between the \Bra luminosity and the accretion rate. Finally, we offer a brief comparison of METIS to CRIRES(+) (Section~\ref{Th:vglCRIRES}) and a glimpse of what MICADO could deliver (Section~\ref{Th:MICADO}).

\subsection{Neglect of magnetic fields}
 \label{Th:MagAkk}

Magnetospheric accretion (e.g., \citealp{koenigl91,hartmann16}) seems logically required even at planetary masses to explain hydrogen-line emission from (effectively) isolated accretors. Their line shapes seem to confirm or at least not contradict magnetospheric accretion as the dominant process, with examples including GQ~Lup~b \citep{demars23}, TWA~27~B/2M1207~b \citep{m23alois,aoyama24twa,patapis25}, Delorme~1~(AB)~b (\citealp{betti22b,betti22c,ringqvist23,demars26}), 2M1115 \citep{viswanath24},
or Cha\,1107-7626 \citep{almendrosabad25}.
Indeed, without truncation of the CPD and nearly-free-fall acceleration of the gas along the accretion columns, it is difficult to imagine how lines could be formed since a (hot) boundary layer connecting CPD and planet (e.g., \citealp{lyndenbell74,kenyon87,kley96,hertfelder17,dong21}) would possibly only lead to continuum emission (depending on the heating of the boundary layer; see also \citealp{mendigut20} in the context of Herbig Ae/Be stars).

However, planets in PPDs can be accreting through magnetospheric accretion from their CPD but also through pure hydrodynamic accretion from the parental PPD. The latter will be the sole mechanism associated with line emission if the magnetic field is weak\footnote{Searches for magnetic fields around relatively old ($\sim0.1$--1~Gyr) planets through their radio emission have returned in part tight upper limits \citep[e.g.,][]{narang24}, but young and therefore bright planets could easily have much stronger, kG-scale magnetic fields \citep{christensen09,zhu15,katarzy16}, similar to low-mass brown dwarfs \citep[e.g.,][]{kao18}.} and thus not able to truncate the CPD, and/or if the radial accretion flow through the CPD is large. If on the contrary there are accretion columns, covering on the order of a percent of the planetary surface, the preshock density at the footpoints on the surface of the planet will simply be higher but the emission mechanism otherwise the same. Such narrow columns might be found on \PDSb (e.g., \citealp{hashimoto20,hasegawa21}).
Thus the present work will apply in those cases too, for appropriate parameter values, within the approximation that we do not include emission by the accretion columns (the `flow model', as in \citealt{kwanfischer11}; see \citealt{hashimoto25}) nor complex interactions of the radiation from the hot spots and/or the accretion column with those columns (e.g., \citealp{muzerolle01,thanathibodee24}).
If the latter do contribute, they might do so on a different timescale, as \citet{demars26} have found for \Dlrmb.

In any case, given the number of different complex mechanisms and free parameters at play, assuming a negligible contribution from magnetospheric processes or possible chromospheric activity (e.g., \citealp{white03,mohanty05a,manara13})
to the line formation is a useful limiting case. It is also likely directly relevant for several planetary-mass objects, at least at epochs of more vigorous accretion. Also, in their analysis of several 20--100~$\MJ$ accretors, \citet{hashimoto25} find that a significant fraction (15/70) could be categorised as shock-dominated as opposed to flow-dominated, at least based on their line-flux ratios. There was no obvious correlation with object parameters but even some of the high-mass objects were well explained by the shock model, which a deeper analysis supports \citep{aoyama26}.  %
Thus the shock model seems even more widely applicable than initially envisioned.

An important implication of the line shapes we obtained, reflecting the input ones from the \citet{aoyama18} grid, is that shock emission leads only to profiles that are smooth, whether Gaussian in the core or wider, and nearly symmetric except at low flux levels. \citet{maea21} needed very high local accretion rates for the accreting gas to leave a significant velocity-gradient-smeared extinction imprint in the line profile. For the framework considered here \citep{m24expeditus}, with accretion from Hill-sphere scales, such a mass flux concentration appears unlikely. On the other hand, magnetospheric accretion can lead to similarly smooth but also to more complex profiles over its parameter space \citep{demars23,thanathibodee24,demars26}. Therefore, a structured, complex line profile would point away from shocks as being the dominant source of emission. In general, the spectral resolution of METIS are sufficient to reveal this (Figs.~\ref{Abb:SRV} and~\ref{Abb:Haupt}).%

\subsection{Neglect of extinction}
 \label{Th:ohneExtinkt}

The visual ISM extinction to \PDS is $\AV\approx0.05$~mag \citep{mueller18} but in general even larger amounts of extinction can be relatively reliably corrected. This is thanks to the large number of studies dealing with extinction, especially in the diffuse ISM or in denser regions such as protostellar cores (e.g., \citealp{ossenkopfhenning94,lutz96,chiar06,wang19,gordon23}).  %

What can be more problematic is in-system extinction. However, we focus on gap-opening gas giants and study near-infrared wavelengths. Thus, the PPD might not attenuate the light emitted by the forming planet very much.
Depending on the atmospheric model they use, \citet{wang21vlti} find a range of in-system extinctions. However, their best-fitting models yield $A_K\approx0.2$--$0.5$~mag at most, which implies a small \Bra extinction $\ABra\approx0.1$--0.2~mag when using the \citet{chiar06} scaling.
This is a different situation from the results reported by \citet{cugno25b} for AS\,209. There, a significant amount of extinction is found in the gap, but it is shallow, while \PDS has a deep gap \citep{keppler18}.

\citet{maea21} tried to estimate how extinguishing the infalling gas and dust are at \Ha. The infall geometries they considered, if anything, overestimated the local density. Therefore, their results should be rather conservative. Despite this, they find that for a wide range of masses and up to high accretion rates, the gas should mostly not lead to significant extinction at \Ha.
If we translate these results to \Bra, they will hold even more securely, since $\ABra/\AV=0.07$ for an extinction parameter $\RV=3.1$ \citep{cardelli89,chiar06}.
Because of their proximity in wavelength, the other $L$- and $M$-band lines will be affected by a similar amount of extinction, with only 7\,\%\ more extinction at \Pfg or 20\,\%\ more at \Pfd than at \Bra \citep{chiar06}.
From their hydrodynamical simulations and adopted dust properties, \citet{alarc24} obtain a \Bra extinction at \PDSb around $\ABra\approx2$~mag for their specific accretion rate, choice of disc viscosity, and so on. We will consider this a moderate value given the large uncertainty in key parameters.  %
For now, we tentatively conclude that neglecting extinction by the infalling gas and dust is an acceptable working approximation for \PDSb.

Nevertheless, since both the photospheric emission of and the line emission at the planet surface dominate over the contributions from the CPD, any non-negligible extinction would likely affect both components in the same way. Therefore, only the total flux would be lowered but the relative contributions not changed, making it easy to apply our framework also to other systems.

\subsection[Relation to the work of Takasao et al. (2021)]{Relation to the work of \citet{takasao21}}
 \label{Th:vglTakasao21}

\citet{takasao21} present predictions of line emission using the \citet{aoyama18} line-emission models, but their purpose and approach differ from ours in several ways. The main difference is in the larger-scale hydrodynamical simulations which their calculations use. Whereas in our case, based on the results from radiation-hydrodynamical simulations, the planet surface and the CPD surface emit, in theirs almost all the line emission comes from where a thin supersonic sub-surface flow hits the planet surface. \citet{m22Schock} compare these scenarios. Which is closer to reality, depends on the physics of the postshock cooling, especially on the ability of the gas to cool, which is set by the density and chemical reactions involving the CO, H$_2$, and OH abundances, involving in turn questions about energy-level populations (see Appendix~B of \citealt{takasao21}).  %

Worth mentioning but not a fundamental difference, \citet{takasao21} look only at the \Ha line and studied only one case ($\MP=10~\MJ$). They perform hydrodynamical simulations to obtain the gas flow, whereas we adopt a semi-analytical approach to make covering a large parameter space tractable.
Finally, to compute the high-resolution line profiles, we take the explicit viewing geometry into account, including the inclination, whereas they seem to take a simpler weighted sum.
Thus our works appear complementary to some extent.

\subsection[Relation to the work of Dom\'inguez-Jamett al. (2025)]{Relation to the work of \citet{dom25}}
 \label{Th:vglDom25}
 
We comment on our modelling in the light of the radio emission detected at \PDSc. Its properties are similar to those of \PDSb \citep{hammond25,trevascus25}, which is not detected at all at radio frequencies\footnote{%
This conundrum could be explained by a difference in how the free--free and the hydrogen-line emission scale with planet mass.}.
To explain the non-detection of \PDSc in Band~9 of ALMA %
despite its detection at Bands~3 (marginal), 4, and~7, \citet{dom25} find that the most likely scenario is for the radio spectrum to be dominated by free--free emission from an accretion shock at the surface of the CPD. Within the assumption of a uniform mass influx, a high accretion rate was needed, crudely $\MPkt\sim10^{-4}$--$10^{-1}~\MPktEinhJ$.
This is several orders of magnitude above the estimate from the maximum observed \Ha luminosity,
$\LHa\approx2\times10^{-7}~\LSonne$ \citep{close25a}, which the \citet{AMIM21L} scaling translates to
$\MPkt\approx10^{-6}~\MPktEinhJ$ when conservatively assuming $\AV=3$~mag of extinction and a planet mass $\MP=2~\MJ$ and radius $\RP=2~\RJ$ on the low and high side, respectively. Without de-reddening and with a higher $\MP=10~\MJ$, the corresponding line-emitting accretion rate is reduced to $\MPkt\sim10^{-8}~\MPktEinhJ$.  %
Therefore, the scenario used in \citet{dom25} is not consistent with the \Ha data, or conversely, our set-up in the present work agrees with the Band~9 non-detection but would lead to too little flux at Bands~3, 4, and~7.

To resolve the tension, it could be sufficient to improve the free--free emission model. One aspect would be to have the gas be already dissociated as it reaches the CPD.
This lowers the critical preshock velocity needed to have free--free and line emission, %
because the shock energy is no longer consumed by dissociation. The required mass accretion rate would still be
$\sim10^{-6}~\MPktEinhJ$, which remains higher than, or at least close to, the estimate from \Ha. Another issue is the source of the free electrons. Given the photospheric temperature of \PDSc ($\Teff\approx1000$~K), thermal ionisation is at best partial some distance away from the planet, and some non-thermal processes would need to be invoked to provide any ionisation. Therefore, the free--free emission model would need to be improved to have a physically consistent picture. Finally, we point out the dust-ring model put forth by \citet{shibaike26} as an alternative or explanation of the ALMA observations. A flux contribution by an optically thick ring could alleviate the tension.

\subsection{Accretion rate from line luminosity}
 \label{Th:MPktausLLinie}

Section~\ref{Th:Parvar} showed that the line shape depends very little on the physical parameters if it comes only from a shock. The converse is that a precise measurement of the profile will yield only loose constraints even on the mass over radius $\MP/\RP$, on which the velocity depends the most \citep{ab22}. Figure~\ref{Abb:Varpar} did not explore the whole parameter space systematically but it suggests that the other parameters likely have a small effect at most.

\begin{figure} %
 \centering
\includegraphics[width=0.45\textwidth]{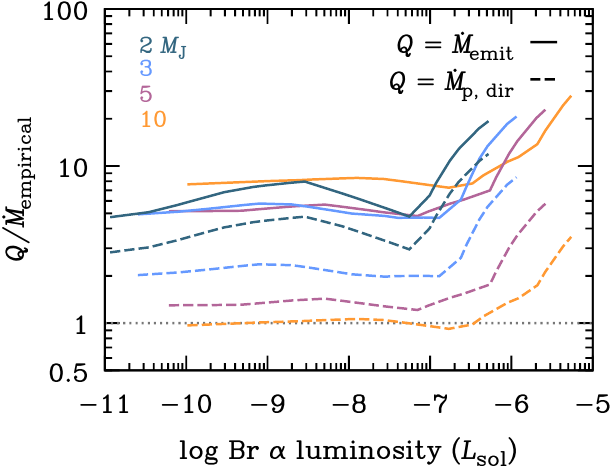}%
\caption{%
Correction factor from the traditional \MPktempir from \LAkk (Eq.~(\ref{Gl:MPktempir})) to the accretion rate $Q$ for different definitions: 
$Q=\MPktEmiss$ (Fig.~\ref{Abb:MPktEmiss}; \textit{solid lines})
and
$Q=\MPktPdir$ (Eq.~(\ref{Gl:MPktkumul}); %
\textit{dashed})
for $\MP=2$--10~$\MJ$ (teal to orange) as a function of \LBra.
The mass growth rate \MPktvekst is likely greater by a large factor.
A dotted line highlights $y=1$. The parameters are fixed as in Figure~\ref{Abb:Haupt}.
The known accretors \PDSbc and \WISPb are predicted to have $\LBra\sim10^{-8}~\LSonne$ (Fig.~\ref{Abb:LLinie}).
}%
\label{Abb:MPktausLLinie}
\end{figure}

However, the integrated line flux can inform about the accretion rate.
Figure~\ref{Abb:MPktausLLinie} in turn compares the line-emitting mass flux
 \MPktEmiss (Fig.~\ref{Abb:MPktEmiss})
 and
 \MPktPdir, which is the accretion rate that reaches the planet surface directly (Eq.~(\ref{Gl:MPktkumul}); \citealp{m24expeditus}),
to `the accretion rate' \MPktempir as derived empirically from the line luminosity:\footnote{%
For stars, most accretion energy is emitted in the continuum, with much less energy in the lines, whereas for planets it is the other way around (see e.g.\ \citealt{hashimoto25,aoyama26} and references therein).} %
\begin{equation}
 \label{Gl:MPktempir}
 \MPktempir \equiv \frac{\LAkk(\LLinie)}{G\MP/\RP},
\end{equation}
here for \Bra and using the \citet{AMIM21L} \LAkk--\LLinie scaling, which is intended to be appropriate at these masses.
In the nominal case, $\LBra=1.8\times10^{-8}~\LSonne$ (Fig.~\ref{Abb:MPktEmiss}), yielding $\LAkk=1.1\times10^{-4}~\LSonne$, which translates into $\MPktempir = 9\times10^{-8}~\MPktEinhJ$.
Using instead the extrapolated relationships of \citet{Komarova+Fischer2020} or \citet{testi25} would yield an \MPktempir smaller by a factor up to 10$^6$ or 100, respectively.
We used the grid of Section~\ref{Th:alleszusammen}, in which only the mass and mass infall rate into the Hill sphere (through the surface density) are varied, while the radius is kept constant, as are the other parameters. For the conversion to \MPktPdir, we also show in Figure~\ref{Abb:MPktausLLinie} a case with $\RP=1.5~\RJ$.

The correction factor from \MPktempir to \MPktEmiss does not depend much on mass, and increases with \MP, but varies also with flux, spanning $\MPktEmiss/\MPktempir\approx5$--10 at line luminosities up to $\LBra\sim10^{-6}~\LSonne$.
The reason is that Equation~(\ref{Gl:MPktempir}) assumes that all the gas is shocking with \vFfinfty on the surface of the planet, while in reality this velocity is the upper bound (Eq.~(\ref{Gl:vFfinfty}) and Fig.~\ref{Abb:Bild}). Therefore, a higher mass flux towards the inner regions is needed to generate the same line luminosity.

Figure~\ref{Abb:MPktausLLinie} predicts the mass flux shocking directly on the planet surface \MPktPdir to be 3--10~times higher than the canonical estimate \MPktempir for $\MP=2~\MJ$ but only 1--3~times higher for $\MP=10~\MJ$, with a dependence on the \Bra luminosity.
In fact, for the fiducial case, $\MPktPdir=2\times10^{-7}~\MPktEinhJ$, which implies a `planet-only' accretion luminosity of
\begin{subequations}
\begin{align}
 \LAkkPdir &=\MPktPdir\left\langle\frac{1}{2}v_r^2\right\rangle\\
      &=\frac{G\MP\MPktPdir}{\RP}\frac{1+\left\langle\mu\right\rangle}{2}\\
      &\approx1.0\times10^{-4}~\LSonne
\end{align}
\end{subequations}
for the appropriate average of incoming kinetic energy density, with $\mu\equiv\cos\theta$ (Eq.~(4a) of \citealp{m24expeditus}).
With their $\LBra\approx10^{-8}~\LSonne$, \PDSb and \WISPb (which seem similar; see beginning of Section~\ref{Th:Disk}) thus likely have $\MPktPdir\approx1.3\MPktempir$ and $\MPktEmiss\approx5\MPktempir$ according to Fig.~\ref{Abb:MPktausLLinie}.

For comparison, the total photospheric\footnote{We avoid the qualifier `bolometric' here because of the ambiguity of whether its meaning of `total' applies to the entire SED, including the CPD emission, or only to the photospheric emission.} luminosity is
$\Lphot=(1.3\pm0.2)\times10^{-4}~\LSonne$ for the nominal parameters in this work or some best-fit combinations of \citet{wang21vlti} or \citet{blakely25}, for example. Thus the photospheric luminosity is comparable to or slightly larger than the accretion luminosity onto the planet surface.  %
Interestingly, from the fit in Eq.~(A4a) of \citet{Aoyama+2020}, the fraction of the incoming accretion energy that is travelling downwards from the shock into the planet is around $\fdown=0.8$--0.9. Thus the shock could be heating up the (top layers of the) planet to some extent, since $\fdown\times\LAkkPdir\sim\Lphot$. The details will depend on the exact spectrum of this radiation and are beyond the scope of this work.

As discussed in Section~\ref{Th:MPkt}, the mass growth rate \MPktvekst, which is often the physical quantity of interest, is in general different from both \MPktPdir and \MPktEmiss.
If, as we assume in this work, no magnetospheric accretion is taking place, even \MPktEmiss is probably a lower bound on \MPktvekst since also the gas hitting the CPD further out could ultimately be accreted onto the planet, by flowing radially through the CPD and passing through a boundary layer. The radius corresponding to \MPktEmiss is essentially $\rvkrit\approx8~\RJ$ (Eq.~(\ref{Gl:rvkrit})), well within the centrifugal radius $\Rzent=\fzent\RHill=167~\RJ$ for the fiducial parameters, while the gas out to approximately \Rzent or a fraction of order unity thereof accretes onto the planet (\citealp{ab25}; see discussion in Sect.~2.4.3 of \citealt{m24expeditus}).
Even at $\MP=20~\MJ$, %
$\rvkrit\approx20~\RJ$
and is thus again much smaller than the corresponding $\Rzent\approx265~\RJ$.
As discussed in Section~\ref{Th:MPkt}, converting \MPktempir into \MPktvekst requires modelling aspects of the CPD that are outside the scope of this work.
While one could formally integrate the mass flow onto the CPD down to its thickness at the Hill sphere, the details of the assumed simplified structure (e.g., thickness and flaring, dynamics) have a disproportionate bearing on the results.

However, if magnetospheric accretion is active, \MPktempir will be much closer to the total mass growth rate because both the gas falling directly onto the planet and the gas first hitting the CPD, even well outside of \rvkrit, ultimately creates shocks on the planet surface. In this case, the correction factors of Figure~\ref{Abb:MPktausLLinie} would not be required.
This strengthens the case for using METIS to study accreting planets since the line shape can point towards which mechanisms govern accretion on the smallest scales.

\subsection{Brief comparison of METIS to CRIRES+}
 \label{Th:vglCRIRES}

The upraded CRyogenic InfraRed Echelle Spectrograph (CRIRES) instrument at VLT, CRIRES+ \citep{dorn23}, offers adaptive-optics-assisted $R=\textrm{50,000}$--100,000 spectroscopy at 0.95--5.3~\mum.
In the background-limited regime, CRIRES+ is approximately 10\,\%\ more sensitive than \mbox{(old-)}CRIRES \citep{grant24}, and based on the sensitivity curves in Figure~3 of \citet{kendrew10}, %
this would make METIS approximately 70~times more sensitive than CRIRES+ at $L$ and 20~times at $M$, for a fixed integration time and required \SNR. In the context of their analysis with cross-correlation spectroscopy, \citet{parker24} highlight that the sky thermal background decreases with $D^4$, where $D$ is the primary-mirror diameter, and they estimate an S/N improvement of $\approx7.5$ at $M$ between CRIRES+ and METIS. However, they caution that several factors can change this number.

Thus an order-of-magnitude increase in sensitivity seems likely, and it could be more. The sensitivity will likely be even more impressive for closer-in planets, which should more numerous (e.g., \citealp{fernandes19}). Compared to Figure~13 of \citet{parker24}, the sensitivity curves for METIS (see Fig.~\ref{Abb:ELT-Kontrascht}) are essentially compressed horizontally by a factor of~4--5, and this would let many planets become visible that were in the contrast-limited regime before, the steep inner part of the sensitivity curves.  %

\subsection{Perspectives for MICADO}
 \label{Th:MICADO}

\begin{figure} %
 \centering
\includegraphics[width=0.24\textwidth]{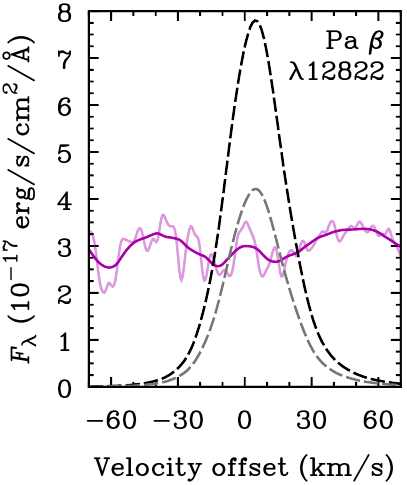}~~~%
\includegraphics[width=0.24\textwidth]{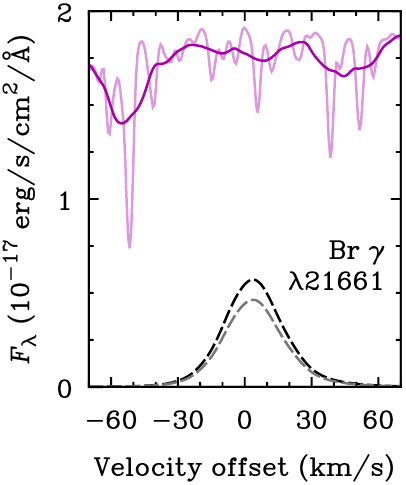}
\caption{%
Shock excess (\textit{black curve}) and resolved continuum (\textit{pink}; at full resolution or broadened to $v\sin i=10~\kms$: \textit{pale or darker}) at the MICADO-relevant accretion lines \Pab and \Brg.
We use the fiducial parameters for the shock and the atmosphere (Table~\ref{Tab:Par}) but scale the model flux by $\fPhot=0.13$ and $\fPhot=0.6$ to approximately match the SPHERE \citep{Stolker+20b} and GRAVITY \citep{wang21vlti} data, respectively (shown in Fig.~\ref{Abb:ProDiMoSpektrum}).
The shock emission is also shown extincted by $\AV=3$~mag (\textit{grey}).
We did not convolve with the spectral resolution of MICADO (up to $R=\textrm{20,000}$; the atmospheric spectrum has a sampling 20~times higher) and did not add any noise.
}
\label{Abb:LinienfuerMICADO}
\end{figure}

Our model can easily be applied to other wavelengths. Another first-generation ELT instrument, the Multi-AO Imaging Camera for Deep Observations (MICADO\footnote{See \url{https://elt.eso.org/instrument/MICADO}.}), will provide imaging and spectroscopy in the NIR at 0.8--2.4~\mum (in the $IYJHK$ bands; e.g., \citealp{baudoz23,palmabifani23,sturm24}). This covers many lines used to study accretion also at low-mass objects, in particular \Brg$\lambda21661$, \Pab$\lambda12822$ to \Pae=\,Pa8\,$\lambda9548$, and the \HeI$\lambda10833$ triplet \citep[e.g.,][]{alcal17,betti22b,betti22c,erkal22}.

Figure~\ref{Abb:LinienfuerMICADO} shows the \Pab and \Brg shock lines computed with our model for the fiducial parameters. The shape is identical to that of the other transitions, as Fig.~\ref{Abb:VarparLinie} shows.
As throughout, we use the same CIFIST atmospheric model with $\Teff=1400$~K with $\RP=2~\RJ$ but additionally scale the flux at \Pab by $\fPhot=0.13$ and at \Brg by $\fPhot=0.6$ (for the photosphere only). These factors were chosen to have a good visual match to the SPHERE \citep{Stolker+20b} $H$-band data and the GRAVITY \citep{wang21vlti} $K$-band data, both at much lower resolution. As a reminder, we use a cloud-free BT-Settl model as a conservative estimate of the amount of structure in the photosphere; a DRIFT-PHOENIX model, which matches the overall SED much better (see Fig.~\ref{Abb:ProDiMoSpektrum}), should be smoother or at least not much more structured. As in Section~\ref{Th:PlanEm}, we also broaden the photosphere to $v\sin i=10~\kms$.
We do not include any instrumental noise.

The amount of extinction has not been constrained securely, with values $\AV\approx0$--3~mag assumed or found in different works \citep[e.g.,][]{Stolker+20b,uyama21b,blakely25,taylor26b}. Therefore, we also also consider the conservative case that the inferred atmospheric properties already include extinction. Then, we still need to dim the line emission from our model. With \citet{wang19} and $\AV\approx3$~mag from the DRIFT-PHOENIX, two-component fit in \citep{wang21vlti}, this gives $A_\lambda=0.7$~mag at \Pab and 0.2~mag and \Brg.

Comparing Figure~\ref{Abb:LinienfuerMICADO} to Figure~\ref{Abb:Atmbeitragnorm} shows that the \Brg emission
(peak excess: $F_\lambda\approx0.5\times10^{-17}$~erg\,s$^{-1}$\,cm$^{-2}$\,\AA$^{-1}$) is weaker than the \Bra emission
($F_\lambda\approx0.9\times10^{-17}$ in the same units), while the \Pab line ($F_\lambda\approx8\times10^{-17}$) is stronger than \Bra. This holds generally speaking for shock emission \citep{aoyama18}, also in empirical scalings \citep{testi25}, and in `preshock emission' models \citep{kwanfischer11}, as easily seen (except for \Bra) in \citet{betti22c}.
The width of the shock-emitted \Pab and \Brg lines is identical to that of the others shown in Fig.~\ref{Abb:Varpar}, with a small difference for \Ha.

The simple comparison to the continuum in Fig.~\ref{Abb:LinienfuerMICADO} suggests that while the \Brg emission might be too weak for easy detection, \Pab is promising thanks to its good contrast with the continuum: its peak is 1.3~(2.7) times brighter than the photosphere with (without) $\AV=3$~mag.
Even with extinction, extracting the \Pab line at least at modest significance should not be very challenging because there are no strong spectral features close to \Pab.

\citet{uyama21b} searched for \Pab emission at \PDSbc using the Keck/OSIRIS spectrograph, with a resolution $R\approx4000$. For \PDSb, they report a 5$\sigma$ upper limit line-integrated flux $\FPab<1.4\times10^{-16}$~erg\,s$^{-1}$\,cm$^{-2}$, which corresponds to $\LPab<5.6\times10^{-8}~\LSonne$, based on the instrumental line spread function (LSF; see e.g.\ around Eq.~(13) of \citealt{wang21vlti}), assumed to be a Gaussian with $\FWHM=70~\kms$.
As shown in Figure~\ref{Abb:LLinie}, our predicted non-extincted line-integrated luminosity is $\LPab=4.6\times10^{-8}~\LSonne$. This agrees very well especially since this line strength comes from our choice of an accretion rate matching the maximum \Ha observed (Section~\ref{Th:Parvar}), which in general was at a different epoch than in \citet{uyama21b}. Also, extinction could possibly reduce the observable line strength.

At \Brg, both \citet{christiaens19b} with VLT/SINFONI and \citet{wang21vlti} with VLTI/GRAVITY derived a 5$\sigma$ upper flux limit $\FBrg<8.4\times10^{-17}$~erg\,s$^{-1}$\,cm$^{-2}$, which translates into an upper limit of $\LBrg<3.3\times10^{-8}~\LSonne$. Our prediction of $\LBrg=1.3\times10^{-8}~\LSonne$ compares well with this.

Thus, the agreement between our prediction and the upper limits seems acceptable. As an aside, one should keep in mind that if the \Ha (which we use to set the accretion rate) originates not only from the shock but is boosted by the preshock emission of a magnetospheric flow, the line strengths we predict are likely on the high side.

Whether lines such as \Pab will have a higher \SNR{} with MICADO relative to \Bra{} with METIS for accreting planets in general, depends to some extent on the continuum level and the strength of the features; on the extinction, which could be higher than $\AV=3$~mag, and the location of the extincting matter (e.g., in the planet atmosphere; only around the planet but not the CPD surface; or between any line-emitting region and the observer); and on the intrinsic relative strength of the lines.
Nevertheless, all these lines, and their ratios \citep[e.g.,][]{gatti06,betti22b,betti22c} will help in the analysis of accretion processes.

We conclude with comments on the practical observability with MICADO.
The thermal background from the atmosphere and the telescope is likely negligible for MICADO, with speckle noise dominating, which is the opposite situation compared to METIS (Figure~\ref{Abb:ELT-Kontrascht}). %
With ADI post-processing, median imaging contrasts around $3\times10^{-6}$ should be achievable in the $H$ band \citep{baudoz23}, making \PDSb in principle easily observable thanks to its contrast $\Delta H\approx9.1~\textrm{mag}\approx2\times10^{-4}$ \citep{keppler18}. \PDSA ($r'=12$~mag; \citealp{henden15}) is bright enough for the SCAO to work well.

However, MICADO has only one focal-plane wheel \citep{kravchenko24}. Therefore, either the coronagraph for high-contrast observations or the long-slit mask for spectroscopy can be used at a given time. %
Consequently, for high-resolution spectroscopy, the whole ADI sequence cannot be used because of the diffraction spikes from the ELT, which are significant even at the separation of \PDSb, around~$10\lambda/D$ (Appendix~\ref{Th:wosindbundc}, now for $\lambda\approx2$~\mum).

Therefore, a small-field IFU in the MICADO bands would likely simplify the observational approach and better match the ADI.  %
Ideally, if this IFU is only on-axis (centred on the star), it would need to be large enough to include \PDSc (with minimum separation $\rho\approx210$~mas; Fig.~\ref{Abb:wosindbundc}) and \WISPb (with currently $\rho\approx320$~mas), that is, at least 640~mas large in at least one dimension. An off-axis IFU, which can be offset from the star by a small amount, with the AO locked on the star, could be smaller and still suited for observing accreting planets.
\section{Summary and conclusions}
 \label{Th:Zus}

In preparation for ELT/METIS coming online, we have developed in \citet{m24expeditus} and this work a framework to predict spectrally resolved hydrogen accretion-line profiles for gas giants.
For the local emission, we used the shock model of \citet{aoyama18}.
While our set-up could accommodate magnetospheric accretion,
we restricted ourselves to the
cases when it is not at play.
We took \PDSb as an example, noting that \WISPb is very similar, and we have predicted how well METIS can detect the \Bra line and reveal its shape.
Our main findings concerning \PDSb and planetary-mass accretors in disc gaps more generally are outlined in the following:
\begin{enumerate}

 \item The range of observed \Ha fluxes, combined with planetary scaling relationships \citep{aoyama18,AMIM21L} predicts an integrated \Bra line luminosity at \PDSb\ of $\LBra\approx10^{-9}$--$10^{-8}~\LSonne$.  %
 We used the maximum as our fiducial case (Fig.~\ref{Abb:LLinie}).
 
 \item The \Bra line profile, coming mostly from the shock on the planet surface with a tens-of-percent contribution from the CPD surface shock (Fig.~\ref{Abb:Varpar}), has a remarkably constant $\FWHM\approx30$~\kms, even up to preshock velocities of $v_0\approx200$~\kms, for low to moderate preshock densities (Fig.~\ref{Abb:DvAo18}). Thus, this is often clearly narrower than the maximum free-fall velocity. The line is Gaussian in its core only,
  with broader wings that are brighter on the red side.
  The peak is slightly offset, by $\Delta v\approx+2$~\kms, which is easily measurable.
   The line shape is very modestly sensitive to the mass and radius and barely sensitive to the other parameters within a reasonable range.
 Magnetospheric accretion can lead to similar profiles, but also to more complex ones \citep{edwards94,thanathibodee24,demars26}. Consequently, a complex shape would imply that shock emission is subdominant.  %

 \item The pure photospheric emission of \PDSb will already be detectable at $\SNR\approx12$ per spectral bin in 4~h (Fig.~\ref{Abb:Komboalles}). If caught in an episode of strong accretion, corresponding to moderate to high literature values, the shock excess will be easily detectable at \Bra with a peak per-bin $\SNR=3$ in 10~min.
 However, the non-Gaussian wings of the photosphere-subtracted spectrum most likely cannot be detected in a reasonable amount of time.
 
 \item At least at \Bra, the detailed spectral appearance of the photosphere depends on the choice of the opacity line list, dominated around $\Teff\approx1400$~K by water (H$_2$O) and hydrogen sulphide (H$_2$S). Recent calculations disagree on the exact
 position and strength of the features (Figs.~\ref{Abb:kappaundFlLinienlisten} and~\ref{Abb:AtmbeitragnormmitgCMCRT}).

 \item A helpful factor for detecting a line excess is the possible rotation of young gas giants, even if only around $v\sin i\approx10~\kms$. It flattens the photospheric features but without affecting the accretion signature (Fig.~\ref{Abb:Atmbeitragnorm}). A moderate inclination $i\approx50\degr$---which is the inclination of the \PDS system---is optimal for this enhancement of the detectability.
 
 \item The PPD emission and the residual speckle noise from the star should be negligible at \Bra (Fig.~\ref{Abb:ELT-Kontrascht}). The CPD emission is likely small, but more importantly it probably does not introduce spectral features that impede a measurement of the shock line, as we verified with ProDiMo (Fig.~\ref{Abb:ProDiMoSpektrum}). However, a more systematic and detailed assessment of the CPD contribution is beyond the scope of this work but seems warranted.
 
 \item The other transitions accessible to METIS (e.g., \Pfb, \Pfg, \Pfd, \Pfe in the Pfund series, several Humphries lines) are likely of limited use for detecting accreting planets because the photospheric signal is much stronger and likely displays strong features (Fig.~\ref{Abb:Komboalles}). These higher-order lines, however, could be useful at accretors with a higher mass or accretion rate.
 
 \item For $\MP\approx0.5$--$1~\MJ$, planets would need to be accreting very vigorously with a line-emission mass flux of $\MPktEmiss\gtrsim10^{-5}~\MPktEinhJ$ for their shock excess to have a chance of being detectable. Lower masses, $\MP\lesssim0.5~\MJ$, do not emit any lines. A large parameter space is shown in Fig.~\ref{Abb:Haupt}.

 \item We distinguished between different definitions of the accretion rate, including the mass growth rate \MPktvekst, the line-emitting \MPktEmiss, and the canonically inferred \MPktempir (Sects.~\ref{Th:MPkt} and~\ref{Th:MPktausLLinie}). Within our assumption of emission only from shocks on the planet and CPD surfaces, the line-emitting flux is larger by a factor of five to 20 than the canonical estimate \MPktempir from \LAkk--\LLinie relations for planets (Fig.~\ref{Abb:MPktausLLinie}).
 If magnetospheric accretion is active, \MPktempir should be closer to the mass growth rate \MPktvekst. %

\end{enumerate}

The broad-brush predictions for \PDSb apply similarly to \PDSc, with a positive nuance. \PDSc has a similar accretion rate \citep{zhou25} and mass (\citealp{hammond25,trevascus25}; however, there is a large range) as \PDSb, while the photosphere of \PDSc seems less bright \citep{wang21vlti}.
Together, this should let the \Bra line stand out better from the photosphere. \PDSc is viewed through some PPD material \citep{Haffert+2019}, but the actual amount of extinction to the line-emitting regions is hardly constrained \citep[e.g.,][]{Aoyama+Ikoma2019,wang21vlti}. It could well be that this extinction is negligible especially at \Bra.
Moreover, the different physical separations from the star are not directly relevant for the local accretion, apart from possibly setting how much of the global radial gas flow is intercepted by each planet \citep{lubow06,bergez23}. Therefore, the models here should apply similarly.

On a related note, the VLT visitor instrument RISTRETTO \citep{chazelas20,lovis22,lovis24}, designed for reflected-light $R=\textrm{140,000}$ spectroscopy at the diffraction limit of the VLT, will be able to study accreting planets in their \Ha line \citep{blackman26}. Thus, it holds an excellent synergy potential, especially if observations are performed simultaneously with METIS. Its current estimated date of first light is 2029, around the same time as METIS.
Our work could be extended in many ways. One would be to incorporate the emission from magnetospheric accretion columns if models predicted them in the NIR. At planetary masses, both this preshock emission and the (post-)shock emission considered so far likely contribute \citep{hashimoto25}.
We have focussed on the line excess, but calculating the continuum in the $L$ and $M$ bands, especially from the CPD, would be an interesting challenge, for which there are currently only a few observational constraints. This would also represent an extension of the calculations by \citet{oberg23metis}.

\begin{acknowledgements}
{\small
We thank the anonymous referee for a helpful, careful report that significantly improved the clarity of central findings of this work.
We are indebted to Olivier Absil for providing us with METIS contrast curves in different modes and patiently exchanging on the topic.
We gratefully acknowledge the insightful comments by Aster Taylor that led to clarifications and improvements.
We thank Connor Robinson for helpful discussions about radiative transfer
and Cade B\"urgy for comments on the manuscript, in particular insightfully asking about the rotation of the line-emitting gas.
We also thank Hiro Takami for discussions of the observability,
Pierre Baudoz, J\"org-Uwe Pott, Markus Feldt, and Wolfgang Brandner for thoughts about MICADO,
Jeff Bary for helpful comments about the line centroids,
Josh Blackman for comments about RISTRETTO,
Dorian Demars for explanations concerning line widths,  %
Anu Dudhia for a quick and helpful clarification, %
Eric Gaidos for discussions about \PDSA,  %
and Elsie Lee for a high-resolution computation with gCMCRT.
The main calculations carried out here were performed on the \texttt{bachelor} cluster of the MPIA.
G-DM and TH 
acknowledge support from the European Research Council (ERC) under the European Union's Horizon 2020 Research and Innovation Programme via the ERC Advanced Grant `ORIGINS' (Nr.~832428; PI: Th.~Henning).
G-DM and MB
received funding from the European Research Council (ERC) under the European Union's Horizon 2020 Research and Innovation Programme via the ERC Consolidator Grant `PROTOPLANETS' (Nr.~101002188; PI: M.~Benisty).
G-DM 
acknowledges the support of the Deutsche Forschungsgemeinschaft (DFG) through grant MA~9185/2-1.
This publication made use of \href{https://www.wolframalpha.com}{\texttt{Wolfram|alpha}} to obtain Equation~(\ref{Gl:avph}).
This publication made use of the `Theoretical Spectra' service of VOSA (\url{https://svo.cab.inta-csic.es}; \citealp{bayo08}), developed under the Spanish Virtual Observatory (SVO) project funded by MCIN/AEI/10.13039/501100011033/ through grant PID2020-112949GB-I00. VOSA has been partially updated by using funding from the European Union's Horizon 2020 Research and Innovation Programme `EXOPLANETS-A' (Nr.~776403).
This research has made use of the SVO Filter Profile Service `Carlos Rodrigo', funded by MCIN/AEI/10.13039/501100011033/ through grant PID2023-146210NB-I00.
This research has used NASA's Astrophysics Data System Bibliographic Services.
}
\end{acknowledgements}

\bibliographystyle{aa_url}
\bibliography{std.bib}{}

\begin{thebibliography}{267}
\expandafter\ifx\csname natexlab\endcsname\relax\def\natexlab#1{#1}\fi

\bibitem[{{Abramowitz} \& {Stegun}(1972)}]{abramowitz72}
{Abramowitz}, M. \& {Stegun}, I.~A. 1972, {Handbook of Mathematical Functions},
  Applied Mathematics Series 55 (U.S.\@ Department of Commerce, National Bureau
  of Standards)

\bibitem[{{Adams} {et~al.}(2025){Adams}, {Zhou}, {Marleau}, {Apai}, {Biller},
  {Carter}, {Vos}, {Whiteford}, {Birkmann}, {Karalidi}, {Tan}, {Wang},
  {Aoyama}, {Bowler}, {Bonnefoy}, \& {Hashimoto}}]{adamszhou25}
{Adams}, A.~D., {Zhou}, Y., {Marleau}, G.-D., {et~al.} 2025,
  \href{http://dx.doi.org/10.3847/1538-3881/ae07d2}{\color{magenta}\aj},
  \href{https://ui.adsabs.harvard.edu/abs/2025AJ....170..289A}{170, 289}

\bibitem[{{Adams} \& {Batygin}(2022)}]{ab22}
{Adams}, F.~C. \& {Batygin}, K. 2022,
  \href{http://dx.doi.org/10.3847/1538-4357/ac7a3e}{\color{magenta}\apj},
  \href{https://ui.adsabs.harvard.edu/abs/2022ApJ...934..111A}{934, 111}

\bibitem[{{Adams} \& {Batygin}(2025)}]{ab25}
{Adams}, F.~C. \& {Batygin}, K. 2025,
  \href{http://dx.doi.org/10.1088/1538-3873/adcf57}{\color{magenta}\pasp},
  \href{https://ui.adsabs.harvard.edu/abs/2025PASP..137e4401A}{137, 054401}

\bibitem[{{Alarc{\'o}n} {et~al.}(2024){Alarc{\'o}n}, {Bergin}, \&
  {Cugno}}]{alarc24}
{Alarc{\'o}n}, F., {Bergin}, E.~A., \& {Cugno}, G. 2024,
  \href{http://dx.doi.org/10.3847/1538-4357/ad3938}{\color{magenta}\apj},
  \href{https://ui.adsabs.harvard.edu/abs/2024ApJ...966..225A}{966, 225}

\bibitem[{{Alcal{\'a}} {et~al.}(2017){Alcal{\'a}}, {Manara}, {Natta}, {Frasca},
  {Testi}, {Nisini}, {Stelzer}, {Williams}, {Antoniucci}, {Biazzo}, {Covino},
  {Esposito}, {Getman}, \& {Rigliaco}}]{alcal17}
{Alcal{\'a}}, J.~M., {Manara}, C.~F., {Natta}, A., {et~al.} 2017,
  \href{http://dx.doi.org/10.1051/0004-6361/201629929}{\color{magenta}\aap},
  \href{https://ui.adsabs.harvard.edu/abs/2017A%26A...600A..20A}{600, A20}

\bibitem[{{Allard} {et~al.}(2012){Allard}, {Homeier}, \&
  {Freytag}}]{allard12philtrans}
{Allard}, F., {Homeier}, D., \& {Freytag}, B. 2012,
  \href{http://dx.doi.org/10.1098/rsta.2011.0269}{\color{magenta}Philosophical
  Transactions of the Royal Society of London Series A},
  \href{http://cdsads.u-strasbg.fr/abs/2012RSPTA.370.2765A}{370, 2765}

\bibitem[{{Allard} {et~al.}(2013){Allard}, {Homeier}, \& {Freytag}}]{allard13}
{Allard}, F., {Homeier}, D., \& {Freytag}, B. 2013, \memsai,
  \href{https://ui.adsabs.harvard.edu/abs/2013MmSAI..84.1053A}{84, 1053}

\bibitem[{{Almendros-Abad} {et~al.}(2025){Almendros-Abad}, {Scholz}, {Damian},
  {Jayawardhana}, {Bayo}, {Flagg}, {Mu{\v{z}}i{\'c}}, {Natta}, {Pinilla}, \&
  {Testi}}]{almendrosabad25}
{Almendros-Abad}, V., {Scholz}, A., {Damian}, B., {et~al.} 2025,
  \href{http://dx.doi.org/10.3847/2041-8213/ae09a8}{\color{magenta}\apjl},
  \href{https://ui.adsabs.harvard.edu/abs/2025ApJ...992L...2A}{992, L2}

\bibitem[{{Aoyama} {et~al.}(2026){Aoyama}, {Hashimoto}, {Marleau}, {Manara},
  {Alcal{\'a}}, {Herczeg}, {Bergez-Casalou}, \& {Ikoma}}]{aoyama26}
{Aoyama}, Y., {Hashimoto}, J., {Marleau}, G.-D., {et~al.} 2026,
  \href{https://ui.adsabs.harvard.edu/abs/2026arXiv260923778A}{\aj,
  arXiv:2609.23778}

\bibitem[{{Aoyama} \& {Ikoma}(2019)}]{Aoyama+Ikoma2019}
{Aoyama}, Y. \& {Ikoma}, M. 2019,
  \href{http://dx.doi.org/10.3847/2041-8213/ab5062}{\color{magenta}\apjl},
  \href{https://ui.adsabs.harvard.edu/abs/2019ApJ...885L..29A}{885, L29}

\bibitem[{{Aoyama} {et~al.}(2018){Aoyama}, {Ikoma}, \& {Tanigawa}}]{aoyama18}
{Aoyama}, Y., {Ikoma}, M., \& {Tanigawa}, T. 2018,
  \href{http://dx.doi.org/10.3847/1538-4357/aadc11}{\color{magenta}\apj},
  \href{https://ui.adsabs.harvard.edu/abs/2018ApJ...866...84A}{866, 84}

\bibitem[{{Aoyama} {et~al.}(2024){Aoyama}, {Marleau}, \&
  {Hashimoto}}]{aoyama24twa}
{Aoyama}, Y., {Marleau}, G.-D., \& {Hashimoto}, J. 2024,
  \href{http://dx.doi.org/10.3847/1538-3881/ad67df}{\color{magenta}\aj},
  \href{https://ui.adsabs.harvard.edu/abs/2024AJ....168..155A}{168, 155}

\bibitem[{{Aoyama} {et~al.}(2021){Aoyama}, {Marleau}, {Ikoma}, \&
  {Mordasini}}]{AMIM21L}
{Aoyama}, Y., {Marleau}, G.-D., {Ikoma}, M., \& {Mordasini}, C. 2021,
  \href{http://dx.doi.org/10.3847/2041-8213/ac19bd}{\color{magenta}\apjl},
  \href{https://ui.adsabs.harvard.edu/abs/2021ApJ...917L..30A}{917, L30}

\bibitem[{{Aoyama} {et~al.}(2020){Aoyama}, {Marleau}, {Mordasini}, \&
  {Ikoma}}]{Aoyama+2020}
{Aoyama}, Y., {Marleau}, G.-D., {Mordasini}, C., \& {Ikoma}, M.
  \href{https://ui.adsabs.harvard.edu/abs/2020arXiv201106608A}{2020,
  arXiv:2011.06608}

\bibitem[{{Ayliffe} \& {Bate}(2009)}]{ab09b}
{Ayliffe}, B.~A. \& {Bate}, M.~R. 2009,
  \href{http://dx.doi.org/10.1111/j.1365-2966.2009.15002.x}{\color{magenta}\mnras},
  \href{https://ui.adsabs.harvard.edu/abs/2009MNRAS.397..657A}{397, 657}

\bibitem[{{Azzam} {et~al.}(2016){Azzam}, {Tennyson}, {Yurchenko}, \&
  {Naumenko}}]{azzam16}
{Azzam}, A. A.~A., {Tennyson}, J., {Yurchenko}, S.~N., \& {Naumenko}, O.~V.
  2016, \href{http://dx.doi.org/10.1093/mnras/stw1133}{\color{magenta}\mnras},
  \href{https://ui.adsabs.harvard.edu/abs/2016MNRAS.460.4063A}{460, 4063}

\bibitem[{{Bae} {et~al.}(2023){Bae}, {Isella}, {Zhu}, {Martin}, {Okuzumi}, \&
  {Suriano}}]{bae23ppvii}
{Bae}, J., {Isella}, A., {Zhu}, Z., {et~al.} 2023, in Astronomical Society of
  the Pacific Conference Series, Vol. 534, Protostars and Planets VII, ed.
  {Inutsuka}, S., {Aikawa}, Y., {Muto}, T., {Tomida}, K., \& {Tamura}, M.,
  \href{https://ui.adsabs.harvard.edu/abs/2023ASPC..534..423B}{423}

\bibitem[{{Bailer-Jones} {et~al.}(2021){Bailer-Jones}, {Rybizki}, {Fouesneau},
  {Demleitner}, \& {Andrae}}]{bj21}
{Bailer-Jones}, C.~A.~L., {Rybizki}, J., {Fouesneau}, M., {Demleitner}, M., \&
  {Andrae}, R. 2021,
  \href{http://dx.doi.org/10.3847/1538-3881/abd806}{\color{magenta}\aj},
  \href{https://ui.adsabs.harvard.edu/abs/2021AJ....161..147B}{161, 147}

\bibitem[{{Balmer} {et~al.}(2026){Balmer}, {Xuan}, {Choksi}, {Hoch},
  {Madurowicz}, {Pueyo}, {Rickman}, {Perrin}, {Sanghi}, {Messier}, {Bruinsma},
  {Follette}, {Marleau}, \& {Ward-Duong}}]{balmer26jwst}
{Balmer}, W., {Xuan}, J., {Choksi}, N., {et~al.} 2026, {Measuring the
  composition and circumplanetary environment of a gap-clearing protoplanet},
  JWST Proposal. Cycle 5, ID. \#9993

\bibitem[{{Balsalobre-Ruza} {et~al.}(2026){Balsalobre-Ruza}, {Christiaens},
  {Hu{\'e}lamo}, {Hammond}, {Benisty}, {Lacour}, {Blakely}, {van Holstein},
  {Latour}, {Lillo-Box}, {Wahhaj}, {Trevascus}, {Absil}, {Bae}, {Charalambous},
  {de Gregorio-Monsalvo}, {Mendigut{\'\i}a}, {Petrovich}, {Ribas}, \&
  {Juillard}}]{balsalobreruza26}
{Balsalobre-Ruza}, O., {Christiaens}, V., {Hu{\'e}lamo}, N., {et~al.}
  \href{https://ui.adsabs.harvard.edu/abs/2026arXiv260626381B}{2026,
  arXiv:2606.26381}

\bibitem[{{Baraffe} {et~al.}(2003){Baraffe}, {Chabrier}, {Barman}, {Allard}, \&
  {Hauschildt}}]{baraffe03}
{Baraffe}, I., {Chabrier}, G., {Barman}, T.~S., {Allard}, F., \& {Hauschildt},
  P.~H. 2003,
  \href{http://dx.doi.org/10.1051/0004-6361:20030252}{\color{magenta}\aap},
  \href{https://ui.adsabs.harvard.edu/abs/2003A%26A...402..701B}{402, 701}

\bibitem[{{Barber} {et~al.}(2006){Barber}, {Tennyson}, {Harris}, \&
  {Tolchenov}}]{barber06}
{Barber}, R.~J., {Tennyson}, J., {Harris}, G.~J., \& {Tolchenov}, R.~N. 2006,
  \href{http://dx.doi.org/10.1111/j.1365-2966.2006.10184.x}{\color{magenta}\mnras},
  \href{https://ui.adsabs.harvard.edu/abs/2006MNRAS.368.1087B}{368, 1087}

\bibitem[{{Batygin} \& {Adams}(2025)}]{batygin25}
{Batygin}, K. \& {Adams}, F.~C. 2025,
  \href{http://dx.doi.org/10.1038/s41550-025-02512-y}{\color{magenta}Nature
  Astronomy}, \href{https://ui.adsabs.harvard.edu/abs/2025NatAs...9..835B}{9,
  835}

\bibitem[{{Batygin} \& {Morbidelli}(2020)}]{batygin20}
{Batygin}, K. \& {Morbidelli}, A. 2020,
  \href{http://dx.doi.org/10.3847/1538-4357/ab8937}{\color{magenta}\apj},
  \href{https://ui.adsabs.harvard.edu/abs/2020ApJ...894..143B}{894, 143}

\bibitem[{{Baudino} {et~al.}(2017){Baudino}, {Molli{\`e}re}, {Venot},
  {Tremblin}, {B{\'e}zard}, \& {Lagage}}]{baudino17}
{Baudino}, J.-L., {Molli{\`e}re}, P., {Venot}, O., {et~al.} 2017,
  \href{http://dx.doi.org/10.3847/1538-4357/aa95be}{\color{magenta}\apj},
  \href{https://ui.adsabs.harvard.edu/abs/2017ApJ...850..150B}{850, 150}

\bibitem[{{Baudoz} {et~al.}(2023){Baudoz}, {Huby}, {Vidal}, {Gendron},
  {Clenet}, \& {Davies}}]{baudoz23}
{Baudoz}, P., {Huby}, E., {Vidal}, F., {et~al.} 2023, in Adaptive Optics for
  Extremely Large Telescopes (AO4ELT7),
  \href{https://ui.adsabs.harvard.edu/abs/2023aoel.confE..90B}{90}

\bibitem[{{Bayo} {et~al.}(2008){Bayo}, {Rodrigo}, {Barrado Y Navascu{\'e}s},
  {Solano}, {Guti{\'e}rrez}, {Morales-Calder{\'o}n}, \& {Allard}}]{bayo08}
{Bayo}, A., {Rodrigo}, C., {Barrado Y Navascu{\'e}s}, D., {et~al.} 2008,
  \href{http://dx.doi.org/10.1051/0004-6361:200810395}{\color{magenta}\aap},
  \href{https://ui.adsabs.harvard.edu/abs/2008A%26A...492..277B}{492, 277}

\bibitem[{{Benisty} {et~al.}(2021){Benisty}, {Bae}, {Facchini}, {Keppler},
  {Teague}, {Isella}, {Kurtovic}, {P{\'e}rez}, {Sierra}, {Andrews},
  {Carpenter}, {Czekala}, {Dominik}, {Henning}, {Menard}, {Pinilla}, \&
  {Zurlo}}]{benisty21}
{Benisty}, M., {Bae}, J., {Facchini}, S., {et~al.} 2021,
  \href{http://dx.doi.org/10.3847/2041-8213/ac0f83}{\color{magenta}\apjl},
  \href{https://ui.adsabs.harvard.edu/abs/2021ApJ...916L...2B}{916, L2}

\bibitem[{{Benisty} {et~al.}(2026){Benisty}, {Facchini}, {Teague}, {Zagaria},
  {Curone}, {Pezzotta}, {Bae}, {Cugno}, \& {Price}}]{benisty26}
{Benisty}, M., {Facchini}, S., {Teague}, R., {et~al.} 2026,
  \href{http://dx.doi.org/10.3847/2041-8213/aea214}{\color{magenta}\apjl},
  \href{https://ui.adsabs.harvard.edu/abs/2026ApJ..1009L..32B}{1009, L32}

\bibitem[{{Bergez-Casalou} {et~al.}(2023){Bergez-Casalou}, {Bitsch}, \&
  {Raymond}}]{bergez23}
{Bergez-Casalou}, C., {Bitsch}, B., \& {Raymond}, S.~N. 2023,
  \href{http://dx.doi.org/10.1051/0004-6361/202244988}{\color{magenta}\aap},
  \href{https://ui.adsabs.harvard.edu/abs/2023A%26A...669A.129B}{669, A129}

\bibitem[{{B{\'e}thune} \& {Rafikov}(2019)}]{b19b}
{B{\'e}thune}, W. \& {Rafikov}, R.~R. 2019,
  \href{http://dx.doi.org/10.1093/mnras/stz1870}{\color{magenta}\mnras},
  \href{https://ui.adsabs.harvard.edu/abs/2019MNRAS.488.2365B}{488, 2365}

\bibitem[{{Betti} {et~al.}(2022{\natexlab{a}}){Betti}, {Follette},
  {Ward-Duong}, {Aoyama}, {Marleau}, {Bary}, {Robinson}, {Janson}, {Balmer},
  {Chauvin}, \& {Palma-Bifani}}]{betti22b}
{Betti}, S.~K., {Follette}, K.~B., {Ward-Duong}, K., {et~al.}
  2022{\natexlab{a}},
  \href{http://dx.doi.org/10.3847/2041-8213/ac85ef}{\color{magenta}\apjl},
  \href{https://ui.adsabs.harvard.edu/abs/2022ApJ...935L..18B}{935, L18}

\bibitem[{{Betti} {et~al.}(2022{\natexlab{b}}){Betti}, {Follette},
  {Ward-Duong}, {Aoyama}, {Marleau}, {Bary}, {Robinson}, {Janson}, {Balmer},
  {Chauvin}, \& {Palma-Bifani}}]{betti22c}
{Betti}, S.~K., {Follette}, K.~B., {Ward-Duong}, K., {et~al.}
  2022{\natexlab{b}},
  \href{http://dx.doi.org/10.3847/2041-8213/aca331}{\color{magenta}\apjl},
  \href{https://ui.adsabs.harvard.edu/abs/2022ApJ...941L..20B}{941, L20}

\bibitem[{{Biller}(2017)}]{biller17}
{Biller}, B. 2017,
  \href{http://dx.doi.org/10.1080/21672857.2017.1303105}{\color{magenta}The
  Astronomical Review},
  \href{https://ui.adsabs.harvard.edu/abs/2017AstRv..13....1B}{13, 1}

\bibitem[{{Biller} {et~al.}(2015){Biller}, {Vos}, {Bonavita}, {Buenzli},
  {Baxter}, {Crossfield}, {Allers}, {Liu}, {Bonnefoy}, {Deacon}, {Brandner},
  {Schlieder}, {Dupuy}, {Kopytova}, {Manjavacas}, {Allard}, {Homeier}, \&
  {Henning}}]{biller15}
{Biller}, B.~A., {Vos}, J., {Bonavita}, M., {et~al.} 2015,
  \href{http://dx.doi.org/10.1088/2041-8205/813/2/L23}{\color{magenta}\apjl},
  \href{https://ui.adsabs.harvard.edu/abs/2015ApJ...813L..23B}{813, L23}

\bibitem[{{Blackman} {et~al.}(2026){Blackman}, {Mordasini}, {Marleau}, {Lovis},
  {Bugatti}, {Aoyama}, \& {Blind}}]{blackman26}
{Blackman}, J.~W., {Mordasini}, C., {Marleau}, G.-D., {et~al.} 2026,
  \href{http://dx.doi.org/10.1051/0004-6361/202659437}{\color{magenta}\aap},
  \href{https://ui.adsabs.harvard.edu/abs/2026A&A...711A.148B}{711, A148}

\bibitem[{{Blakely} {et~al.}(2025){Blakely}, {Johnstone}, {Cugno},
  {Sivaramakrishnan}, {Tuthill}, {Dong}, {Pope}, {Albert}, {Charles}, {Cooper},
  {Furio}, {Desdoigts}, {Doyon}, {Francis}, {Greenbaum}, {Lafreni{\'e}re},
  {Lloyd}, {Meyer}, {Pueyo}, {Ray}, {S{\'a}nchez-Berm{\'u}dez}, {Soulain},
  {Thatte}, {Thompson}, \& {Vandal}}]{blakely25}
{Blakely}, D., {Johnstone}, D., {Cugno}, G., {et~al.} 2025,
  \href{http://dx.doi.org/10.3847/1538-3881/ad9b94}{\color{magenta}\aj},
  \href{https://ui.adsabs.harvard.edu/abs/2025AJ....169..137B}{169, 137}

\bibitem[{{Bohr}(1913)}]{bohr13}
{Bohr}, N. 1913,
  \href{http://dx.doi.org/10.1080/14786441308634955}{\color{magenta}Philosophical
  Magazine}, \href{https://ui.adsabs.harvard.edu/abs/1913PMag...26....1B}{26,
  1}

\bibitem[{{Bowler} {et~al.}(2011){Bowler}, {Liu}, {Kraus}, {Mann}, \&
  {Ireland}}]{bowler11}
{Bowler}, B.~P., {Liu}, M.~C., {Kraus}, A.~L., {Mann}, A.~W., \& {Ireland},
  M.~J. 2011,
  \href{http://dx.doi.org/10.1088/0004-637X/743/2/148}{\color{magenta}\apj},
  \href{https://ui.adsabs.harvard.edu/abs/2011ApJ...743..148B}{743, 148}

\bibitem[{{Brandl} {et~al.}(2021){Brandl}, {Bettonvil}, {van Boekel},
  {Glauser}, {Quanz}, {Absil}, {Amorim}, {Feldt}, {Glasse}, {G{\"u}del}, {Ho},
  {Labadie}, {Meyer}, {Pantin}, {van Winckel}, \& {METIS
  Consortium}}]{brandl21}
{Brandl}, B., {Bettonvil}, F., {van Boekel}, R., {et~al.} 2021,
  \href{http://dx.doi.org/10.18727/0722-6691/5218}{\color{magenta}The
  Messenger}, \href{https://ui.adsabs.harvard.edu/abs/2021Msngr.182...22B}{182,
  22}

\bibitem[{{Brandl} {et~al.}(2022){Brandl}, {Bettonvil}, {van Boekel},
  {Glauser}, {Quanz}, {Absil}, {Feldt}, {Garcia}, {Glasse}, {Guedel},
  {Labadie}, {Meyer}, {Pantin}, {Wang}, {van Winckel}, {Ag{\'o}cs}, {Amorim},
  {Bertram}, {Burtscher}, {Delacroix}, {Laun}, {Lesman}, {Raskin}, {Salo},
  {Scheithauer}, {Stuik}, {Todd}, {Haupt}, \& {Siebenmorgen}}]{brandl22}
{Brandl}, B.~R., {Bettonvil}, F., {van Boekel}, R., {et~al.} 2022, in Society
  of Photo-Optical Instrumentation Engineers (SPIE) Conference Series, Vol.
  12184, Ground-based and Airborne Instrumentation for Astronomy IX, ed.
  {Evans}, C.~J., {Bryant}, J.~J., \& {Motohara}, K.,
  \href{https://ui.adsabs.harvard.edu/abs/2022SPIE12184E..21B}{1218421}

\bibitem[{{Bryan} {et~al.}(2020){Bryan}, {Ginzburg}, {Chiang}, {Morley},
  {Bowler}, {Xuan}, \& {Knutson}}]{bryan20}
{Bryan}, M.~L., {Ginzburg}, S., {Chiang}, E., {et~al.} 2020,
  \href{http://dx.doi.org/10.3847/1538-4357/abc0ef}{\color{magenta}\apj},
  \href{https://ui.adsabs.harvard.edu/abs/2020ApJ...905...37B}{905, 37}

\bibitem[{{B{\"u}rgy} {et~al.}(2026){B{\"u}rgy}, {Benisty}, {Alqubelat},
  {Manara}, \& {Facchini}}]{buergy26}
{B{\"u}rgy}, C.~J., {Benisty}, M., {Alqubelat}, H., {Manara}, C.~F., \&
  {Facchini}, S. 2026,
  \href{http://dx.doi.org/10.1051/0004-6361/202662086}{\color{magenta}\aap},
  \href{https://ui.adsabs.harvard.edu/abs/2026A&A...713L...1B}{713, L1}

\bibitem[{{Campbell-White} {et~al.}(2023){Campbell-White}, {Manara}, {Benisty},
  {Natta}, {Claes}, {Frasca}, {Bae}, {Facchini}, {Isella}, {P{\'e}rez},
  {Pinilla}, {Sicilia-Aguilar}, \& {Teague}}]{campbell23}
{Campbell-White}, J., {Manara}, C.~F., {Benisty}, M., {et~al.} 2023,
  \href{http://dx.doi.org/10.3847/1538-4357/acf0c0}{\color{magenta}\apj},
  \href{https://ui.adsabs.harvard.edu/abs/2023ApJ...956...25C}{956, 25}

\bibitem[{{Cardelli} {et~al.}(1989){Cardelli}, {Clayton}, \&
  {Mathis}}]{cardelli89}
{Cardelli}, J.~A., {Clayton}, G.~C., \& {Mathis}, J.~S. 1989,
  \href{http://dx.doi.org/10.1086/167900}{\color{magenta}\apj},
  \href{https://ui.adsabs.harvard.edu/abs/1989ApJ...345..245C}{345, 245}

\bibitem[{{Carlomagno} {et~al.}(2020{\natexlab{a}}){Carlomagno}, {Delacroix},
  {Absil}, {Cantalloube}, {de Xivry}, {Pathak}, {Agocs}, {Bertram}, {Brandl},
  {Burtscher}, {Doelman}, {Feldt}, {Glauser}, {Hippler}, {Kenworthy}, {Por},
  {Snik}, {Stuik}, \& {van Boekel}}]{carlomagno20err}
{Carlomagno}, B., {Delacroix}, C., {Absil}, O., {et~al.} 2020{\natexlab{a}},
  \href{http://dx.doi.org/10.1117/1.JATIS.6.4.049801}{\color{magenta}\jatis},
  \href{https://ui.adsabs.harvard.edu/abs/2020JATIS...6d9801C}{6, 049801}

\bibitem[{{Carlomagno} {et~al.}(2020{\natexlab{b}}){Carlomagno}, {Delacroix},
  {Absil}, {Cantalloube}, {Orban de Xivry}, {Pathak}, {Agocs}, {Bertram},
  {Brandl}, {Burtscher}, {Feldt}, {Glauser}, {Hippler}, {Kenworthy}, {Stuik},
  \& {van Boekel}}]{carlomagno20}
{Carlomagno}, B., {Delacroix}, C., {Absil}, O., {et~al.} 2020{\natexlab{b}},
  \href{http://dx.doi.org/10.1117/1.JATIS.6.3.035005}{\color{magenta}\jatis},
  \href{https://ui.adsabs.harvard.edu/abs/2020JATIS...6c5005C}{6, 035005}

\bibitem[{{Carvalho} \& {Johns-Krull}(2023)}]{carvalho23}
{Carvalho}, A. \& {Johns-Krull}, C.~M. 2023,
  \href{http://dx.doi.org/10.3847/2515-5172/acd37e}{\color{magenta}\rnaas},
  \href{https://ui.adsabs.harvard.edu/abs/2023RNAAS...7...91C}{7, 91}

\bibitem[{{Casassus} {et~al.}(2026){Casassus}, {C{\'a}rcamo},
  {Dom{\'\i}nguez-Jamett}, {Aoyama}, {Marleau}, {Chrenko}, {Baobab Liu}, \&
  {Ercolano}}]{casassus26}
{Casassus}, S., {C{\'a}rcamo}, M., {Dom{\'\i}nguez-Jamett}, O., {et~al.} 2026,
  \href{http://dx.doi.org/10.1051/0004-6361/202659553}{\color{magenta}\aap},
  \href{https://ui.adsabs.harvard.edu/abs/2026A&A...710L..10C}{710, L10}

\bibitem[{{Chazelas} {et~al.}(2020){Chazelas}, {Lovis}, {Blind}, {K{\"u}hn},
  {Genolet}, {Hughes}, {Turbet}, {Hagelberg}, {Restori}, {Kasper}, \& {Cerpa
  Urra}}]{chazelas20}
{Chazelas}, B., {Lovis}, C., {Blind}, N., {et~al.} 2020, in SPIE Conf.\ Ser.\@,
  Vol. 11448, SPIE Conf.\ Ser.\@,
  \href{https://ui.adsabs.harvard.edu/abs/2020SPIE11448E..75C}{1144875}

\bibitem[{{Chen} \& {Szul{\'a}gyi}(2022)}]{chen22}
{Chen}, X. \& {Szul{\'a}gyi}, J. 2022,
  \href{http://dx.doi.org/10.1093/mnras/stac1976}{\color{magenta}\mnras},
  \href{https://ui.adsabs.harvard.edu/abs/2022MNRAS.516..506C}{516, 506}

\bibitem[{{Chiar} \& {Tielens}(2006)}]{chiar06}
{Chiar}, J.~E. \& {Tielens}, A.~G.~G.~M. 2006,
  \href{http://dx.doi.org/10.1086/498406}{\color{magenta}\apj},
  \href{https://ui.adsabs.harvard.edu/abs/2006ApJ...637..774C}{637, 774}

\bibitem[{{Christensen} {et~al.}(2009){Christensen}, {Holzwarth}, \&
  {Reiners}}]{christensen09}
{Christensen}, U.~R., {Holzwarth}, V., \& {Reiners}, A. 2009,
  \href{http://dx.doi.org/10.1038/nature07626}{\color{magenta}\nat},
  \href{https://ui.adsabs.harvard.edu/abs/2009Natur.457..167C}{457, 167}

\bibitem[{{Christiaens} {et~al.}(2019{\natexlab{a}}){Christiaens},
  {Cantalloube}, {Casassus}, {Price}, {Absil}, {Pinte}, {Girard}, \&
  {Montesinos}}]{christiaens19a}
{Christiaens}, V., {Cantalloube}, F., {Casassus}, S., {et~al.}
  2019{\natexlab{a}},
  \href{http://dx.doi.org/10.3847/2041-8213/ab212b}{\color{magenta}\apjl},
  \href{https://ui.adsabs.harvard.edu/abs/2019ApJ...877L..33C}{877, L33}

\bibitem[{{Christiaens} {et~al.}(2019{\natexlab{b}}){Christiaens}, {Casassus},
  {Absil}, {Cantalloube}, {Gomez Gonzalez}, {Girard}, {Ram{\'\i}rez}, {Pairet},
  {Salinas}, {Price}, {Pinte}, {Quanz}, {Jord{\'a}n}, {Mawet}, \&
  {Wahhaj}}]{christiaens19b}
{Christiaens}, V., {Casassus}, S., {Absil}, O., {et~al.} 2019{\natexlab{b}},
  \href{http://dx.doi.org/10.1093/mnras/stz1232}{\color{magenta}\mnras},
  \href{https://ui.adsabs.harvard.edu/abs/2019MNRAS.486.5819C}{486, 5819}

\bibitem[{{Christiaens} {et~al.}(2024){Christiaens}, {Samland}, {Henning},
  {Portilla-Revelo}, {Perotti}, {Matthews}, {Absil}, {Decin}, {Kamp},
  {Boccaletti}, {Tabone}, {Marleau}, {van Dishoeck}, {G{\"u}del}, {Lagage},
  {Barrado}, {Caratti o Garatti}, {Glauser}, {Olofsson}, {Ray}, {Scheithauer},
  {Vandenbussche}, {Waters}, {Arabhavi}, {Grant}, {Jang}, {Kanwar},
  {Schreiber}, {Schwarz}, {Temmink}, \& {{\"O}stlin}}]{christiaens24}
{Christiaens}, V., {Samland}, M., {Henning}, T., {et~al.} 2024,
  \href{http://dx.doi.org/10.1051/0004-6361/202349089}{\color{magenta}\aap},
  \href{https://ui.adsabs.harvard.edu/abs/2024A%26A...685L...1C}{685, L1}

\bibitem[{{Christiaens} {et~al.}(2025){Christiaens}, {Samland}, {Henning},
  {Portilla-Revelo}, {Perotti}, {Matthews}, {Absil}, {Decin}, {Kamp},
  {Boccaletti}, {Tabone}, {Marleau}, {van Dishoeck}, {G{\"u}del}, {Lagage},
  {Barrado}, {Caratti o Garatti}, {Glauser}, {Olofsson}, {Ray}, {Scheithauer},
  {Vandenbussche}, {Waters}, {Arabhavi}, {Grant}, {Jang}, {Kanwar},
  {Schreiber}, {Schwarz}, {Temmink}, \& {{\"O}stlin}}]{christiaens25}
{Christiaens}, V., {Samland}, M., {Henning}, T., {et~al.} 2025,
  \href{http://dx.doi.org/10.1051/0004-6361/202451943e}{\color{magenta}\aap},
  \href{https://ui.adsabs.harvard.edu/abs/2025A%26A...694C...2C}{694, C2}

\bibitem[{{Claret}(2000)}]{claret00}
{Claret}, A. 2000, \aap,
  \href{https://ui.adsabs.harvard.edu/abs/2000A&A...363.1081C}{363, 1081}

\bibitem[{{Close} {et~al.}(2025{\natexlab{a}}){Close}, {Males}, {Li},
  {Haffert}, {Long}, {Hedglen}, {Weinberger}, {Follette}, {Apai}, {Doyon},
  {Foster}, {Gasho}, {Van Gorkom}, {Guyon}, {Kautz}, {Kueny}, {Lumbres},
  {McLeod}, {McEwen}, {Pavao}, {Pearce}, {Perez}, {Schatz}, {Szul{\'a}gyi},
  {Wagner}, \& {Wu}}]{close25a}
{Close}, L.~M., {Males}, J.~R., {Li}, J., {et~al.} 2025{\natexlab{a}},
  \href{http://dx.doi.org/10.3847/1538-3881/ad8648}{\color{magenta}\aj},
  \href{https://ui.adsabs.harvard.edu/abs/2025AJ....169...35C}{169, 35}

\bibitem[{{Close} {et~al.}(2025{\natexlab{b}}){Close}, {van Capelleveen},
  {Weible}, {Wagner}, {Haffert}, {Males}, {Ilyin}, {Kenworthy}, {Li}, {Long},
  {Ertel}, {Ginski}, {Weinberger}, {Follette}, {Liberman}, {Twitchell},
  {Johnson}, {Kueny}, {Apai}, {Doyon}, {Foster}, {Gasho}, {Van Gorkom},
  {Guyon}, {Kautz}, {McLeod}, {McEwen}, {Pearce}, {Schatz}, {Hedglen}, {Wu},
  {Isbell}, {Power}, {Carlson}, {Close}, {Tonucci}, \& {Mars}}]{close25b}
{Close}, L.~M., {van Capelleveen}, R.~F., {Weible}, G., {et~al.}
  2025{\natexlab{b}},
  \href{http://dx.doi.org/10.3847/2041-8213/adf7a5}{\color{magenta}\apjl},
  \href{https://ui.adsabs.harvard.edu/abs/2025ApJ...990L...9C}{990, L9}

\bibitem[{{Cugno} {et~al.}(2025){Cugno}, {Facchini}, {Alarcon}, {Bae},
  {Benisty}, {Eilers}, {Leung}, {Meyer}, {Pueyo}, {Teague}, {Bergin}, {Girard},
  {Helled}, {Huang}, \& {Leisenring}}]{cugno25b}
{Cugno}, G., {Facchini}, S., {Alarcon}, F., {et~al.} 2025,
  \href{http://dx.doi.org/10.3847/1538-3881/ae0acd}{\color{magenta}\aj},
  \href{https://ui.adsabs.harvard.edu/abs/2025AJ....170..317C}{170, 317}

\bibitem[{{Cugno} {et~al.}(2019){Cugno}, {Quanz}, {Hunziker}, {Stolker},
  {Schmid}, {Avenhaus}, {Baudoz}, {Bohn}, {Bonnefoy}, {Buenzli}, {Chauvin},
  {Cheetham}, {Desidera}, {Dominik}, {Feautrier}, {Feldt}, {Ginski}, {Girard},
  {Gratton}, {Hagelberg}, {Hugot}, {Janson}, {Lagrange}, {Langlois}, {Magnard},
  {Maire}, {Menard}, {Meyer}, {Milli}, {Mordasini}, {Pinte}, {Pragt},
  {Roelfsema}, {Rigal}, {Szul{\'a}gyi}, {van Boekel}, {van der Plas}, {Vigan},
  {Wahhaj}, \& {Zurlo}}]{Cugno+2019}
{Cugno}, G., {Quanz}, S.~P., {Hunziker}, S., {et~al.} 2019,
  \href{http://dx.doi.org/10.1051/0004-6361/201834170}{\color{magenta}\aap},
  \href{https://ui.adsabs.harvard.edu/abs/2019A%26A...622A.156C}{622, A156}

\bibitem[{{Currie} {et~al.}(2025){Currie}, {Hashimoto}, {Aoyama}, {Dong},
  {Fukagawa}, {Muto}, {Dykes}, {El Morsy}, \& {Tamura}}]{currie25b}
{Currie}, T., {Hashimoto}, J., {Aoyama}, Y., {et~al.} 2025,
  \href{http://dx.doi.org/10.3847/2041-8213/adf7a0}{\color{magenta}\apjl},
  \href{https://ui.adsabs.harvard.edu/abs/2025ApJ...990L..42C}{990, L42}

\bibitem[{{Cushing} {et~al.}(2026){Cushing}, {Trucks}, {Hardegree-Ullman},
  {Burgasser}, {Carey}, {Fortney}, {Gelino}, {Gizis}, {Kirkpatrick}, {Leggett},
  {Mace}, {Marley}, \& {Morley}}]{cushing26}
{Cushing}, M.~C., {Trucks}, J.~L., {Hardegree-Ullman}, K.~K., {et~al.} 2026,
  \href{http://dx.doi.org/10.3847/1538-3881/ae8e7c}{\color{magenta}\aj},
  \href{https://ui.adsabs.harvard.edu/abs/2026AJ....172..166C}{172, 166}

\bibitem[{{de Regt} {et~al.}(2025){de Regt}, {Snellen}, {Allard}, {Gonz{\'a}lez
  Picos}, {Gandhi}, {Grasser}, {Landman}, {Molli{\`e}re}, {Nasedkin},
  {Stolker}, \& {Zhang}}]{deregt25}
{de Regt}, S., {Snellen}, I.~A.~G., {Allard}, N.~F., {et~al.} 2025,
  \href{http://dx.doi.org/10.1051/0004-6361/202453190}{\color{magenta}\aap},
  \href{https://ui.adsabs.harvard.edu/abs/2025A%26A...696A.225D}{696, A225}

\bibitem[{{Delacroix} {et~al.}(2022){Delacroix}, {Absil}, {Orban de Xivry},
  {Shinde}, {Pathak}, {Cantalloube}, {Carlomagno}, {Christiaens}, {Bon{\'e}},
  {Dolkens}, {Kenworthy}, \& {Doelman}}]{delacroix22}
{Delacroix}, C., {Absil}, O., {Orban de Xivry}, G., {et~al.} 2022, in Society
  of Photo-Optical Instrumentation Engineers (SPIE) Conference Series, Vol.
  12187, Modeling, Systems Engineering, and Project Management for Astronomy X,
  ed. {Angeli}, G.~Z. \& {Dierickx}, P.,
  \href{https://ui.adsabs.harvard.edu/abs/2022SPIE12187E..0FD}{121870F}

\bibitem[{{Demars} {et~al.}(2023){Demars}, {Bonnefoy}, {Dougados}, {Aoyama},
  {Thanathibodee}, {Marleau}, {Tremblin}, {Delorme}, {Palma-Bifani}, {Petrus},
  {Bowler}, {Chauvin}, \& {Lagrange}}]{demars23}
{Demars}, D., {Bonnefoy}, M., {Dougados}, C., {et~al.} 2023,
  \href{http://dx.doi.org/10.1051/0004-6361/202346221}{\color{magenta}\aap},
  \href{https://ui.adsabs.harvard.edu/abs/2023A%26A...676A.123D}{676, A123}

\bibitem[{{Demars} {et~al.}(2026){Demars}, {Bonnefoy}, {Dougados}, {Viswanath},
  {Ringqvist}, {Janson}, {Aoyama}, {Thanathibodee}, {Marleau}, {Manara},
  {Rigliaco}, {Szul{\'a}gyi}, {Sicilia-Aguilar}, {Bouvier}, {Alecian},
  {Petrus}, \& {Houll{\'e}}}]{demars26}
{Demars}, D., {Bonnefoy}, M., {Dougados}, C., {et~al.} 2026,
  \href{http://dx.doi.org/10.1051/0004-6361/202554644}{\color{magenta}\aap},
  \href{https://ui.adsabs.harvard.edu/abs/2026A%26A...706A..57D}{706, A57}

\bibitem[{{\relax DLMF}(2026)}]{dlmf126}
{\relax DLMF}. 2026, {\it NIST Digital Library of Mathematical Functions},
  \url{https://dlmf.nist.gov/}, Release 1.2.6 of 2026-03-15, f.~W.~J. Olver,
  A.~B. {Olde Daalhuis}, D.~W. Lozier, B.~I. Schneider, R.~F. Boisvert, C.~W.
  Clark, B.~R. Miller, B.~V. Saunders, H.~S. Cohl, and M.~A. McClain, eds.

\bibitem[{{Do {\'O}} {et~al.}(2025){Do {\'O}}, {Bae}, {Konopacky}, {Nguyen},
  {Diamond}, {Go{\'z}dziewski}, \& {Jankowski}}]{do25}
{Do {\'O}}, C.~R., {Bae}, J., {Konopacky}, Q.~M., {et~al.} 2025,
  \href{http://dx.doi.org/10.3847/1538-4357/ae12ec}{\color{magenta}\apj},
  \href{https://ui.adsabs.harvard.edu/abs/2025ApJ...995..190D}{995, 190}

\bibitem[{{Doi} {et~al.}(2024){Doi}, {Kataoka}, {Liu}, {Yoshida}, {Benisty},
  {Dong}, {Yamato}, \& {Hashimoto}}]{doi24}
{Doi}, K., {Kataoka}, A., {Liu}, H.~B., {et~al.} 2024,
  \href{http://dx.doi.org/10.3847/2041-8213/ad7f51}{\color{magenta}\apjl},
  \href{https://ui.adsabs.harvard.edu/abs/2024ApJ...974L..25D}{974, L25}

\bibitem[{{Dom{\'\i}nguez-Jamett} {et~al.}(2025){Dom{\'\i}nguez-Jamett},
  {Casassus}, {Baobab Liu}, {Aoyama}, {C{\'a}rcamo}, {Weber}, {Chrenko},
  {Marleau}, {Ercolano}, \& {Szul{\'a}gyi}}]{dom25}
{Dom{\'\i}nguez-Jamett}, O., {Casassus}, S., {Baobab Liu}, H., {et~al.} 2025,
  \href{http://dx.doi.org/10.1051/0004-6361/202554485}{\color{magenta}\aap},
  \href{https://ui.adsabs.harvard.edu/abs/2025A%26A...702A..18D}{702, A18}

\bibitem[{{Donati} {et~al.}(2024){Donati}, {Cristofari}, {Alencar},
  {K{\'o}sp{\'a}l}, {Bouvier}, {Moutou}, {Carmona}, {Gregorio-Hetem},
  {M{\'e}nard}, {Artigau}, {Doyon}, {Takami}, {Shang}, {Dias do Nascimento},
  {M{\'e}nard}, {Gaidos}, \& {SPIRou Science Team}}]{donati24}
{Donati}, J.~F., {Cristofari}, P.~I., {Alencar}, S.~H.~P., {et~al.} 2024,
  \href{http://dx.doi.org/10.1093/mnras/stae2506}{\color{magenta}\mnras},
  \href{https://ui.adsabs.harvard.edu/abs/2024MNRAS.535.3363D}{535, 3363}

\bibitem[{{Dong} {et~al.}(2021){Dong}, {Jiang}, \& {Armitage}}]{dong21}
{Dong}, J., {Jiang}, Y.-F., \& {Armitage}, P.~J. 2021,
  \href{http://dx.doi.org/10.3847/1538-4357/ac1941}{\color{magenta}\apj},
  \href{https://ui.adsabs.harvard.edu/abs/2021ApJ...921...54D}{921, 54}

\bibitem[{{Dorn} {et~al.}(2023){Dorn}, {Bristow}, {Smoker}, {Rodler}, {Lavail},
  {Accardo}, {van den Ancker}, {Baade}, {Baruffolo}, {Courtney-Barrer},
  {Blanco}, {Brucalassi}, {Cumani}, {Follert}, {Haimerl}, {Hatzes}, {Haug},
  {Heiter}, {Hinterschuster}, {Hubin}, {Ives}, {Jung}, {Jones}, {Kaeufl},
  {Kirchbauer}, {Klein}, {Kochukhov}, {Korhonen}, {K{\"o}hler}, {Lizon},
  {Moins}, {Molina-Conde}, {Marquart}, {Neeser}, {Oliva}, {Pallanca},
  {Pasquini}, {Paufique}, {Piskunov}, {Reiners}, {Schneller}, {Schmutzer},
  {Seemann}, {Slumstrup}, {Smette}, {Stegmeier}, {Stempels}, {Tordo},
  {Valenti}, {Valenzuela}, {Vernet}, {Vinther}, \& {Wehrhahn}}]{dorn23}
{Dorn}, R.~J., {Bristow}, P., {Smoker}, J.~V., {et~al.} 2023,
  \href{http://dx.doi.org/10.1051/0004-6361/202245217}{\color{magenta}\aap},
  \href{https://ui.adsabs.harvard.edu/abs/2023A&A...671A..24D}{671, A24}

\bibitem[{{Draine} \& {Bertoldi}(1996)}]{draine96}
{Draine}, B.~T. \& {Bertoldi}, F. 1996,
  \href{http://dx.doi.org/10.1086/177689}{\color{magenta}\apj},
  \href{https://ui.adsabs.harvard.edu/abs/1996ApJ...468..269D}{468, 269}

\bibitem[{{Dullemond} {et~al.}(2012){Dullemond}, {Juhasz}, {Pohl}, {Sereshti},
  {Shetty}, {Peters}, {Commercon}, \& {Flock}}]{dullemond12}
{Dullemond}, C.~P., {Juhasz}, A., {Pohl}, A., {et~al.} 2012, {RADMC-3D: A
  multi-purpose radiative transfer tool},
  \href{https://ascl.net/1202.015}{\color{magenta}{Astrophysics Source Code
  Library}},
  \href{https://ui.adsabs.harvard.edu/abs/2012ascl.soft02015D}{record
  ascl:1202.015}

\bibitem[{{Edwards} {et~al.}(1994){Edwards}, {Hartigan}, {Ghandour}, \&
  {Andrulis}}]{edwards94}
{Edwards}, S., {Hartigan}, P., {Ghandour}, L., \& {Andrulis}, C. 1994,
  \href{http://dx.doi.org/10.1086/117134}{\color{magenta}\aj},
  \href{https://ui.adsabs.harvard.edu/abs/1994AJ....108.1056E}{108, 1056}

\bibitem[{{Eriksson} {et~al.}(2020){Eriksson}, {Asensio Torres}, {Janson},
  {Aoyama}, {Marleau}, {Bonnefoy}, \& {Petrus}}]{eriksson20}
{Eriksson}, S.~C., {Asensio Torres}, R., {Janson}, M., {et~al.} 2020,
  \href{http://dx.doi.org/10.1051/0004-6361/202038131}{\color{magenta}\aap},
  \href{https://ui.adsabs.harvard.edu/abs/2020A%26A...638L...6E}{638, L6}

\bibitem[{{Erkal} {et~al.}(2022){Erkal}, {Manara}, {Schneider}, {Vincenzi},
  {Nisini}, {Coffey}, {Alcal{\'a}}, {Fedele}, \& {Antoniucci}}]{erkal22}
{Erkal}, J., {Manara}, C.~F., {Schneider}, P.~C., {et~al.} 2022,
  \href{http://dx.doi.org/10.1051/0004-6361/202244254}{\color{magenta}\aap},
  \href{https://ui.adsabs.harvard.edu/abs/2022A%26A...666A.188E}{666, A188}

\bibitem[{{Faruqi} {et~al.}(2026){Faruqi}, {Speedie}, {Pudritz}, \&
  {Meru}}]{faruqi26}
{Faruqi}, A., {Speedie}, J., {Pudritz}, R.~E., \& {Meru}, F. 2026,
  \href{http://dx.doi.org/10.3847/1538-4357/ae6272}{\color{magenta}\apj},
  \href{https://ui.adsabs.harvard.edu/abs/2026ApJ..1003..215F}{1003, 215}

\bibitem[{{Feldt} {et~al.}(2024){Feldt}, {Bertram}, {Correia}, {Absil},
  {C{\'a}rdenas V{\'a}zquez}, {Coppejans}, {Kulas}, {Obereder}, {Orban de
  Xivry}, {Scheithauer}, \& {Steuer}}]{feldt24}
{Feldt}, M., {Bertram}, T., {Correia}, C., {et~al.} 2024,
  \href{http://dx.doi.org/10.1007/s10686-024-09968-2}{\color{magenta}Experimental
  Astronomy}, \href{https://ui.adsabs.harvard.edu/abs/2024ExA....58...20F}{58,
  20}

\bibitem[{{Fendt}(2003)}]{fendt03}
{Fendt}, C. 2003,
  \href{http://dx.doi.org/10.1051/0004-6361:20034154}{\color{magenta}\aap},
  \href{https://ui.adsabs.harvard.edu/abs/2003A%26A...411..623F}{411, 623}

\bibitem[{{Fernandes} {et~al.}(2019){Fernandes}, {Mulders}, {Pascucci},
  {Mordasini}, \& {Emsenhuber}}]{fernandes19}
{Fernandes}, R.~B., {Mulders}, G.~D., {Pascucci}, I., {Mordasini}, C., \&
  {Emsenhuber}, A. 2019,
  \href{http://dx.doi.org/10.3847/1538-4357/ab0300}{\color{magenta}\apj},
  \href{https://ui.adsabs.harvard.edu/abs/2019ApJ...874...81F}{874, 81}

\bibitem[{{Fiorellino} {et~al.}(2025){Fiorellino}, {Alcal{\'a}}, {Manara},
  {Pittman}, {{\'A}brah{\'a}m}, {Venuti}, {Cabrit}, {Claes}, {Fang},
  {K{\'o}sp{\'a}l}, {Lodato}, {Mauco}, \& {Tychoniec}}]{fiorellino25}
{Fiorellino}, E., {Alcal{\'a}}, J.~M., {Manara}, C.~F., {et~al.} 2025,
  \href{http://dx.doi.org/10.1051/0004-6361/202556603}{\color{magenta}\aap},
  \href{https://ui.adsabs.harvard.edu/abs/2025A&A...704A..42F}{704, A42}

\bibitem[{{Flores-Rivera} {et~al.}(2023){Flores-Rivera}, {Flock}, {Kurtovic},
  {Husemann}, {Banzatti}, {Ringqvist}, {Kamann}, {M{\"u}ller}, {Fendt},
  {Garc{\'\i}a Lopez}, {Marleau}, {Henning}, {Carrasco-Gonz{\'a}lez}, {van
  Boekel}, {Keppler}, {Launhardt}, \& {Aoyama}}]{floresrivera23}
{Flores-Rivera}, L., {Flock}, M., {Kurtovic}, N.~T., {et~al.} 2023,
  \href{http://dx.doi.org/10.1051/0004-6361/202141664}{\color{magenta}\aap},
  \href{https://ui.adsabs.harvard.edu/abs/2023A%26A...670A.126F}{670, A126}

\bibitem[{{Follette} {et~al.}(2023){Follette}, {Close}, {Males}, {Ward-Duong},
  {Balmer}, {Redai}, {Morales}, {Sarosi}, {Dacus}, {De Rosa}, {Garcia Toro},
  {Leonard}, {Macintosh}, {Morzinski}, {Mullen}, {Palmo}, {Saitoti}, {Spiro},
  {Treiber}, {Wagner}, {Wang}, {Wang}, {Watson}, \& {Weinberger}}]{follette23}
{Follette}, K.~B., {Close}, L.~M., {Males}, J.~R., {et~al.} 2023,
  \href{http://dx.doi.org/10.3847/1538-3881/acc183}{\color{magenta}\aj},
  \href{https://ui.adsabs.harvard.edu/abs/2023AJ....165..225F}{165, 225}

\bibitem[{{Fortney} {et~al.}(2005){Fortney}, {Marley}, {Hubickyj},
  {Bodenheimer}, \& {Lissauer}}]{fort05}
{Fortney}, J.~J., {Marley}, M.~S., {Hubickyj}, O., {Bodenheimer}, P., \&
  {Lissauer}, J.~J. 2005,
  \href{http://dx.doi.org/10.1002/asna.200510465}{\color{magenta}Astron.\
  Nachrichten},
  \href{https://ui.adsabs.harvard.edu/abs/2005AN....326..925F}{326, 925}

\bibitem[{{Fulton} {et~al.}(2021){Fulton}, {Rosenthal}, {Hirsch}, {Isaacson},
  {Howard}, {Dedrick}, {Sherstyuk}, {Blunt}, {Petigura}, {Knutson}, {Behmard},
  {Chontos}, {Crepp}, {Crossfield}, {Dalba}, {Fischer}, {Henry}, {Kane},
  {Kosiarek}, {Marcy}, {Rubenzahl}, {Weiss}, \& {Wright}}]{fulton21}
{Fulton}, B.~J., {Rosenthal}, L.~J., {Hirsch}, L.~A., {et~al.} 2021,
  \href{http://dx.doi.org/10.3847/1538-4365/abfcc1}{\color{magenta}\apjs},
  \href{https://ui.adsabs.harvard.edu/abs/2021ApJS..255...14F}{255, 14}

\bibitem[{{Fung} {et~al.}(2019){Fung}, {Zhu}, \& {Chiang}}]{fung19}
{Fung}, J., {Zhu}, Z., \& {Chiang}, E. 2019,
  \href{http://dx.doi.org/10.3847/1538-4357/ab53da}{\color{magenta}\apj},
  \href{https://ui.adsabs.harvard.edu/abs/2019ApJ...887..152F}{887, 152}

\bibitem[{{Gaia Collaboration} {et~al.}(2018){Gaia Collaboration}, {Brown},
  {Vallenari}, {Prusti}, {de Bruijne}, {Babusiaux}, {Bailer-Jones}, {Biermann},
  {Evans}, {Eyer}, {Jansen}, {Jordi}, {Klioner}, {Lammers}, {Lindegren},
  {Luri}, {Mignard}, {Panem}, {Pourbaix}, {Randich}, {Sartoretti}, {Siddiqui},
  {Soubiran}, {van Leeuwen}, {Walton}, {Arenou}, {Bastian}, {Cropper},
  {Drimmel}, {Katz}, {Lattanzi}, {Bakker}, {Cacciari}, {Casta{\~n}eda},
  {Chaoul}, {Cheek}, {De Angeli}, {Fabricius}, {Guerra}, {Holl}, {Masana},
  {Messineo}, {Mowlavi}, {Nienartowicz}, {Panuzzo}, {Portell}, {Riello},
  {Seabroke}, {Tanga}, {Th{\'e}venin}, {Gracia-Abril}, {Comoretto},
  {Garcia-Reinaldos}, {Teyssier}, {Altmann}, {Andrae}, {Audard},
  {Bellas-Velidis}, {Benson}, {Berthier}, {Blomme}, {Burgess}, {Busso},
  {Carry}, {Cellino}, {Clementini}, {Clotet}, {Creevey}, {Davidson}, {De
  Ridder}, {Delchambre}, {Dell'Oro}, {Ducourant},
  {Fern{\'a}ndez-Hern{\'a}ndez}, {Fouesneau}, {Fr{\'e}mat}, {Galluccio},
  {Garc{\'\i}a-Torres}, {Gonz{\'a}lez-N{\'u}{\~n}ez}, {Gonz{\'a}lez-Vidal},
  {Gosset}, {Guy}, {Halbwachs}, {Hambly}, {Harrison}, {Hern{\'a}ndez},
  {Hestroffer}, {Hodgkin}, {Hutton}, {Jasniewicz}, {Jean-Antoine-Piccolo},
  {Jordan}, {Korn}, {Krone-Martins}, {Lanzafame}, {Lebzelter}, {L{\"o}ffler},
  {Manteiga}, {Marrese}, {Mart{\'\i}n-Fleitas}, {Moitinho}, {Mora}, {Muinonen},
  {Osinde}, {Pancino}, {Pauwels}, {Petit}, {Recio-Blanco}, {Richards},
  {Rimoldini}, {Robin}, {Sarro}, {Siopis}, {Smith}, {Sozzetti}, {S{\"u}veges},
  {Torra}, {van Reeven}, {Abbas}, {Abreu Aramburu}, {Accart}, {Aerts},
  {Altavilla}, {{\'A}lvarez}, {Alvarez}, {Alves}, {Anderson}, {Andrei},
  {Anglada Varela}, {Antiche}, {Antoja}, {Arcay}, {Astraatmadja}, {Bach},
  {Baker}, {Balaguer-N{\'u}{\~n}ez}, {Balm}, {Barache}, {Barata}, {Barbato},
  {Barblan}, {Barklem}, {Barrado}, {Barros}, {Barstow}, {Bartholom{\'e}
  Mu{\~n}oz}, {Bassilana}, {Becciani}, {Bellazzini}, {Berihuete}, {Bertone},
  {Bianchi}, {Bienaym{\'e}}, {Blanco-Cuaresma}, {Boch}, {Boeche}, {Bombrun},
  {Borrachero}, {Bossini}, {Bouquillon}, {Bourda}, {Bragaglia}, {Bramante},
  {Breddels}, {Bressan}, {Brouillet}, {Br{\"u}semeister}, {Brugaletta},
  {Bucciarelli}, {Burlacu}, {Busonero}, {Butkevich}, {Buzzi}, {Caffau},
  {Cancelliere}, {Cannizzaro}, {Cantat-Gaudin}, {Carballo}, {Carlucci},
  {Carrasco}, {Casamiquela}, {Castellani}, {Castro-Ginard}, {Charlot},
  {Chemin}, {Chiavassa}, {Cocozza}, {Costigan}, {Cowell}, {Crifo}, {Crosta},
  {Crowley}, {Cuypers}, {Dafonte}, {Damerdji}, {Dapergolas}, {David}, {David},
  {de Laverny}, \& {De Luise}}]{gDR2}
{Gaia Collaboration}, {Brown}, A.~G.~A., {Vallenari}, A., {et~al.} 2018,
  \href{http://dx.doi.org/10.1051/0004-6361/201833051}{\color{magenta}\aap},
  \href{https://ui.adsabs.harvard.edu/abs/2018A%26A...616A...1G}{616, A1}

\bibitem[{{Gaia Collaboration} {et~al.}(2023){Gaia Collaboration}, {Vallenari},
  {Brown}, {Prusti}, {de Bruijne}, {Arenou}, {Babusiaux}, {Biermann},
  {Creevey}, {Ducourant}, {Evans}, {Eyer}, {Guerra}, {Hutton}, {Jordi},
  {Klioner}, {Lammers}, {Lindegren}, {Luri}, {Mignard}, {Panem}, {Pourbaix},
  {Randich}, {Sartoretti}, {Soubiran}, {Tanga}, {Walton}, {Bailer-Jones},
  {Bastian}, {Drimmel}, {Jansen}, {Katz}, {Lattanzi}, {van Leeuwen}, {Bakker},
  {Cacciari}, {Casta{\~n}eda}, {De Angeli}, {Fabricius}, {Fouesneau},
  {Fr{\'e}mat}, {Galluccio}, {Guerrier}, {Heiter}, {Masana}, {Messineo},
  {Mowlavi}, {Nicolas}, {Nienartowicz}, {Pailler}, {Panuzzo}, {Riclet}, {Roux},
  {Seabroke}, {Sordo}, {Th{\'e}venin}, {Gracia-Abril}, {Portell}, {Teyssier},
  {Altmann}, {Andrae}, {Audard}, {Bellas-Velidis}, {Benson}, {Berthier},
  {Blomme}, {Burgess}, {Busonero}, {Busso}, {C{\'a}novas}, {Carry}, {Cellino},
  {Cheek}, {Clementini}, {Damerdji}, {Davidson}, {de Teodoro}, {Nu{\~n}ez
  Campos}, {Delchambre}, {Dell'Oro}, {Esquej}, {Fern{\'a}ndez-Hern{\'a}ndez},
  {Fraile}, {Garabato}, {Garc{\'\i}a-Lario}, {Gosset}, {Haigron}, {Halbwachs},
  {Hambly}, {Harrison}, {Hern{\'a}ndez}, {Hestroffer}, {Hodgkin}, {Holl},
  {Jan{\ss}en}, {Jevardat de Fombelle}, {Jordan}, {Krone-Martins}, {Lanzafame},
  {L{\"o}ffler}, {Marchal}, {Marrese}, {Moitinho}, {Muinonen}, {Osborne},
  {Pancino}, {Pauwels}, {Recio-Blanco}, {Reyl{\'e}}, {Riello}, {Rimoldini},
  {Roegiers}, {Rybizki}, {Sarro}, {Siopis}, {Smith}, {Sozzetti}, {Utrilla},
  {van Leeuwen}, {Abbas}, {{\'A}brah{\'a}m}, {Abreu Aramburu}, {Aerts},
  {Aguado}, {Ajaj}, {Aldea-Montero}, {Altavilla}, {{\'A}lvarez}, {Alves},
  {Anders}, {Anderson}, {Anglada Varela}, {Antoja}, {Baines}, {Baker},
  {Balaguer-N{\'u}{\~n}ez}, {Balbinot}, {Balog}, {Barache}, {Barbato},
  {Barros}, {Barstow}, {Bartolom{\'e}}, {Bassilana}, {Bauchet}, {Becciani},
  {Bellazzini}, {Berihuete}, {Bernet}, {Bertone}, {Bianchi}, {Binnenfeld},
  {Blanco-Cuaresma}, {Blazere}, {Boch}, {Bombrun}, {Bossini}, {Bouquillon},
  {Bragaglia}, {Bramante}, {Breedt}, {Bressan}, {Brouillet}, {Brugaletta},
  {Bucciarelli}, {Burlacu}, {Butkevich}, {Buzzi}, {Caffau}, {Cancelliere},
  {Cantat-Gaudin}, {Carballo}, {Carlucci}, {Carnerero}, {Carrasco},
  {Casamiquela}, {Castellani}, {Castro-Ginard}, {Chaoul}, {Charlot}, {Chemin},
  {Chiaramida}, {Chiavassa}, {Chornay}, {Comoretto}, {Contursi}, {Cooper},
  {Cornez}, {Cowell}, {Crifo}, {Cropper}, {Crosta}, {Crowley}, {Dafonte},
  {Dapergolas}, {David}, {David}, {de Laverny}, {De Luise}, {De March}, {De
  Ridder}, {de Souza}, {de Torres}, {del Peloso}, {del Pozo}, {Delbo},
  {Delgado}, {Delisle}, {Demouchy}, {Dharmawardena}, {Di Matteo}, {Diakite},
  {Diener}, {Distefano}, {Dolding}, {Edvardsson}, {Enke}, {Fabre}, {Fabrizio},
  {Faigler}, {Fedorets}, {Fernique}, {Fienga}, {Figueras}, {Fournier},
  {Fouron}, {Fragkoudi}, {Gai}, {Garcia-Gutierrez}, {Garcia-Reinaldos},
  {Garc{\'\i}a-Torres}, {Garofalo}, {Gavel}, {Gavras}, {Gerlach}, {Geyer},
  {Giacobbe}, {Gilmore}, {Girona}, {Giuffrida}, {Gomel}, {Gomez},
  {Gonz{\'a}lez-N{\'u}{\~n}ez}, {Gonz{\'a}lez-Santamar{\'\i}a},
  {Gonz{\'a}lez-Vidal}, {Granvik}, {Guillout}, {Guiraud},
  {Guti{\'e}rrez-S{\'a}nchez}, {Guy}, {Hatzidimitriou}, {Hauser}, {Haywood},
  {Helmer}, {Helmi}, {Sarmiento}, {Hidalgo}, {Hilger}, {H{\l}adczuk}, {Hobbs},
  {Holland}, {Huckle}, {Jardine}, {Jasniewicz}, {Jean-Antoine Piccolo},
  {Jim{\'e}nez-Arranz}, {Jorissen}, {Juaristi Campillo}, {Julbe}, {Karbevska},
  {Kervella}, {Khanna}, {Kontizas}, {Kordopatis}, {Korn}, {K{\'o}sp{\'a}l},
  {Kostrzewa-Rutkowska}, {Kruszy{\'n}ska}, {Kun}, {Laizeau}, {Lambert},
  {Lanza}, {Lasne}, {Le Campion}, {Lebreton}, {Lebzelter}, {Leccia}, {Leclerc},
  {Lecoeur-Taibi}, {Liao}, {Licata}, {Lindstr{\o}m}, {Lister}, {Livanou},
  {Lobel}, {Lorca}, {Loup}, {Madrero Pardo}, {Magdaleno Romeo}, {Managau},
  {Mann}, {Manteiga}, {Marchant}, {Marconi}, {Marcos}, {Marcos Santos},
  {Mar{\'\i}n Pina}, {Marinoni}, {Marocco}, {Marshall}, {Martin Polo},
  {Mart{\'\i}n-Fleitas}, {Marton}, {Mary}, {Masip}, {Massari},
  {Mastrobuono-Battisti}, {Mazeh}, {McMillan}, {Messina}, {Michalik}, {Millar},
  {Mints}, {Molina}, {Molinaro}, {Moln{\'a}r}, {Monari}, {Mongui{\'o}},
  {Montegriffo}, {Montero}, {Mor}, {Mora}, {Morbidelli}, {Morel}, {Morris},
  {Muraveva}, {Murphy}, {Musella}, {Nagy}, {Noval}, {Oca{\~n}a}, {Ogden},
  {Ordenovic}, {Osinde}, {Pagani}, {Pagano}, {Palaversa}, {Palicio},
  {Pallas-Quintela}, {Panahi}, {Payne-Wardenaar}, {Pe{\~n}alosa Esteller},
  {Penttil{\"a}}, {Pichon}, {Piersimoni}, {Pineau}, {Plachy}, {Plum}, {Poggio},
  {Pr{\v{s}}a}, {Pulone}, {Racero}, {Ragaini}, {Rainer}, {Raiteri}, {Rambaux},
  {Ramos}, {Ramos-Lerate}, {Re Fiorentin}, {Regibo}, {Richards}, {Rios Diaz},
  {Ripepi}, {Riva}, {Rix}, {Rixon}, {Robichon}, {Robin}, {Robin}, {Roelens},
  {Rogues}, {Rohrbasser}, {Romero-G{\'o}mez}, {Rowell}, {Royer}, {Ruz Mieres},
  {Rybicki}, {Sadowski}, {S{\'a}ez N{\'u}{\~n}ez}, {Sagrist{\`a} Sell{\'e}s},
  {Sahlmann}, {Salguero}, {Samaras}, {Sanchez Gimenez}, {Sanna},
  {Santove{\~n}a}, {Sarasso}, {Schultheis}, {Sciacca}, {Segol}, {Segovia},
  {S{\'e}gransan}, {Semeux}, {Shahaf}, {Siddiqui}, {Siebert}, {Siltala},
  {Silvelo}, {Slezak}, {Slezak}, {Smart}, {Snaith}, {Solano}, {Solitro},
  {Souami}, {Souchay}, {Spagna}, {Spina}, {Spoto}, {Steele},
  {Steidelm{\"u}ller}, {Stephenson}, {S{\"u}veges}, {Surdej}, {Szabados},
  {Szegedi-Elek}, {Taris}, {Taylor}, {Teixeira}, {Tolomei}, {Tonello}, {Torra},
  {Torra}, {Torralba Elipe}, {Trabucchi}, {Tsounis}, {Turon}, {Ulla}, {Unger},
  {Vaillant}, {van Dillen}, {van Reeven}, {Vanel}, {Vecchiato}, {Viala},
  {Vicente}, {Voutsinas}, {Weiler}, {Wevers}, {Wyrzykowski}, {Yoldas}, {Yvard},
  {Zhao}, {Zorec}, {Zucker}, \& {Zwitter}}]{gDR3}
{Gaia Collaboration}, {Vallenari}, A., {Brown}, A.~G.~A., {et~al.} 2023,
  \href{http://dx.doi.org/10.1051/0004-6361/202243940}{\color{magenta}\aap},
  \href{https://ui.adsabs.harvard.edu/abs/2023A%26A...674A...1G}{674, A1}

\bibitem[{{Gaidos} {et~al.}(2024){Gaidos}, {Thanathibodee}, {Hoffman}, {Ong},
  {Hinkle}, {Shappee}, \& {Banzatti}}]{gaidos24}
{Gaidos}, E., {Thanathibodee}, T., {Hoffman}, A., {et~al.} 2024,
  \href{http://dx.doi.org/10.3847/1538-4357/ad3447}{\color{magenta}\apj},
  \href{https://ui.adsabs.harvard.edu/abs/2024ApJ...966..167G}{966, 167}

\bibitem[{{Gatti} {et~al.}(2006){Gatti}, {Testi}, {Natta}, {Randich}, \&
  {Muzerolle}}]{gatti06}
{Gatti}, T., {Testi}, L., {Natta}, A., {Randich}, S., \& {Muzerolle}, J. 2006,
  \href{http://dx.doi.org/10.1051/0004-6361:20066095}{\color{magenta}\aap},
  \href{https://ui.adsabs.harvard.edu/abs/2006A&A...460..547G}{460, 547}

\bibitem[{{Gharib-Nezhad} {et~al.}(2024){Gharib-Nezhad}, {Batalha}, {Chubb},
  {Freedman}, {Gordon}, {Gamache}, {Hargreaves}, {Lewis}, {Tennyson}, \&
  {Yurchenko}}]{gn24}
{Gharib-Nezhad}, E.~S., {Batalha}, N.~E., {Chubb}, K., {et~al.} 2024,
  \href{http://dx.doi.org/10.1093/rasti/rzad058}{\color{magenta}RAS Techniques
  and Instruments},
  \href{https://ui.adsabs.harvard.edu/abs/2024RASTI...3...44G}{3, 44}

\bibitem[{{Ginski} {et~al.}(2025){Ginski}, {Pinilla}, {Benisty}, {Pinte},
  {Claes}, {Mamajek}, {Kenworthy}, {Murphy}, {Manara}, {Bae}, {Birnstiel},
  {Byrne}, {Dominik}, {Facchini}, {Garufi}, {Gratton}, {Hogerheijde}, {van
  Holstein}, {Huang}, {Langlois}, {Lawlor}, {Ma}, {McLachlan}, {Menard},
  {Rigliaco}, {Ribas}, {Schmidt}, {Sierra}, {Tazaki}, {Williams}, \&
  {Zurlo}}]{ginski25}
{Ginski}, C., {Pinilla}, P., {Benisty}, M., {et~al.} 2025,
  \href{http://dx.doi.org/10.1051/0004-6361/202451647}{\color{magenta}\aap},
  \href{https://ui.adsabs.harvard.edu/abs/2025A%26A...699A.237G}{699, A237}

\bibitem[{{Gordon} {et~al.}(2022){Gordon}, {Rothman}, {Hargreaves}, {Hashemi},
  {Karlovets}, {Skinner}, {Conway}, {Hill}, {Kochanov}, {Tan}, {Wcis{\l}o},
  {Finenko}, {Nelson}, {Bernath}, {Birk}, {Boudon}, {Campargue}, {Chance},
  {Coustenis}, {Drouin}, {Flaud}, {Gamache}, {Hodges}, {Jacquemart}, {Mlawer},
  {Nikitin}, {Perevalov}, {Rotger}, {Tennyson}, {Toon}, {Tran}, {Tyuterev},
  {Adkins}, {Baker}, {Barbe}, {Can{\`e}}, {Cs{\'a}sz{\'a}r}, {Dudaryonok},
  {Egorov}, {Fleisher}, {Fleurbaey}, {Foltynowicz}, {Furtenbacher}, {Harrison},
  {Hartmann}, {Horneman}, {Huang}, {Karman}, {Karns}, {Kassi}, {Kleiner},
  {Kofman}, {Kwabia-Tchana}, {Lavrentieva}, {Lee}, {Long}, {Lukashevskaya},
  {Lyulin}, {Makhnev}, {Matt}, {Massie}, {Melosso}, {Mikhailenko}, {Mondelain},
  {M{\"u}ller}, {Naumenko}, {Perrin}, {Polyansky}, {Raddaoui}, {Raston},
  {Reed}, {Rey}, {Richard}, {T{\'o}bi{\'a}s}, {Sadiek}, {Schwenke},
  {Starikova}, {Sung}, {Tamassia}, {Tashkun}, {Vander Auwera}, {Vasilenko},
  {Vigasin}, {Villanueva}, {Vispoel}, {Wagner}, {Yachmenev}, \&
  {Yurchenko}}]{gordon22}
{Gordon}, I.~E., {Rothman}, L.~S., {Hargreaves}, R.~J., {et~al.} 2022,
  \href{http://dx.doi.org/10.1016/j.jqsrt.2021.107949}{\color{magenta}\jqsrt},
  \href{https://ui.adsabs.harvard.edu/abs/2022JQSRT.27707949G}{277, 107949}

\bibitem[{{Gordon} {et~al.}(2023){Gordon}, {Clayton}, {Decleir}, {Fitzpatrick},
  {Massa}, {Misselt}, \& {Tollerud}}]{gordon23}
{Gordon}, K.~D., {Clayton}, G.~C., {Decleir}, M., {et~al.} 2023,
  \href{http://dx.doi.org/10.3847/1538-4357/accb59}{\color{magenta}\apj},
  \href{https://ui.adsabs.harvard.edu/abs/2023ApJ...950...86G}{950, 86}

\bibitem[{{Grant} {et~al.}(2024){Grant}, {Bettoni}, {Banzatti}, {van Dishoeck},
  {Brittain}, {Fedele}, {Henning}, {Manara}, {Semenov}, \& {Whelan}}]{grant24}
{Grant}, S.~L., {Bettoni}, G., {Banzatti}, A., {et~al.} 2024,
  \href{http://dx.doi.org/10.1051/0004-6361/202347905}{\color{magenta}\aap},
  \href{https://ui.adsabs.harvard.edu/abs/2024A%26A...684A.213G}{684, A213}

\bibitem[{{Gray}(1992)}]{gray92}
{Gray}, D.~F. 1992, {The observation and analysis of stellar photospheres.},
  Vol.~20 (Cambridge, UK: Cambridge University Press)

\bibitem[{{Haffert} {et~al.}(2019){Haffert}, {Bohn}, {de Boer}, {Snellen},
  {Brinchmann}, {Girard}, {Keller}, \& {Bacon}}]{Haffert+2019}
{Haffert}, S.~Y., {Bohn}, A.~J., {de Boer}, J., {et~al.} 2019,
  \href{http://dx.doi.org/10.1038/s41550-019-0780-5}{\color{magenta}\natas},
  \href{https://ui.adsabs.harvard.edu/abs/2019NatAs...3..749H}{3, 749}

\bibitem[{{Hammond} {et~al.}(2025){Hammond}, {Christiaens}, {Price}, {Blakely},
  {Trevascus}, {Bonse}, {Cantalloube}, {Marleau}, {Pinte}, {Juillard},
  {Samland}, {Thompson}, \& {Wallace}}]{hammond25}
{Hammond}, I., {Christiaens}, V., {Price}, D.~J., {et~al.} 2025,
  \href{http://dx.doi.org/10.1093/mnras/staf586}{\color{magenta}\mnras},
  \href{https://ui.adsabs.harvard.edu/abs/2025MNRAS.539.1613H}{539, 1613}

\bibitem[{{Hartmann} {et~al.}(2016){Hartmann}, {Herczeg}, \&
  {Calvet}}]{hartmann16}
{Hartmann}, L., {Herczeg}, G., \& {Calvet}, N. 2016,
  \href{http://dx.doi.org/10.1146/annurev-astro-081915-023347}{\color{magenta}\araa},
  \href{https://ui.adsabs.harvard.edu/abs/2016ARA%26A..54..135H}{54, 135}

\bibitem[{{Hartmann} {et~al.}(1994){Hartmann}, {Hewett}, \&
  {Calvet}}]{hartmann94}
{Hartmann}, L., {Hewett}, R., \& {Calvet}, N. 1994,
  \href{http://dx.doi.org/10.1086/174104}{\color{magenta}\apj},
  \href{https://ui.adsabs.harvard.edu/abs/1994ApJ...426..669H}{426, 669}

\bibitem[{{Hasegawa} {et~al.}(2021){Hasegawa}, {Kanagawa}, \&
  {Turner}}]{hasegawa21}
{Hasegawa}, Y., {Kanagawa}, K.~D., \& {Turner}, N.~J. 2021,
  \href{http://dx.doi.org/10.3847/1538-4357/ac257b}{\color{magenta}\apj},
  \href{https://ui.adsabs.harvard.edu/abs/2021ApJ...923...27H}{923, 27}

\bibitem[{{Hashimoto} \& {Aoyama}(2025)}]{hashimoto25}
{Hashimoto}, J. \& {Aoyama}, Y. 2025,
  \href{http://dx.doi.org/10.3847/1538-3881/ad957e}{\color{magenta}\aj},
  \href{https://ui.adsabs.harvard.edu/abs/2025AJ....169...93H}{169, 93}

\bibitem[{{Hashimoto} {et~al.}(2020){Hashimoto}, {Aoyama}, {Konishi}, {Uyama},
  {Takasao}, {Ikoma}, \& {Tanigawa}}]{hashimoto20}
{Hashimoto}, J., {Aoyama}, Y., {Konishi}, M., {et~al.} 2020,
  \href{http://dx.doi.org/10.3847/1538-3881/ab811e}{\color{magenta}\aj},
  \href{https://ui.adsabs.harvard.edu/abs/2020AJ....159..222H}{159, 222}

\bibitem[{{Hashimoto} {et~al.}(2026){Hashimoto}, {Aoyama}, {Takami}, \&
  {Takasao}}]{hashimoto26}
{Hashimoto}, J., {Aoyama}, Y., {Takami}, M., \& {Takasao}, S. 2026,
  \href{http://dx.doi.org/10.3847/1538-3881/ae93ad}{\color{magenta}\aj},
  \href{https://ui.adsabs.harvard.edu/abs/2026AJ....172..189H}{172, 189}

\bibitem[{{Helling} {et~al.}(2008){Helling}, {Dehn}, {Woitke}, \&
  {Hauschildt}}]{helling08}
{Helling}, C., {Dehn}, M., {Woitke}, P., \& {Hauschildt}, P.~H. 2008,
  \href{http://dx.doi.org/10.1086/533462}{\color{magenta}\apjl},
  \href{https://ui.adsabs.harvard.edu/abs/2008ApJ...675L.105H}{675, L105}

\bibitem[{{Henden} {et~al.}(2015){Henden}, {Levine}, {Terrell}, \&
  {Welch}}]{henden15}
{Henden}, A.~A., {Levine}, S., {Terrell}, D., \& {Welch}, D.~L. 2015, in
  American Astronomical Society Meeting Abstracts, Vol. 225, American
  Astronomical Society Meeting Abstracts \#225,
  \href{https://ui.adsabs.harvard.edu/abs/2015AAS...22533616H}{336.16}

\bibitem[{{Heng} \& {McCray}(2007)}]{heng07}
{Heng}, K. \& {McCray}, R. 2007,
  \href{http://dx.doi.org/10.1086/509601}{\color{magenta}\apj},
  \href{https://ui.adsabs.harvard.edu/abs/2007ApJ...654..923H}{654, 923}

\bibitem[{{Hertfelder} \& {Kley}(2017)}]{hertfelder17}
{Hertfelder}, M. \& {Kley}, W. 2017,
  \href{http://dx.doi.org/10.1051/0004-6361/201730847}{\color{magenta}\aap},
  \href{https://ui.adsabs.harvard.edu/abs/2017A%26A...605A..24H}{605, A24}

\bibitem[{{Horne} \& {Marsh}(1986)}]{horne86}
{Horne}, K. \& {Marsh}, T.~R. 1986,
  \href{http://dx.doi.org/10.1093/mnras/218.4.761}{\color{magenta}\mnras},
  \href{https://ui.adsabs.harvard.edu/abs/1986MNRAS.218..761H}{218, 761}

\bibitem[{{Hsu} {et~al.}(2021{\natexlab{a}}){Hsu}, {Burgasser}, {Theissen},
  {Gelino}, {Birky}, {Diamant}, {Bardalez Gagliuffi}, {Aganze}, {Blake}, \&
  {Faherty}}]{hsu21}
{Hsu}, C.-C., {Burgasser}, A.~J., {Theissen}, C.~A., {et~al.}
  2021{\natexlab{a}},
  \href{http://dx.doi.org/10.3847/1538-4365/ac1c7d}{\color{magenta}\apjs},
  \href{https://ui.adsabs.harvard.edu/abs/2021ApJS..257...45H}{257, 45}

\bibitem[{{Hsu} {et~al.}(2021{\natexlab{b}}){Hsu}, {Theissen}, {Burgasser}, \&
  {Birky}}]{hsu21zndo}
{Hsu}, C.-C., {Theissen}, C., {Burgasser}, A., \& {Birky}, J.
  2021{\natexlab{b}}, {SMART: The Spectral Modeling Analysis and RV Tool},
  \href{https://zenodo.org/records/4765258}{\color{magenta}Zenodo},
  \href{https://ui.adsabs.harvard.edu/abs/2021zndo...4765258H}{record 4765258}

\bibitem[{{Hsu} {et~al.}(2024){Hsu}, {Wang}, {Blake}, {Xuan}, {Zhang},
  {Ruffio}, {Horstman}, {Cronin}, {Sappey}, {Xin}, {Finnerty}, {Echeverri},
  {Mawet}, {Jovanovic}, {Do {\'O}}, {Baker}, {Bartos}, {Calvin}, {Cetre},
  {Delorme}, {Doppmann}, {Fitzgerald}, {Liberman}, {L{\'o}pez}, {Morris},
  {Pezzato-Rovner}, {Schofield}, {Skemer}, {Wallace}, \& {Wang}}]{hsu24c}
{Hsu}, C.-C., {Wang}, J.~J., {Blake}, G.~A., {et~al.} 2024,
  \href{http://dx.doi.org/10.3847/2041-8213/ad95e8}{\color{magenta}\apjl},
  \href{https://ui.adsabs.harvard.edu/abs/2024ApJ...977L..47H}{977, L47}

\bibitem[{{Hsu} {et~al.}(2026){Hsu}, {Wang}, {Xuan}, {Zhang}, {Ruffio},
  {Mawet}, {Finnerty}, {Horstman}, {Cronin}, {Xin}, {Sappey}, {Echeverri},
  {Jovanovic}, {Baker}, {Bartos}, {Blake}, {Calvin}, {Cetre}, {Delorme},
  {Doppmann}, {Fitzgerald}, {Konopacky}, {Liberman}, {L{\'o}pez}, {Morris},
  {Pezzato}, {Schofield}, {Skemer}, {Wallace}, \& {Wang}}]{hsu26}
{Hsu}, C.-C., {Wang}, J.~J., {Xuan}, J.~W., {et~al.} 2026,
  \href{http://dx.doi.org/10.3847/1538-3881/ae434b}{\color{magenta}\aj},
  \href{https://ui.adsabs.harvard.edu/abs/2026AJ....171..224H}{171, 224}

\bibitem[{{Isella} {et~al.}(2019){Isella}, {Benisty}, {Teague}, {Bae},
  {Keppler}, {Facchini}, \& {P{\'e}rez}}]{isella19}
{Isella}, A., {Benisty}, M., {Teague}, R., {et~al.} 2019,
  \href{http://dx.doi.org/10.3847/2041-8213/ab2a12}{\color{magenta}\apjl},
  \href{https://ui.adsabs.harvard.edu/abs/2019ApJ...879L..25I}{879, L25}

\bibitem[{{Jang} {et~al.}(2025){Jang}, {Arabhavi}, {Kaeufer}, {Waters}, {Kamp},
  {Henning}, {Caratti o Garatti}, {van Dishoeck}, {Perotti}, {Kanwar},
  {G{\"u}del}, {Morales-Calder{\'o}n}, {Grant}, \& {Christiaens}}]{jang25}
{Jang}, H., {Arabhavi}, A.~M., {Kaeufer}, T., {et~al.} 2025,
  \href{http://dx.doi.org/10.1051/0004-6361/202556193}{\color{magenta}\aap},
  \href{https://ui.adsabs.harvard.edu/abs/2025A&A...703A..53J}{703, A53}

\bibitem[{{Jang} {et~al.}(2024){Jang}, {Waters}, {Kaeufer}, {Tamanai},
  {Perotti}, {Christiaens}, {Kamp}, {Henning}, {Min}, {Arabhavi}, {Barrado},
  {van Dishoeck}, {Gasman}, {Grant}, {G{\"u}del}, {Lagage}, {Lahuis},
  {Schwarz}, {Tabone}, \& {Temmink}}]{jang24}
{Jang}, H., {Waters}, R., {Kaeufer}, T., {et~al.} 2024,
  \href{http://dx.doi.org/10.1051/0004-6361/202451589}{\color{magenta}\aap},
  \href{https://ui.adsabs.harvard.edu/abs/2024A%26A...691A.148J}{691, A148}

\bibitem[{{Jones} {et~al.}(2013){Jones}, {Noll}, {Kausch}, {Szyszka}, \&
  {Kimeswenger}}]{jones13}
{Jones}, A., {Noll}, S., {Kausch}, W., {Szyszka}, C., \& {Kimeswenger}, S.
  2013,
  \href{http://dx.doi.org/10.1051/0004-6361/201322433}{\color{magenta}\aap},
  \href{https://ui.adsabs.harvard.edu/abs/2013A&A...560A..91J}{560, A91}

\bibitem[{{Kaeufer} {et~al.}(2024{\natexlab{a}}){Kaeufer}, {Min}, {Woitke},
  {Kamp}, \& {Arabhavi}}]{k24}
{Kaeufer}, T., {Min}, M., {Woitke}, P., {Kamp}, I., \& {Arabhavi}, A.~M.
  2024{\natexlab{a}},
  \href{http://dx.doi.org/10.1051/0004-6361/202449936}{\color{magenta}\aap},
  \href{https://ui.adsabs.harvard.edu/abs/2024A%26A...687A.209K}{687, A209}

\bibitem[{{Kaeufer} {et~al.}(2024{\natexlab{b}}){Kaeufer}, {Woitke}, {Kamp},
  {Kanwar}, \& {Min}}]{k24b}
{Kaeufer}, T., {Woitke}, P., {Kamp}, I., {Kanwar}, J., \& {Min}, M.
  2024{\natexlab{b}},
  \href{http://dx.doi.org/10.1051/0004-6361/202450891}{\color{magenta}\aap},
  \href{https://ui.adsabs.harvard.edu/abs/2024A%26A...690A.100K}{690, A100}

\bibitem[{{Kammerer} {et~al.}(2025){Kammerer}, {Winterhalder}, {Lacour},
  {Stolker}, {Marleau}, {Balmer}, {Moore}, {Piscarreta}, {Toci}, {M{\'e}rand},
  {Nowak}, {Rickman}, {Pueyo}, {Pourr{\'e}}, {Nasedkin}, {Wang}, {Bourdarot},
  {Eisenhauer}, {Henning}, {Garcia Lopez}, {van Dishoeck}, {Forveille},
  {Monnier}, {Abuter}, {Amorim}, {Benisty}, {Berger}, {Beust}, {Blunt},
  {Boccaletti}, {Bonnefoy}, {Bonnet}, {Bordoni}, {Brandner}, {Cantalloube},
  {Caselli}, {Ceva}, {Charnay}, {Chauvin}, {Chavez}, {Chomez}, {Choquet},
  {Christiaens}, {Cl{\'e}net}, {Coud{\'e} du Foresto}, {Cridland}, {Davies},
  {Dembet}, {Dexter}, {Drescher}, {Duvert}, {Eckart}, {Fontanive}, {F{\"o}rster
  Schreiber}, {Garcia}, {Gendron}, {Genzel}, {Gillessen}, {Girard}, {Grant},
  {Hagelberg}, {Haubois}, {Hei{\ss}el}, {Hinkley}, {Hippler}, {Houll{\'e}},
  {Hubert}, {Jocou}, {Keppler}, {Kervella}, {Kreidberg}, {Kurtovic},
  {Lagrange}, {Lapeyr{\`e}re}, {Le Bouquin}, {Lutz}, {Maire}, {Mang},
  {Matthews}, {Molli{\`e}re}, {Mordasini}, {Mouillet}, {Ott}, {Otten},
  {Paladini}, {Paumard}, {Perraut}, {Perrin}, {Pfuhl}, {Ribeiro},
  {Rustamkulov}, {S{\'e}gransan}, {Shangguan}, {Shimizu}, {Samland}, {Sing},
  {Stadler}, {Straub}, {Straubmeier}, {Sturm}, {Tacconi}, {Udry}, {Vigan},
  {Vincent}, {von Fellenberg}, {Widmann}, {Woillez}, \& {Yazici}}]{kammerer25}
{Kammerer}, J., {Winterhalder}, T.~O., {Lacour}, S., {et~al.} 2025,
  \href{http://dx.doi.org/10.1051/0004-6361/202556860}{\color{magenta}\aap},
  \href{https://ui.adsabs.harvard.edu/abs/2025A&A...704A.318K}{704, A318}

\bibitem[{{Kamp} {et~al.}(2017){Kamp}, {Thi}, {Woitke}, {Rab}, {Bouma}, \&
  {M{\'e}nard}}]{kamp17}
{Kamp}, I., {Thi}, W.-F., {Woitke}, P., {et~al.} 2017,
  \href{http://dx.doi.org/10.1051/0004-6361/201730388}{\color{magenta}\aap},
  \href{https://ui.adsabs.harvard.edu/abs/2017A&A...607A..41K}{607, A41}

\bibitem[{{Kanagawa} {et~al.}(2017){Kanagawa}, {Tanaka}, {Muto}, \&
  {Tanigawa}}]{kanagawa17}
{Kanagawa}, K.~D., {Tanaka}, H., {Muto}, T., \& {Tanigawa}, T. 2017,
  \href{http://dx.doi.org/10.1093/pasj/psx114}{\color{magenta}\pasj},
  \href{https://ui.adsabs.harvard.edu/abs/2017PASJ...69...97K}{69, 97}

\bibitem[{{Kao} {et~al.}(2018){Kao}, {Hallinan}, {Pineda}, {Stevenson}, \&
  {Burgasser}}]{kao18}
{Kao}, M.~M., {Hallinan}, G., {Pineda}, J.~S., {Stevenson}, D., \& {Burgasser},
  A. 2018,
  \href{http://dx.doi.org/10.3847/1538-4365/aac2d5}{\color{magenta}\apjs},
  \href{https://ui.adsabs.harvard.edu/abs/2018ApJS..237...25K}{237, 25}

\bibitem[{{Katarzy{\'n}ski} {et~al.}(2016){Katarzy{\'n}ski}, {Gawro{\'n}ski},
  \& {Go{\'z}dziewski}}]{katarzy16}
{Katarzy{\'n}ski}, K., {Gawro{\'n}ski}, M., \& {Go{\'z}dziewski}, K. 2016,
  \href{http://dx.doi.org/10.1093/mnras/stw1354}{\color{magenta}\mnras},
  \href{https://ui.adsabs.harvard.edu/abs/2016MNRAS.461..929K}{461, 929}

\bibitem[{{Kendrew} {et~al.}(2010){Kendrew}, {Jolissaint}, {Brandl}, {Lenzen},
  {Pantin}, {Glasse}, {Blommaert}, {Venema}, {Siebenmorgen}, \&
  {Molster}}]{kendrew10}
{Kendrew}, S., {Jolissaint}, L., {Brandl}, B., {et~al.} 2010, in Society of
  Photo-Optical Instrumentation Engineers (SPIE) Conference Series, Vol. 7735,
  Ground-based and Airborne Instrumentation for Astronomy III, ed. {McLean},
  I.~S., {Ramsay}, S.~K., \& {Takami}, H.,
  \href{https://ui.adsabs.harvard.edu/abs/2010SPIE.7735E..5FK}{77355F}

\bibitem[{{Kenyon} \& {Hartmann}(1987)}]{kenyon87}
{Kenyon}, S.~J. \& {Hartmann}, L. 1987,
  \href{http://dx.doi.org/10.1086/165866}{\color{magenta}\apj},
  \href{https://ui.adsabs.harvard.edu/abs/1987ApJ...323..714K}{323, 714}

\bibitem[{{Keppler} {et~al.}(2018){Keppler}, {Benisty}, {M{\"u}ller},
  {Henning}, {van Boekel}, {Cantalloube}, {Ginski}, {van Holstein}, {Maire},
  {Pohl}, {Samland}, {Avenhaus}, {Baudino}, {Boccaletti}, {de Boer},
  {Bonnefoy}, {Chauvin}, {Desidera}, {Langlois}, {Lazzoni}, {Marleau},
  {Mordasini}, {Pawellek}, {Stolker}, {Vigan}, {Zurlo}, {Birnstiel},
  {Brandner}, {Feldt}, {Flock}, {Girard}, {Gratton}, {Hagelberg}, {Isella},
  {Janson}, {Juhasz}, {Kemmer}, {Kral}, {Lagrange}, {Launhardt}, {Matter},
  {M{\'e}nard}, {Milli}, {Molli{\`e}re}, {Olofsson}, {P{\'e}rez}, {Pinilla},
  {Pinte}, {Quanz}, {Schmidt}, {Udry}, {Wahhaj}, {Williams}, {Buenzli},
  {Cudel}, {Dominik}, {Galicher}, {Kasper}, {Lannier}, {Mesa}, {Mouillet},
  {Peretti}, {Perrot}, {Salter}, {Sissa}, {Wildi}, {Abe}, {Antichi},
  {Augereau}, {Baruffolo}, {Baudoz}, {Bazzon}, {Beuzit}, {Blanchard}, {Brems},
  {Buey}, {De Caprio}, {Carbillet}, {Carle}, {Cascone}, {Cheetham}, {Claudi},
  {Costille}, {Delboulb{\'e}}, {Dohlen}, {Fantinel}, {Feautrier}, {Fusco},
  {Giro}, {Gluck}, {Gry}, {Hubin}, {Hugot}, {Jaquet}, {Le Mignant}, {Llored},
  {Madec}, {Magnard}, {Martinez}, {Maurel}, {Meyer}, {M{\"o}ller-Nilsson},
  {Moulin}, {Mugnier}, {Orign{\'e}}, {Pavlov}, {Perret}, {Petit}, {Pragt},
  {Puget}, {Rabou}, {Ramos}, {Rigal}, {Rochat}, {Roelfsema}, {Rousset}, {Roux},
  {Salasnich}, {Sauvage}, {Sevin}, {Soenke}, {Stadler}, {Suarez}, {Turatto}, \&
  {Weber}}]{keppler18}
{Keppler}, M., {Benisty}, M., {M{\"u}ller}, A., {et~al.} 2018,
  \href{http://dx.doi.org/10.1051/0004-6361/201832957}{\color{magenta}\aap},
  \href{https://ui.adsabs.harvard.edu/abs/2018A%26A...617A..44K}{617, A44}

\bibitem[{{Kley} \& {Lin}(1996)}]{kley96}
{Kley}, W. \& {Lin}, D.~N.~C. 1996,
  \href{http://dx.doi.org/10.1086/177115}{\color{magenta}\apj},
  \href{https://ui.adsabs.harvard.edu/abs/1996ApJ...461..933K}{461, 933}

\bibitem[{{Knierim} {et~al.}(2026){Knierim}, {Batygin}, {Helled}, {Morf}, \&
  {Adams}}]{knierim26a}
{Knierim}, H., {Batygin}, K., {Helled}, R., {Morf}, L., \& {Adams}, F.~C. 2026,
  \href{http://dx.doi.org/10.1051/0004-6361/202556984}{\color{magenta}\aap},
  \href{https://ui.adsabs.harvard.edu/abs/2026A&A...706A..51K}{706, A51}

\bibitem[{{Koenigl}(1991)}]{koenigl91}
{Koenigl}, A. 1991,
  \href{http://dx.doi.org/10.1086/185972}{\color{magenta}\apjl},
  \href{https://ui.adsabs.harvard.edu/abs/1991ApJ...370L..39K}{370, L39}

\bibitem[{{Komarova} \& {Fischer}(2020)}]{Komarova+Fischer2020}
{Komarova}, O. \& {Fischer}, W.~J. 2020,
  \href{http://dx.doi.org/10.3847/2515-5172/ab67bb}{\color{magenta}RNAAS},
  \href{https://ui.adsabs.harvard.edu/abs/2020RNAAS...4....6K}{4, 6}

\bibitem[{{Kramida} {et~al.}(2022){Kramida}, {Ralchenko}, {Reader}, \& {and
  NIST ASD Team}}]{NIST_ASD20230624}
{Kramida}, A., {Ralchenko}, Y., {Reader}, J., \& {and NIST ASD Team}. 2022,
  {NIST Atomic Spectra Database (ver. 5.10), [Online]. Available:
  {\url{https://physics.nist.gov/asd}} [2023, June 24]. National Institute of
  Standards and Technology, Gaithersburg, MD}

\bibitem[{{Krapp} {et~al.}(2024){Krapp}, {Kratter}, {Youdin},
  {Ben{\'\i}tez-Llambay}, {Masset}, \& {Armitage}}]{krapp24}
{Krapp}, L., {Kratter}, K.~M., {Youdin}, A.~N., {et~al.} 2024,
  \href{http://dx.doi.org/10.3847/1538-4357/ad644a}{\color{magenta}\apj},
  \href{https://ui.adsabs.harvard.edu/abs/2024ApJ...973..153K}{973, 153}

\bibitem[{{Kravchenko} {et~al.}(2024){Kravchenko}, {Rabien}, {Deysenroth},
  {Neumeier}, {Spallek}, {Honsberg}, {Barl}, {Ziegleder}, {Sturm}, \&
  {Davies}}]{kravchenko24}
{Kravchenko}, K., {Rabien}, S., {Deysenroth}, M., {et~al.} 2024,
  \href{https://ui.adsabs.harvard.edu/abs/2024arXiv240901714K}{\href{http://dx.doi.org/10.48550/arXiv.2409.01714}{\color{magenta}SPIE},
  arXiv:2409.01714}

\bibitem[{{Kwan} \& {Fischer}(2011)}]{kwanfischer11}
{Kwan}, J. \& {Fischer}, W. 2011,
  \href{http://dx.doi.org/10.1111/j.1365-2966.2010.17863.x}{\color{magenta}\mnras},
  \href{https://ui.adsabs.harvard.edu/abs/2011MNRAS.411.2383K}{411, 2383}

\bibitem[{{Lagrange} {et~al.}(2023){Lagrange}, {Philipot}, {Rubini}, {Meunier},
  {Kiefer}, {Kervella}, {Delorme}, \& {Beust}}]{lagrange23}
{Lagrange}, A.-M., {Philipot}, F., {Rubini}, P., {et~al.} 2023,
  \href{http://dx.doi.org/10.1051/0004-6361/202346165}{\color{magenta}\aap},
  \href{https://ui.adsabs.harvard.edu/abs/2023A&A...677A..71L}{677, A71}

\bibitem[{{Lawlor} {et~al.}(2026){Lawlor}, {van Capelleveen}, {Bourdarot},
  {Ginski}, {Kenworthy}, {Stolker}, {Close}, {Bohn}, {Eisenhauer}, {Garcia},
  {H{\"o}nig}, {Kammerer}, {Kreidberg}, {Lacour}, {Le Bouquin}, {Mamajek},
  {Nowak}, {Paumard}, {Straubmeier}, {van der Marel}, \& {The Exogravity
  Collaboration}}]{lawlor26}
{Lawlor}, C., {van Capelleveen}, R.~F., {Bourdarot}, G., {et~al.} 2026,
  \href{http://dx.doi.org/10.3847/2041-8213/ae4b3b}{\color{magenta}\apjl},
  \href{https://ui.adsabs.harvard.edu/abs/2026ApJ..1000L..38L}{1000, L38}

\bibitem[{{Lee} {et~al.}(2022){Lee}, {Wardenier}, {Prinoth}, {Parmentier},
  {Grimm}, {Baeyens}, {Carone}, {Christie}, {Deitrick}, {Kitzmann}, {Mayne},
  {Roman}, \& {Thorsbro}}]{lee22}
{Lee}, E. K.~H., {Wardenier}, J.~P., {Prinoth}, B., {et~al.} 2022,
  \href{http://dx.doi.org/10.3847/1538-4357/ac61d6}{\color{magenta}\apj},
  \href{https://ui.adsabs.harvard.edu/abs/2022ApJ...929..180L}{929, 180}

\bibitem[{{Lega} {et~al.}(2024){Lega}, {Benisty}, {Cridland}, {Morbidelli},
  {Schulik}, \& {Lambrechts}}]{lega24}
{Lega}, E., {Benisty}, M., {Cridland}, A., {et~al.} 2024,
  \href{http://dx.doi.org/10.1051/0004-6361/202450899}{\color{magenta}\aap},
  \href{https://ui.adsabs.harvard.edu/abs/2024A%26A...690A.183L}{690, A183}

\bibitem[{{Lenzen} {et~al.}(2003){Lenzen}, {Hartung}, {Brandner}, {Finger},
  {Hubin}, {Lacombe}, {Lagrange}, {Lehnert}, {Moorwood}, \&
  {Mouillet}}]{lenzen03}
{Lenzen}, R., {Hartung}, M., {Brandner}, W., {et~al.} 2003, in Society of
  Photo-Optical Instrumentation Engineers (SPIE) Conference Series, Vol. 4841,
  Instrument Design and Performance for Optical/Infrared Ground-based
  Telescopes, ed. {Iye}, M. \& {Moorwood}, A. F.~M.,
  \href{https://ui.adsabs.harvard.edu/abs/2003SPIE.4841..944L}{944--952}

\bibitem[{{Leschinski} {et~al.}(2020){Leschinski}, {Buddelmeijer}, {Czoske},
  {Verdugo}, {Verdoes-Kleijn}, \& {Zeilinger}}]{leschinski20}
{Leschinski}, K., {Buddelmeijer}, H., {Czoske}, O., {et~al.} 2020, in Society
  of Photo-Optical Instrumentation Engineers (SPIE) Conference Series, Vol.
  11452, Software and Cyberinfrastructure for Astronomy VI, ed. {Guzman}, J.~C.
  \& {Ibsen}, J.,
  \href{https://ui.adsabs.harvard.edu/abs/2020SPIE11452E..1ZL}{114521Z}

\bibitem[{{Li} {et~al.}(2025){Li}, {Close}, {Long}, {Males}, {Haffert},
  {Weinberger}, {Follette}, {Andrews}, {Carpenter}, {Foster}, {Van Gorkom},
  {Hedglen}, {Herczeg}, {Johnson}, {Kautz}, {Kueny}, {Li}, {Liberman}, {Long},
  {Lumbres}, {Marino}, {Matr{\`a}}, {McEwen}, {Guyon}, {Pearce}, {P{\'e}rez},
  {Pinilla}, {Schatz}, {Shi}, {Twitchell}, {Wagner}, {Wilner}, {Wu}, {Zhang},
  \& {Zhu}}]{li25}
{Li}, J., {Close}, L.~M., {Long}, F., {et~al.} 2025,
  \href{http://dx.doi.org/10.3847/2041-8213/adfcbd}{\color{magenta}\apjl},
  \href{https://ui.adsabs.harvard.edu/abs/2025ApJ...990L..70L}{990, L70}

\bibitem[{{Linder} {et~al.}(2019){Linder}, {Mordasini}, {Molli{\`e}re},
  {Marleau}, {Malik}, {Quanz}, \& {Meyer}}]{linder19}
{Linder}, E.~F., {Mordasini}, C., {Molli{\`e}re}, P., {et~al.} 2019,
  \href{http://dx.doi.org/10.1051/0004-6361/201833873}{\color{magenta}\aap},
  \href{https://ui.adsabs.harvard.edu/abs/2019A%26A...623A..85L}{623, A85}

\bibitem[{{Lovelace} {et~al.}(2011){Lovelace}, {Covey}, \&
  {Lloyd}}]{lovelace11}
{Lovelace}, R.~V.~E., {Covey}, K.~R., \& {Lloyd}, J.~P. 2011,
  \href{http://dx.doi.org/10.1088/0004-6256/141/2/51}{\color{magenta}\aj},
  \href{https://ui.adsabs.harvard.edu/abs/2011AJ....141...51L}{141, 51}

\bibitem[{{Lovis} {et~al.}(2022){Lovis}, {Blind}, {Chazelas}, {K{\"u}hn},
  {Genolet}, {Hughes}, {Sordet}, {Schnell}, {Turbet}, {Fusco}, {Sauvage},
  {Bugatti}, {Billot}, {Hagelberg}, {Hocini}, \& {Guyon}}]{lovis22}
{Lovis}, C., {Blind}, N., {Chazelas}, B., {et~al.} 2022, in SPIE Conf.\ Ser.\@,
  Vol. 12184, Ground-based and Airborne Instrumentation for Astronomy IX, ed.
  {Evans}, C.~J., {Bryant}, J.~J., \& {Motohara}, K.,
  \href{https://ui.adsabs.harvard.edu/abs/2022SPIE12184E..1QL}{121841Q}

\bibitem[{{Lovis} {et~al.}(2024){Lovis}, {Blind}, {Chazelas}, {Shinde},
  {Bugatti}, {Restori}, {Dinis}, {Genolet}, {Hughes}, {Sordet}, {Schnell},
  {Rihs}, {Crausaz}, {Turbet}, {Billot}, {Fusco}, {Neichel}, {Sauvage}, {Santos
  Diaz}, {Houelle}, {Blackman}, {Lanotte}, {K{\"u}hn}, {Hagelberg}, {Guyon},
  {Martinez}, {Spang}, {Mordasini}, {Ehrenreich}, {Demory}, \&
  {Bolmont}}]{lovis24}
{Lovis}, C., {Blind}, N., {Chazelas}, B., {et~al.} 2024,
  \href{https://ui.adsabs.harvard.edu/abs/2024arXiv240902875L}{\href{http://dx.doi.org/10.48550/arXiv.2409.02875}{\color{magenta}SPIE},
  arXiv:2409.02875}

\bibitem[{{Lubow} \& {D'Angelo}(2006)}]{lubow06}
{Lubow}, S.~H. \& {D'Angelo}, G. 2006,
  \href{http://dx.doi.org/10.1086/500356}{\color{magenta}\apj},
  \href{https://ui.adsabs.harvard.edu/abs/2006ApJ...641..526L}{641, 526}

\bibitem[{{Luhman} {et~al.}(2023){Luhman}, {Tremblin}, {Birkmann},
  {Manjavacas}, {Valenti}, {Alves de Oliveira}, {Beck}, {Giardino},
  {L{\"u}tzgendorf}, {Rauscher}, \& {Sirianni}}]{luhman23c}
{Luhman}, K.~L., {Tremblin}, P., {Birkmann}, S.~M., {et~al.} 2023,
  \href{http://dx.doi.org/10.3847/2041-8213/acd635}{\color{magenta}\apjl},
  \href{https://ui.adsabs.harvard.edu/abs/2023ApJ...949L..36L}{949, L36}

\bibitem[{{Lutz} {et~al.}(1996){Lutz}, {Feuchtgruber}, {Genzel}, {Kunze},
  {Rigopoulou}, {Spoon}, {Wright}, {Egami}, {Katterloher}, {Sturm},
  {Wieprecht}, {Sternberg}, {Moorwood}, \& {de Graauw}}]{lutz96}
{Lutz}, D., {Feuchtgruber}, H., {Genzel}, R., {et~al.} 1996, \aap,
  \href{https://ui.adsabs.harvard.edu/abs/1996A%26A...315L.269L}{315, L269}

\bibitem[{{Lynden-Bell} \& {Pringle}(1974)}]{lyndenbell74}
{Lynden-Bell}, D. \& {Pringle}, J.~E. 1974,
  \href{http://dx.doi.org/10.1093/mnras/168.3.603}{\color{magenta}\mnras},
  \href{https://ui.adsabs.harvard.edu/abs/1974MNRAS.168..603L}{168, 603}

\bibitem[{{Malik} {et~al.}(2015){Malik}, {Meru}, {Mayer}, \& {Meyer}}]{malik15}
{Malik}, M., {Meru}, F., {Mayer}, L., \& {Meyer}, M. 2015,
  \href{http://dx.doi.org/10.1088/0004-637X/802/1/56}{\color{magenta}\apj},
  \href{https://ui.adsabs.harvard.edu/abs/2015ApJ...802...56M}{802, 56}

\bibitem[{{Manara} {et~al.}(2013){Manara}, {Testi}, {Rigliaco}, {Alcal{\'a}},
  {Natta}, {Stelzer}, {Biazzo}, {Covino}, {Covino}, {Cupani}, {D'Elia}, \&
  {Randich}}]{manara13}
{Manara}, C.~F., {Testi}, L., {Rigliaco}, E., {et~al.} 2013,
  \href{http://dx.doi.org/10.1051/0004-6361/201220921}{\color{magenta}\aap},
  \href{https://ui.adsabs.harvard.edu/abs/2013A%26A...551A.107M}{551, A107}

\bibitem[{{Marleau}(2026)}]{m24expeditus}
{Marleau}, G.-D. 2026,
  \href{http://dx.doi.org/10.3847/1538-4357/ae2604}{\color{magenta}\apj},
  \href{https://ui.adsabs.harvard.edu/abs/2026ApJ..1000..153M}{1000, 153}

\bibitem[{{Marleau} {et~al.}(2024){Marleau}, {Aoyama}, {Hashimoto}, \&
  {Zhou}}]{m23alois}
{Marleau}, G.-D., {Aoyama}, Y., {Hashimoto}, J., \& {Zhou}, Y. 2024,
  \href{http://dx.doi.org/10.3847/1538-4357/ad1ee9}{\color{magenta}\apj},
  \href{https://ui.adsabs.harvard.edu/abs/2024ApJ...964...70M}{964, 70}

\bibitem[{{Marleau} {et~al.}(2022){Marleau}, {Aoyama}, {Kuiper}, {Follette},
  {Turner}, {Cugno}, {Manara}, {Haffert}, {Kitzmann}, {Ringqvist}, {Wagner},
  {van Boekel}, {Sallum}, {Janson}, {Schmidt}, {Venuti}, {Lovis}, \&
  {Mordasini}}]{maea21}
{Marleau}, G.-D., {Aoyama}, Y., {Kuiper}, R., {et~al.} 2022,
  \href{http://dx.doi.org/10.1051/0004-6361/202037494}{\color{magenta}\aap},
  \href{https://ui.adsabs.harvard.edu/abs/2022A%26A...657A..38M}{657, A38}

\bibitem[{{Marleau} {et~al.}(2019){Marleau}, {Coleman}, {Leleu}, \&
  {Mordasini}}]{m18hip}
{Marleau}, G.-D., {Coleman}, G.~A.~L., {Leleu}, A., \& {Mordasini}, C. 2019,
  \href{http://dx.doi.org/10.1051/0004-6361/201833597}{\color{magenta}\aap},
  \href{https://ui.adsabs.harvard.edu/abs/2019A%26A...624A..20M}{624, A20}

\bibitem[{{Marleau} \& {Cumming}(2014)}]{mc14}
{Marleau}, G.-D. \& {Cumming}, A. 2014,
  \href{http://dx.doi.org/10.1093/mnras/stt1967}{\color{magenta}\mnras},
  \href{https://ui.adsabs.harvard.edu/abs/2014MNRAS.437.1378M}{437, 1378}

\bibitem[{{Marleau} {et~al.}(2023){Marleau}, {Kuiper}, {B{\'e}thune}, \&
  {Mordasini}}]{m22Schock}
{Marleau}, G.-D., {Kuiper}, R., {B{\'e}thune}, W., \& {Mordasini}, C. 2023,
  \href{http://dx.doi.org/10.3847/1538-4357/accf12}{\color{magenta}\apj},
  \href{https://ui.adsabs.harvard.edu/abs/2023ApJ...952...89M}{952, 89}

\bibitem[{{Marley} {et~al.}(2007){Marley}, {Fortney}, {Hubickyj},
  {Bodenheimer}, \& {Lissauer}}]{marl07}
{Marley}, M.~S., {Fortney}, J.~J., {Hubickyj}, O., {Bodenheimer}, P., \&
  {Lissauer}, J.~J. 2007,
  \href{http://dx.doi.org/10.1086/509759}{\color{magenta}\apj},
  \href{https://ui.adsabs.harvard.edu/abs/2007ApJ...655..541M}{655, 541}

\bibitem[{{Marley} {et~al.}(2021){Marley}, {Saumon}, {Visscher}, {Lupu},
  {Freedman}, {Morley}, {Fortney}, {Seay}, {Smith}, {Teal}, \&
  {Wang}}]{marley21}
{Marley}, M.~S., {Saumon}, D., {Visscher}, C., {et~al.} 2021,
  \href{http://dx.doi.org/10.3847/1538-4357/ac141d}{\color{magenta}\apj},
  \href{https://ui.adsabs.harvard.edu/abs/2021ApJ...920...85M}{920, 85}

\bibitem[{{Marois} {et~al.}(2006){Marois}, {Lafreni{\`e}re}, {Doyon},
  {Macintosh}, \& {Nadeau}}]{marois06a}
{Marois}, C., {Lafreni{\`e}re}, D., {Doyon}, R., {Macintosh}, B., \& {Nadeau},
  D. 2006, \href{http://dx.doi.org/10.1086/500401}{\color{magenta}\apj},
  \href{https://ui.adsabs.harvard.edu/abs/2006ApJ...641..556M}{641, 556}

\bibitem[{{Mendigut{\'\i}a}(2020)}]{mendigut20}
{Mendigut{\'\i}a}, I. 2020,
  \href{http://dx.doi.org/10.3390/galaxies8020039}{\color{magenta}Galaxies},
  \href{https://ui.adsabs.harvard.edu/abs/2020Galax...8...39M}{8, 39}

\bibitem[{{Mesa} {et~al.}(2019){Mesa}, {Keppler}, {Cantalloube}, {Rodet},
  {Charnay}, {Gratton}, {Langlois}, {Boccaletti}, {Bonnefoy}, {Vigan},
  {Flasseur}, {Bae}, {Benisty}, {Chauvin}, {de Boer}, {Desidera}, {Henning},
  {Lagrange}, {Meyer}, {Milli}, {M{\"u}ller}, {Pairet}, {Zurlo}, {Antoniucci},
  {Baudino}, {Brown Sevilla}, {Cascone}, {Cheetham}, {Claudi}, {Delorme},
  {D'Orazi}, {Feldt}, {Hagelberg}, {Janson}, {Kral}, {Lagadec}, {Lazzoni},
  {Ligi}, {Maire}, {Martinez}, {Menard}, {Meunier}, {Perrot}, {Petrus},
  {Pinte}, {Rickman}, {Rochat}, {Rouan}, {Samland}, {Sauvage}, {Schmidt},
  {Udry}, {Weber}, \& {Wildi}}]{mesa19}
{Mesa}, D., {Keppler}, M., {Cantalloube}, F., {et~al.} 2019,
  \href{http://dx.doi.org/10.1051/0004-6361/201936764}{\color{magenta}\aap},
  \href{https://ui.adsabs.harvard.edu/abs/2019A%26A...632A..25M}{632, A25}

\bibitem[{{Mihalas} \& {Mihalas}(1984)}]{mihalas84}
{Mihalas}, D. \& {Mihalas}, B.~W. 1984, Foundations of radiation hydrodynamics
  (Oxford University Press)

\bibitem[{{Mohanty} {et~al.}(2005){Mohanty}, {Jayawardhana}, \&
  {Basri}}]{mohanty05a}
{Mohanty}, S., {Jayawardhana}, R., \& {Basri}, G. 2005,
  \href{http://dx.doi.org/10.1086/429794}{\color{magenta}\apj},
  \href{https://ui.adsabs.harvard.edu/abs/2005ApJ...626..498M}{626, 498}

\bibitem[{{Molli{\`e}re} {et~al.}(2019){Molli{\`e}re}, {Wardenier}, {van
  Boekel}, {Henning}, {Molaverdikhani}, \& {Snellen}}]{moll19}
{Molli{\`e}re}, P., {Wardenier}, J.~P., {van Boekel}, R., {et~al.} 2019,
  \href{http://dx.doi.org/10.1051/0004-6361/201935470}{\color{magenta}\aap},
  \href{https://ui.adsabs.harvard.edu/abs/2019A%26A...627A..67M}{627, A67}

\bibitem[{{Mordasini} {et~al.}(2017){Mordasini}, {Marleau}, \&
  {Molli{\`e}re}}]{mordasini17}
{Mordasini}, C., {Marleau}, G.-D., \& {Molli{\`e}re}, P. 2017,
  \href{http://dx.doi.org/10.1051/0004-6361/201630077}{\color{magenta}\aap},
  \href{https://ui.adsabs.harvard.edu/abs/2017A%26A...608A..72M}{608, A72}

\bibitem[{{Morley} {et~al.}(2024){Morley}, {Mukherjee}, {Marley}, {Fortney},
  {Visscher}, {Lupu}, {Gharib-Nezhad}, {Thorngren}, {Freedman}, \&
  {Batalha}}]{morley24}
{Morley}, C.~V., {Mukherjee}, S., {Marley}, M.~S., {et~al.} 2024,
  \href{http://dx.doi.org/10.3847/1538-4357/ad71d5}{\color{magenta}\apj},
  \href{https://ui.adsabs.harvard.edu/abs/2024ApJ...975...59M}{975, 59}

\bibitem[{{M{\"u}ller} {et~al.}(2018){M{\"u}ller}, {Keppler}, {Henning},
  {Samland}, {Chauvin}, {Beust}, {Maire}, {Molaverdikhani}, {van Boekel},
  {Benisty}, {Boccaletti}, {Bonnefoy}, {Cantalloube}, {Charnay}, {Baudino},
  {Gennaro}, {Long}, {Cheetham}, {Desidera}, {Feldt}, {Fusco}, {Girard},
  {Gratton}, {Hagelberg}, {Janson}, {Lagrange}, {Langlois}, {Lazzoni}, {Ligi},
  {M{\'e}nard}, {Mesa}, {Meyer}, {Molli{\`e}re}, {Mordasini}, {Moulin},
  {Pavlov}, {Pawellek}, {Quanz}, {Ramos}, {Rouan}, {Sissa}, {Stadler}, {Vigan},
  {Wahhaj}, {Weber}, \& {Zurlo}}]{mueller18}
{M{\"u}ller}, A., {Keppler}, M., {Henning}, T., {et~al.} 2018,
  \href{http://dx.doi.org/10.1051/0004-6361/201833584}{\color{magenta}\aap},
  \href{https://ui.adsabs.harvard.edu/abs/2018A%26A...617L...2M}{617, L2}

\bibitem[{{Muzerolle} {et~al.}(2001){Muzerolle}, {Calvet}, \&
  {Hartmann}}]{muzerolle01}
{Muzerolle}, J., {Calvet}, N., \& {Hartmann}, L. 2001,
  \href{http://dx.doi.org/10.1086/319779}{\color{magenta}\apj},
  \href{https://ui.adsabs.harvard.edu/abs/2001ApJ...550..944M}{550, 944}

\bibitem[{{Narang} {et~al.}(2024){Narang}, {Manoj}, {Chandra}, {Banerjee},
  {Tyagi}, {Tamura}, {Henning}, {Mathew}, {Lazio}, {Surya}, \&
  {Nayak}}]{narang24}
{Narang}, M., {Manoj}, P., {Chandra}, C.~H.~I., {et~al.} 2024,
  \href{http://dx.doi.org/10.1093/mnras/stae536}{\color{magenta}\mnras},
  \href{https://ui.adsabs.harvard.edu/abs/2024MNRAS.529.1161N}{529, 1161}

\bibitem[{{Noll} {et~al.}(2012){Noll}, {Kausch}, {Barden}, {Jones}, {Szyszka},
  {Kimeswenger}, \& {Vinther}}]{noll12}
{Noll}, S., {Kausch}, W., {Barden}, M., {et~al.} 2012,
  \href{http://dx.doi.org/10.1051/0004-6361/201219040}{\color{magenta}\aap},
  \href{https://ui.adsabs.harvard.edu/abs/2012A&A...543A..92N}{543, A92}

\bibitem[{{Oberg} {et~al.}(2023){Oberg}, {Kamp}, {Cazaux}, {Rab}, \&
  {Czoske}}]{oberg23metis}
{Oberg}, N., {Kamp}, I., {Cazaux}, S., {Rab}, C., \& {Czoske}, O. 2023,
  \href{http://dx.doi.org/10.1051/0004-6361/202244845}{\color{magenta}\aap},
  \href{https://ui.adsabs.harvard.edu/abs/2023A%26A...670A..74O}{670, A74}

\bibitem[{{Okuzumi} {et~al.}(2026){Okuzumi}, {Muto}, {Tominaga}, \&
  {Shimizu}}]{okuzumi26}
{Okuzumi}, S., {Muto}, T., {Tominaga}, R.~T., \& {Shimizu}, S. 2026,
  \href{http://dx.doi.org/10.1093/pasj/psag010}{\color{magenta}\pasj},
  \href{https://ui.adsabs.harvard.edu/abs/2026PASJ...78..673O}{78, 673}

\bibitem[{{Ossenkopf} \& {Henning}(1994)}]{ossenkopfhenning94}
{Ossenkopf}, V. \& {Henning}, T. 1994, \aap,
  \href{https://ui.adsabs.harvard.edu/abs/1994A%26A...291..943O}{291, 943}

\bibitem[{{Palma-Bifani} {et~al.}(2023){Palma-Bifani}, {Baudoz}, {Huby}, \&
  {Chauvin}}]{palmabifani23}
{Palma-Bifani}, P., {Baudoz}, P., {Huby}, E., \& {Chauvin}, G. 2023, in
  SF2A-2023: Proceedings of the Annual meeting of the French Society of
  Astronomy and Astrophysics, ed. {N'Diaye}, M., {Siebert}, A., {Lagarde}, N.,
  {et~al.},
  \href{https://ui.adsabs.harvard.edu/abs/2023sf2a.conf..239P}{239--242}

\bibitem[{{Papaloizou} \& {Nelson}(2005)}]{papnel05}
{Papaloizou}, J.~C.~B. \& {Nelson}, R.~P. 2005,
  \href{http://dx.doi.org/10.1051/0004-6361:20042029}{\color{magenta}\aap},
  \href{https://ui.adsabs.harvard.edu/abs/2005A%26A...433..247P}{433, 247}

\bibitem[{{Parker} {et~al.}(2024){Parker}, {Birkby}, {Landman}, {Wardenier},
  {Young}, {Vaughan}, {van Sluijs}, {Brogi}, {Parmentier}, \&
  {Line}}]{parker24}
{Parker}, L.~T., {Birkby}, J.~L., {Landman}, R., {et~al.} 2024,
  \href{http://dx.doi.org/10.1093/mnras/stae1277}{\color{magenta}\mnras},
  \href{https://ui.adsabs.harvard.edu/abs/2024MNRAS.531.2356P}{531, 2356}

\bibitem[{{Partridge} \& {Schwenke}(1997)}]{partridge97}
{Partridge}, H. \& {Schwenke}, D.~W. 1997,
  \href{http://dx.doi.org/10.1063/1.473987}{\color{magenta}\jcp},
  \href{https://ui.adsabs.harvard.edu/abs/1997JChPh.106.4618P}{106, 4618}

\bibitem[{{Patapis} {et~al.}(2025){Patapis}, {Morales-Calder{\'o}n},
  {Arabhavi}, {K{\"u}hnle}, {Gasman}, {Cugno}, {Molli{\`e}re}, {Matthews},
  {M{\^a}lin}, {Whiteford}, {Lagage}, {Waters}, {Guedel}, {Henning},
  {Vandenbussche}, {Absil}, {Argyriou}, {Barrado}, {Baudoz}, {Boccaletti},
  {Bouwman}, {Cossou}, {Coulais}, {Decin}, {Gastaud}, {Glasse}, {Glauser},
  {Grant}, {Min}, {Kamp}, {Olofsson}, {Pye}, {Rouan}, {Royer}, {Scheithauer},
  {Sun}, {Tremblin}, {Colina}, {Ray}, {{\"O}stlin}, {van Dishoeck}, \&
  {Wright}}]{patapis25}
{Patapis}, P., {Morales-Calder{\'o}n}, M., {Arabhavi}, A.~M., {et~al.} 2025,
  \href{http://dx.doi.org/10.1051/0004-6361/202556296}{\color{magenta}\aap},
  \href{https://ui.adsabs.harvard.edu/abs/2025A&A...704A...5P}{704, A5}

\bibitem[{{Plunkett} {et~al.}(2025){Plunkett}, {Follette}, {Marleau}, \&
  {Nielsen}}]{plunkett25}
{Plunkett}, C., {Follette}, K.~B., {Marleau}, G.-D., \& {Nielsen}, E.~L. 2025,
  \href{http://dx.doi.org/10.3847/1538-3881/adc09d}{\color{magenta}\aj},
  \href{https://ui.adsabs.harvard.edu/abs/2025AJ....169..262P}{169, 262}

\bibitem[{{Polyansky} {et~al.}(2018){Polyansky}, {Kyuberis}, {Zobov},
  {Tennyson}, {Yurchenko}, \& {Lodi}}]{polyansky18}
{Polyansky}, O.~L., {Kyuberis}, A.~A., {Zobov}, N.~F., {et~al.} 2018,
  \href{http://dx.doi.org/10.1093/mnras/sty1877}{\color{magenta}\mnras},
  \href{https://ui.adsabs.harvard.edu/abs/2018MNRAS.480.2597P}{480, 2597}

\bibitem[{{Portilla-Revelo} {et~al.}(2023){Portilla-Revelo}, {Kamp},
  {Facchini}, {van Dishoeck}, {Law}, {Rab}, {Bae}, {Benisty}, {{\"O}berg}, \&
  {Teague}}]{pr23}
{Portilla-Revelo}, B., {Kamp}, I., {Facchini}, S., {et~al.} 2023,
  \href{http://dx.doi.org/10.1051/0004-6361/202346607}{\color{magenta}\aap},
  \href{https://ui.adsabs.harvard.edu/abs/2023A%26A...677A..76P}{677, A76}

\bibitem[{{Quillen} \& {Trilling}(1998)}]{quillen98}
{Quillen}, A.~C. \& {Trilling}, D.~E. 1998,
  \href{http://dx.doi.org/10.1086/306421}{\color{magenta}\apj},
  \href{https://ui.adsabs.harvard.edu/abs/1998ApJ...508..707Q}{508, 707}

\bibitem[{{Rab} {et~al.}(2019){Rab}, {Kamp}, {Ginski}, {Oberg}, {Muro-Arena},
  {Dominik}, {Waters}, {Thi}, \& {Woitke}}]{rab19}
{Rab}, C., {Kamp}, I., {Ginski}, C., {et~al.} 2019,
  \href{http://dx.doi.org/10.1051/0004-6361/201834899}{\color{magenta}\aap},
  \href{https://ui.adsabs.harvard.edu/abs/2019A%26A...624A..16R}{624, A16}

\bibitem[{{Radcliffe} {et~al.}(2026){Radcliffe}, {Charnay}, {Lagrange},
  {Kiefer}, {B{\'e}zard}, {Petrus}, {Palma-Bifani}, {Ravet}, {Leconte}, \&
  {Marleau}}]{radcliffe26}
{Radcliffe}, A., {Charnay}, B., {Lagrange}, A.-M., {et~al.} 2026,
  \href{http://dx.doi.org/10.1051/0004-6361/202659006}{\color{magenta}\aap},
  \href{https://ui.adsabs.harvard.edu/abs/2026A&A...711A.257R}{711, A257}

\bibitem[{{Radigan} {et~al.}(2014){Radigan}, {Lafreni{\`e}re}, {Jayawardhana},
  \& {Artigau}}]{radigan14}
{Radigan}, J., {Lafreni{\`e}re}, D., {Jayawardhana}, R., \& {Artigau}, E. 2014,
  \href{http://dx.doi.org/10.1088/0004-637X/793/2/75}{\color{magenta}\apj},
  \href{https://ui.adsabs.harvard.edu/abs/2014ApJ...793...75R}{793, 75}

\bibitem[{{Ramsay} {et~al.}(2018){Ramsay}, {Casali}, {Amico}, {Bezawada},
  {Cirasuolo}, {Conzelmann}, {Egner}, {Frank}, {George}, {Gonz{\'a}lez
  Herrera}, {Hammersley}, {Haupt}, {Heijmans}, {Ives}, {Jakob}, {Kerber},
  {Koehler}, {Mainieri}, {Manescau}, {Marchetti}, {Oberti}, {Padovani},
  {Schmid}, {Schimpelsberger}, {Siebenmorgen}, {Tamai}, \& {Vernet}}]{ramsay18}
{Ramsay}, S., {Casali}, M., {Amico}, P., {et~al.} 2018, in Society of
  Photo-Optical Instrumentation Engineers (SPIE) Conference Series, Vol. 10702,
  Ground-based and Airborne Instrumentation for Astronomy VII, ed. {Evans},
  C.~J., {Simard}, L., \& {Takami}, H.,
  \href{https://ui.adsabs.harvard.edu/abs/2018SPIE10702E..1PR}{107021P}

\bibitem[{{Razumovskiy} {et~al.}(2025){Razumovskiy}, {Fomin}, \&
  {Astanin}}]{razumovskiy24}
{Razumovskiy}, M., {Fomin}, B., \& {Astanin}, D. 2025,
  \href{http://dx.doi.org/10.1016/j.jqsrt.2025.109599}{\color{magenta}\jqsrt},
  \href{https://ui.adsabs.harvard.edu/abs/2025JQSRT.34609599R}{346, 109599}

\bibitem[{{Rigliaco} {et~al.}(2015){Rigliaco}, {Pascucci}, {Duchene},
  {Edwards}, {Ardila}, {Grady}, {Mendigut{\'\i}a}, {Montesinos}, {Mulders},
  {Najita}, {Carpenter}, {Furlan}, {Gorti}, {Meijerink}, \&
  {Meyer}}]{rigliaco15}
{Rigliaco}, E., {Pascucci}, I., {Duchene}, G., {et~al.} 2015,
  \href{http://dx.doi.org/10.1088/0004-637X/801/1/31}{\color{magenta}\apj},
  \href{https://ui.adsabs.harvard.edu/abs/2015ApJ...801...31R}{801, 31}

\bibitem[{{Ringqvist} {et~al.}(2023){Ringqvist}, {Viswanath}, {Aoyama},
  {Janson}, {Marleau}, \& {Brandeker}}]{ringqvist23}
{Ringqvist}, S.~C., {Viswanath}, G., {Aoyama}, Y., {et~al.} 2023,
  \href{http://dx.doi.org/10.1051/0004-6361/202245424}{\color{magenta}\aap},
  \href{https://ui.adsabs.harvard.edu/abs/2023A%26A...669L..12R}{669, L12}

\bibitem[{{Riols} \& {Lesur}(2018)}]{riols18}
{Riols}, A. \& {Lesur}, G. 2018,
  \href{http://dx.doi.org/10.1051/0004-6361/201833212}{\color{magenta}\aap},
  \href{https://ui.adsabs.harvard.edu/abs/2018A&A...617A.117R}{617, A117}

\bibitem[{{Rogers} {et~al.}(2024){Rogers}, {de Marchi}, \& {Brandl}}]{rogers24}
{Rogers}, C., {de Marchi}, G., \& {Brandl}, B. 2024,
  \href{http://dx.doi.org/10.1051/0004-6361/202449282}{\color{magenta}\aap},
  \href{https://ui.adsabs.harvard.edu/abs/2024A%26A...684L...8R}{684, L8}

\bibitem[{{Rothman} {et~al.}(2013){Rothman}, {Gordon}, {Babikov}, {Barbe},
  {Chris Benner}, {Bernath}, {Birk}, {Bizzocchi}, {Boudon}, {Brown},
  {Campargue}, {Chance}, {Cohen}, {Coudert}, {Devi}, {Drouin}, {Fayt}, {Flaud},
  {Gamache}, {Harrison}, {Hartmann}, {Hill}, {Hodges}, {Jacquemart}, {Jolly},
  {Lamouroux}, {Le Roy}, {Li}, {Long}, {Lyulin}, {Mackie}, {Massie},
  {Mikhailenko}, {M{\"u}ller}, {Naumenko}, {Nikitin}, {Orphal}, {Perevalov},
  {Perrin}, {Polovtseva}, {Richard}, {Smith}, {Starikova}, {Sung}, {Tashkun},
  {Tennyson}, {Toon}, {Tyuterev}, \& {Wagner}}]{rothman13}
{Rothman}, L.~S., {Gordon}, I.~E., {Babikov}, Y., {et~al.} 2013,
  \href{http://dx.doi.org/10.1016/j.jqsrt.2013.07.002}{\color{magenta}\jqsrt},
  \href{https://ui.adsabs.harvard.edu/abs/2013JQSRT.130....4R}{130, 4}

\bibitem[{{Rotman} {et~al.}(2025){Rotman}, {Welbanks}, {Line}, {McGill},
  {Radica}, \& {Nixon}}]{rotman25}
{Rotman}, Y., {Welbanks}, L., {Line}, M.~R., {et~al.} 2025,
  \href{http://dx.doi.org/10.3847/1538-4357/adef04}{\color{magenta}\apj},
  \href{https://ui.adsabs.harvard.edu/abs/2025ApJ...989..201R}{989, 201}

\bibitem[{{Rousset} {et~al.}(2003){Rousset}, {Lacombe}, {Puget}, {Hubin},
  {Gendron}, {Fusco}, {Arsenault}, {Charton}, {Feautrier}, {Gigan}, {Kern},
  {Lagrange}, {Madec}, {Mouillet}, {Rabaud}, {Rabou}, {Stadler}, \&
  {Zins}}]{rousset03}
{Rousset}, G., {Lacombe}, F., {Puget}, P., {et~al.} 2003, in Society of
  Photo-Optical Instrumentation Engineers (SPIE) Conference Series, Vol. 4839,
  Adaptive Optical System Technologies II, ed. {Wizinowich}, P.~L. \&
  {Bonaccini}, D.,
  \href{https://ui.adsabs.harvard.edu/abs/2003SPIE.4839..140R}{140--149}

\bibitem[{{Rybicki} \& {Lightman}(1979)}]{rl79}
{Rybicki}, G.~B. \& {Lightman}, A.~P. 1979, {Radiative processes in
  astrophysics} (New York: Wiley)

\bibitem[{{Sagynbayeva} {et~al.}(2025){Sagynbayeva}, {Li}, {Kuznetsova}, {Zhu},
  {Jiang}, \& {Armitage}}]{sagynbayeva25}
{Sagynbayeva}, S., {Li}, R., {Kuznetsova}, A., {et~al.} 2025,
  \href{http://dx.doi.org/10.3847/1538-4357/add934}{\color{magenta}\apj},
  \href{https://ui.adsabs.harvard.edu/abs/2025ApJ...987..216S}{987, 216}

\bibitem[{{Salyk} {et~al.}(2013){Salyk}, {Herczeg}, {Brown}, {Blake},
  {Pontoppidan}, \& {van Dishoeck}}]{salyk13}
{Salyk}, C., {Herczeg}, G.~J., {Brown}, J.~M., {et~al.} 2013,
  \href{http://dx.doi.org/10.1088/0004-637X/769/1/21}{\color{magenta}\apj},
  \href{https://ui.adsabs.harvard.edu/abs/2013ApJ...769...21S}{769, 21}

\bibitem[{{Sanghi} {et~al.}(2022){Sanghi}, {Zhou}, \& {Bowler}}]{sanghi22}
{Sanghi}, A., {Zhou}, Y., \& {Bowler}, B.~P. 2022,
  \href{http://dx.doi.org/10.3847/1538-3881/ac477e}{\color{magenta}\aj},
  \href{https://ui.adsabs.harvard.edu/abs/2022AJ....163..119S}{163, 119}

\bibitem[{{Schmidt} {et~al.}(2008){Schmidt}, {Neuh{\"a}user}, {Seifahrt},
  {Vogt}, {Bedalov}, {Helling}, {Witte}, \& {Hauschildt}}]{schmidt08}
{Schmidt}, T.~O.~B., {Neuh{\"a}user}, R., {Seifahrt}, A., {et~al.} 2008,
  \href{http://dx.doi.org/10.1051/0004-6361:20078840}{\color{magenta}\aap},
  \href{https://ui.adsabs.harvard.edu/abs/2008A%26A...491..311S}{491, 311}

\bibitem[{{Schulik} {et~al.}(2019){Schulik}, {Johansen}, {Bitsch}, \&
  {Lega}}]{schulik19}
{Schulik}, M., {Johansen}, A., {Bitsch}, B., \& {Lega}, E. 2019,
  \href{http://dx.doi.org/10.1051/0004-6361/201935473}{\color{magenta}\aap},
  \href{https://ui.adsabs.harvard.edu/abs/2019A%26A...632A.118S}{632, A118}

\bibitem[{{Schulik} {et~al.}(2020){Schulik}, {Johansen}, {Bitsch}, {Lega}, \&
  {Lambrechts}}]{schulik20}
{Schulik}, M., {Johansen}, A., {Bitsch}, B., {Lega}, E., \& {Lambrechts}, M.
  2020,
  \href{http://dx.doi.org/10.1051/0004-6361/202037556}{\color{magenta}\aap},
  \href{https://ui.adsabs.harvard.edu/abs/2020A%26A...642A.187S}{642, A187}

\bibitem[{{Shibaike} \& {Mordasini}(2024)}]{shibaike24}
{Shibaike}, Y. \& {Mordasini}, C. 2024,
  \href{http://dx.doi.org/10.1051/0004-6361/202449522}{\color{magenta}\aap},
  \href{https://ui.adsabs.harvard.edu/abs/2024A%26A...687A.166S}{687, A166}

\bibitem[{{Shibaike} {et~al.}(2026){Shibaike}, {Okuzumi}, {Ueda}, {Doi}, \&
  {Fukagawa}}]{shibaike26}
{Shibaike}, Y., {Okuzumi}, S., {Ueda}, T., {Doi}, K., \& {Fukagawa}, M. 2026,
  \href{http://dx.doi.org/10.3847/2041-8213/ae86f5}{\color{magenta}\apjl},
  \href{https://ui.adsabs.harvard.edu/abs/2026ApJ..1006L..26S}{1006, L26}

\bibitem[{{Shridharan} {et~al.}(2026){Shridharan}, {Manoj}, {Pathak}, {Caratti
  o Garatti}, {Banerjee}, {Henning}, {Kamp}, {van Dishoeck}, {Tyagi}, {Arun},
  {Mathew}, {G{\"u}del}, \& {Lagage}}]{shridharan26}
{Shridharan}, B., {Manoj}, P., {Pathak}, V.~C., {et~al.} 2026,
  \href{http://dx.doi.org/10.1051/0004-6361/202556384}{\color{magenta}\aap},
  \href{https://ui.adsabs.harvard.edu/abs/2026A&A...708A..22S}{708, A22}

\bibitem[{{Sierra} {et~al.}(2026){Sierra}, {Izquierdo}, {Pinilla}, {Chen},
  {Benisty}, {P{\'e}rez}, {Paneque-Carre{\~n}o}, {Agurto-Gangas}, {Bae},
  {Carpenter}, {Cieza}, {Deng}, {Facchini}, {Gonz{\'a}lez-Ruilova}, {Kurtovic},
  {Kuznetsova}, {Manara}, {Miley}, {Ribas}, {Rosotti}, \& {Vioque}}]{sierra26}
{Sierra}, A., {Izquierdo}, A.~F., {Pinilla}, P., {et~al.} 2026,
  \href{http://dx.doi.org/10.1093/mnras/stag1700}{\color{magenta}\mnras},
  \href{https://ui.adsabs.harvard.edu/abs/2026MNRAS.552g1700S}{552, stag1700}

\bibitem[{{Skinner} \& {Audard}(2022)}]{skinner22}
{Skinner}, S.~L. \& {Audard}, M. 2022,
  \href{http://dx.doi.org/10.3847/1538-4357/ac892f}{\color{magenta}\apj},
  \href{https://ui.adsabs.harvard.edu/abs/2022ApJ...938..134S}{938, 134}

\bibitem[{{Smak}(1969)}]{smak69}
{Smak}, J. 1969, \actaa,
  \href{https://ui.adsabs.harvard.edu/abs/1969AcA....19..155S}{19, 155}

\bibitem[{{Smak}(1981)}]{smak81}
{Smak}, J. 1981, \actaa,
  \href{https://ui.adsabs.harvard.edu/abs/1981AcA....31..395S}{31, 395}

\bibitem[{{Smith}(1994)}]{smith94}
{Smith}, M.~D. 1994, \aap,
  \href{https://ui.adsabs.harvard.edu/abs/1994A&A...287..523S}{287, 523}

\bibitem[{{Snellen}(2025)}]{snellen25}
{Snellen}, I. A.~G. 2025,
  \href{http://dx.doi.org/10.1146/annurev-astro-052622-031342}{\color{magenta}\araa},
  \href{https://ui.adsabs.harvard.edu/abs/2025ARA&A..63...83S}{63, 83}

\bibitem[{{Spiegel} \& {Burrows}(2012)}]{sb12}
{Spiegel}, D.~S. \& {Burrows}, A. 2012,
  \href{http://dx.doi.org/10.1088/0004-637X/745/2/174}{\color{magenta}\apj},
  \href{https://ui.adsabs.harvard.edu/abs/2012ApJ...745..174S}{745, 174}

\bibitem[{{Stolker} {et~al.}(2020){Stolker}, {Marleau}, {Cugno},
  {Molli{\`e}re}, {Quanz}, {Todorov}, \& {K{\"u}hn}}]{Stolker+20b}
{Stolker}, T., {Marleau}, G.-D., {Cugno}, G., {et~al.} 2020,
  \href{http://dx.doi.org/10.1051/0004-6361/202038878}{\color{magenta}\aap},
  \href{https://ui.adsabs.harvard.edu/abs/2020A%26A...644A..13S}{644, A13}

\bibitem[{{Sturm} {et~al.}(2024){Sturm}, {Davies}, {Alves}, {Cl{\'e}net},
  {Kotilainen}, {Monna}, {Nicklas}, {Pott}, {Tolstoy}, {Vulcani}, {Achren},
  {Annadevara}, {Anwand-Heerwart}, {Arcidiacono}, {Barboza}, {Barl}, {Baudoz},
  {Bender}, {Bezawada}, {Biondi}, {Bizenberger}, {Blin}, {Bon{\'e}},
  {Bonifacio}, {Borgo}, {Born}, {Buey}, {Cao}, {Chapron}, {Chauvin}, {Chemla},
  {Cloiseau}, {Cohen}, {Colin}, {Czoske}, {Dette}, {Deysenroth}, {Dijkstra},
  {Dreizler}, {Dupuis}, {Egmond}, {Eisenhauer}, {Elswijk}, {Emslander},
  {Fabricius}, {Fasola}, {Ferreira}, {F{\"o}rster Schreiber}, {Fontana},
  {Gaudemard}, {Gautherot}, {Gendron}, {Gennet}, {Genzel}, {Ghouchou},
  {Gillessen}, {Gratadour}, {Grazian}, {Grupp}, {Guieu}, {Gullieuszik}, {Haan},
  {Hartke}, {Hartl}, {Haussmann}, {Helin}, {Hess}, {Hofferbert}, {Huber},
  {Huby}, {Huet}, {Ives}, {Janssen}, {Jaufmann}, {Jilg}, {Jodlbauer}, {Jost},
  {Kausch}, {Kellermann}, {Kerber}, {Kravcar}, {Kravchenko}, {Kulcs{\'a}r},
  {Kuncarayakti}, {Kunst}, {Kwast}, {Lang}, {Lange}, {Lapeyrere}, {Le Ruyet},
  {Leschinski}, {Locatelli}, {Massari}, {Mattila}, {Mei}, {Merlin}, {Meyer},
  {Michel}, {Mohr}, {Montarg{\`e}s}, {M{\"u}ller}, {M{\"u}nch}, {Navarro},
  {Neumann}, {Neumayer}, {Neumeier}, {Pedichini}, {Pfl{\"u}ger}, {Piazzesi},
  {Pinard}, {Porras}, {Portulari}, {Przybilla}, {Rabien}, {Raffard},
  {Ragazzoni}, {Ramlau}, {Ramos}, {Ramsay}, {Raynaud}, {Rhode}, {Richter},
  {Rix}, {Rodenhuis}, {Rohloff}, {Romp}, {Rousselot}, {Sabha}, {Sassolas},
  {Schlichter}, {Schuil}, {Schweitzer}, {Seemann}, {Sevin}, {Simioni},
  {Spallek}, {S{\"o}nmez}, {Suuronen}, {Taburet}, {Thomas}, {Tisserand},
  {Vaccari}, {Valenti}, {Verdoes Kleijn}, {Verdugo}, {Vidal}, {Wagner},
  {Wegner}, {Winden}, {Witschel}, {Zanella}, {Zeilinger}, {Ziegleder}, \&
  {Ziegler}}]{sturm24}
{Sturm}, E., {Davies}, R., {Alves}, J., {et~al.} 2024, in Society of
  Photo-Optical Instrumentation Engineers (SPIE) Conference Series, Vol. 13096,
  Ground-based and Airborne Instrumentation for Astronomy X, ed. {Bryant},
  J.~J., {Motohara}, K., \& {Vernet}, J. R.~D.,
  \href{https://ui.adsabs.harvard.edu/abs/2024SPIE13096E..11S}{1309611}

\bibitem[{{Sun} {et~al.}(2024){Sun}, {Huang}, {Dong}, \& {Liu}}]{sun24}
{Sun}, X., {Huang}, P., {Dong}, R., \& {Liu}, S.-F. 2024,
  \href{http://dx.doi.org/10.3847/1538-4357/ad57c2}{\color{magenta}\apj},
  \href{https://ui.adsabs.harvard.edu/abs/2024ApJ...972...25S}{972, 25}

\bibitem[{{Sun} {et~al.}(2026){Sun}, {Marleau}, \& {Liu}}]{sun26}
{Sun}, X., {Marleau}, G.-D., \& {Liu}, S.-F. 2026,
  \href{http://dx.doi.org/10.1051/0004-6361/202659308}{\color{magenta}\aap},
  \href{https://ui.adsabs.harvard.edu/abs/2026A&A...712A.221S}{712, A221}

\bibitem[{{Szul{\'a}gyi} \& {Ercolano}(2020)}]{szul20}
{Szul{\'a}gyi}, J. \& {Ercolano}, B. 2020,
  \href{http://dx.doi.org/10.3847/1538-4357/abb5a2}{\color{magenta}\apj},
  \href{https://ui.adsabs.harvard.edu/abs/2020ApJ...902..126S}{902, 126}

\bibitem[{{Szul{\'a}gyi} {et~al.}(2016){Szul{\'a}gyi}, {Masset}, {Lega},
  {Crida}, {Morbidelli}, \& {Guillot}}]{szul16}
{Szul{\'a}gyi}, J., {Masset}, F., {Lega}, E., {et~al.} 2016,
  \href{http://dx.doi.org/10.1093/mnras/stw1160}{\color{magenta}\mnras},
  \href{https://ui.adsabs.harvard.edu/abs/2016MNRAS.460.2853S}{460, 2853}

\bibitem[{{Takami} {et~al.}(2025){Takami}, {Otten}, {Absil}, {Delacroix},
  {Karr}, \& {Wang}}]{takami25}
{Takami}, M., {Otten}, G., {Absil}, O., {et~al.} 2025,
  \href{http://dx.doi.org/10.1088/1538-3873/adbbc4}{\color{magenta}\pasp},
  \href{https://ui.adsabs.harvard.edu/abs/2025PASP..137c4504T}{137, 034504}

\bibitem[{{Takasao} {et~al.}(2021){Takasao}, {Aoyama}, \& {Ikoma}}]{takasao21}
{Takasao}, S., {Aoyama}, Y., \& {Ikoma}, M. 2021,
  \href{http://dx.doi.org/10.3847/1538-4357/ac0f7e}{\color{magenta}\apj},
  \href{https://ui.adsabs.harvard.edu/abs/2021ApJ...921...10T}{921, 10}

\bibitem[{{Tanigawa} {et~al.}(2012){Tanigawa}, {Ohtsuki}, \&
  {Machida}}]{tanigawa12}
{Tanigawa}, T., {Ohtsuki}, K., \& {Machida}, M.~N. 2012,
  \href{http://dx.doi.org/10.1088/0004-637X/747/1/47}{\color{magenta}\apj},
  \href{https://ui.adsabs.harvard.edu/abs/2012ApJ...747...47T}{747, 47}

\bibitem[{{Taylor} \& {Adams}(2024)}]{taylor24}
{Taylor}, A.~G. \& {Adams}, F.~C. 2024,
  \href{http://dx.doi.org/10.1016/j.icarus.2024.116044}{\color{magenta}\icarus},
  \href{https://ui.adsabs.harvard.edu/abs/2024Icar..41516044T}{415, 116044}

\bibitem[{{Taylor} \& {Adams}(2025)}]{taylor25}
{Taylor}, A.~G. \& {Adams}, F.~C. 2025,
  \href{http://dx.doi.org/10.1016/j.icarus.2024.116327}{\color{magenta}\icarus},
  \href{https://ui.adsabs.harvard.edu/abs/2025Icar..42516327T}{425, 116327}

\bibitem[{{Taylor} \& {Adams}(2026)}]{taylor26b}
{Taylor}, A.~G. \& {Adams}, F.~C. 2026,
  \href{http://dx.doi.org/10.1016/j.icarus.2026.117264}{\color{magenta}\icarus},
  \href{https://ui.adsabs.harvard.edu/abs/2026Icar..46017264T}{460, 117264}

\bibitem[{{Teague} \& {Foreman-Mackey}(2018)}]{teague18centroid}
{Teague}, R. \& {Foreman-Mackey}, D. 2018,
  \href{http://dx.doi.org/10.3847/2515-5172/aae265}{\color{magenta}RNAAS},
  \href{https://ui.adsabs.harvard.edu/abs/2018RNAAS...2..173T}{2, 173}

\bibitem[{{Tennyson} {et~al.}(2024){Tennyson}, {Yurchenko}, {Zhang},
  {Bowesman}, {Brady}, {Buldyreva}, {Chubb}, {Gamache}, {Gorman}, {Guest},
  {Hill}, {Kefala}, {Lynas-Gray}, {Mellor}, {McKemmish}, {Mitev}, {Mizus},
  {Owens}, {Peng}, {Perri}, {Pezzella}, {Polyansky}, {Qu}, {Semenov}, {Smola},
  {Solokov}, {Somogyi}, {Upadhyay}, {Wright}, \& {Zobov}}]{tennyson24}
{Tennyson}, J., {Yurchenko}, S.~N., {Zhang}, J., {et~al.} 2024,
  \href{http://dx.doi.org/10.1016/j.jqsrt.2024.109083}{\color{magenta}\jqsrt},
  \href{https://ui.adsabs.harvard.edu/abs/2024JQSRT.32609083T}{326, 109083}

\bibitem[{{Testi} {et~al.}(2025){Testi}, {Natta}, {Gozzi}, {Manara},
  {Williams}, {Claes}, {Lebreuilly}, {Hennebelle}, {Klessen}, \&
  {Molinari}}]{testi25}
{Testi}, L., {Natta}, A., {Gozzi}, S., {et~al.} 2025,
  \href{http://dx.doi.org/10.1051/0004-6361/202554149}{\color{magenta}\aap},
  \href{https://ui.adsabs.harvard.edu/abs/2025A&A...703A.277T}{703, A277}

\bibitem[{{Thanathibodee} {et~al.}(2019){Thanathibodee}, {Calvet}, {Bae},
  {Muzerolle}, \& {Hern{\'a}ndez}}]{thanathibodee19}
{Thanathibodee}, T., {Calvet}, N., {Bae}, J., {Muzerolle}, J., \&
  {Hern{\'a}ndez}, R.~F. 2019,
  \href{http://dx.doi.org/10.3847/1538-4357/ab44c1}{\color{magenta}\apj},
  \href{https://ui.adsabs.harvard.edu/abs/2019ApJ...885...94T}{885, 94}

\bibitem[{{Thanathibodee} {et~al.}(2020){Thanathibodee}, {Molina}, {Calvet},
  {Serna}, {Bae}, {Reynolds}, {Hern{\'a}ndez}, {Muzerolle}, \&
  {Hern{\'a}ndez}}]{Thanathibodee+2020}
{Thanathibodee}, T., {Molina}, B., {Calvet}, N., {et~al.} 2020,
  \href{http://dx.doi.org/10.3847/1538-4357/ab77c1}{\color{magenta}\apj},
  \href{https://ui.adsabs.harvard.edu/abs/2020ApJ...892...81T}{892, 81}

\bibitem[{{Thanathibodee} {et~al.}(2024){Thanathibodee}, {Robinson}, {Calvet},
  {Espaillat}, {Pittman}, {Arulanantham}, {France}, {G{\"u}nther}, {Chang}, \&
  {Schneider}}]{thanathibodee24}
{Thanathibodee}, T., {Robinson}, C.~E., {Calvet}, N., {et~al.} 2024,
  \href{http://dx.doi.org/10.3847/1538-4357/ad7b2d}{\color{magenta}\apj},
  \href{https://ui.adsabs.harvard.edu/abs/2024ApJ...975..193T}{975, 193}

\bibitem[{{Tofflemire} {et~al.}(2025){Tofflemire}, {Manara}, {Banzatti},
  {Pontoppidan}, {Najita}, {Nisini}, {Whelan}, {Campbell-White}, {Alqubelat},
  {Kraus}, {Rab}, {Houge}, {Krijt}, {Muzerolle}, {Fiorellino}, {Benisty},
  {Tychoniec}, {Salyk}, {Bourdarot}, \& {Hyden}}]{tofflemire25}
{Tofflemire}, B.~M., {Manara}, C.~F., {Banzatti}, A., {et~al.} 2025,
  \href{http://dx.doi.org/10.3847/1538-4357/adcc23}{\color{magenta}\apj},
  \href{https://ui.adsabs.harvard.edu/abs/2025ApJ...985..224T}{985, 224}

\bibitem[{{Trevascus} {et~al.}(2025){Trevascus}, {Blunt}, {Christiaens},
  {Matthews}, {Hammond}, {Brandner}, {Wang}, {Lacour}, {Vigan}, {Balmer},
  {Bonnefoy}, {Burn}, {Chauvin}, {Gratton}, {Houll{\'e}}, {Hinkley},
  {Kammerer}, {Kreidberg}, {Marleau}, {Mesa}, {Otten}, {Nowak}, {Rickman},
  {Sanchez-Bermudez}, \& {Sauter}}]{trevascus25}
{Trevascus}, D., {Blunt}, S., {Christiaens}, V., {et~al.} 2025,
  \href{http://dx.doi.org/10.1051/0004-6361/202553936}{\color{magenta}\aap},
  \href{https://ui.adsabs.harvard.edu/abs/2025A%26A...698A..19T}{698, A19}

\bibitem[{{Trevascus} {et~al.}(2026){Trevascus}, {Brandner}, {Balsalobre-Ruza},
  {Lacour}, {Abd El Dayem}, {Aimar}, {Berdeu}, {Berger}, {Bourdarot},
  {Christiaens}, {Correia}, {Davies}, {Defr{\`e}re}, {Drescher}, {Eckart},
  {Eisenhauer}, {Fabricius}, {Feuchtgruber}, {Flesch}, {F{\"o}rster Schreiber},
  {Foschi}, {Fournier}, {Garcia}, {Garcia Lopez}, {Genzel}, {Gillessen},
  {Hammond}, {H{\"o}nig}, {Houll{\'e}}, {Joharle}, {Kervella}, {Kreidberg},
  {Labadie}, {Lai}, {Laugier}, {Le Bouquin}, {Leftley}, {Li}, {Lopez}, {Lutz},
  {Marleau}, {Mang}, {M{\'e}rand}, {Millour}, {Montarg{\`e}s}, {Moruj{\~a}o},
  {Nowacki}, {Nowak}, {Osorno}, {Ott}, {Pappert}, {Paumard}, {Perraut},
  {Perrin}, {Petrov}, {Petrucci}, {Pourr{\'e}}, {Rabien}, {Ribeiro},
  {Robbe-Dubois}, {Sadun Bordoni}, {Sanchez Bermudez}, {Santos}, {Sauter},
  {Scigliuto}, {Shangguan}, {Shimizu}, {Soulez}, {Straubmeier}, {Sturm},
  {Subroweit}, {Sykes}, {Tacconi}, {Th{\'e}venet}, {Urso}, {Vincent},
  {Woillez}, \& {The Gravity+ Collaboration}}]{trevascus26}
{Trevascus}, D., {Brandner}, W., {Balsalobre-Ruza}, O., {et~al.} 2026,
  \href{http://dx.doi.org/10.1051/0004-6361/202660732}{\color{magenta}\aap},
  \href{https://ui.adsabs.harvard.edu/abs/2026A&A...713L..16T}{713, L16}

\bibitem[{{Ulrich}(1976)}]{ulrich76}
{Ulrich}, R.~K. 1976,
  \href{http://dx.doi.org/10.1086/154840}{\color{magenta}\apj},
  \href{https://ui.adsabs.harvard.edu/abs/1976ApJ...210..377U}{210, 377}

\bibitem[{{Uyama} {et~al.}(2021){Uyama}, {Xie}, {Aoyama}, {Beichman},
  {Hashimoto}, {Dong}, {Hasegawa}, {Ikoma}, {Mawet}, {McElwain}, {Ruffio},
  {Wagner}, {Wang}, \& {Zhou}}]{uyama21b}
{Uyama}, T., {Xie}, C., {Aoyama}, Y., {et~al.} 2021,
  \href{http://dx.doi.org/10.3847/1538-3881/ac2739}{\color{magenta}\aj},
  \href{https://ui.adsabs.harvard.edu/abs/2021AJ....162..214U}{162, 214}

\bibitem[{{van Capelleveen} {et~al.}(2025){van Capelleveen}, {Ginski},
  {Kenworthy}, {Byrne}, {Lawlor}, {McLachlan}, {Mamajek}, {Stolker}, {Benisty},
  {Bohn}, {Close}, {Dominik}, {Haffert}, {Landman}, {Ma}, {Snellen}, {Tazaki},
  {van der Marel}, {Welzel}, \& {Zhang}}]{vancapelleveen25a}
{van Capelleveen}, R.~F., {Ginski}, C., {Kenworthy}, M.~A., {et~al.} 2025,
  \href{http://dx.doi.org/10.3847/2041-8213/adf721}{\color{magenta}\apjl},
  \href{https://ui.adsabs.harvard.edu/abs/2025ApJ...990L...8V}{990, L8}

\bibitem[{{Viswanath} {et~al.}(2026){Viswanath}, {Bonnefoy}, {Dougados},
  {Ringqvist}, {Janson}, {Demars}, {Sicilia-Aguilar}, {Bouvier}, {Marleau},
  {Alecian}, \& {Chauvin}}]{viswanath26}
{Viswanath}, G., {Bonnefoy}, M., {Dougados}, C., {et~al.} 2026,
  \href{http://dx.doi.org/10.1051/0004-6361/202558444}{\color{magenta}\aap},
  \href{https://ui.adsabs.harvard.edu/abs/2026A%26A...710A..69V}{710, A69}

\bibitem[{{Viswanath} {et~al.}(2024){Viswanath}, {Ringqvist}, {Demars},
  {Janson}, {Bonnefoy}, {Aoyama}, {Marleau}, {Dougados}, {Szul{\'a}gyi}, \&
  {Thanathibodee}}]{viswanath24}
{Viswanath}, G., {Ringqvist}, S.~C., {Demars}, D., {et~al.} 2024,
  \href{http://dx.doi.org/10.1051/0004-6361/202450881}{\color{magenta}\aap},
  \href{https://ui.adsabs.harvard.edu/abs/2024A%26A...691A..64V}{691, A64}

\bibitem[{{Vos} {et~al.}(2022){Vos}, {Faherty}, {Gagn{\'e}}, {Marley},
  {Metchev}, {Gizis}, {Rice}, \& {Cruz}}]{vos22}
{Vos}, J.~M., {Faherty}, J.~K., {Gagn{\'e}}, J., {et~al.} 2022,
  \href{http://dx.doi.org/10.3847/1538-4357/ac4502}{\color{magenta}\apj},
  \href{https://ui.adsabs.harvard.edu/abs/2022ApJ...924...68V}{924, 68}

\bibitem[{{Wahhaj} {et~al.}(2024){Wahhaj}, {Benisty}, {Ginski}, {Swastik},
  {Arora}, {van Holstein}, {De Rosa}, {Yang}, {Bae}, \& {Ren}}]{wahhaj24}
{Wahhaj}, Z., {Benisty}, M., {Ginski}, C., {et~al.} 2024,
  \href{http://dx.doi.org/10.1051/0004-6361/202349018}{\color{magenta}\aap},
  \href{https://ui.adsabs.harvard.edu/abs/2024A%26A...687A.257W}{687, A257}

\bibitem[{{Wang} {et~al.}(2020){Wang}, {Ginzburg}, {Ren}, {Wallack}, {Gao},
  {Mawet}, {Bond}, {Cetre}, {Wizinowich}, {De Rosa}, {Ruane}, {Liu}, {Absil},
  {Alvarez}, {Baranec}, {Choquet}, {Chun}, {Defr{\`e}re}, {Delorme},
  {Duch{\^e}ne}, {Forsberg}, {Ghez}, {Guyon}, {Hall}, {Huby}, {Jolivet},
  {Jensen-Clem}, {Jovanovic}, {Karlsson}, {Lilley}, {Matthews}, {M{\'e}nard},
  {Meshkat}, {Millar-Blanchaer}, {Ngo}, {Orban de Xivry}, {Pinte}, {Ragland},
  {Serabyn}, {Catal{\'a}n}, {Wang}, {Wetherell}, {Williams}, {Ygouf}, \&
  {Zuckerman}}]{wang20}
{Wang}, J.~J., {Ginzburg}, S., {Ren}, B., {et~al.} 2020,
  \href{http://dx.doi.org/10.3847/1538-3881/ab8aef}{\color{magenta}\aj},
  \href{https://ui.adsabs.harvard.edu/abs/2020AJ....159..263W}{159, 263}

\bibitem[{{Wang} {et~al.}(2021{\natexlab{a}}){Wang}, {Kulikauskas}, \&
  {Blunt}}]{wang21withp}
{Wang}, J.~J., {Kulikauskas}, M., \& {Blunt}, S. 2021{\natexlab{a}},
  {whereistheplanet: Predicting positions of directly imaged companions},
  \href{https://ascl.net/2101.003}{\color{magenta}{Astrophysics Source Code
  Library}},
  \href{https://ui.adsabs.harvard.edu/abs/2021ascl.soft01003W}{record
  ascl:2101.003}

\bibitem[{{Wang} {et~al.}(2021{\natexlab{b}}){Wang}, {Vigan}, {Lacour},
  {Nowak}, {Stolker}, {De Rosa}, {Ginzburg}, {Gao}, {Abuter}, {Amorim},
  {Asensio-Torres}, {Baub{\"o}ck}, {Benisty}, {Berger}, {Beust}, {Beuzit},
  {Blunt}, {Boccaletti}, {Bohn}, {Bonnefoy}, {Bonnet}, {Brandner},
  {Cantalloube}, {Caselli}, {Charnay}, {Chauvin}, {Choquet}, {Christiaens},
  {Cl{\'e}net}, {Coud{\'e} Du Foresto}, {Cridland}, {de Zeeuw}, {Dembet},
  {Dexter}, {Drescher}, {Duvert}, {Eckart}, {Eisenhauer}, {Facchini}, {Gao},
  {Garcia}, {Garcia Lopez}, {Gardner}, {Gendron}, {Genzel}, {Gillessen},
  {Girard}, {Haubois}, {Hei{\ss}el}, {Henning}, {Hinkley}, {Hippler},
  {Horrobin}, {Houll{\'e}}, {Hubert}, {Jim{\'e}nez-Rosales}, {Jocou},
  {Kammerer}, {Keppler}, {Kervella}, {Meyer}, {Kreidberg}, {Lagrange},
  {Lapeyr{\`e}re}, {Le Bouquin}, {L{\'e}na}, {Lutz}, {Maire}, {M{\'e}nard},
  {M{\'e}rand}, {Molli{\`e}re}, {Monnier}, {Mouillet}, {M{\"u}ller},
  {Nasedkin}, {Ott}, {Otten}, {Paladini}, {Paumard}, {Perraut}, {Perrin},
  {Pfuhl}, {Pueyo}, {Rameau}, {Rodet}, {Rodr{\'\i}guez-Coira}, {Rousset},
  {Scheithauer}, {Shangguan}, {Shimizu}, {Stadler}, {Straub}, {Straubmeier},
  {Sturm}, {Tacconi}, {van Dishoeck}, {Vincent}, {von Fellenberg},
  {Ward-Duong}, {Widmann}, {Wieprecht}, {Wiezorrek}, {Woillez}, \& {Gravity
  Collaboration}}]{wang21vlti}
{Wang}, J.~J., {Vigan}, A., {Lacour}, S., {et~al.} 2021{\natexlab{b}},
  \href{http://dx.doi.org/10.3847/1538-3881/abdb2d}{\color{magenta}\aj},
  \href{https://ui.adsabs.harvard.edu/abs/2021AJ....161..148W}{161, 148}

\bibitem[{{Wang} \& {Chen}(2019)}]{wang19}
{Wang}, S. \& {Chen}, X. 2019,
  \href{http://dx.doi.org/10.3847/1538-4357/ab1c61}{\color{magenta}\apj},
  \href{https://ui.adsabs.harvard.edu/abs/2019ApJ...877..116W}{877, 116}

\bibitem[{{Ward} \& {Canup}(2010)}]{ward10}
{Ward}, W.~R. \& {Canup}, R.~M. 2010,
  \href{http://dx.doi.org/10.1088/0004-6256/140/5/1168}{\color{magenta}\aj},
  \href{https://ui.adsabs.harvard.edu/abs/2010AJ....140.1168W}{140, 1168}

\bibitem[{{White} \& {Basri}(2003)}]{white03}
{White}, R.~J. \& {Basri}, G. 2003,
  \href{http://dx.doi.org/10.1086/344673}{\color{magenta}\apj},
  \href{https://ui.adsabs.harvard.edu/abs/2003ApJ...582.1109W}{582, 1109}

\bibitem[{{Witte} {et~al.}(2011){Witte}, {Helling}, {Barman}, {Heidrich}, \&
  {Hauschildt}}]{witte11}
{Witte}, S., {Helling}, C., {Barman}, T., {Heidrich}, N., \& {Hauschildt},
  P.~H. 2011,
  \href{http://dx.doi.org/10.1051/0004-6361/201014105}{\color{magenta}\aap},
  \href{https://ui.adsabs.harvard.edu/abs/2011A%26A...529A..44W}{529, A44}

\bibitem[{{Witte} {et~al.}(2009){Witte}, {Helling}, \& {Hauschildt}}]{witte09}
{Witte}, S., {Helling}, C., \& {Hauschildt}, P.~H. 2009,
  \href{http://dx.doi.org/10.1051/0004-6361/200811501}{\color{magenta}\aap},
  \href{https://ui.adsabs.harvard.edu/abs/2009A&A...506.1367W}{506, 1367}

\bibitem[{{Wittenmyer} {et~al.}(2020){Wittenmyer}, {Wang}, {Horner}, {Butler},
  {Tinney}, {Carter}, {Wright}, {Jones}, {Bailey}, {O'Toole}, \&
  {Johns}}]{wittenmyer20}
{Wittenmyer}, R.~A., {Wang}, S., {Horner}, J., {et~al.} 2020,
  \href{http://dx.doi.org/10.1093/mnras/stz3436}{\color{magenta}\mnras},
  \href{https://ui.adsabs.harvard.edu/abs/2020MNRAS.492..377W}{492, 377}

\bibitem[{{Woitke} {et~al.}(2009){Woitke}, {Kamp}, \& {Thi}}]{woitke09}
{Woitke}, P., {Kamp}, I., \& {Thi}, W.-F. 2009,
  \href{http://dx.doi.org/10.1051/0004-6361/200911821}{\color{magenta}\aap},
  \href{https://ui.adsabs.harvard.edu/abs/2009A&A...501..383W}{501, 383}

\bibitem[{{Woitke} {et~al.}(2016){Woitke}, {Min}, {Pinte}, {Thi}, {Kamp},
  {Rab}, {Anthonioz}, {Antonellini}, {Baldovin-Saavedra}, {Carmona}, {Dominik},
  {Dionatos}, {Greaves}, {G{\"u}del}, {Ilee}, {Liebhart}, {M{\'e}nard},
  {Rigon}, {Waters}, {Aresu}, {Meijerink}, \& {Spaans}}]{woitke16}
{Woitke}, P., {Min}, M., {Pinte}, C., {et~al.} 2016,
  \href{http://dx.doi.org/10.1051/0004-6361/201526538}{\color{magenta}\aap},
  \href{https://ui.adsabs.harvard.edu/abs/2016A%26A...586A.103W}{586, A103}

\bibitem[{{Woitke} {et~al.}(2018){Woitke}, {Min}, {Thi}, {Roberts}, {Carmona},
  {Kamp}, {M{\'e}nard}, \& {Pinte}}]{woitke18}
{Woitke}, P., {Min}, M., {Thi}, W.~F., {et~al.} 2018,
  \href{http://dx.doi.org/10.1051/0004-6361/201731460}{\color{magenta}\aap},
  \href{https://ui.adsabs.harvard.edu/abs/2018A%26A...618A..57W}{618, A57}

\bibitem[{{Woitke} {et~al.}(2024){Woitke}, {Thi}, {Arabhavi}, {Kamp},
  {K{\'o}sp{\'a}l}, \& {{\'A}brah{\'a}m}}]{woitke24}
{Woitke}, P., {Thi}, W.~F., {Arabhavi}, A.~M., {et~al.} 2024,
  \href{http://dx.doi.org/10.1051/0004-6361/202347730}{\color{magenta}\aap},
  \href{https://ui.adsabs.harvard.edu/abs/2024A%26A...683A.219W}{683, A219}

\bibitem[{{Xie} {et~al.}(2020){Xie}, {Haffert}, {de Boer}, {Kenworthy},
  {Brinchmann}, {Girard}, {Snellen}, \& {Keller}}]{xie20}
{Xie}, C., {Haffert}, S.~Y., {de Boer}, J., {et~al.} 2020,
  \href{http://dx.doi.org/10.1051/0004-6361/202038242}{\color{magenta}\aap},
  \href{https://ui.adsabs.harvard.edu/abs/2020A%26A...644A.149X}{644, A149}

\bibitem[{{Zhang}(2024)}]{zhang24}
{Zhang}, Z. 2024,
  \href{http://dx.doi.org/10.3847/2515-5172/ad4481}{\color{magenta}\rnaas},
  \href{https://ui.adsabs.harvard.edu/abs/2024RNAAS...8..114Z}{8, 114}

\bibitem[{{Zhang} {et~al.}(2025){Zhang}, {Molli{\`e}re}, {Fortney}, \&
  {Marley}}]{zhangmoll25}
{Zhang}, Z., {Molli{\`e}re}, P., {Fortney}, J.~J., \& {Marley}, M.~S. 2025,
  \href{http://dx.doi.org/10.3847/1538-3881/addfcb}{\color{magenta}\aj},
  \href{https://ui.adsabs.harvard.edu/abs/2025AJ....170...64Z}{170, 64}

\bibitem[{{Zhou} {et~al.}(2026){Zhou}, {Biller}, {Carter}, {Perrin}, {Poon},
  {Su{\'a}rez}, {Ben J. Sutlieff}, {Vos}, {Wang}, {Balmer}, {Bryan},
  {Boccaletti}, {Girard}, {Gonzales}, {Kammerer}, {Leisenring}, {Palma-Bifani},
  {Wagner}, {Apai}, {Bonnefoy}, {Bowler}, {Franson}, {Liu}, {Mel{\'e}ndez},
  {Metchev}, {Petrus}, {Pueyo}, {Rebollido}, {Skemer}, {Tan}, \&
  {Whiteford}}]{zhou26}
{Zhou}, Y., {Biller}, B.~A., {Carter}, A.~L., {et~al.} 2026,
  \href{http://dx.doi.org/10.3847/1538-3881/ae8bab}{\color{magenta}\aj},
  \href{https://ui.adsabs.harvard.edu/abs/2026AJ....172..152Z}{172, 152}

\bibitem[{{Zhou} {et~al.}(2025){Zhou}, {Bowler}, {Sanghi}, {Marleau},
  {Takasao}, {Aoyama}, {Hasegawa}, {Thanathibodee}, {Uyama}, {Hashimoto},
  {Wagner}, {Calvet}, {Demars}, {Wu}, {Biddle}, {Haffert}, \& {Bryan}}]{zhou25}
{Zhou}, Y., {Bowler}, B.~P., {Sanghi}, A., {et~al.} 2025,
  \href{http://dx.doi.org/10.3847/2041-8213/adb134}{\color{magenta}\apjl},
  \href{https://ui.adsabs.harvard.edu/abs/2025ApJ...980L..39Z}{980, L39}

\bibitem[{{Zhou} {et~al.}(2021){Zhou}, {Bowler}, {Wagner}, {Schneider}, {Apai},
  {Kraus}, {Close}, {Herczeg}, \& {Fang}}]{zhou21}
{Zhou}, Y., {Bowler}, B.~P., {Wagner}, K.~R., {et~al.} 2021,
  \href{http://dx.doi.org/10.3847/1538-3881/abeb7a}{\color{magenta}\aj},
  \href{https://ui.adsabs.harvard.edu/abs/2021AJ....161..244Z}{161, 244}

\bibitem[{{Zhu}(2015)}]{zhu15}
{Zhu}, Z. 2015,
  \href{http://dx.doi.org/10.1088/0004-637X/799/1/16}{\color{magenta}\apj},
  \href{https://ui.adsabs.harvard.edu/abs/2015ApJ...799...16Z}{799, 16}

\bibitem[{{Zurlo} {et~al.}(2020){Zurlo}, {Cugno}, {Montesinos}, {Perez},
  {Canovas}, {Casassus}, {Christiaens}, {Cieza}, \& {Huelamo}}]{Zurlo+2020}
{Zurlo}, A., {Cugno}, G., {Montesinos}, M., {et~al.} 2020,
  \href{http://dx.doi.org/10.1051/0004-6361/201936891}{\color{magenta}\aap},
  \href{https://ui.adsabs.harvard.edu/abs/2020A%26A...633A.119Z}{633, A119}

\end{thebibliography}

\begin{appendix}

\section{Flux at observer}
 \label{Th:FlbeimBeob}

\subsection{Emission from the planetary surface}

\begin{figure}[hp] %
 \centering
 \includegraphics[width=0.47\textwidth]{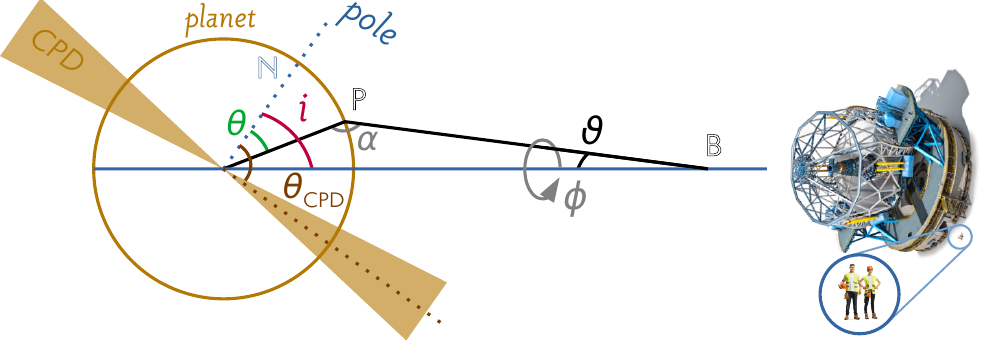}
\caption{%
Geometry for the calculation of the emission, with some important angles indicated (see text too). The point $\Vekt{P}$ on the planet is in general not in the plane of the paper, and $\alpha$ is the (true, non-projected) angle. The azimuthal angle around the planet--observer axis is denoted by $\phi$ and the azimuthal angle around the pole of the planet is denoted by $\varphi$ (`varphi').
Figure only partially to scale.
}
\label{Abb:Winkelabb}
\end{figure}

We wished to obtain the flux at the observer from the surface of the planet.
For this, we first ignored the CPD and later integrated over the appropriate surface.
We let the observer be at distance $d$ from the planet of radius \RP. Whether bolometric or monochromatic,  %
the (radial) flux  %
at the observer is the integral of the intensity $I$ from the projected visible area of the planet (e.g., \citealp{rl79} or Sect.~65 of \citealp{mihalas84}):
\begin{equation}
 \label{Gl:F}
 F = \int_{0}^{2\pi}\!\!\int_0^{\varthmaxP} I \cos\vartheta\,\sin\vartheta\,{\rm d}\vartheta\, {\rm d}\phi,
\end{equation}
where $\vartheta$ is the angle measured from the axis connecting the observer and a point on the surface, $\phi$ is the angle around this axis, and $\varthmaxP=\arcsin(\RP/d)$ is half the maximum angular extent of the planet as seen from the observer.
One assumption here is that the intensity $I$ is a constant function of the angle from the local normal at every point of the surface of the emitter.
Figure~\ref{Abb:Winkelabb} illustrates the geometry. The spherical angles centred on the planet are denoted by $\theta$ for the polar angle and $\varphi$ for the azimuthal one.
We further define the inclination of the planet, the angle between its pole, and the direction to the observer, as $i$. 
For a generic inclination, $\phi\neq\varphi$; the two are the same\footnote{More precisely, offset by an arbitrarily-chosen but constant phase.} only for $i=0$, a pole-on viewing geometry.
The `pole' here refers to the direction of the orbital angular momentum of the planet since this is the one relevant for the infall of gas.
The spin axis of the planet could differ 
but this is not of concern for our work.

Since $\RP/d=4.6\times10^{-11}(\RP/2~\RJ)/(d/100~\mathrm{pc})\ll1$, it will be convenient when integrating numerically to rescale the $\vartheta$ variable.
We do it at first exactly by defining
\begin{subequations}
\label{Gl:varmuDef}
\begin{align}
 \varmu & \equiv\cos\vartheta,\eqsep\varmumin\equiv\cos\varthmaxP\\
 \varmu & \equiv\varmumin + m (1-\varmumin),
\end{align}
\end{subequations}
so that Equation~(\ref{Gl:F}) becomes  %
\begin{equation}
 \label{Gl:Fvarmu}
 F = \left(1-\varmumin\right) \int_{0}^{2\pi}\int_0^1 I \left[\varmumin+m\left(1-\varmumin\right)\right]  \,{\rm d}m\, {\rm d}\phi.
\end{equation}
If the intensity is constant over the (visible) surface, Equation~(\ref{Gl:F}) or~(\ref{Gl:Fvarmu}) yields
\begin{equation}
 F_{\mathrm{const.}~I} = \pi I \left(1-\varmumin^2\right)
\end{equation}
exactly. Evaluated at the surface ($d\rightarrow\RP$, i.e., $\varmumin=0$),
this relates the intensity to the surface flux density $\FOb$ by
\begin{equation}
\label{Gl:I2F}
 I = \frac{1}{\pi}\FOb.
\end{equation}
One can check %
that this holds in fact also for an arbitrary $\FOb(\theta,\varphi)$,
since an observer at $d$ sees only the local surface in the limit $d\rightarrow\RP$.
Going back to the general case, but in the far-field limit, that is, with $\varmumin=\cos\varthmaxP\approx1-\frac{1}{2}(\RP/d)^2$, Equation~(\ref{Gl:Fvarmu}) becomes to second order in $\RP/d$ the usual inverse-square law, $F = F_0\times(\RP/d)^2$, with $F_0=\iint \FOb\, {\rm d}m\,{\rm d}\phi$ the surface-integrated flux.

To compute the integral in Equation~(\ref{Gl:Fvarmu}), we need to relate the two sets of angles $(\vartheta,\,\phi)$ and $(\theta,\,\varphi)$.
By our assumption of azimuthally symmetric infall onto the planet, the emission will depend only on the angle from the pole: $\FOb=\FOb(\theta)$. Therefore, we only need to obtain $\theta=\theta(\vartheta,\phi)$.

We let $\alpha$ be the angle between the planet--observer axis and the surface point as seen from the centre of the planet; if the point is in the plane of Figure~\ref{Abb:Winkelabb}, $\alpha=i-\theta$, but this does not hold in general. Applying the law of sines to the complementary angle of that triangle yields in general
\begin{equation}
 \alpha = \arcsin\left(\frac{\sin\vartheta}{\RP/d}\right) - \vartheta.
\end{equation}
Again, if the surface point is in the plane of Figure~\ref{Abb:Winkelabb} (more exactly defined by the planet--observer and the pole vectors), we obtain:
\begin{equation}
 \label{Gl:thetaPolebene}
 \thetaPolebene = \left|i-\alpha\right| = \left|i-\arcsin\left(\frac{\sin\vartheta}{\RP/d}\right) + \vartheta\right|.
\end{equation}
It is easy to check that \thetaPolebene fulfils the intuitive limiting cases.

To obtain the $\theta$ coordinate of a generic point on the surface of the planet, the simplest approach is to express its position and that of the pole in Cartesian coordinates, calculate the distance between the pole at the surface and the point, and use the equation for a chord (the law of cosines) to find the angle $\theta$. In Figure~\ref{Abb:Winkelabb}, we set the origin at the centre of the planet, $x$ is out of the page, $y$ is towards the observer, and $z$ is upwards on the page. Then, the observer is at $\Vekt{B}=(0,\,d,\,0)$, the pole at $\Vekt{N}=(0,\,\RP\cos i,\,\RP\sin i)$, and the arbitrary point at $\Vekt{P}=(\RP\sin\phi\sin\alpha,\,\RP\cos\alpha,\,\RP\cos\phi\sin\alpha)$. The expression for $\Vekt{P}$ uses effectively a coordinate system centred on the planet with the planet--observer axis as pole and the `observer angle' $\phi$ as azimuth.
Then, the distance from the point to the previously defined (the `true') pole of the planet is
\begin{align}
 \left\Vert\Vekt{P}-\Vekt{N}\right\Vert =\RP\Big[\sin^2\phi\sin^2\alpha
   \,+&\left(\cos\alpha-\cos i\right)^2  \notag\\
     & +\left(\cos\phi\sin\alpha-\sin i\right)^2\Big]^{1/2},
\end{align}
with the sought angle
\begin{equation}
 \label{Gl:thetavonvarthphi}
 \theta(\vartheta,\,\phi) = 2\arcsin\left(\frac{\left\Vert\Vekt{P}-\Vekt{N}\right\Vert}{2\RP}\right).
\end{equation}
With this, we can integrate Equation~(\ref{Gl:Fvarmu}) numerically by evaluating $I=I(\theta(\vartheta,\,\phi))$, Taylor-expanding to first order in the $\RP\ll d$, $\varmumin\approx1$ limit since we use double-precision arithmetic: we write the prefactor in Equation~(\ref{Gl:Fvarmu}) as $1-\varmumin\approx1-\frac{1}{2}(\RP/d)^2$, and we write the integration angle $\vartheta$ as $\vartheta\approx(\RP/d)\sqrt{1-m}$ from Equation~(\ref{Gl:varmuDef}).
The $\cos\vartheta$ factor in the integrand, however, remains equal to unity to machine precision.
The integral has to be calculated at every wavelength.

We use the \texttt{int\_2d} routine from \texttt{IDL}/\texttt{GDL} \citep{abramowitz72},
which uses iterated Gaussian quadrature with a user-chosen number of transformation points: 6, 10, 20, 48, or 96.
Simply integrating over the whole CPD can lead to a massive underestimation of the integral for large CPDs because the emission is strongly concentrated close to the planet, which even 96~points may not capture. Therefore, we divide up the integral over $m$ for the CPD into several zones: from what corresponds to $r=\RP$ to $2\RP$, from $r=2\RP$ to $5\RP$, and so on. We verified carefully for the fiducial values and in what should be challenging cases (large CPD because of high mass and/or larger \fzent) that adding many more intermediate sectors still leads to very nearly the same answer (to tenths of a percent). We also controlled the number of transformation points and found that while $N=20$ is often enough to obtain the same answer as with $N=96$ within less than a few percent, some parameter combinations require $N=48$. We therefore used this for all calculations.

The CPD will block the emission from the planetary surface for angles $\theta>\thzpSch$, which defines an implicit function of $(\vartheta,\phi)$.
Since there is no simple solution, we keep the integration bounds as in Equation~(\ref{Gl:Fvarmu}) but simply discard the contribution of those angles to the integrand by multiplying the contribution by zero before summing.

\subsection{Emission from the CPD surface}

We let the CPD extend from a spherical radius (i.e., measured along the surface, which is assumed to be at a constant angle from the pole) $r=\rmin$ to \rmax.  %
Ignoring at first the planet, the radial flux from the CPD is
again given by Equation~(\ref{Gl:F}). However, $\theta=\thzpSch$ is fixed and it is now $r$ along the CPD surface that varies with $\vartheta$. Also, the flux is a function only of $r$. Therefore, we need $r=r(\vartheta)$.

Consider the two spherical coordinate systems centred on the planet introduced above: $(r,\,\vartheta,\phi)$ with its $z$ axis pointing towards the observer and $x$ pointing up in the plane of the page, and $(r,\,\theta,\,\varphi)$ with its $\tilde{z}$ axis along the pole and its $\tilde{x}$ at 90\degr\ also in the plane of the page. The $y$ and $\tilde{y}$ axes are the same and both systems are righthanded; they are related by a rotation around $y$ or $\tilde{y}$ by an angle $i$. Therefore, considering an arbitrary point with coordinates $(x,\,y,\,z)\entspr(\tilde{x},\,\tilde{y},\,\tilde{z})$ in the respective systems, one can write:%
\begin{subequations}
\label{Gl:Koordrot}
\begin{align}
 x  &= r\sin\alpha\,\cos\phi, \eqsep \tilde{x} = x\cos i-z\sin i = r\cos\varphi\sin\theta,\\
 z  &= r\cos\alpha,\phantom{\cos\phi} \eqsep \;\tilde{z} = x\sin i + z\cos i = r\cos\theta.
\end{align}
\end{subequations}
This leads to:
\begin{align}
 \sin\alpha\,\cos\phi\sin i + \cos\alpha\,\cos i &=\cos\theta \label{Gl:vonztilde}\\
 \sin\alpha\,\cos\phi\cos i - \cos\alpha\,\sin i &=\sin\theta\,\cos\varphi. \label{Gl:vonxtilde}
\end{align}
Solving Equation~(\ref{Gl:vonztilde}) yields, with $n$ an arbitrary integer,%
\begin{subequations}
 \label{Gl:avph}  %
\begin{align}
\frac{\alpha}{2} &= n\pi+\arctan\left(\frac{\sin i\cos\phi\,\pm\!\sqrt{\Delta}}{\cos\theta+\cos i}\right),\\
 \Delta &\equiv \cos^2i-\cos^2\theta+\cos^2i\,\cos^2\phi,
\end{align}
\end{subequations}
where taking $n=0$ and the positive root yields the correct solution as one can easily check by plotting Equation~(\ref{Gl:vonztilde}).
In principle, one could have instead solved Equation~(\ref{Gl:vonztilde}) for $\sin\alpha$ or $\sin^2\alpha$ but one would have needed to take special care to choose the correct, different, roots for different values of $\phi$; thus solving for $\alpha$ directly was simpler.

Next, we need to relate the `observer angles' $(\vartheta,\,\phi)$ that we will need for the integration to the `source angles' $(r,\,\varphi)$. We let $D$ be the distance from the observer to a point (the right black segment in Figure~\ref{Abb:Winkelabb} for a point on the sphere but the definitions are general) and the other angles be defined as up to now in Figure~\ref{Abb:Winkelabb}. With the inverse rotation from Equations~(\ref{Gl:Koordrot}), the distance in Cartesian coordinates yields
\begin{equation}
\label{Gl:D2Gl}
 D^2 = r^2 + d^2 -2 d r \left(\cos\theta\,\cos i - \sin\theta\,\sin i \,\cos\phi\right).
\end{equation}
With Equation~(\ref{Gl:vonxtilde}) and the appropriate law of sines, which is
\begin{equation}
 \frac{\sin\vartheta}{r} = \frac{\sin\alpha}{D},
\end{equation}
we can reexpress Equation~(\ref{Gl:D2Gl}) as%
\begin{subequations}
\label{Gl:athr}
\begin{align}
 0 &= \left[\frac{\sin^2\alpha}{\sin^2\vartheta}-1\right] \left(\frac{r}{d}\right)^2 + 2A\frac{r}{d} -1,\\
   A &\equiv\cos\theta\,\cos i-\left(\sin\alpha\,\cos\phi\,\cos i-\cos\alpha\,\sin i\right)\sin i.
\end{align}
\end{subequations}
This is a simple quadratic equation for $r/d$ and is general; we did not assume $r/d$ to be small, nor does the point $\Vekt{P}$ have to be on the CPD.
We now have everything at hand and can obtain $\FOb(r(\vartheta))$ in the integrand in Equation~(\ref{Gl:Fvarmu}).

The only remaining step is to determine the region of integration for the CPD, which is an ellipse with a `hole' defined by the planet. Writing the bounds in general is not easy, so we take an easier approach. We integrate in observer coordinates $(\vartheta,\,\phi)$ only over the region where (1)~$r\leqslant\rmax$ and the angle~$\theta(\vartheta,\,\phi)$ in Equation~(\ref{Gl:thetavonvarthphi})
either (2a)~has no solution %
or (2b)~is smaller than \thzpSch.
It is easiest to see that this works by considering the converse: the planet surface contributes (is visible) where $\theta$ exists according to Equation~(\ref{Gl:thetavonvarthphi}) and is smaller than where the CPD begins, at \thzpSch.
\section[Line widths and centres in the model of Aoyama et al. (2018)]{Line widths and centres in the model of \citet{aoyama18}}
 \label{Th:DvundmuAo18}

\begin{figure*}[!ht]
\sidecaption 
\includegraphics[width=12cm]{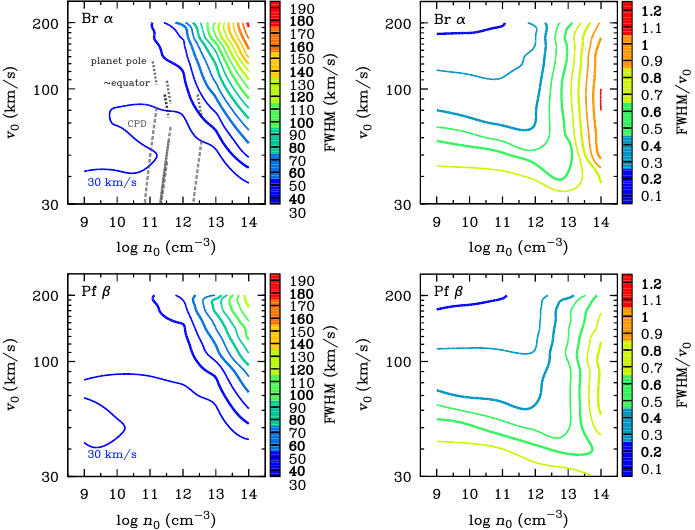} 
\caption{%
Line widths of the \citet{aoyama18} model, that is locally as a function of $(n_0,\,v_0)$, for \Bra (\textit{top row}) and \Pfb (\textit{bottom}), as the Doppler FWHM (\textit{left column}) or the FWHM divided by $v_0$ (\textit{right}).
Thick contours correspond to the bold numbers along the colourbars.
The lowest widths at $v_0=30~\kms$ are $\FWHM\approx23~\kms$ for both lines.
The input model grid has a step of 10~\kms in $v_0$ and 1~dex in $n_0$.
In the top row, we show for the fiducial case 
and some variations from Figure~\ref{Abb:Varpar}
the shock conditions on the planet and CPD surfaces (\textit{dark} and \textit{pale grey dotted lines}, respectively), with the `$\sim$equator' actually referring to the intersection of the planet surface with the CPD, at \thzpSch.%
}
\label{Abb:DvAo18}
\end{figure*}

For reference, we plot in Figure~\ref{Abb:DvAo18} the true FWHM in the model of \citet{aoyama18}. Similar plots can be found in \citet{Aoyama+Ikoma2019} or elsewhere but as a smooth colour scale. Here, we use contours for increased legibility and plot both the Doppler width of the line and the Doppler width relative to the input parameter $v_0$. We show the same plots for \Ha in \citet{blackman26}. Appendix~A of \citet{takasao21} also discusses the width as a function of number density for the \Ha line.

At low preshock number densities\footnote{This quantity is in fact directly proportional to the mass volume density: $X\rho_0=n_0 \mH$ (\citealt{aoyama18}, from \citealt{kwanfischer11}).} $n_0\lesssim10^{12}$~cm$^{-3}$, the relative \Bra line width $\FWHM/v_0$ \textit{decreases} almost only with $v_0$, from $\FWHM/v_0\approx0.8$ to 0.2 as $v_0$ goes from 30~to 200~\kms.
At high $n_0$, the line begins to saturate and the relative width increases only with $n_0$, going from 0.3 to 1 at the highest densities.
The \Pfb, which offers some hope of being observable (Fig.~\ref{Abb:Komboalles}), shows a similar qualitative behaviour but is less saturated and thus less wide.

This comparison shows that the line width is usually tens of percent of $v_0$, and not of order unity times $v_0$ as one could naively think.
This is discussed in Section~\ref{Th:EmLiProfNormalfall}.
The line shapes from accreting planets, as computed in this work, are a superposition of many such profiles, but this conclusion will still hold qualitatively.

Another important fact is that for a given accretor, the effective (flux-averaged) $v_0$ is essentially fixed, whereas the effective $n_0$ will vary with the accretion rate. When an object is accreting less vigorously, $n_0$ is lower, and $\FWHM\approx25$--$40~\kms$ for any to the extent that preshock velocity. 
During episodes of strong(er) accretion (higher $n_0$), however, the FWHM becomes directly proportional to the preshock velocity.
Therefore, specifically for \Bra, even a low-spectral-resolution determination of the FWHM will reveal the preshock density if $v_0\propto\surd[\MP/\RP]$ is known and $\FWHM\gtrsim40~\kms$, assuming shock emission.

\begin{figure*}[!ht]
\sidecaption
\includegraphics[width=12cm]{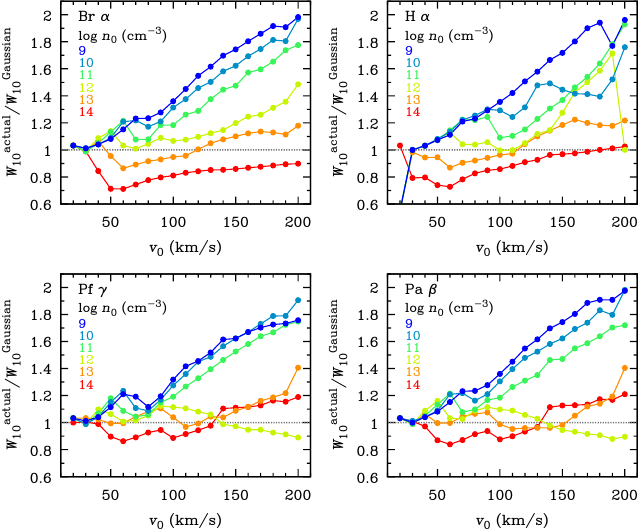}
\caption{%
In the \citet{aoyama18} model,
ratio \WVerhaeltnis (Eq.~(\ref{Gl:WV})) between the actual ten-percent width $W_{10}$ of the emission line and the $W_{10}$ it would have if it were a Gaussian corresponding to the FWHM of the line.
A flat-topped line has a ratio $\WVerhaeltnis<1$, a Gaussian has $\WVerhaeltnis=1$, and a broad-winged profile has $\WVerhaeltnis>1$.
Each panel shows this for a different transition (see label), as a function of $n_0$ (\textit{colour}) and $v_0$ (\textit{abscissae}).
}
\label{Abb:W10relGauAo18}
\end{figure*}

In Figure~\ref{Abb:W10relGauAo18} we assess to what extent the emission line profile in the \citet{aoyama18} model are Gaussian. For every line, we measure the FWHM and the actual ten-percent width $W_{10}$. We then compute and plot
\begin{equation}
 \label{Gl:WV}
 \WVerhaeltnis \equiv \frac{W_{10}}{\FWHM}\sqrt{\frac{\ln 2}{\ln 10}},
\end{equation}
which compares the actual $W_{10}$ to what it would be if the line shape were Gaussian with a width equal to the FWHM; a Gaussian has $\WVerhaeltnis=1$.
If the core of a line is saturated, that is, if it has a flat top or even a central dip \citep{aoyama18}, the half-maximum is reached at a lower flux value, where the line is wider. Then, its $W_{10}$ width is smaller than would be expected based on the FWHM,  %
leading to $\WVerhaeltnis<1$. Conversely, lines made up of a strong narrow component from deeper and cooler region, and a weaker wider component from shallower (closer to the shock front) and hotter regions, have a larger $W_{10}$ than their FWHM sugests, leading to $\WVerhaeltnis>1$.
In general, \WVerhaeltnis grows towards higher $v_0$ or lower $n_0$.

Finally, we comment on the position of the line centre in the \citet{aoyama18} models, which we extracted with the quadratic peak-fitting method (e.g., as used in \citealt{teague18centroid}), even though simply using the peak yields the same results thanks to the high resolution of the models.
We measure the shift relative to the line centres used by \citet{aoyama18}, given by the Rydberg formula\footnote{The model of \citet{aoyama18} currently assumes a perfect Coulomb potential, which considers only the principal quantum number.
This yields values that differ by $\Delta v=0.18~\kms$ at \Bra (and up to $\Delta v\approx2.5$~\kms over the main lines down even to $\HepsM=\mathrm{H}7$) from the mean values reported for example in the NIST database as quoted in Section~\ref{Th:lokEm}, which are experimental weighted averages over the spin states and finer splittings.
Taking the fine structure into account could affect slightly the line profile and would shift the line centres. However, the important point is that the computations of \citet{aoyama18} are done relative to the line centres, so their precise values do not matter beyond shifting the line profiles.}
\citep{bohr13} %
taking the reduced mass of the hydrogen nucleus into account:
\begin{equation}
 \label{Gl:Rydberg}
 \frac{1}{\lambda_0} = \Rydinfty\frac{\mp}{\mp+\me}\left(\frac{1}{\nl^2}-\frac{1}{\nupp^2}\right),
\end{equation}
where $\Rydinfty=10.9737316~\mumM^{-1}$ is the Rydberg constant, $\mp$ ($\me$) the proton (electron) mass, and $\nupp$ ($\nl$) the upper (lower) energy level of the transition.
We found that in the \citet{aoyama18} grid over $(n_0,v_0)$, the \Bra lines\footnote{At least in Figure~\ref{Abb:Varpar}, the shift is the same for all transitions shown there but we have not examined this systematically.} are shifted by $\mu\approx1$--4~\kms, usually around $\mu\approx2~\kms$.
They are never blueshifted, making a blueshift a tell-tale sign for a non-shock origin, most likely magnetospheric accretion, as \citet{demars23} discuss for Balmer series lines. However, in general, line profiles do not deliver a clear diagnostic of their origin \citep{edwards94,hashimoto25}.   %

\section{Photospheric brightness as a function of effective temperature}
 \label{Th:Phothelligkeitmehr}

\begin{figure}[!ht]
 \centering
 \includegraphics[width=0.4\textwidth]{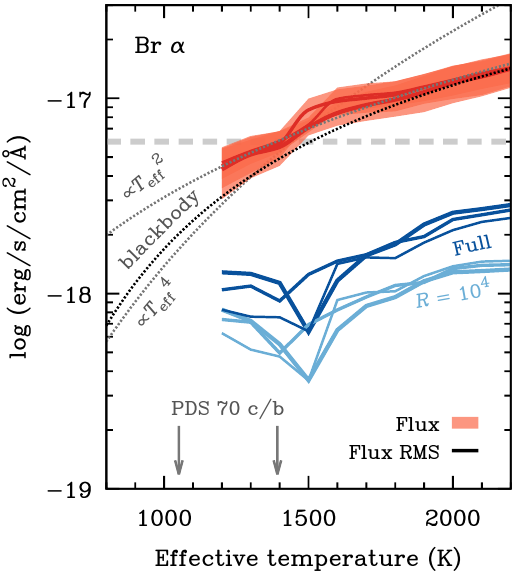}~~~
\caption{%
Photospheric signal against which the emission lines compete for detection at \Bra.
Average flux of the planet's photosphere using CIFIST models (\textit{red shaded bands}), at \Teff equal to integer multiples of $100$~K.
We take $d=113.4$~pc and fix $\RP=2~\RJ$.
The RMS over a window of $\Delta v=2200$~\kms around the line centre
(e.g., at \Bra, for a total of 30~nm) is shown for a non-rotating planet
(full resolution; \textit{dark blue lines})  %
and for $v\sin i=30~\kms$ as an extreme case (poor man's broadening: $R=10^4$; \textit{light blue}).
The three curves in each case are for $\log g=3.5$, 4.0, 4.5 (\textit{thick to thin}).  %
The blackbody flux and scalings of $y\propto\Teff^2$ and $\propto\Teff^4$ are given for reference (\textit{grey and black dotted lines}).
The \Teff values of the \PDS planets are highlighted.
The sensitivity limit of Equation~(\ref{Gl:Empf}) is shown (\textit{thick dashed grey line}). %
}
\label{Abb:AtmoFundRMS}
\end{figure}

For an accretion line to be easily detectable, whether generated in a shock or a magnetospheric scenario, it should be brighter especially than the RMS, and preferably also than the average flux from the atmosphere.
Figure~\ref{Abb:AtmoFundRMS} shows the average and the RMS in the CIFIST models at \Bra for a range of \Teff and $\log g$ values.
The RMS is not very sensitive to the chosen size of the window.
This figure is meant as a reference for any prediction of emission-line strength; once scaled to the object of interest, it makes it easy to estimate whether the atmospheric emission might be a concern or not.

\begin{figure*}[!ht]
\sidecaption
\includegraphics[width=12cm]{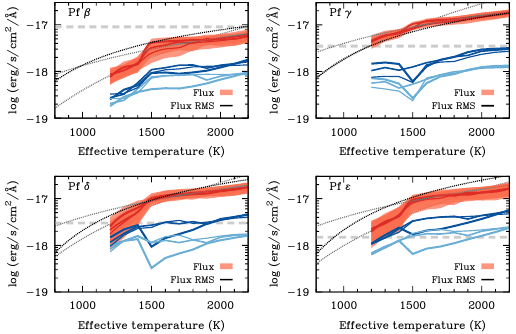}
\caption{%
As in Figure~\ref{Abb:AtmoFundRMS}
but
for \Pfb ($M$ band) and \Pfg, \Pfd, and \Pfe ($L$).
}
\label{Abb:AtmoFundRMSextra}
\end{figure*}

In Figure~\ref{Abb:AtmoFundRMSextra}, we show the brightness of the photosphere as in Figure~\ref{Abb:AtmoFundRMS} but now for the Pfund lines in the $M$ (\Pfb) and $L$ (\Pfg, \Pfd, \Pfe$=$\,Pf\,10) bands.
Whether the \Pfb line is still better detectable depends on the line strength ratio.
In particular, the photosphere at \Pfb is less detectable because it turns out to be less bright than at \Bra or \Pfg by about 0.5~dex, while the sensitivity in the $M$ band is less good by a factor of five to ten (Fig.~\ref{Abb:Linien}).

In general, the RMS is very dependent on the model, especially on the choice of line lists (see Appendix~\ref{Th:Linienlisteneffekt}). This calculation should therefore be repeated with other atmospheric models, and, importantly, they too should be resolved. Currently, such models do not seem publicly available at $\lambda\approx4$~\mum (see Section~\ref{Th:PlanEm}).

As a tentative assessment of the importance of metallicity and also of clouds with a model family, however, we investigated the Sonora Diamondback models \citep{morley24}.
At $\lambda=2$--20~\mum, they are sampled (not convolved) with $\lambda/\Delta \lambda=\textrm{60,000}$ (i.e., a sampling of 50~\kms, leaving at most two bins over the shock line width), and the spectral features are not resolved, but the trends should apply qualitatively. Therefore we computed the average and the RMS at the spacing provided, calling these a pseudo-average and a pseudo-RMS, to be analysed with caution. We inspected the models around the emission lines and verified that varying the surface gravity, metallicity, or cloud parameters did not change the average flux nor the pseudo-RMS significantly (not shown).

\section{Estimate of the CPD spectrum with ProDiMo}
 \label{Th:zpSchProDiMo}

\subsection{Set-up}

We used the radiation thermo-chemical disc modelling code ProDiMo \citep{woitke09,woitke16,kamp17}
to predict the continuum and near- and mid-IR emission line spectrum of the CPD of \PDSb.
We adopted the same fiducial planet properties as in Table~\ref{Tab:Par}.
However, as in \citet{rab19}, we used a DRIFT-PHOENIX \citep{helling08,witte09,witte11} model as input to irradiate the CPD because the spectral shape reproduces more clearly the VLT/SPHERE \textit{YJH}-band and VLTI/GRAVITY $K$-band spectroscopy \citep{mueller18,wang21vlti}. We do not expect the spectral shape to matter very much for the emission from the CPD, as \citet{sun26} confirm.  %
We took a spectrum with lower spectral resolution than what we used in the main text (Section~\ref{Th:PlanEm}), which is sufficient and appropriate for ProDiMo.

For the viscous heating in the CPD midplane, we assumed a radial inward mass flow $\MPktzpSchin=6\times10^{-8}~\MPktEinhJ$,
which comes from assuming only for these purposes $\MPktzpSchin\approx\MPktempir$, with \MPktempir from Eq.~(\ref{Gl:MPktempir}) with the \citet{AMIM21L} scaling applied to the highest \Ha flux which \citet{close25a} report, $\FHa=1.6\times10^{-16}$~erg\,s$^{-1}$\,cm$^{-2}$.
Our adopted $\MPktzpSchin$ value is somewhat smaller than but comparable to the fiducial case of \citet{shibaike26} for \PDSc, who use $\MPktzpSchin=2\times10^{-7}~\MPktEinhJ$.
We ran two CPD models with different inner radii of $\rin=0.0019~\mathrm{au}=4~\RJ$ or $\rin=3~\RJ$ for comparison; in both cases we adopted $\rout=2.4~\mathrm{au}=5020~\RJ$ and a CPD gas mass of $\MzpSch=2.1\times10^{-5}~\MSonne$. %
We assumed a canonical dust-to-gas ratio $\fpg=0.01$ and the typical T Tauri disc dust composition and grain sizes from \citet{woitke16}. We calculated dust settling according to \citet{riols18} using $\alpha=10^{-3}$.  %

ProDiMo models cannot model the disk all the way down to the planetary surface. Therefore, countary to the modelling in the main paper, here we needed to include an inner rim.
The inner rim was rounded using $\mathrm{reduc} = 10^{-5}$ and $\mathrm{raduc} = 2$ (see Equation~(71) of \citealp{woitke24})
and the radial gas surface density profile was set to be relatively flat,
with a power law exponent $\epsilon=0.1$ and a tapering radius of $\Rtap=1$~au (tapering exponent $\gamma=0.5$).
While these parameter choices are reasonable, they are specific, and the interested reader can refer to the recent work of \citet{sun26} for a systematic analysis of the effect of varying twelve CPD model parameters, including those controlling the inner gap. Those CPD models use RADMC-3D \citep{dullemond12} to compute the thermal structure and resulting spectra self-consistently but without including the gas chemistry and thermal balance to predict gas emission lines. Their treatment of the irradiation is also simpler.

The CPD is embedded in the thermal radiation of the protoplanetary disc. Therefore, we assumed a background radiation temperature of 19~K following \citet{pr23}. In addition, the UV radiation field of \PDSA illuminates the CPD from the outside, which we approximated by using $\chi=65.7$ in units of the Draine field\footnote{That is, we set a local effective ISM field, with $\chi$ denoting the factor by which the Draine field integrated over 912--2050~\AA{} is scaled.} \citep{draine96}, appropriate for the separation of \PDSb{} (21~au; \citealp{wang21vlti}). 
The dust temperature in the CPD was calculated from continuum radiative transfer taking the dust properties listed above. We included a detailed heating--cooling balance to find the gas temperature iteratively with the gas chemistry. For the latter, we assumed the large DIANA chemical network \citep{kamp17}. The gas emission line fluxes were calculated using escape probability \citep{woitke24}.
Section~\ref{Th:ProDiMohoheAufl} describes our estimate of the line profiles.

\subsection{Global view of the results}
 \label{Th:ProDiMoglob}

\begin{figure*}[!ht]
\sidecaption
\includegraphics[width=12cm]{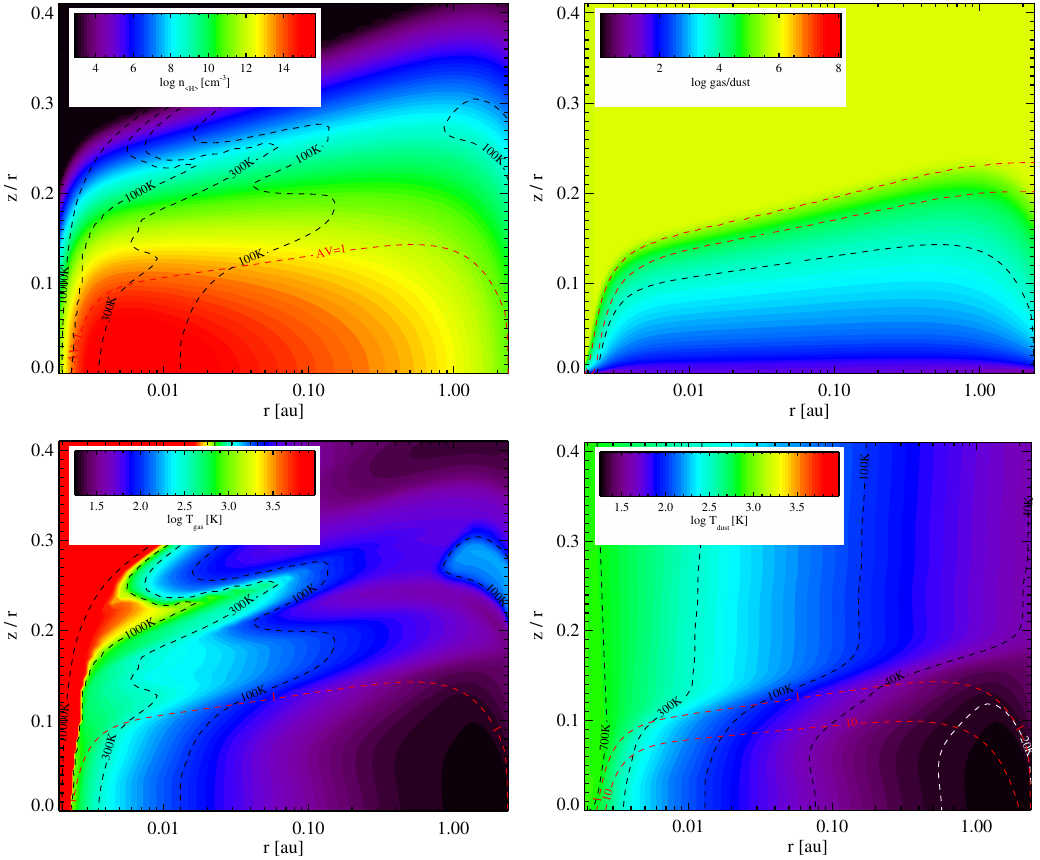}
 \caption{%
Structure of the CPD calculated with ProDiMo. Contrary to our set-up in the main text, we introduce here (out of technical necessity) an inner rim at $\rin=4~\RJ$. However, we repeat the calculations with $\rin=3~\RJ$ to assess the robustness of the results (see text).
The panels, across and down, show as a colour scale the `total' number density of hydrogen $\langle\mathrm{H}\rangle\equiv\mathrm{H}+\frac12\mathrm{H}_2$, the inverse of the dust-to-gas ratio \fpg, the gas temperature, and the dust temperature.
Dashed lines show contour contours of gas temperature, dust temperature, or $\AV$.
}
\label{Abb:ProDiMoBild}
\end{figure*}

Figure~\ref{Abb:ProDiMoBild} displays for reference the gas density, inverse of the dust-to-gas ratio, and gas and dust temperatures of the CPD.
The most striking feature, beyond the dust settling, is the decoupling between the gas and dust temperatures above $z/r\sim0.1$, especially within $r\lesssim0.01~\textrm{au}\approx20~\RJ$. The gas temperature reaches 3000--10,000~K due to the irradiation by the planet. Therefore, including this energy input is important.  %

\begin{figure}[!ht]
 \centering
\includegraphics[width=0.45\textwidth]{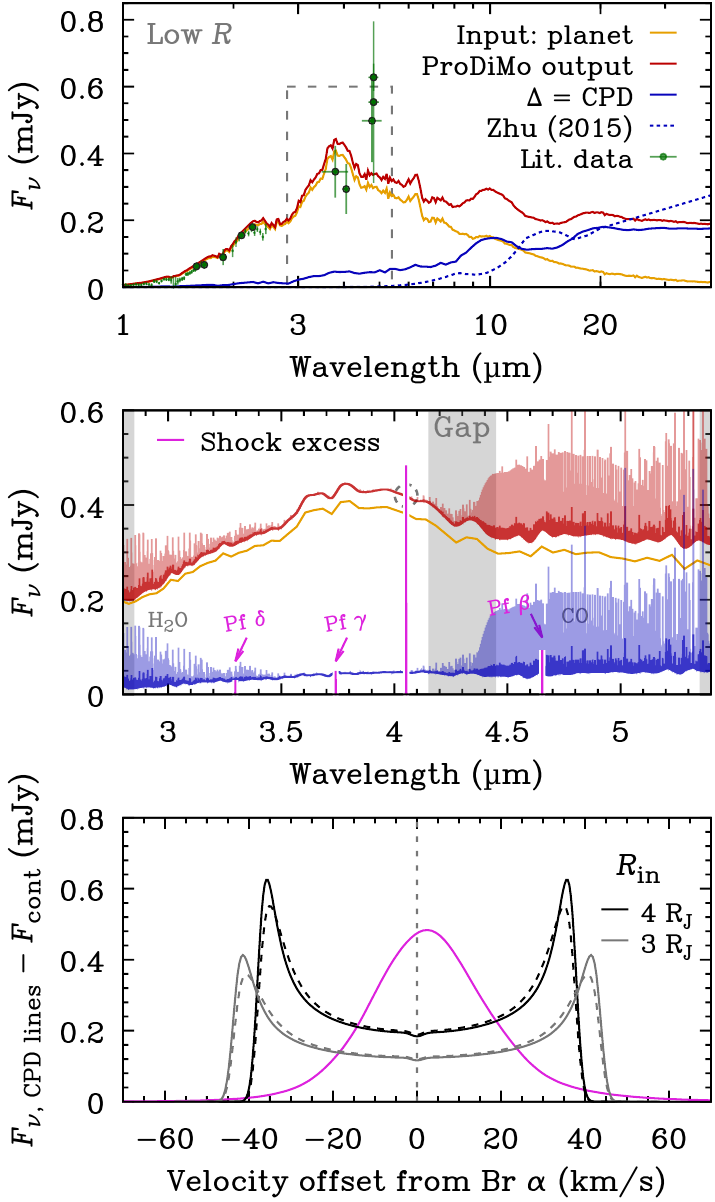}~~~
\caption{%
ProDiMo modelling of the CPD of \PDSb to estimate its contribution at \Bra.
\textit{Top panel}: Input planetary DRIFT-PHOENIX spectrum (\textit{gold curve}),
total spectrum from ProDiMo at $i\approx50\degr$ %
(\textit{red}), and their difference (\textit{blue}), that is the effective contribution from the CPD.
For clarity, we display smoothed spectra calculated without line emission,
and do not add the shock emission.
We compare with, but do not try to fit, literature data
(\textit{green}; \citealp{mueller18,Stolker+20b,wang21vlti,blakely25,christiaens24,christiaens25}).
We also show a model by \citet{zhu15} for the same $\MP\MPkt=3\times10^{-7}~\MMPktEinhJ$ and $\rin=4~\RJ$ as in the ProDiMo modelling (\textit{dotted blue}; details in text).
\textit{Middle panel}:  %
Zoom-in on the $L$ and $M$ bands (grey shaded region: non-sensitivity gaps from Fig.~\ref{Abb:Linien}, due to H$_2$O). %
Contrary to what we use in the main text,
the input planetary spectrum is at a low resolution, so all lines seen in the output come from the CPD. Their real peaks are likely much lower because no Keplerian broadening is applied (see text).
We overlay the resolved \Bra, \Pfb, \Pfg, \Pfd lines in the nominal case (\textit{pink}).
\textit{Bottom panel}: 
Zoom-in around the \Bra line, showing the Keplerian-broadened lines from ProDiMo.
Only the \Bra line itself from the CPD is noticeable.
We subtract the continuum from the CPD and do not include the pseudocontinuum nor lines from the planetary photosphere.
We show the result for $\rin=4$ ($3~\RJ$) in \textit{black} (\textit{grey}), and for an intensity powerlaw $I_\nu\propto r^{-\alpha}$ with $\alpha=15$ ($\alpha=10$) as \textit{solid} (\textit{dotted}) lines.
We also show the shock excess in the nominal case (\textit{pink}).
}
\label{Abb:ProDiMoSpektrum}
\end{figure}

The resulting spectrum is shown in Figure~\ref{Abb:ProDiMoSpektrum}.
The top panel gives a global view from 1~to 30~\mum.
The partial spectroscopy is matched rather well even though we took approximate values, nevertheless directly inspired by the data (Section~\ref{Th:Par}).
We subtract the input planetary spectrum from the ProDiMo output to reveal the contribution of the CPD (blue line), comprised of scattered photons from the planet and of emission from the CPD itself.
Around 4~\mum, on broad spectral scales, the CPD contributes only at most ten percent to the total flux. To first order, this justifies our focus in the main text on the planetary photosphere as a source of `astrophysical noise' with respect to the detection and measurement of the shock line (Section~\ref{Th:PlanEm}). The results from the scan of a large parameter space by \citet{sun26} suggest that this likely holds generally.

We compare briefly our ProDiMo spectrum to a model\footnote{See \url{https://www.physics.unlv.edu/~zhzhu/CPD.html}.} from the pioneering work by \citet{zhu15}.
We used $\MP\MPktzpSchin=3\times10^{-7}~\MMPktEinhJ$ (interpolating logarithmically between the \texttt{1em10} and \texttt{1em9} cases) and $\rin=4~\RJ$ from our adopted values, and took a non-truncated model, with $\rout=1000~\RJ$ (instead of only $\rout=50\rin$ if truncated).
This broadly agrees with \citet{christiaens19a} and the update by \citet{wang21vlti}, who find best-fit values $\MP\MPkt\approx(1$--$9)\times10^{-7}~\MMPktEinhJ$ and $\rin\approx3~\RJ$ when fitting the data simultaneously to a DRIFT-PHOENIX photosphere and a \citet{zhu15} model.

Figure~\ref{Abb:ProDiMoSpektrum} shows that the \citet{zhu15} model is much fainter below 12~\mum, with in fact (not visible on this scale) $F_\nu\approx10^{-4}$~mJy at 4~\mum.
Most striking is that the 10- and 20-\mum silicate resonance features are in absorption, whereas in the ProDiMo model they are in emission, as for very-low-mass stars or brown dwarfs \citep[e.g.,][]{jang25}.
This difference is presumably because the models of \citet{zhu15} do not include the irradiation of the CPD by the planet, whereas ProDiMo does, which leads to a temperature inversion, as Fig.~\ref{Abb:ProDiMoBild} shows.
Empirical constraints in the mid-infrared (MIR) would be valuable but are challenging due to the projected proximity of \PDSb to its host star.

\subsection{High-resolution results}
 \label{Th:ProDiMohoheAufl}

A further important check is whether the CPD adds features on the spectral scales probed by METIS. The 
middle 
panel of Figure~\ref{Abb:ProDiMoSpektrum} shows that it does past 4.3~\mum, for example, where there is CO band emission from the inner parts of the CPD, because in the CPD surface the gas is hotter than the dust.
However, between the H$_2$O grove ending at around 3.5~\mum and a broad CO thicket beginning at 4.3~\mum, the lines which the CPD contributes are essentially not noticeable in this view of the spectral landscape.
The line heights are overestimated because, for this panel, we simply use the line fluxes from the line-escape probability computation and convolve them with the resolution of METIS. This would be accurate only if all lines originated far from the planet, in which case the Keplerian broadening would be negligible.

To obtain a more realistic high-resolution spectrum from the CPD, we convolved the estimated spectrum with the Keplerian rotation profile of the CPD.
This requires knowledge of the radial dependence of the (wavelength-dependent) intensity of the emission from the CPD.
Firstly, we verified that within 500~\kms of \Bra, only the \Bra line is significant. The closest other notable features are three roughly 1000--10000~times fainter water lines $-100$~to $-400~\kms$ away, respectively.
Secondly, we approximated the radial intensity profile for the hydrogen line in the ProDiMo CPD model by a power law, which yielded
$I_\nu\propto r^{-\alpha}$ with $\alpha\approx15$; the emission is heavily dominated by the innner regions of the CPD.
As we verified (dotted versus full lines in Fig.~\ref{Abb:ProDiMoSpektrum}), the exact value of $|\alpha|$ does not matter if it is large.
In the innermost regions where the \Bra is emitted, the temperature reaches at most $T\sim1000$~K (Fig.~\ref{Abb:ProDiMoBild}). Therefore, the intrinsic line width is of the order of the sound speed
$\cs\approx2~\surd(T/1000~\mathrm{K})~\kms$, which is similar to the resolution of METIS.
Therefore, we considered for simplicity only the instrumental broadening.

To obtain the Keplerian broadening kernel, we assumed optically thin emission from a geometrically thin disc, and integrated over the inclined CPD, using the projected line-of-sight velocity.
We verified our integration routine by comparing to exact analytical results, which exist for a few values of $\alpha$ or otherwise could be obtained as the numerical evaluation of Gau\ss{}'s hypergeometrical function \citep{smak69,smak81}.
As shown in \citet{horne86}, assuming optically thin or thick emission changes only modestly the broadening kernel (especially the central part), which is beyond the level of our estimate.
The main features of the broadening kernel and thus, given the narrowness of the line, of the resulting line, are the separation of the `Keplerian horns' $\Delta v\approx2\vKep\sin i$ where $\vKep$ is the Keplerian velocity (here, at the inner edge of the CPD because $\alpha$ is large; cf.\ \citealp{horne86}),   %
and the steepness of the wings, which in velocity space go as $I_\nu\sim (\Delta v)^{\pm(2\alpha+5)}$.  %

The bottom panel of Fig.~\ref{Abb:ProDiMoSpektrum} shows the result of the estimate. In the ProDiMo model with $\rin=4~\RJ$, the line-integrated, continuum-subtracted CPD \Bra flux is
$\FBra=5.5\times10^{-17}$~erg\,s$^{-1}$\,cm$^{-2}$,
and for $\rin=3~\RJ$, it turns out to be smaller but is very similar, with $4.0\times10^{-17}$~erg\,s$^{-1}$\,cm$^{-2}$.
Thanks to the strong Keplerian broadening and the narrowness of the shock lines, the two Keplerian horns of the \Bra emission from the CPD do not interfere with the shock emission.
A comparison to more precide results from the `Fast Line Tracer' (FLiTs), developed by M.~Min \citep{woitke18}, show broad qualitative agreement (not shown).
Thus we conclude that, for our purposes, the \Bra emission of the CPD can be securely ignored to first order, but that it could possibly be studied with METIS.

\section[Predicted positions of PDS 70 b and c]{Predicted positions of \PDSbc}
 \label{Th:wosindbundc}

\begin{figure}[!ht]
 \centering
\includegraphics[width=0.44\textwidth]{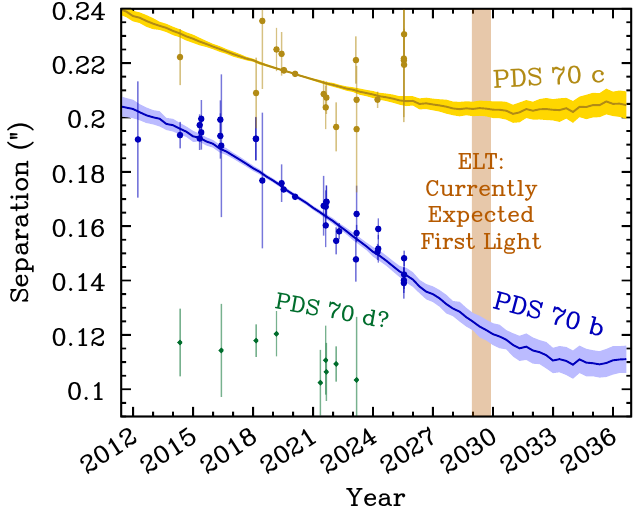}
\caption{%
Measured (from the literature as compiled in \citealt{trevascus25}, additionally with the new epochs in \citealt{balsalobreruza26}) and predicted (thanks to \url{www.whereistheplanet.com}, up to the data in \citealt{wang21vlti}) separation of \PDSbc as a function of date.
As a reference, a vertical band highlights the current expected first-light epoch for the ELT.
The separation of candidate \PDSd is also shown \citep{hammond25}
but its nature remains unclear \citep{trevascus26,balsalobreruza26}.
}
\label{Abb:wosindbundc}
\end{figure}
In Figure~\ref{Abb:wosindbundc}, we show the predicted separation of \PDSbc from the central star. We used\footnote{In the last stages of this work, we noticed the convenient wrapper \url{https://github.com/vandalt/querytheplanet}.} \url{https://github.com/semaphoreP/whereistheplanet} \citep{wang21withp}, which bases its fit on the data summarised in \citet{wang21vlti}. This thus does not include newer astrometric data
but we overplot the data from the compilation by \citet{trevascus25}: data originally from that work in addition to
\citet{keppler18,mueller18,christiaens19b,Haffert+2019,mesa19,wang20,wang21vlti,christiaens24,wahhaj24,christiaens25,blakely25,close25a,hammond25,balsalobreruza26}.
The fit still seems sufficiently accurate at this level. We also display the separation of candidate \PDSd \citep{mesa19,christiaens24,hammond25,trevascus25}, whose presence does not have to affect the stability of the system \citep{do25}.
Its nature, as well as that of other point sources close in, remains a complicated question \citep{trevascus26,balsalobreruza26}.

At the current expected epoch of first light of the ELT (2029), the separation of \PDSb will be only $\rho=0\farcs12$, while for \PDSc it should still be $\rho\approx0\farcs21$.
Also, \PDSd will likely be around $\rho\approx0\farcs1$--$0\farcs2$ away from the central star.
Past around 2040, the separation of \PDSb should start increasing appreciably.
The diffraction-limited beam size of the ELT is considerably smaller, with $\sigma\approx0\farcs026$,
and we discuss in Section~\ref{Th:RauschenvonPrimEm} the suppression of the stellar light at that separation.

Finally, we do not show it but report here that between 2025 and 2037, the radial velocity of \PDSb, in the usual sense of a velocity away from or towards the observer, relative to the systemic velocity, goes from $\mbox{RV}\approx+2$ to $-2$~\kms, while that of \PDSc goes from $\mbox{RV}\approx-0.5$ to $+0.5~\kms$.
Relative to the barycentre of the solar system, this is in addition to
the systemic radial velocity of only $\vSyst\approx3.1\pm1.4$~\kms \citep{gDR2}
(or $\vSyst=0.7\pm3.2~\kms$ in \citealt{gDR3}).  %

\section{Atmospheric extinction}
 \label{Th:tauAtm}

\begin{figure*}[!ht]
 \centering
\includegraphics[width=0.32\textwidth]{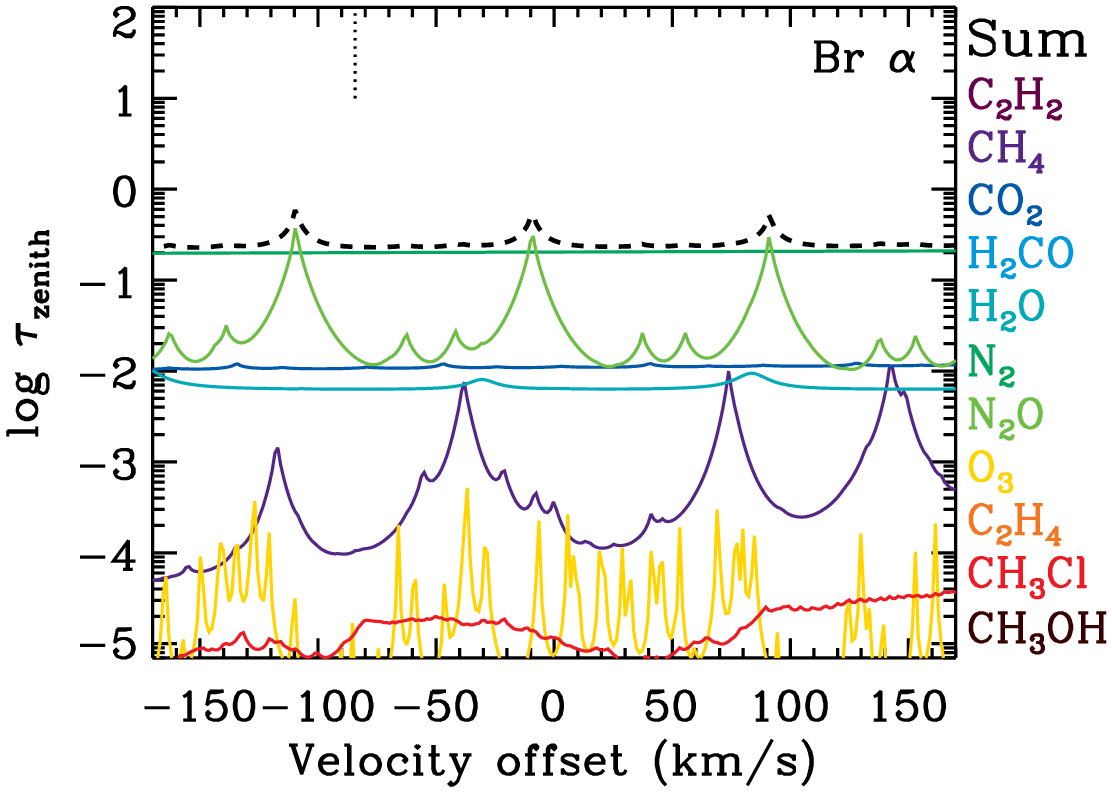}~~%
\includegraphics[width=0.32\textwidth]{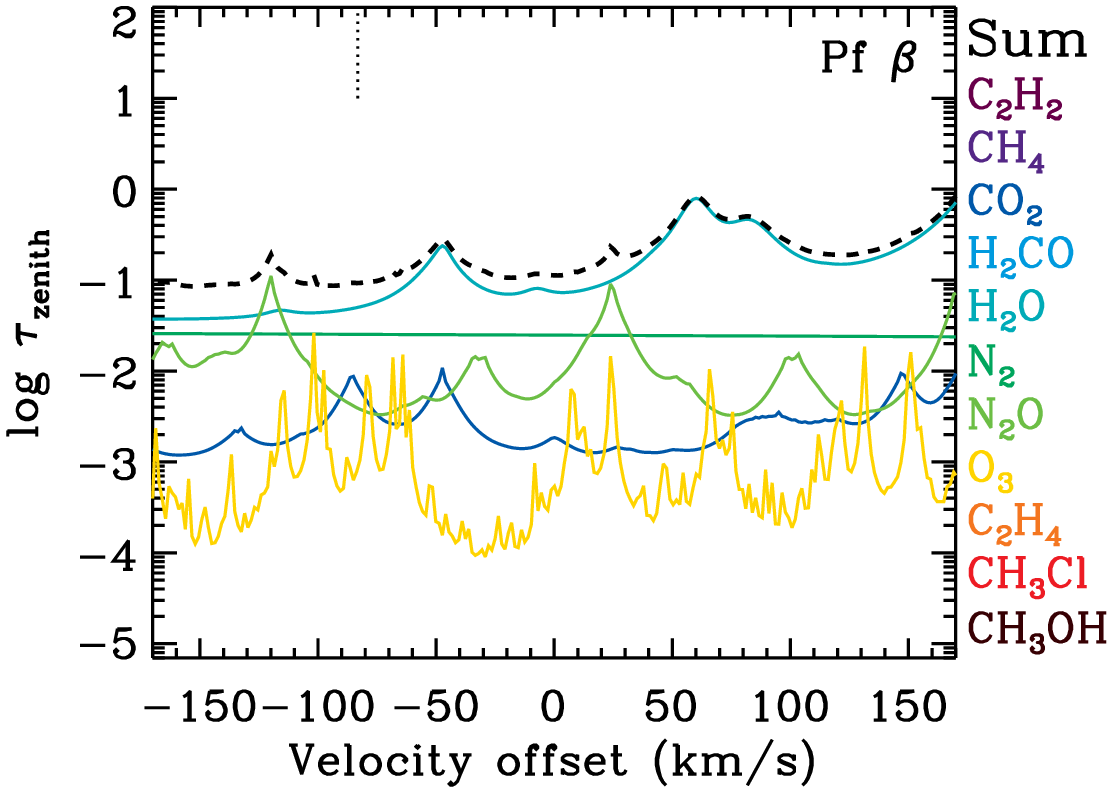}~~%
\includegraphics[width=0.32\textwidth]{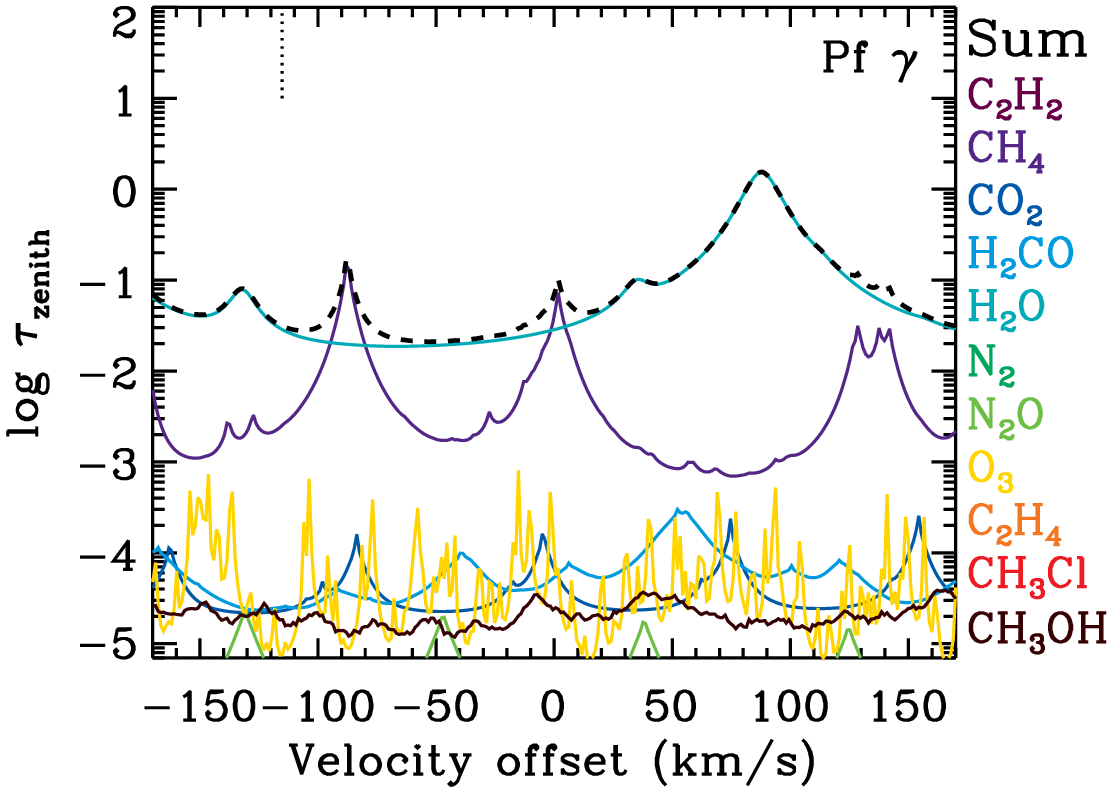}\\
\addvspace{0.5em}
\includegraphics[width=0.32\textwidth]{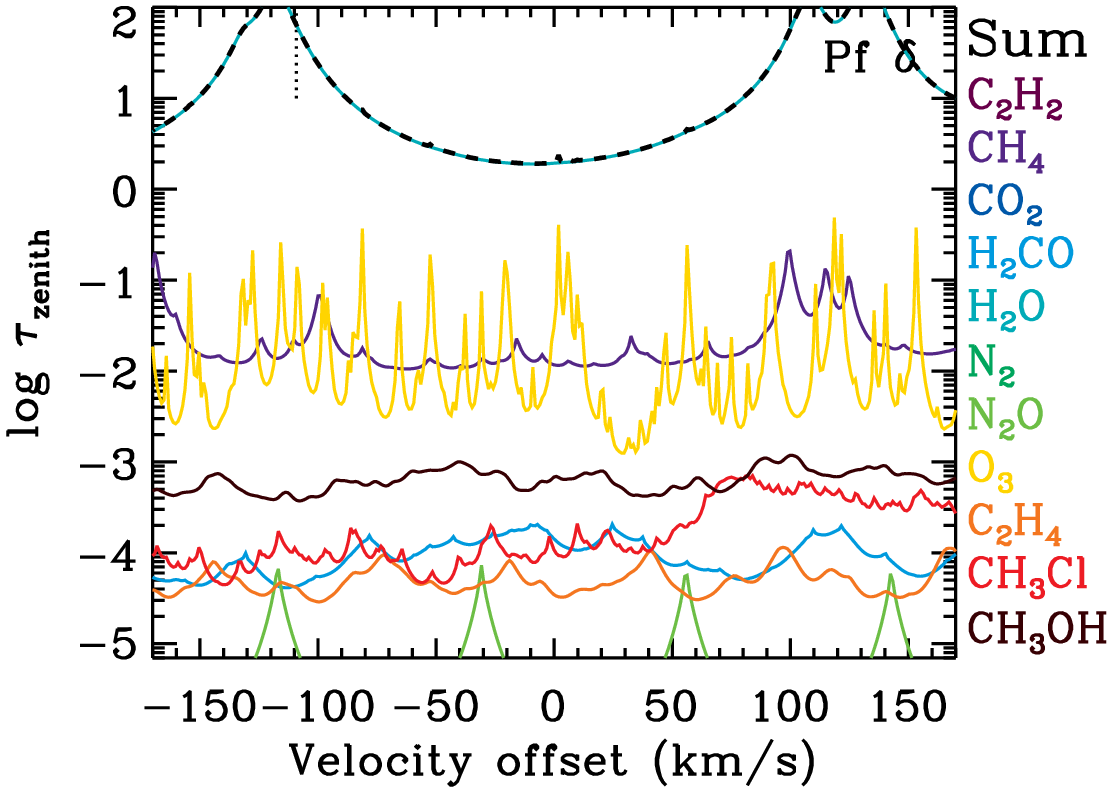}~~~%
\includegraphics[width=0.32\textwidth]{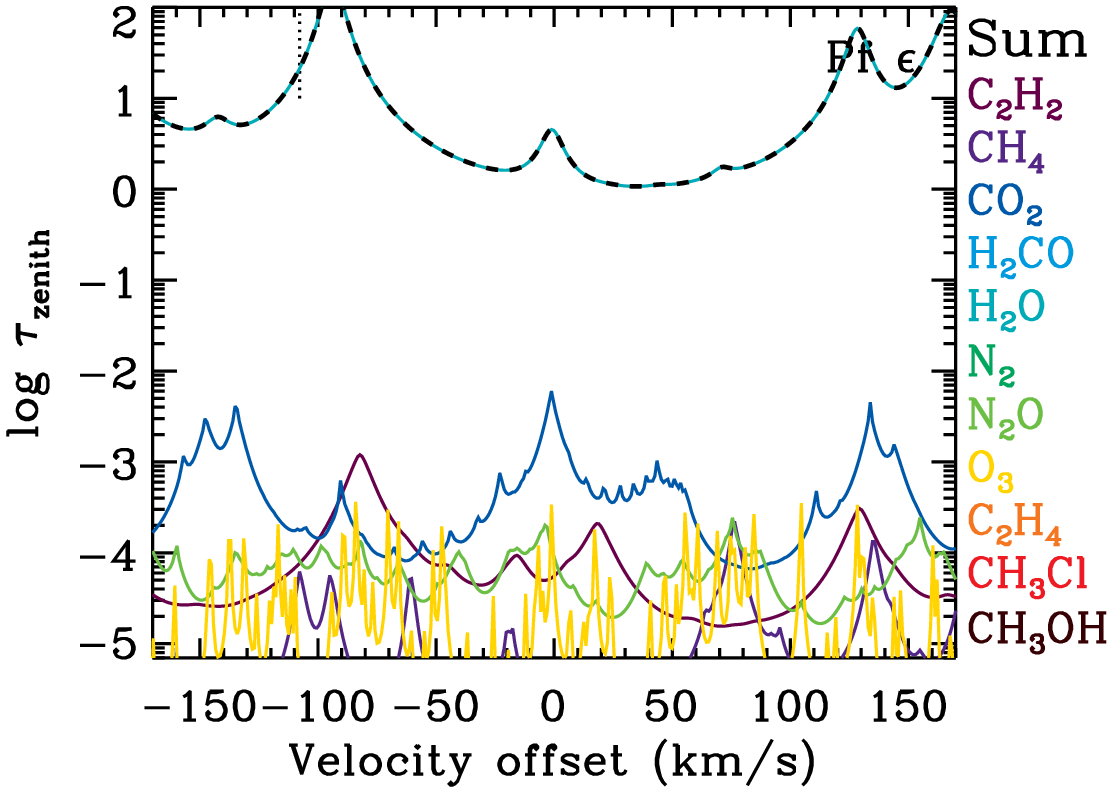}~~~%
%
%
\includegraphics[width=0.29\textwidth]{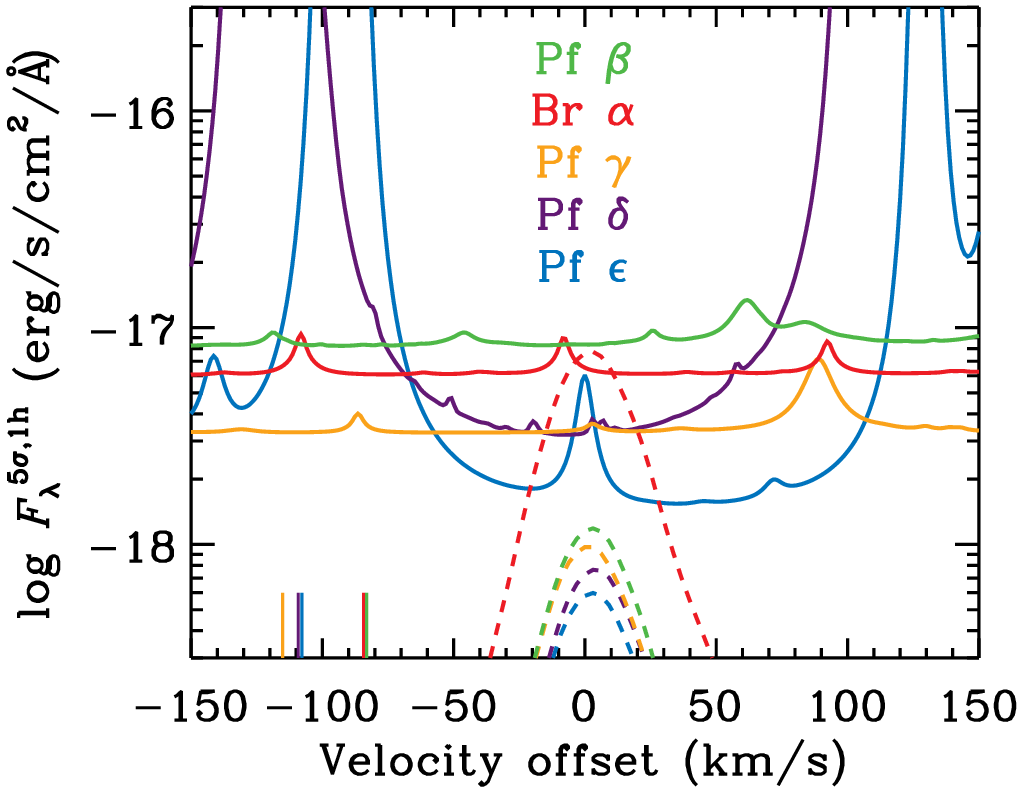}~~~~~~~%
\caption{%
Around all Brackett- and Pfund series hydrogen lines observable with METIS,
optical depth at zenith in the atmosphere of the Earth for each of the strongest components (\textit{coloured lines}) and for their sum (\textit{thick black}).
For each hydrogen line, the central wavelength of the transition in air at STP is shown (from the NIST database; \textit{vertical dotted line}).
(\textit{Last panel})~Summary of the noise curves, as in Figure~\ref{Abb:DetektorempfpaarL} but over a narrower velocity range.
The curves are dominated, though not solely set, by the atmosphere.
}
\label{Abb:tauAtmundDetektorempf}
\end{figure*}

We used the data from the `Atmospheric Infrared Spectrum Atlas' from Oxford University\footnote{See \url{https://eodg.atm.ox.ac.uk/ATLAS}.},  %
 which gives the zenith optical depth as a function of vacuum wavelength (A.~Dudhia 2025, priv.\ comm.),
to identify and plot the strongest absorbers in the atmosphere of Earth in Figure~\ref{Abb:tauAtmundDetektorempf}.
The `Atlas' was built using the HITRAN 2012 data release \citep{rothman13}.

In the last panel of Figure~\ref{Abb:tauAtmundDetektorempf}, we give an overview of the spectroscopic sensitivity curve of METIS in the background-limited regime for median observing conditions
\citep{lovis22}, as calculated by one of us (RvB),
for segments centred on the main $L$- and $M$-band transitions.
Details are given in Section~\ref{Th:Detektorempf} and a wider view for \Bra, \Pfb, and \Pfg is shown in Figure~\ref{Abb:DetektorempfpaarL}.

The small-scale features of the sensitivity curves are set  %
basically by the atmospheric transmission, as can be easily seen by comparing the shapes, because this controls the number of photons reaching the detector. The larger wavelength scales, seen in Fig.~\ref{Abb:Linien}, are set by several factors including the continuum opacity of the telescope.

\section{Effect of the choice of opacity line list}
 \label{Th:Linienlisteneffekt}
 
In Section~\ref{Th:PlanEm}, we estimated the possible photospheric contribution of \PDSb at \Bra. A strong or feature-rich continuum would make the detection of the line more difficult.
We used the CIFIST \citep{allard12philtrans,allard13} models, which for a range of \Teff values show a series of troughs at \Bra (see Figure~\ref{Abb:Atmbeitragnorm}) and the Pfund lines (see Figure~\ref{Abb:Komboalles}).
Therefore, we wanted to assess the robustness of this prediction, which depends on the abundance of the responsible element(s) and on their line shapes.

We examined the thermal structure and (depth-dependent) chemical composition of a few models, also from the literature, and identified several relatively abundant molecules which are candidates to explain these features. We downloaded single-species opacities from the DACE\footnote{See \url{https://dace.unige.ch/opacityDatabase}.} database and from the MARFA\footnote{See \url{https://marfa.app/}.} opacity service \citep{razumovskiy24} at a few pressures $P\approx10^{-4}$--1~bar and temperatures $T=1000$, 1500~K close to the adopted $\Teff=1400$~K value.

\begin{figure} %
 \centering
\includegraphics[width=0.47\textwidth]{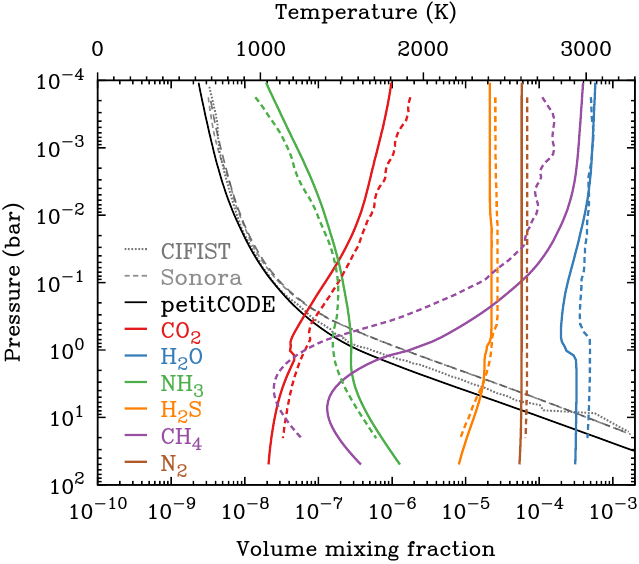}
\caption{%
Pressure--temperature (against the top horizontal axis) and chemical-composition (bottom) structure of cloud-free, solar-composition, $\Teff=1400$~K, $\log g=3.5$ models from petitCODE (\textit{solid lines}) and Sonora Diamondback (\textit{partially transparent grey dashed lines}).
Only the most abundant constituents are shown.
We also show the $P$--$T$ structure for CIFIST (\textit{dotted grey line}) and Sonora Bobcat (essentially identical to Diamondback except at the smallest pressures; also \textit{grey dashed}).
}
\label{Abb:PTStrukt}
\end{figure}

We saw that the opacities depend only little on temperature in that range and that the widths of the opacity features are similar to the atmospheric ones for $P\approx0.1$~bar. We therefore adopted $T=1000$~K and $P=0.1$~bar for the opacity curves. The atmospheric $P$--$T$ structure, which we show in Figure~\ref{Abb:PTStrukt}, passes through this point.
MARFA lets the user choose the wing cut-off
but choosing different values over a wide range led to essentially identical curves at $P=0.1$~bar %
so we kept the default value of $\nuabschn=125$~cm$^{-1}$, while noting that \citet{gn24} recommend $\nuabschn=25$~cm$^{-1}$.

Since the per-layer composition of the CIFIST models is not available, we compared the abundances between the cloud-free models from petitCODE (\citealp{moll19,linder19}; structures: P.~Molli\`ere, priv.\ comm.\ 2022) and Sonora Diamondback \citep{morley24}.
Figure~\ref{Abb:PTStrukt} shows that
the two are very similar to 0.1~dex or better over the whole atmosphere for the main constituents
(CO$_2$, H$_2$O, NH$_3$, CH$_4$, N$_2$) except for CH$_4$, which has an almost constant offset around 0.5~dex. The good overall agreement suggests that this is an appropriate estimate for the CIFIST model.
The $T(P)$ profiles of all three models, also of the corresponding Sonora Bobcat model \citep{marley21}, are essentially identical.

\begin{figure} %
 \centering
\includegraphics[width=0.47\textwidth]{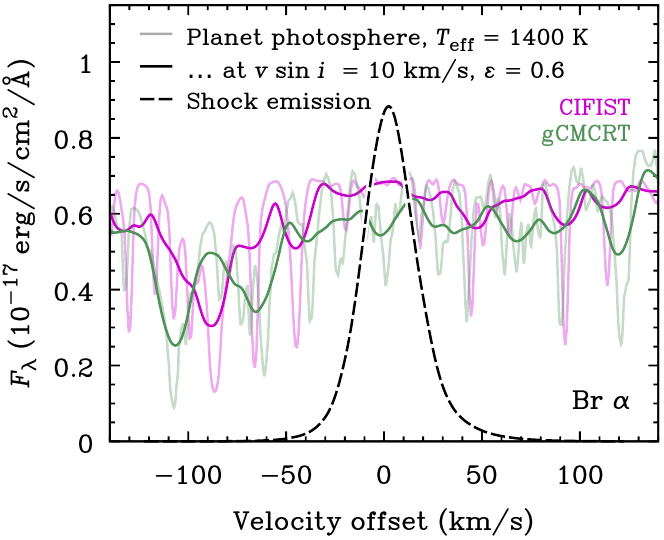}
\caption{%
As in Figure~\ref{Abb:Atmbeitragnorm} but comparing the CIFIST spectrum with the output of the independent code gCMCRT \citep{lee22}, only for $\Teff=1400$~K (details in text), and for double the wavelength range.%
}
\label{Abb:AtmbeitragnormmitgCMCRT}
\end{figure}


Before examing the individual opacity curves of the consistuents of the atmosphere, we show
in Figure~\ref{Abb:AtmbeitragnormmitgCMCRT} a spectrum obtained with the Monte Carlo atmospheric radiative transfer code gCMCRT \citep{lee22} at a resolution $R=10^6$ (kindly computed by Elsie Lee). We used this independent tool to assess   the effect of varying details of the radiation transfer. As input, we used the $P$--$T$ structure of the Sonora Bobcat model \citep{marley21} with 
$\Teff = 1400$~K, $\log g = 3.5$, and solar metallicity and C/O, shown in Figure~\ref{Abb:PTStrukt}.
The gCMCRT spectrum includes the contributions from H$_2$O, CH$_4$, CO, CO$_2$, NH$_3$, and H$_2$S.

On the small scales, the gCMCRT and CIFIST spectra differ in the position of most of the more significant absorption troughs but the qualitative appearance---the approximate number and width of the throughs---is overall similar. Also, the larger-scale structure is comparable.
%

\begin{figure*} %
 \centering
\includegraphics[width=0.94\textwidth]{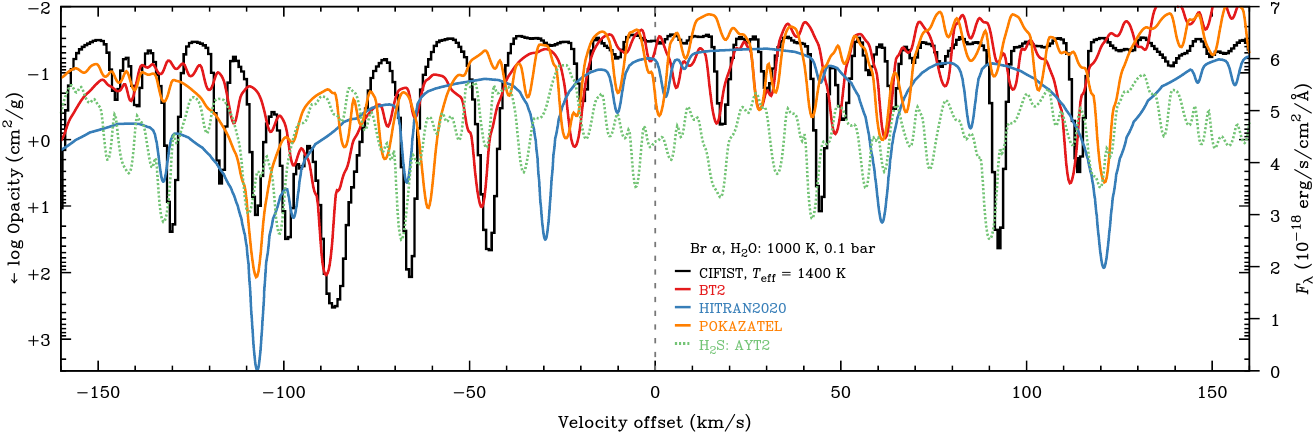}\\
\includegraphics[width=0.94\textwidth]{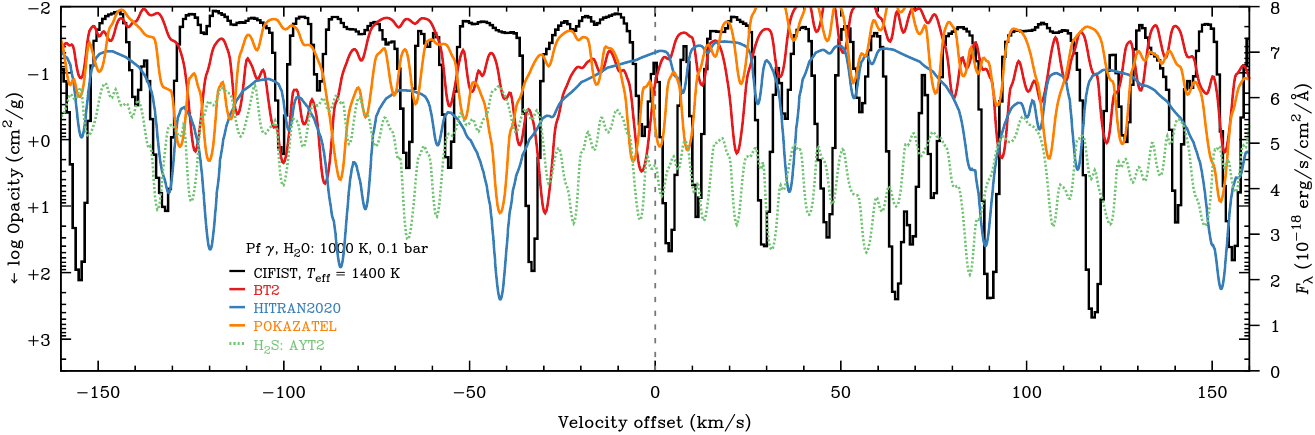}
\caption{%
Opacity 
(cross-section per pure-substance mass) of H$_2$O at $T=1000$~K and $P=0.1$~bar, on a logarithmic scale increasing downwards (\textit{coloured curves; against the left axis}), and flux (at the observer) of the fiducial atmospheric model on a linear scale (\textit{black curve; right axis}). We show regions around \Bra (\textit{top panel}) and \Pfg (\textit{bottom}). The opacities were calculated from the line lists of \citet[][BT2]{barber06}, \citet[][HITRAN2020]{gordon22}, and \citet[POKAZATEL][]{polyansky18} (see legend). The 2024 release of the ExoMol database \citep{tennyson24} recommends the POKAZATEL list. We add the \citet[][AYT2]{azzam16} opacity (from DACE) for H$_2$S (\textit{dashed line}).
}
\label{Abb:kappaundFlLinienlisten}
\end{figure*}

Finally, we display single-substance opacity curves in Fig.~\ref{Abb:kappaundFlLinienlisten}.
Visual comparison of these curves to the spectrum suggests that almost all troughs at \Bra (Figures~\ref{Abb:Atmbeitragnorm} and~\ref{Abb:Komboalles}) can be explained by H$_2$O, with H$_2$S solely or additionally responsible for most the rest.
We indicate approximately the positions of the H$_2$O and H$_2$S features in Figure~\ref{Abb:Komboalles}.
The choice of the line list, however, makes a major difference,
as Figure~\ref{Abb:kappaundFlLinienlisten} demonstrates. 
The CIFIST models use the `BT2' list \citep{barber06} (from DACE) and indeed, the troughs align very well. The shape of the opacity based on HITRAN2020 (\citealt{gordon22}; obtained from MARFA) and on POKAZATEL (\citealt{polyansky18}; from DACE) is quite different in its details, both around \Bra and \Pfg. Only a few troughs are at the same position in all the opacity curve versions within $\Delta v=\pm150~\kms$ of \Bra: only at $\Delta v=+60$ and $\Delta v\approx120$--140~\kms securely, at a few other locations with even more different amplitudes.
This, in addition to any uncertainty in the abundance of water, highlights the need for more theoretical and empirical studies of the opacity of water at these wavelengths.
However, the larger-scale trends (for example, the slope from $\Delta v=-100$ to $+50~\kms$) are similar, as are approximately the qualitative appearance of the throughs.


\end{appendix}

\end{document}